%% file: lumicalib_PAPER.tex
\pdfoutput=1

\documentclass[12pt,a4paper]{article}
\usepackage{ifthen} %
\newboolean{pdflatex}
\setboolean{pdflatex}{true} %

\newboolean{articletitles}
\setboolean{articletitles}{true} %

\newboolean{uprightparticles}
\setboolean{uprightparticles}{false} %

\newboolean{inbibliography}
\setboolean{inbibliography}{false} %

\input{preamble}
\usepackage{longtable} %

\begin{document}

\renewcommand{\thefootnote}{\fnsymbol{footnote}}
\setcounter{footnote}{1}
\setcounter{secnumdepth}{2}

\input{title-LHCb-PAPER}

\renewcommand{\thefootnote}{\arabic{footnote}}
\setcounter{footnote}{0}

\tableofcontents
\cleardoublepage

\pagestyle{plain} %
\setcounter{page}{1}
\pagenumbering{arabic}

\def\subsubsubsection#1{\noindent{\bf #1}\newline}
\input{introduction}

\input{setup}

\input{normalisation}

\input{interactionrate}
\input{formalism}
\input{bgi}

\input{vdm}

\input{combination}

\input{acknowledgements}

\FloatBarrier
\cleardoublepage

\addcontentsline{toc}{section}{References}
\setboolean{inbibliography}{true}
\bibliographystyle{LHCb}
\bibliography{main,LHCb-PAPER,LHCb-CONF,LHCb-DP,lumirefs} %

\FloatBarrier
\cleardoublepage

\FloatBarrier
\cleardoublepage
\input{LHCb_HD_authorlist_2014-07-09}
\end{document}

%% file: preamble.tex
\usepackage{microtype}
\usepackage{lineno}  %
\usepackage{xspace} %
\usepackage{caption} %

\usepackage{graphicx}  %
\usepackage{color}
\usepackage{colortbl}
\graphicspath{{./figs/}{./figsBGI/}{./figsVDM/}} %

\usepackage{amsmath} %
\usepackage{amssymb}
\usepackage{amsfonts}
\usepackage{upgreek} %
\usepackage{textcomp}
\usepackage{gensymb} %
\usepackage{mathtools}

\usepackage{booktabs}
\usepackage{multirow}

\newcommand*\patchAmsMathEnvironmentForLineno[1]{%
\expandafter\let\csname old#1\expandafter\endcsname\csname #1\endcsname
\expandafter\let\csname oldend#1\expandafter\endcsname\csname
end#1\endcsname
 \renewenvironment{#1}%
   {\linenomath\csname old#1\endcsname}%
   {\csname oldend#1\endcsname\endlinenomath}%
}
\newcommand*\patchBothAmsMathEnvironmentsForLineno[1]{%
  \patchAmsMathEnvironmentForLineno{#1}%
  \patchAmsMathEnvironmentForLineno{#1*}%
}
\AtBeginDocument{%
\patchBothAmsMathEnvironmentsForLineno{equation}%
\patchBothAmsMathEnvironmentsForLineno{align}%
\patchBothAmsMathEnvironmentsForLineno{flalign}%
\patchBothAmsMathEnvironmentsForLineno{alignat}%
\patchBothAmsMathEnvironmentsForLineno{gather}%
\patchBothAmsMathEnvironmentsForLineno{multline}%
}

\usepackage{hyperref}    %
\usepackage[all]{hypcap} %

\input{lhcb-symbols-def} %
\input{_commands}

\usepackage{tabulary}

\usepackage{cite} %
\usepackage{mciteplus}

%% file: lhcb-symbols-def.tex
\def\lhcb {\mbox{LHCb}\xspace}

\def\lhc    {\mbox{LHC}\xspace}

\def\velo   {VELO\xspace}

\ifthenelse{\boolean{uprightparticles}}%
{

 \def\PDelta      {\ensuremath{\Delta}\xspace}                 
 \def\PXi      {\ensuremath{\Xi}\xspace}                 
 \def\PLambda      {\ensuremath{\Lambda}\xspace}                 
 \def\PSigma      {\ensuremath{\Sigma}\xspace}                 
 \def\POmega      {\ensuremath{\Omega}\xspace}                 
 \def\PUpsilon      {\ensuremath{\Upsilon}\xspace}

 \def\PB      {\ensuremath{\mathrm{B}}\xspace}                 
                  
 \def\PD      {\ensuremath{\mathrm{D}}\xspace}

 \def\PK      {\ensuremath{\mathrm{K}}\xspace}

 \def\PW      {\ensuremath{\mathrm{W}}\xspace}

 \def\PZ      {\ensuremath{\mathrm{Z}}\xspace}

 \def\Pi      {\ensuremath{\mathrm{i}}\xspace}

}
{

 \mathchardef\PDelta="7101
 \mathchardef\PXi="7104
 \mathchardef\PLambda="7103
 \mathchardef\PSigma="7106
 \mathchardef\POmega="710A
 \mathchardef\PUpsilon="7107
                  
 \def\PB      {\ensuremath{B}\xspace}                 
                  
 \def\PD      {\ensuremath{D}\xspace}

 \def\PK      {\ensuremath{K}\xspace}

 \def\PW      {\ensuremath{W}\xspace}

 \def\PZ      {\ensuremath{Z}\xspace}

 \def\Pi      {\ensuremath{i}\xspace}

}

\def\W      {\ensuremath{\PW}\xspace}

\def\Z      {\ensuremath{\PZ}\xspace}

  \def\Kbar  {\kern 0.2em\overline{\kern -0.2em \PK}{}\xspace}

  \def\Dbar    {\kern 0.2em\overline{\kern -0.2em \PD}{}\xspace}
\def\Bbar    {\ensuremath{\kern 0.18em\overline{\kern -0.18em \PB}{}}\xspace}

  \def\Y#1S{\ensuremath{\PUpsilon{(#1S)}}\xspace}%

\def\Lbar {\ensuremath{\kern 0.1em\overline{\kern -0.1em\PLambda}}\xspace}

\def\AT#1     {\ensuremath{A_{\mathrm{T}}^{#1}}\xspace}           %

\def\C#1      {\ensuremath{\mathcal{C}_{#1}}\xspace}                       %
\def\Cp#1     {\ensuremath{\mathcal{C}_{#1}^{'}}\xspace}                    %
\def\Ceff#1   {\ensuremath{\mathcal{C}_{#1}^{\mathrm{(eff)}}}\xspace}        %
\def\Cpeff#1  {\ensuremath{\mathcal{C}_{#1}^{'\mathrm{(eff)}}}\xspace}       %
\def\Ope#1    {\ensuremath{\mathcal{O}_{#1}}\xspace}                       %
\def\Opep#1   {\ensuremath{\mathcal{O}_{#1}^{'}}\xspace}                    %

\newcommand{\unit}[1]{\ensuremath{\rm\,#1}\xspace}          %

\newcommand{\tev}{\ifthenelse{\boolean{inbibliography}}{\ensuremath{~T\kern -0.05em eV}\xspace}{\ensuremath{\mathrm{\,Te\kern -0.1em V}}\xspace}}
\newcommand{\tevcap}{\ensuremath{\mathrm{\,Te\kern -0.1em V}}\xspace}
\newcommand{\gev}{\ensuremath{\mathrm{\,Ge\kern -0.1em V}}\xspace}
\newcommand{\mev}{\ensuremath{\mathrm{\,Me\kern -0.1em V}}\xspace}
\newcommand{\kev}{\ensuremath{\mathrm{\,ke\kern -0.1em V}}\xspace}
\newcommand{\ev}{\ensuremath{\mathrm{\,e\kern -0.1em V}}\xspace}
\newcommand{\gevc}{\ensuremath{{\mathrm{\,Ge\kern -0.1em V\!/}c}}\xspace}
\newcommand{\mevc}{\ensuremath{{\mathrm{\,Me\kern -0.1em V\!/}c}}\xspace}
\newcommand{\gevcc}{\ensuremath{{\mathrm{\,Ge\kern -0.1em V\!/}c^2}}\xspace}
\newcommand{\gevgevcccc}{\ensuremath{{\mathrm{\,Ge\kern -0.1em V^2\!/}c^4}}\xspace}
\newcommand{\mevcc}{\ensuremath{{\mathrm{\,Me\kern -0.1em V\!/}c^2}}\xspace}

\def\m    {\ensuremath{\rm \,m}\xspace}
\def\cm   {\ensuremath{\rm \,cm}\xspace}

\def\mm   {\ensuremath{\rm \,mm}\xspace}

\def\mum  {\ensuremath{{\,\micro\rm m}}\xspace}

\def\barn{\ensuremath{\rm \,b}\xspace}

\def\mus  {\ensuremath{{\,\micro{\rm s}}}\xspace}
\def\ns   {\ensuremath{{\rm \,ns}}\xspace}
\def\ps   {\ensuremath{{\rm \,ps}}\xspace}

\def\MHz  {\ensuremath{{\rm \,MHz}}\xspace}

\def\kHz  {\ensuremath{{\rm \,kHz}}\xspace}
\def\Hz   {\ensuremath{{\rm \,Hz}}\xspace}

\newcommand{\chisqndf}{\ensuremath{\chi^2/\mathrm{ndf}}\xspace}

\def\gsim{{~\raise.15em\hbox{$>$}\kern-.85em
          \lower.35em\hbox{$\sim$}~}\xspace}

\newcommand{\mean}[1]{\ensuremath{\left\langle #1 \right\rangle}} %

\def\degrees{\ensuremath{\degree}\xspace}

\def\mrad{\ensuremath{\rm \,mrad}\xspace}
\def\rad{\ensuremath{\rm \,rad}\xspace}

\def\betastar {\ensuremath{\beta^*}\xspace}

\def\tell1  {TELL1\xspace}
\def\ukl1   {UKL1\xspace}

\newcommand{\eg}{\mbox{\itshape e.g.}\xspace}
\newcommand{\ie}{\mbox{\itshape i.e.}\xspace}

%% file: _commands.tex
\usepackage{placeins}

\newcommand{\includefig}[4]{
\begin{figure}[tbp]
\centering
\includegraphics[width=#1\textwidth]{#2}
\caption[#3]{#4}
\label{#2}
\end{figure} 
}

\newcommand{\PZero}[1][]{\ensuremath{{P_0^{\mathrm{#1}}}}\xspace}
\newcommand{\sigeff}[2][]{\ensuremath{{\sigma^{#1}_{\mathrm{#2}}}}\xspace}
\newcommand{\mueff}[2][]{\ensuremath{{\mu^{#1}_{\mathrm{#2}}}}\xspace}
\newcommand{\eff}{\ensuremath{\mathrm{vis}}}
\newcommand{\bx}[1]{\ensuremath{{\mathsf{#1}}}\xspace}

\newcommand{\EffVertex}{\ensuremath{\eta_\mathrm{\mbox{\scriptsize Vertex,LHCb}}}\xspace}

\newcommand{\mul}{\multicolumn}
\newcommand{\dyfr}{\displaystyle\frac}
\newcommand{\sect}{Sec.\xspace}

\newcommand{\Ni}[1]{\ensuremath{N_{#1,i}}}
\newcommand{\Si}[1]{\ensuremath{S_{\textnormal{FBCT},#1,i}}}

\newcommand{\fghost}[1]{\ensuremath{f_{\textnormal{ghost},#1}}}
\newcommand{\fsat}[1]{\ensuremath{f_{\textnormal{sat},#1}}}

\newcommand{\mulrow}{\multirow}
\newcommand{\mrw}{\multirow}

\newcommand{\instL}{L}          %
\newcommand{\rate}{R}
\newcommand{\ovint}{\Omega} %
\newcommand{\xsec}{\sigma}

\newcommand{\frev}{\ensuremath{\nu_{\mathrm{rev}}\xspace}}  %
\newcommand{\fr}{\ensuremath{f}\xspace}  %
\newcommand{\beamone}{\mbox{beam\,1}\xspace}
\newcommand{\beamtwo}{\mbox{beam\,2}\xspace}
\newcommand{\Beamone}{\mbox{Beam\,1}\xspace}
\newcommand{\Beamtwo}{\mbox{Beam\,2}\xspace}

\newcommand{\vdm}{VDM\xspace}

\newcommand{\Five}{5}
\newcommand{\PPP}{$pp$\xspace}
\newcommand{\PPB}{$p$Pb\xspace}
\newcommand{\PBP}{Pb$p$\xspace}

\def\hx{\hat{x}}
\def\hy{\hat{y}}
\def\hz{\hat{z}}

\newcommand{\hca}{\ensuremath{\phi}\xspace}   %
\def\sina{\sin\!\hca}
\def\cosa{\cos\!\hca}

\def\ssqa{\sin^2\!\hca}
\def\csqa{\cos^2\!\hca}

\newcommand{\Dx}{\ensuremath{\Delta x}\xspace}
\newcommand{\Dy}{\ensuremath{\Delta y}\xspace}
\newcommand{\Dm}{\ensuremath{\Delta m}\xspace}
\newcommand{\Dxz}{\ensuremath{\Delta x_0}\xspace}
\newcommand{\Dyz}{\ensuremath{\Delta y_0}\xspace}

\renewcommand{\Vec}[1]{\mathbf{#1}}  %
\def\zrf{z_{\rm rf}}   %
\newcommand{\Ntrks}{\ensuremath{N_{\mathrm{Tr}}\xspace}}
\newcommand{\Nbwd}{\ensuremath{N_{\mathrm{bwd}}\xspace}}
\newcommand{\Nfwd}{\ensuremath{N_{\mathrm{fwd}}\xspace}}

\newcommand{\oneD}{{1-d}\xspace}
\newcommand{\twoD}{{2-d}\xspace}

\newcommand{\sNN}{s_{\mbox{\tiny{\it NN}}}}
\newcommand{\rms}{RMS\xspace}

\newcommand{\pp}{\ensuremath{pp}\xspace}
\newcommand{\pA}{\ensuremath{p\text{Pb}}\xspace}
\newcommand{\Ap}{\ensuremath{\text{Pb}p}\xspace}

\newcommand{\rmn}{\ensuremath{\mathrm{n}}\xspace}
\newcommand{\rmw}{\ensuremath{\mathrm{w}}\xspace}

\newcommand{\corcoef}{\lambda}

\newcommand{\roc}{\emph{Track} to \emph{Vertex}}

\newcommand{\xilz}{\ensuremath{\xi_{lz}}}
\newcommand{\siglz}{\ensuremath{\sigma_{lz}}}

\newcommand{\sign}{\ensuremath{\sigma_\mathrm{n}}\xspace}

%% file: title-LHCb-PAPER.tex
\begin{titlepage}
\pagenumbering{roman}

\vspace*{-1.5cm}
\centerline{\large EUROPEAN ORGANIZATION FOR NUCLEAR RESEARCH (CERN)}
\vspace*{1.5cm}
\hspace*{-0.5cm}
\begin{tabular*}{\linewidth}{lc@{\extracolsep{\fill}}r}
\ifthenelse{\boolean{pdflatex}}%
{\vspace*{-2.7cm}\mbox{\!\!\!\includegraphics[width=.14\textwidth]{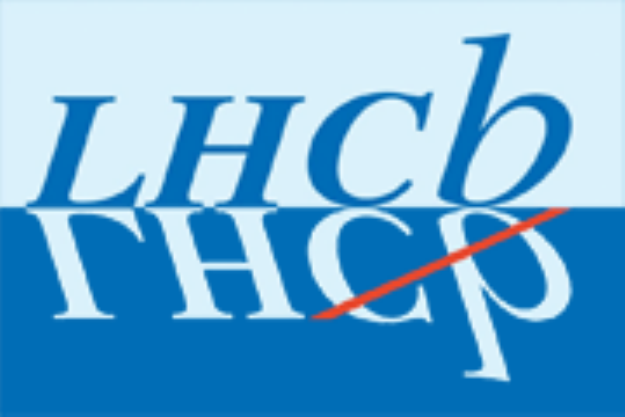}} & &}%
{\vspace*{-1.2cm}\mbox{\!\!\!\includegraphics[width=.12\textwidth]{lhcb-logo.eps}} & &}%
\\
 & & CERN-PH-EP-2014-221 \\  %
 & & LHCb-PAPER-2014-047 \\  %
 & & December 5, 2014 \\ %
 & & \\
\end{tabular*}

\vspace*{4.0cm}

{\bf\boldmath\huge
\begin{center}
  Precision luminosity measurements at LHCb 
\end{center}
}

\vspace*{1.0cm}

\begin{center}
The LHCb collaboration\footnote{Authors are listed at the end of the article.}
\end{center}

\vspace{\fill}

\input{abstract}

\vspace*{1.0cm}

\begin{center}
Published in \href{http://dx.doi.org/10.1088/1748-0221/9/12/P12005}{JINST \textbf{9} (2014) P12005}
\end{center}

\vspace{\fill}

{\footnotesize 
\centerline{\copyright~CERN on behalf of the \lhcb collaboration, license \href{http://creativecommons.org/licenses/by/4.0/}{CC-BY-4.0}.}}
\vspace*{2mm}

\end{titlepage}

\newpage
\setcounter{page}{2}
\mbox{~}
\newpage

\cleardoublepage

%% file: abstract.tex
\begin{abstract}
\noindent
Measuring cross-sections at the LHC requires
the luminosity to be determined accurately at each
centre-of-mass energy $\sqrt{s}$.
In this paper results are reported from the luminosity calibrations carried out
at the LHC interaction point 8 with the LHCb detector
for $\sqrt{s} = 2.76,~7$ and $8\tev$ (proton-proton collisions) and
for $\sqrt{\sNN} = \Five\tev$ (proton-lead collisions).
Both the ``van der Meer scan" and  ``beam-gas imaging" luminosity calibration methods were employed.
It is observed that the beam density profile cannot always be described 
by a function that is factorizable in the two transverse coordinates.
The introduction of a two-dimensional description of the beams improves significantly the consistency of the results.
For proton-proton interactions at $\sqrt{s} = 8\tev$
a relative precision of the luminosity calibration of 1.47\% is obtained using van der Meer scans
and 1.43\% using beam-gas imaging, resulting in a combined precision of 1.12\%.
Applying the calibration to the full data set determines the luminosity with a precision of 1.16\%.
This represents the most precise luminosity measurement achieved so far at a bunched-beam hadron collider.
\end{abstract}

%% file: introduction.tex
\section{Introduction}
\label{sec:Introduction}

The determination of the cross-section of a given subatomic process at high energy colliding-beam 
experiments is generally performed by the measurement of an interaction rate.
To determine such a cross-section on an absolute scale, a measurement of the colliding-beam 
luminosity must be performed.
The requirement for the accuracy on the value of the cross-section is usually driven by the
precision of theoretical predictions for the process. 
At the LHCb experiment~\cite{Alves:2008zz} %
the cross-section measurements for the production of 
vector bosons ($\Z$ and $\W$)~\cite{LHCb-PAPER-2012-008,LHCb-PAPER-2014-033} and the exclusive 
two-photon production of muon pairs~\cite{LHCb-PAPER-2013-059} %
motivate an accuracy of order 1--2\% for the luminosity calibration.

The instantaneous luminosity $\instL$ is defined by the relation between the reaction rate $\rate$ and 
the process cross-section $\xsec$ 
\begin{equation}
\rate = \instL \, \xsec \, .
\label{eq:cross-section}
\end{equation}
The instantaneous luminosity for a colliding bunch pair can be written as~\cite{ref:moller,Napoly:1992kn,Herr:941318}
\begin{equation}
 \label{eq:luminosity}
  L =  N_1 \, N_2 \, \frev \, \ovint \, ,
\end{equation}
where $N_1$ and $N_2$ are the populations of the colliding bunches of {\beamone} and {\beamtwo},
$\frev$ is the revolution frequency
and the beam overlap integral $\ovint$ embodies the passage of the two bunches with spatial particle density distributions 
$\rho_{1}(x,y,z,t)$ and $\rho_{2}(x,y,z,t)$ accross each other.
In the limit of ultra-relativistic particles (velocity close to the speed of light, $v\approx c$),
crossing at small angle, the beam overlap integral is given by
\begin{equation}
 \label{eq:overlapintegral}
  \ovint = 2c \, \displaystyle\int\!\rho_{1}(x,y,z,t) \, \rho_{2}(x,y,z,t) \, dx\, dy\, dz\, dt \, .
\end{equation}

Methods for absolute luminosity determination can be classified 
as being either direct or indirect.
Indirect methods are {\em e.g.} the use of the optical theorem 
to make a simultaneous measurement of the elastic and total
cross-sections~\cite{Antchev:1495764,ATLAS-CONF-2014-040}, %
or the comparison with a
process for which the absolute cross-section is known, either from theory or by a
previous direct measurement. 
Direct methods derive the luminosity from the measurements of the 
colliding beam parameters. 
The analysis described in this paper relies on two direct methods to
determine the absolute luminosity calibration:  
the ``van der Meer scan'' method (VDM)~\cite{LHCb-PAPER-2011-015,vanderMeer:296752,Rubbia:1025746,2011NIMPA.654..634B,ref:vdm-LHC} and 
the ``beam-gas imaging'' method (BGI)~\cite{FerroLuzzi:2005em,LHCb-PAPER-2011-015},
the latter making use of unique capabilities of the LHCb experiment. 
The VDM method exploits the ability to move the beams in both transverse coordinates with high precision and thus to scan 
the overlap integral of the colliding beams at different relative beam positions while measuring a relative rate.
This method, which was first applied at the CERN ISR~\cite{vanderMeer:296752},
is also being used by the other LHC experiments~\cite{Aad:1517411,CMS:2013gfa,Abelev:1700665}. %
The BGI method is based on reconstructing vertices of interactions between beam particles
and gas nuclei in the beam vacuum to measure the angles, positions and shapes of the individual beams
without displacing them.
The shapes obtained with these data are constrained by the distribution of vertices measured with beam-beam interactions.
In both methods, data taken with the LHCb detector located at interaction point (IP) 8
are used in conjunction with data from the LHC beam instrumentation. 

At the LHC, from 2009 to 2013, several luminosity calibration measurements 
were performed with a gradually improving precision.
Different nucleon-nucleon centre-of-mass energies $\sqrt{s}$ and 
different beam species were used: protons on protons (\PPP), 
lead on lead (Pb-Pb) and protons on lead (\PPB or \PBP, where the
first/second beam species applies to {\beamone}/{\beamtwo} in the standard LHC definition \cite{LHCMachine}, 
see Fig.~\ref{fig:lhcb-sketch}).
First LHC luminosity calibrations were obtained by LHCb 
using  \PPP collision data collected at the end of 2009 at
$\sqrt{s} = 900\gev$~\cite{LHCb-PAPER-2010-001} and in 2010 at 
\mbox{$\sqrt{s} = 7\tev$}~\cite{ref:balagura-moriond,ref:plamen-moriond,LHCb-PAPER-2011-015} 
with an accuracy that was limited by the systematic uncertainties associated 
with the normalization of the colliding bunch populations~\cite{BCNWG1,BCNWG2}.  %
Recent detailed studies of the LHC beam current transformers (BCTs) %
significantly reduced these uncertainties~\cite{Barschel:1425904,BCNWG3,BCNWG4},
thus facilitating an improvement of the final precision of the luminosity calibration.
In this paper results are reported from luminosity calibration experiments carried out
at the LHC IP8 with the LHCb detector from 2011 to 2013, %
for $\sqrt{s} = 2.76,~7$ and $8\tev$ in \PPP collisions and
for $\sqrt{\sNN} = \Five\tev$ in \PPB and \PBP collisions.
In addition to performing luminosity calibration measurements, LHCb provided related 
beam-gas interaction measurements as a service to the other LHC experiments.
This included the measurement of the total charge outside the nominally filled slots 
(``ghost charge'', see \sect~\ref{sec:Bunch current normalization}) and of
the single beam size as a function of time during the VDM scans of these other experiments. %

The precision of the luminosity calibration in the LHCb experiment is now limited by the 
systematic uncertainties of the beam overlap determination.
These systematic uncertainties are different, to a large extent, for the VDM and BGI methods.
Therefore, the comparison provides an important cross
check of the results.
The calibration measurements obtained with the VDM and BGI methods are found to be
consistent and are averaged for the final result. 

Since the absolute calibration can only be performed during specific
running periods, a relative normalization method 
is needed to transport the results of the absolute luminosity calibration to the complete data-taking period. 
  To this end, several observables are used, each one corresponding to an 
  effective visible cross-section $\sigma_\mathrm{vis}$. The
  corresponding cross-section is calibrated for each variable using the
  measurements of the absolute luminosity during specific data-taking
  periods.  The integrated luminosity for an arbitrary period of data
  taking is then obtained from the accumulated counts of a calibrated
  visible cross-section.

In the present paper we first describe briefly the LHCb experimental setup 
and data-taking conditions in \sect~\ref{sec:Experimental setup and data-taking conditions},
emphasizing the aspects relevant to the analysis presented here. 
Section~\ref{sec:Bunch current normalization} %
is devoted to the normalization
of the bunch population, while the methods used for the relative normalization technique 
are given in \sect~\ref{sec:InteractionRate}. 
In \sect~\ref{sec:Luminosity formalism for colliding beams} we introduce 
the luminosity formalism for colliding beams.
The determination of the luminosity with the BGI method 
is detailed in \sect~\ref{sec:Beam-gas imaging method} and
with the VDM scan method in \sect~\ref{sec:Van der Meer scan method}.
The combination of the results and conclusions are  given in \sect~\ref{sec:Conclusion}.

%% file: setup.tex
\section{Experimental setup and data-taking conditions}
\label{sec:Experimental setup and data-taking conditions}

\begin{figure}[tbp]
 \centering
 \includegraphics*[width=0.9\textwidth]{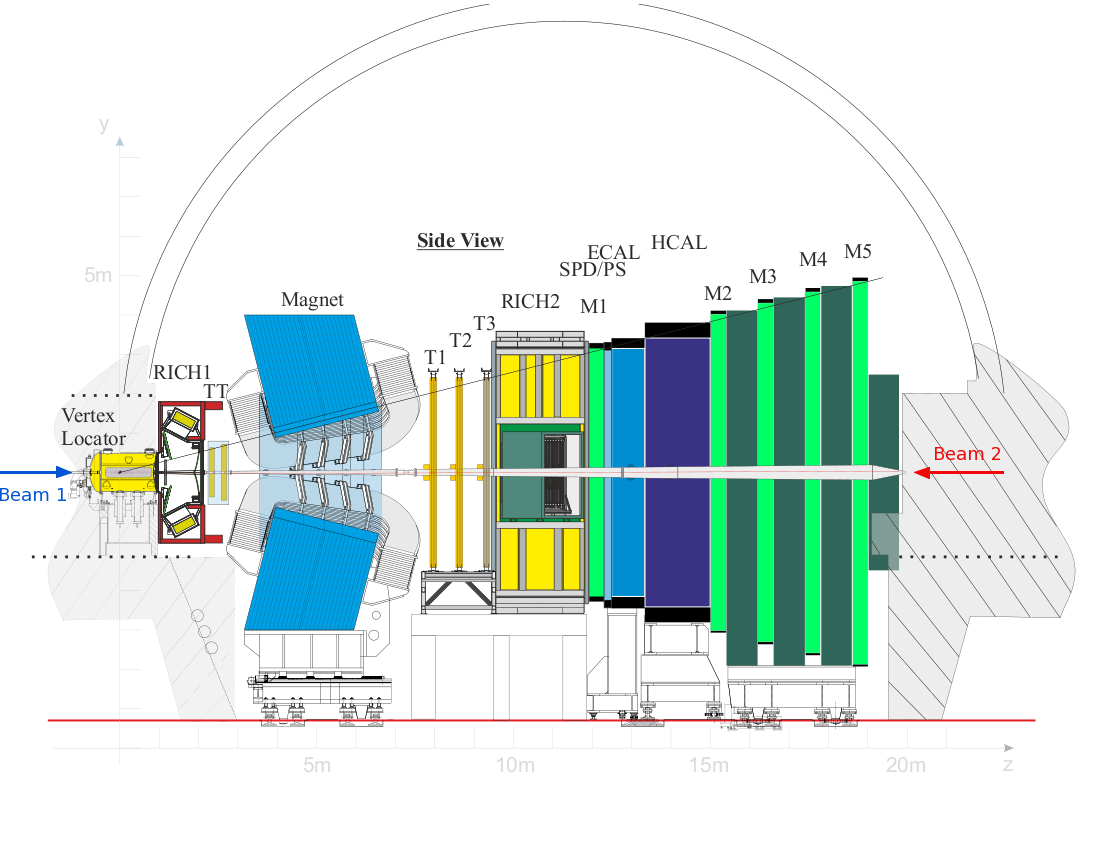}
                \caption{Schematic view of the current LHCb detector.
                         LHC {\beamone} ({\beamtwo}) enters from the left (right) side of the figure.
                         The labels indicate sub-detectors:
                         vertex locator (VELO), 
                         RICH1, RICH2 (ring imaging Cherenkov detectors 1 and 2),
                         TT (tracker Turicensis),
                         T1, T2, T3, (tracking stations 1, 2 and 3),
                         SPD/PS (scintillating pad detector / preshower detector), 
                         ECAL (electromagnetic calorimeter),
                         HCAL (hadron calorimeter), and
                         M1, M2, M3, M4, M5 (muon stations 1, 2, 3, 4, and 5) (drawing from Ref.~\cite{Antunes-Nobrega:630827}).
 }
 \label{fig:lhcb-sketch}
\end{figure}

The LHCb detector (Fig.~\ref{fig:lhcb-sketch}) is a single-arm forward spectrometer with a polar angular coverage of 
approximately 15 to 300\mrad in the horizontal (bending) plane, and 15 to 250\mrad in the vertical plane. 
It is designed for the study of particles containing $b$ or $c$ quarks
and is described in detail elsewhere~\cite{Alves:2008zz}. %

The apparatus contains tracking detectors, ring-imaging Cherenkov
detectors, calorimeters, and a muon identification system. 
The tracking system comprises the vertex locator (VELO) 
surrounding the beam interaction region, a tracking station upstream of
the dipole magnet and three tracking stations located
downstream of the magnet.  
Particles traversing the spectrometer experience a bending-field integral
of around 4\unit{Tm}.

\begin{figure}[tbp]
 \centering
 \includegraphics*[width=0.9\textwidth]{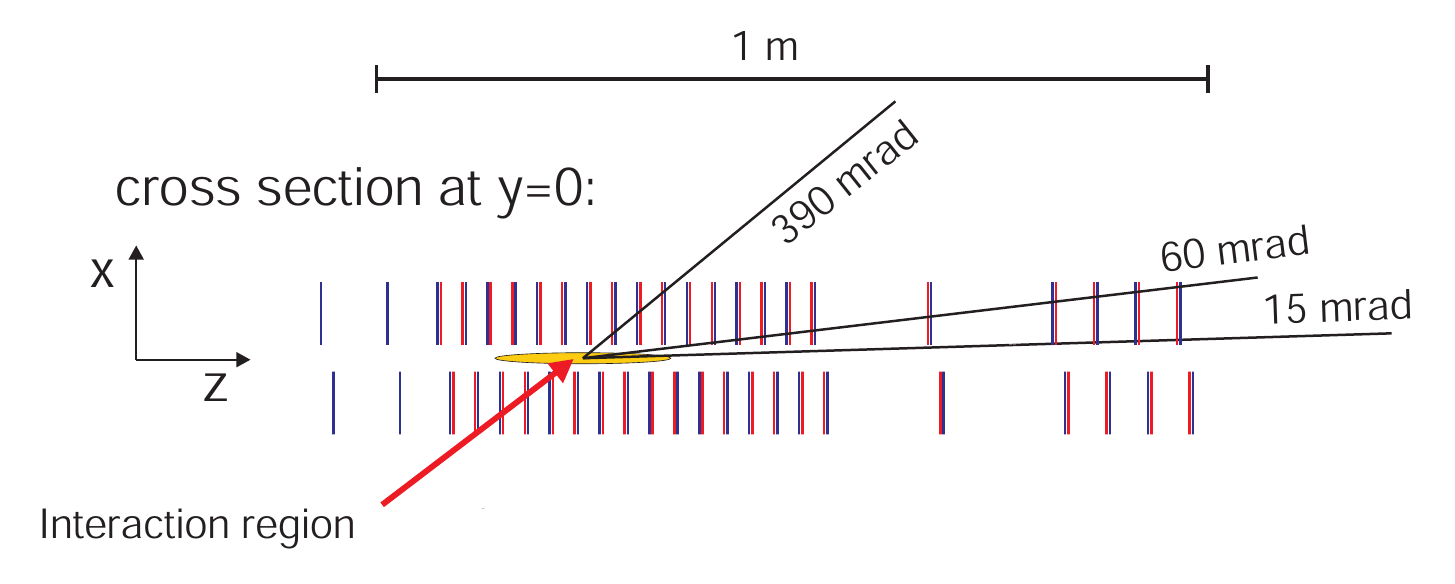} 
 \caption{Sketch of the VELO sensor positions.
  The luminous region is schematically depicted with a filled ellipse.
  Its longitudinal extent, \rms $\sigma = 53\mm$, is indicative.
  Sensors measuring the $R$~($\phi$) coordinates are shown as blue (red) lines.
 The LHC beam of ring 1 (2) enters from the left (right) on this sketch.
 The coordinate system is defined in \sect~\ref{sec:Luminosity formalism for colliding beams} (drawing from Ref.~\cite{Antunes-Nobrega:630827}).
 \label{fig:velo-sketch}
 }
\end{figure}

The VELO plays an essential role in the application of the
VDM and BGI luminosity calibration methods at LHCb. 
It consists of two horizontally retractable halves, each
having 21 modules of radial and azimuthal silicon-strip sensors in a
half-circle shape (Fig.~\ref{fig:velo-sketch}). 
Two additional stations ({\it Pile-Up System}, PU) upstream of the VELO
tracking stations are mainly used in the hardware trigger. 
The VELO has a large acceptance for beam-beam interactions 
owing to its many layers of silicon sensors and their close proximity to
the beam line. 
During nominal operation, the distance between the closest sensor strip and the beams
 is only 8.2\mm. 
During injection and beam adjustments, the two VELO halves are kept apart
in a retracted position 30\mm away from the beams. 
They are brought to their
nominal position close to the beams during stable beam periods only. 
More details about the VELO can be found in Ref.~\cite{LHCb-DP-2014-001}.

The LHCb trigger system~\cite{LHCb-DP-2012-004} consists of two separate
levels: a hardware trigger, which is implemented in custom electronics, and a
software trigger, executed on a farm of commercial processors.  
The hardware trigger is designed to have an accept rate of 1\MHz and uses information from the PU sensors of
the VELO, the calorimeters and the muon system.
These detectors send information to the hardware decision unit, 
where selection algorithms are run synchronously with the
40\MHz LHC bunch crossing. 
For every nominal bunch-crossing slot (\ie each 25\ns) the hardware decision unit
sends its information to the LHCb readout supervisor, which distributes the
synchronous hardware trigger decision to all front-end electronics.  
For every positive hardware decision the full event information of all sub-detectors is
sent to the processor farm and is made available to the software trigger algorithms. 

For luminosity calibration and monitoring,
a trigger strategy is adopted to select beam-beam inelastic interactions and
interactions of the beams with the residual gas in the vacuum chamber.
Events are collected for the four bunch-crossing types:
two colliding bunches (\bx{bb}), one {\beamone} bunch with no {\beamtwo} bunch (\bx{be}),
one {\beamtwo} bunch with no {\beamone} bunch (\bx{eb}) and nominally empty bunch slots
(\bx{ee}).  
Here ``\bx{b}'' stands for ``bunch'' and ``\bx{e}'' stands for ``empty''.
The first two categories of crossings produce particles in the
forward direction and are triggered using calorimeter information.
An additional PU veto is applied for \bx{be} crossings.
Crossings of the type \bx{eb} produce particles 
in the backward direction, are triggered by demanding a minimal hit
multiplicity in the PU, and are vetoed by calorimeter activity.
The trigger for \bx{ee} crossings is defined as the logical OR of the
conditions used for the \bx{be} and \bx{eb} crossings in order to be
sensitive to background from both beams.
In addition to these specific triggers, a decision based on a 
hardware trigger sensitive to any activity in the PU and calorimeter is available.
The latter hardware trigger configuration is used for most measurements 
described in this paper.
Events are then further selected by the software  trigger 
based on track and vertex reconstruction using VELO hits.
During VDM scans specialized trigger configurations are defined that 
optimize the data taking for these measurements 
(see \sect~\ref{sec:Van der Meer scan method}).

The reconstruction of interaction vertices (also called ``primary vertices'', PVs) 
is performed using standard LHCb algorithms \cite{LHCb-PUB-2014-044}.
The initial estimate of the PV position is based on an iterative
clustering of tracks.  
For each track the distance of closest approach (DOCA) with respect to all
other tracks is calculated and tracks are clustered into a PV candidate if their 
DOCA is less than 1\mm. 
An initial position of the PV is obtained from the weighted average of the points of closest approach between all track pairs, after removing outliers.
The final PV coordinates are determined by iteratively improving the
position determination with an adaptive, weighted, least-squares fit.
Participating tracks are %
assigned weights depending on their impact parameter with respect to the PV.
The procedure is repeated for all possible track clusters, excluding tracks from
previously reconstructed PVs, retaining only those with at 
least five tracks. 
For the analysis described here only PVs with a larger number of tracks are used since they
provide better position resolution. 
For the study of beam-gas interactions only PVs with at least ten tracks are used and at least 25 tracks are
required for the study of beam-beam interactions.  
For specific studies different criteria are applied as described below.

\def\mrx{ }
\begin{table}[ptb]
\centering
\caption[Dedicated LHC luminosity calibration fills used for VDM or BGI analysis
in this work.]{%
Dedicated LHC calibration fills during which LHCb performed  the luminosity calibrations
described in this paper or ghost charge and beam size measurements for other LHC experiments.
In most calibration measurements the number of bunches per beam was the same 
for {\beamone} and {\beamtwo}. For the \PPB and  \PBP fills where this was not the case
two numbers are given, the first for the number of {\beamone} bunches, 
the second for the number of {\beamtwo} bunches.
The number of colliding bunches at LHCb is indicated in parentheses (fifth column).
Half crossing angles $\phi_x$ and $\phi_y$, and $\beta^*$ are given as nominal values. 
The VELO vacuum state during BGI measurements is indicated in the column ``Gas injection''.
A state ``off'' means that gas injection was turned off and the VELO ion pumps were turned
off, which resulted in a residual vacuum pressure about a factor four higher than nominal.
A state ``on'' indicates that neon gas was being injected into the beam vacuum. 
During VDM measurements the state was always ``off''.
}
\label{tab:fills}
\begin{tabular}{@{}llllllll@{}}\toprule
Period & Fill & $\phi_x$ ($\phi_y$) &  $\beta^*$   & Bunches & Gas      & Luminosity  \\  %
       &      & ({\micro}rad)        & (m)          & per beam & injection & calibration \\  %
\midrule
\multicolumn{7}{l}{Fills with \PPP at $\sqrt{s}=8\tev$}\\
\mrx{~~Apr 2012} & \mrx{2520} & \mrx{236 (90)} & \mrx{3}& \mrx{48 (6)} & \mrx{on} & BGI                \\  %
\mrx{~~Apr 2012} & \mrx{2523} & \mrx{236 (90)} & \mrx{3}& \mrx{52 (24)} & \mrx{on} & BGI,    VDM     \\  %
\mrx{~~Jul 2012} & \mrx{2852} & \mrx{456 (0)} & \mrx{10}& \mrx{50 (16)} & \mrx{on} & BGI,            \\  %
\mrx{~~Jul 2012} & \mrx{2853} & \mrx{456 (0)} & \mrx{10}& \mrx{35 (16)} & \mrx{on} & BGI,    VDM     \\  %
\mrx{~~Jul 2012} & \mrx{2855} & \mrx{456 (0)} & \mrx{10}& \mrx{48 (6)} & \mrx{on} & BGI                \\  %
\mrx{~~Jul 2012} & \mrx{2856} & \mrx{456 (0)} & \mrx{10}& \mrx{48 (6)} & \mrx{on} & BGI                \\  %
\mrx{~~Nov 2012} & \mrx{3311} & \mrx{456 (0)} & \mrx{10}& \mrx{39 (6)} & \mrx{on} & BGI                \\  %
\mrx{~~Nov 2012} & \mrx{3316} & \mrx{456 (0)} & \mrx{10}& \mrx{39 (6)} & \mrx{on} & BGI                \\  %
\midrule
\multicolumn{7}{l}{Fills with \PPP at $\sqrt{s}=7\tev$}\\
\mrx{~~Oct 2011} & \mrx{2234} & \mrx{270 (0)} & \mrx{10}& \mrx{36 (16)} & \mrx{off} & BGI,      VDM   \\  %
\midrule
\multicolumn{7}{l}{Fills with \PPP at $\sqrt{s}=2.76\tev$}\\
\mrx{~~Feb 2013} & \mrx{3555} & \mrx{855 (0)} & \mrx{10}& \mrx{100 (22)} & \mrx{on} & BGI              \\  %
\mrx{~~Feb 2013} & \mrx{3562} & \mrx{855 (0)} & \mrx{10}& \mrx{39 (6)} & \mrx{on} & BGI                \\  %
\mrx{~~Feb 2013} & \mrx{3563} & \mrx{855 (0)} & \mrx{10}& \mrx{39 (6)} & \mrx{on} & BGI                \\  %
\midrule
\multicolumn{7}{l}{Fills with \PPB at $\sqrt{\sNN}=\Five\tev$}\\
\mrx{~~Jan 2013} & \mrx{3503} & \mrx{456 (0)} & \mrx{2}& \mrx{272+338 (38)}& \mrx{off}& other experiments\\  %
\mrx{~~Jan 2013} & \mrx{3505} & \mrx{456 (0)} & \mrx{2}& \mrx{272+338 (38)}& \mrx{off}&           VDM  \\  %
\midrule
\multicolumn{7}{l}{Fills with \PBP at $\sqrt{\sNN}=\Five\tev$}\\
\mrx{~~Feb 2013} & \mrx{3537} & \mrx{456 (0)} & \mrx{2}& \mrx{314+272 (22)}& \mrx{off}& other experiments \\  %
\mrx{~~Feb 2013} & \mrx{3540} & \mrx{456 (0)} & \mrx{2}& \mrx{314+272 (22)}& \mrx{off}& other experiments \\  %
\mrx{~~Feb 2013} & \mrx{3542} & \mrx{456 (0)} & \mrx{2}& \mrx{338 (39)}    & \mrx{off}&            VDM \\  %
\bottomrule
\end{tabular}
\end{table}

The full list of luminosity calibrations discussed in this paper is summarized
in Table~\ref{tab:fills}.
The table is divided into five sections following the different nucleon-nucleon centre-of-mass 
energies and beam species involved.
A first measurement with intentionally enlarged beta functions at the IP ($\beta^* = 10\m$) was performed
in October 2011 with \PPP collisions at $\sqrt{s}=7\tev$.
Several fills in 2012 were dedicated to luminosity calibration for \PPP collisions 
at $\sqrt{s}=8\tev$,
although only the measurements in July and November were performed with large $\beta^*$. 
The April measurements were performed in non-optimal conditions, 
with focused beams ($\beta^* = 3\m$) and with a tilted crossing plane (a non-zero 
vertical half crossing angle $\phi_y$),
and are therefore primarily used for the VDM calibration method and
to cross-check the effects on the BGI method of the finite vertex resolution.
Calibrations for \PPB and  \PBP were conducted in January 2013 at $\sqrt{\sNN}=\Five\tev$
with VDM scans only.
Further \PPP calibrations were performed at $\sqrt{s}=2.76\tev$ in February 2013, exclusively using the BGI method. 
The number of bunches per beam is also given in the table.
No active gas injection was used to enhance the beam-gas rates and the end of 2011, though
a first rate increase was obtained in October 2011 by degrading the beam vacuum by switching 
off the VELO ion pumps.
Thus, three configurations of the VELO vacuum state have been used, one where the vacuum pumps are
operating (normal state), one where the VELO ion pumps were switched off, and one where, in 
addition to running with pumps off, neon gas was injected 
into the VELO vacuum chamber (see \sect~\ref{sec:Beam-gas imaging method}).
All \PPP BGI calibration measurements of 2012 and 2013 took advantage of gas injection.
During VDM calibration scans, gas injection was always off.
In all \PPP calibration runs discussed here the initial bunch populations ranged between 
0.6 and $1.1\times 10^{11}$ particles. 
For the \PPB and \PBP runs they varied between 1 and $2\times 10^{10}$ elementary charges
(for both beam species).
Calibration experiments with the VDM method included a variety of beam displacement sequences.
The details of these individual experiments are given in the section devoted to 
the VDM analysis (\sect~\ref{sec:Van der Meer scan method}).
In fills 3503, 3537 and 3540, no luminosity calibration was performed at IP8, 
though the LHCb experiment provided ghost charge and beam size measurements for the benefit of
the luminosity calibrations conducted in other LHC experiments.

%% file: normalisation.tex
\section{Bunch current normalization}
\label{sec:Bunch current normalization}

\def\fr{\ensuremath{f}\xspace}

Various detector systems are used to determine with high precision the population of particles in 
each colliding and non-colliding bunch in the LHC.
The longitudinal structure of the LHC beams is shaped by the 400\MHz radio frequency (RF) system.
Both LHC rings are filled with bunches at locations (``RF buckets'') defined by 
the RF system and are organized in ``slots'', which contain each ten consecutive buckets.
Ideally, only one of these buckets is filled with a bunch, called the 
``main bunch'', the other nine are nominally empty.
Only a subset of the slots are filled in a given filling scheme.
In each filled slot, the main bunch occupies the same bucket number.
In practice, a small fraction (typically $<\,10^{-3}$) of the charge in a 
slot occupies nominally empty buckets and are called ``satellite'' bunches.
Additionally, also the nominally empty slots may contain charges.
The total charge outside the filled slots is called ``ghost charge''.

\subsection{Bunch population measurement}
\label{Bunch population measurement}

To measure the population in the main bunches, specific instruments are used
to determine the overall circulating charge, the relative charge in the filled
bunches, the fraction of the charge in the satellite buckets of the filled slots
and the fraction of ghost charge.
Four independent direct-current current-transformers (DCCTs), 
two per ring, 
are used to measure the total beam current circulating in each LHC ring.
The DCCT is designed to be insensitive to the time structure of the beam~\cite{P.Odier.LHC.DCCT}.
Two fast bunch current transformers (FBCTs), one per ring, provide a relative measure of
the individual charges on a slot-by-slot basis~\cite{ref:fbct}. 
The FBCT is designed to produce a signal proportional to the charge in each 25~ns LHC bunch slot.
The captured particles of an LHC bunch are contained within an RF
bucket of 1--1.5\ns length at $\pm2$~standard deviations~\cite{ref:LHC-RF}.
Since 2012, one longitudinal density monitor (LDM)~\cite{Jeff:2012zz,Jeff:1513180} per LHC ring is
available for detecting synchrotron radiation photons emitted by particles deflected in 
a magnetic field. The LDMs are used to obtain the longitudinal beam charge distribution with
a time resolution of about 90\ps to resolve the charge distribution in individual RF buckets.
Finally, the ghost charge fraction is obtained by counting beam-gas interactions with the 
LHCb detector in nominally empty (\bx{ee}) compared to the rates in nominally filled (\bx{bb}, \bx{be} and \bx{eb})
bunch crossings.

Previous LHC luminosity calibration experiments showed that one of the 
dominant uncertainties 
arises from the normalization of the bunch population product $ N_1 \, N_2$.
As a consequence, a detailed study of the normalization was carried out
using data from the LHC beam current transformers (BCTs) and from the LHC 
experiments.
A dedicated analysis procedure was defined and bunch population uncertainties were quantified
for the 2010 LHC luminosity calibration measurements~\cite{BCNWG1,BCNWG2}. 
The precision was limited by the understanding of the BCT data at that stage.
Since then, a number of additional tests were carried out that significantly 
improved the understanding of the bunch current measurements.
Careful calibration measurements and systematic studies of the DCCTs
improved the dominant uncertainty by an order of magnitude~\cite{Barschel:1425904,Colin-thesis}.
Uncertainties on the beam current product for the 2011--2013 measurements are 
well below 1\% and are given in more detail below. 

The accuracy of the relative bunch populations determined with the FBCT is cross-checked
against results from other measurements, such as those obtained from the ATLAS BPTX 
button pick-up~\cite{ATLAS-BPTX} and those derived from the LHCb beam-gas interaction 
rates~\cite{BCNWG3}.
The sum of the FBCT signals of all nominally filled bunch slots is normalized 
to the total number of particles measured by the DCCTs after subtraction of 
the ghost charge and satellite charges, 
\begin{equation}
   \Ni{j} =  \dyfr{I_{{\rm DCCT},j}}{\frev\, Z_j\,e} 
             \cdot (1 - \fghost{j}) 
             \cdot \frac{\Si{j}}{\sum_i \Si{j}}  
             \cdot (1-\fsat{j,i}) \, ,
\end{equation}
defining $\Ni{j}$ as the bunch population of the nominally filled RF bucket of bunch slot $i$
of beam $j$, and
$I_{{\rm DCCT},j}$ as the current measured by the DCCTs and $Z_j\,e$ the charge of
a beam particle ($82\,e$ for Pb beams).
The sum runs over all nominally filled slots and the $\Si{j}$ are the signals measured 
by the FBCT of ring $j$.
The ghost charge fraction is denoted $\fghost{j}$ and 
the fraction of the charge in satellite bunches $\fsat{j,i}$ for beam $j$ and slot $i$.

Ghost charge fractions for the 2011--2013 LHC luminosity calibration fills range 
up to about 2.5\%. 
As mentioned above, these measurements are performed with the LHCb detector.
The results and methods are described in detail in \sect~\ref{sec:ghost_charges}.

Satellite charges have been observed in various  ways with the LHC detectors 
by detecting longitudinally displaced collisions (see for example Ref.~\cite{BCNWG1}).  
The total satellite population fraction ($\fsat{j,i}$) in a bunch slot is usually less than a percent compared
to the associated main bunch population. Nevertheless, it needs to be 
quantified to obtain a precise measurement of the bunch population that actually
contributes to the luminosity.

\subsection{Ghost charge}
\label{sec:ghost_charges}

The determination of the ghost charge from the beam-gas
interaction rate measurements was pioneered in a previous LHCb luminosity calibration~\cite{Plamen-thesis}.
The results presented here benefit from the larger number of beam-gas events obtained 
with neon gas injection in the beam vacuum chamber, which
allows the uncertainty to be reduced
and provides a more detailed determination of the charge distribution
over the LHC ring in a shorter time. 
Systematic uncertainties are further reduced by a better trigger efficiency calibration.
The ghost charge measurement is based on the same data sample as used for the BGI analysis.
The trigger requirements are described in \sect~\ref{sec:Beam-gas imaging method}.
 
To ensure that each vertex is a result of a beam-gas interaction and is assigned to 
the correct beam, several selection criteria are applied~\cite{Colin-thesis} that are based 
on the track directions (all forward for {\beamone}, all backward for {\beamtwo}),
on the transverse position (to exclude interactions with material in the vicinity of the beams), 
on the longitudinal position and on the vertex track multiplicity.

The LHCb data acquisition is synchronised with the LHC RF system with a granularity of 25\ns.
The sampling phase of the detectors relative to the LHC clock is optimized to provide 
the highest efficiency for nominally filled RF buckets,
but the trigger efficiency may vary across the 25\ns bunch slot.
Since the ghost charge is distributed over all RF buckets inside the 25\ns slots,
the trigger efficiency must be known for all possible phases.
A first efficiency measurement was performed in 2010~\cite{Plamen-thesis,LHCb-PAPER-2011-015},
resulting in a ghost charge uncertainty of about 20\% per beam.
A new dedicated measurement was performed in 2012 with the aim of reducing this 
uncertainty by acquiring data for more clock phases and by using neon gas injection 
to increase the statistical accuracy.
The efficiency is determined by measuring dead-time corrected beam-gas interaction rates from 
non-colliding bunches at different clock phases and comparing them
with the standard phase (zero clock shift).
The absolute rate is measured as function of clock shift in 2.5\ns steps.
The beam intensity decay observed during the measurement %
is taken into account.

\begin{figure}[tbp]
  \centering
  \begin{minipage}[c]{0.5\textwidth}
    \includegraphics[width=\textwidth]{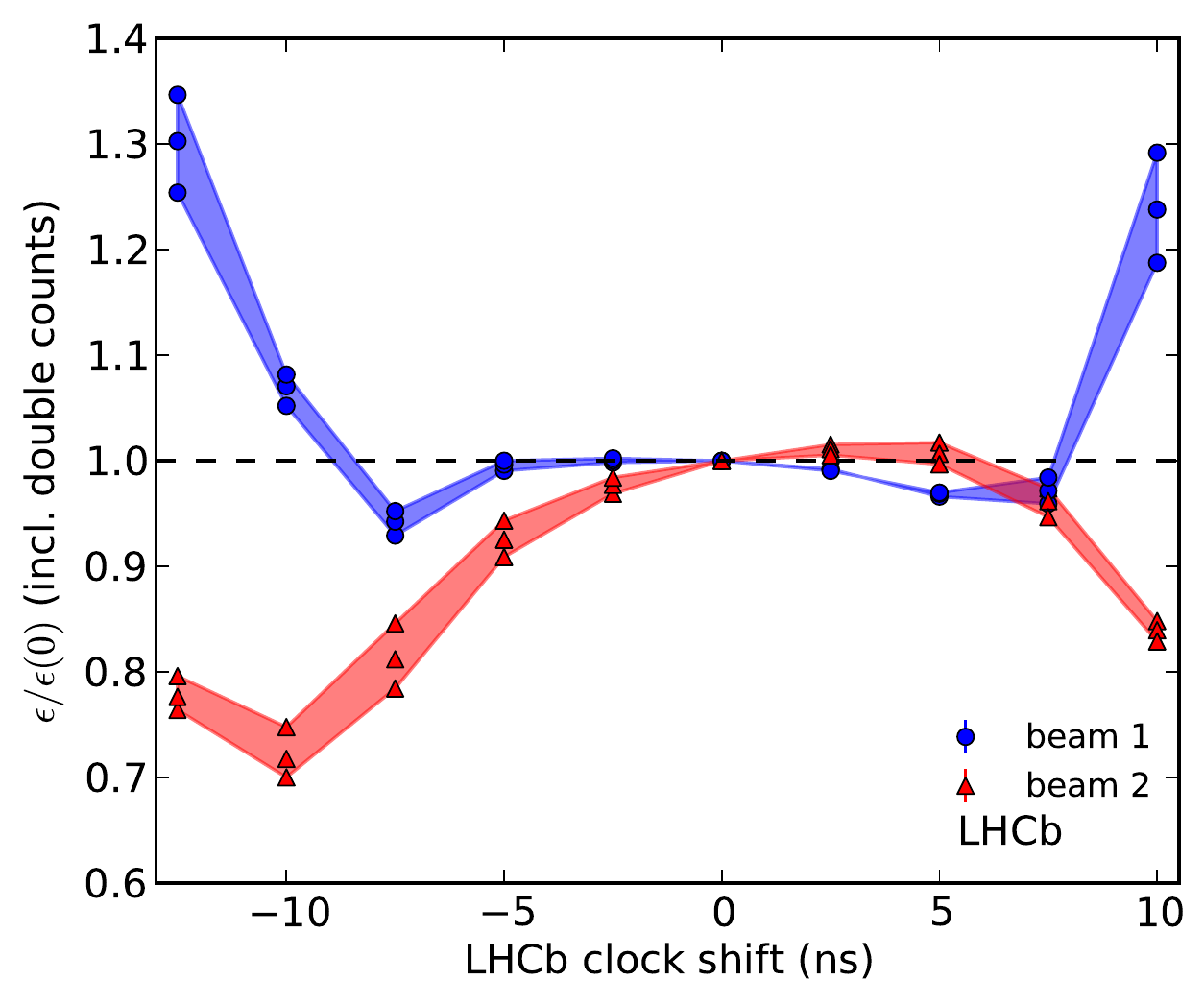}
  \end{minipage}\hfill
  \begin{minipage}[c]{0.5\textwidth}
    \includegraphics[width=\textwidth]{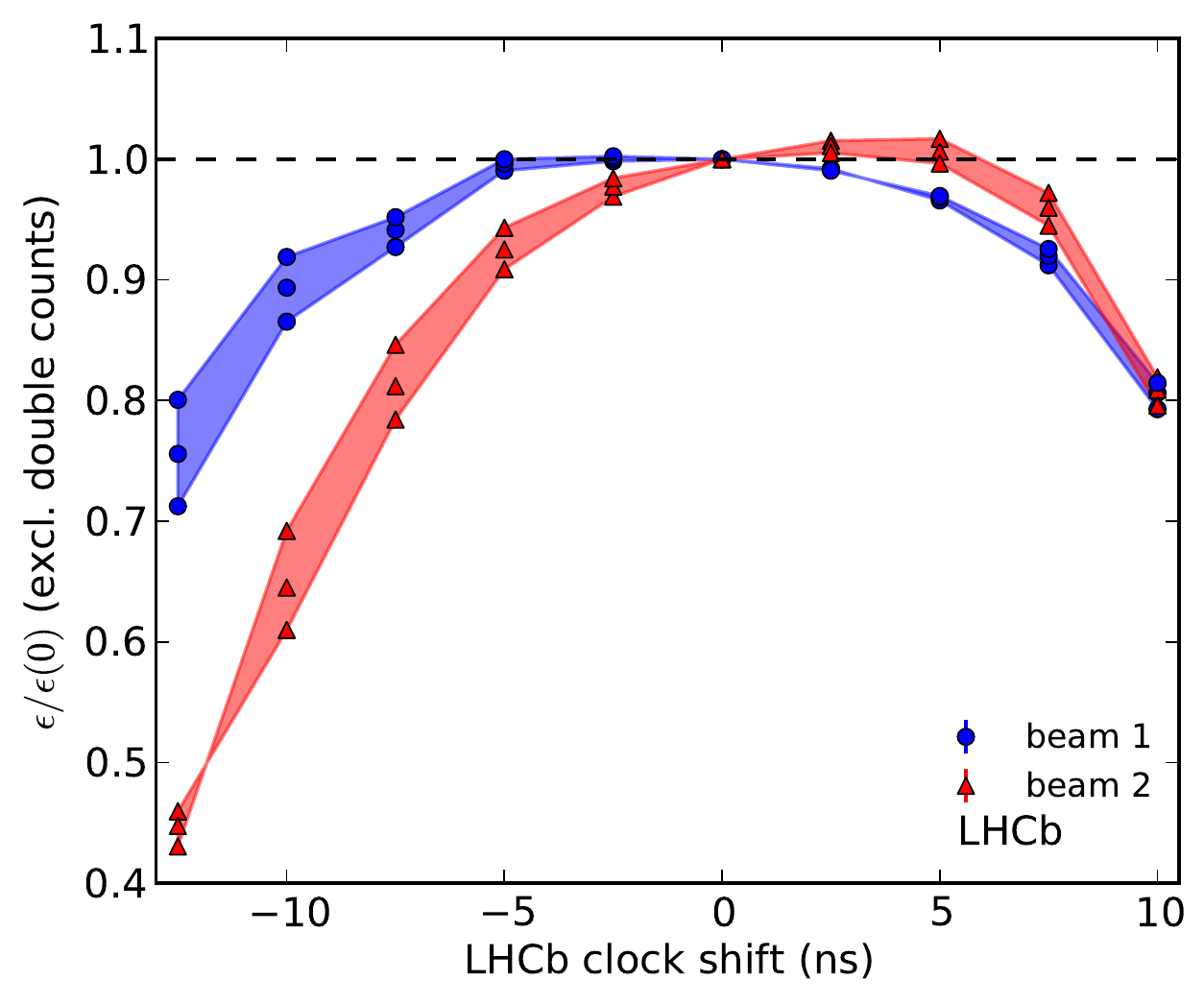}
  \end{minipage}\hfill
  \begin{minipage}[c]{1\linewidth}
    \caption{\small Relative beam-gas trigger efficiency as function of LHCb detector clock shift with respect to the LHC reference
      timing, (left) including or (right) excluding double-counted beam-gas interaction vertices. 
      The efficiency is shown relative to the value at the nominal clock setting (\ie zero shift).
      The shaded areas indicate the variation between the results for thresholds corresponding to 8 and 12 tracks.
      The data points, appearing in groups of three, indicate measurements applying the 8, 10 and 12 track thresholds.
    }
    \label{trigger_efficiency}
  \end{minipage}
\end{figure}

If a beam-gas interaction occurs near the bunch slot edges, that is, the originating 
charge is near the previous or next clock cycle, the resulting VELO sensor signals may be sufficiently long that they are also seen in the neighbouring clock cycle.
Therefore, depending on where the charges are located within the 25\ns bunch slot, 
some vertices are counted twice and thus bias the ghost charge or trigger efficiency 
measurement.
To take this double-counting effect into account, the efficiency is measured 
including all beam-gas events or, alternatively, excluding double-counted vertices.
In addition, the efficiency is measured for different vertex track multiplicity thresholds
(from 8 to 12 tracks) to account for the slightly different 
trigger conditions used for this measurement as compared to later BGI measurements.
The results of the trigger efficiency calibration are shown in Fig.~\ref{trigger_efficiency} and the
values averaged over the 25\ns clock cycle are summarized in Table~\ref{tab:trigger_efficiency}.
\begin{table}[tbp]
  \centering
  \caption{\small Relative beam-gas 
    trigger efficiency for the ghost charge measurement assuming a constant charge distribution within a bunch slot.
  }
  \label{tab:trigger_efficiency}
  \begin{tabular}{@{}lccllll@{}}\toprule
    Beam&\multicolumn{2}{c}{Efficiency average $\epsilon_j$}\\
    & including double-counting & excluding double-counting\\
    \midrule
    1&$1.05\pm 0.03$ & $0.93\pm 0.02$\\
    2&$0.90\pm 0.01$ & $0.86 \pm 0.01$\\
    \bottomrule
  \end{tabular}
\end{table}

\begin{figure}[tbp]
  \centering
  \begin{minipage}[c]{1\textwidth}
    \includegraphics[width=\textwidth]{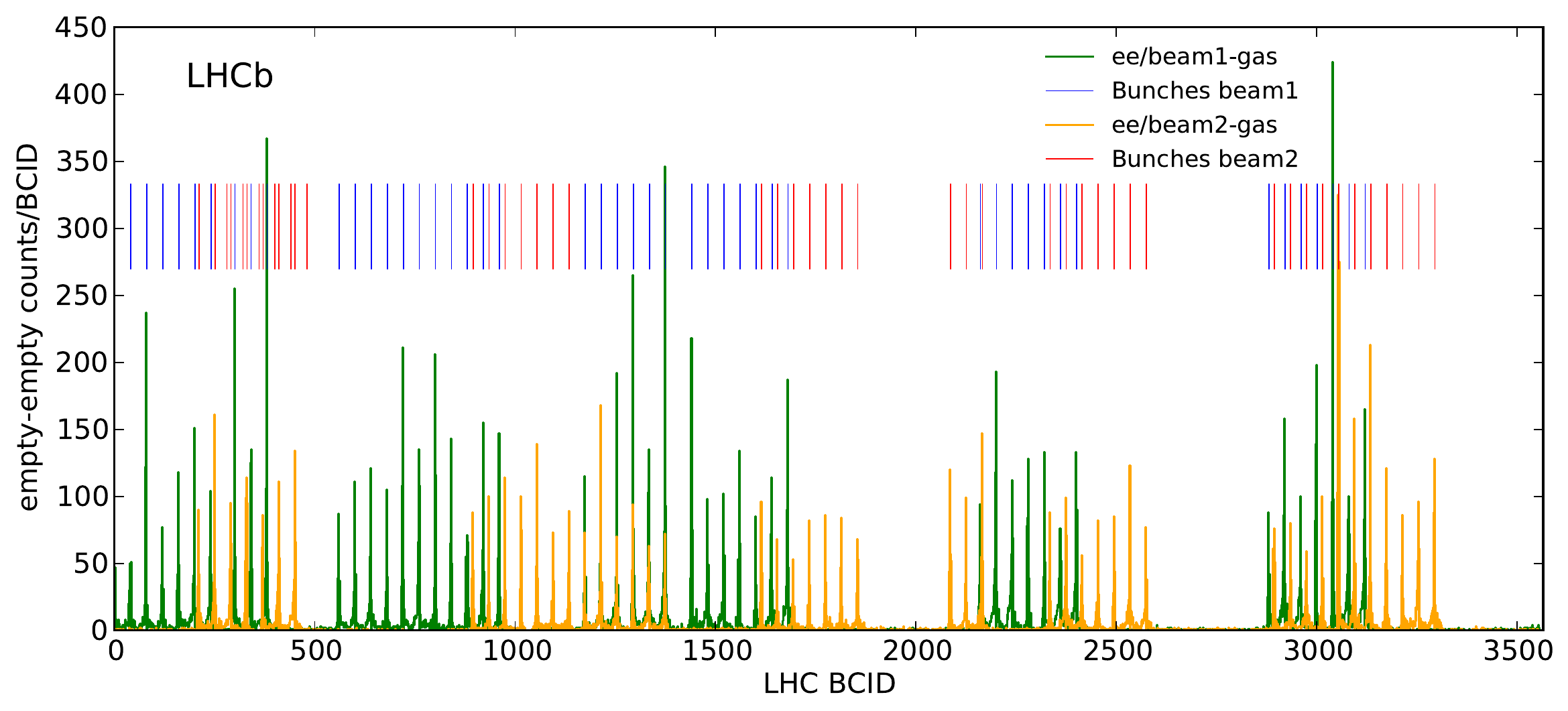}
  \end{minipage}\hfill
  \begin{minipage}[c]{0.5\textwidth}
    \includegraphics[width=\textwidth]{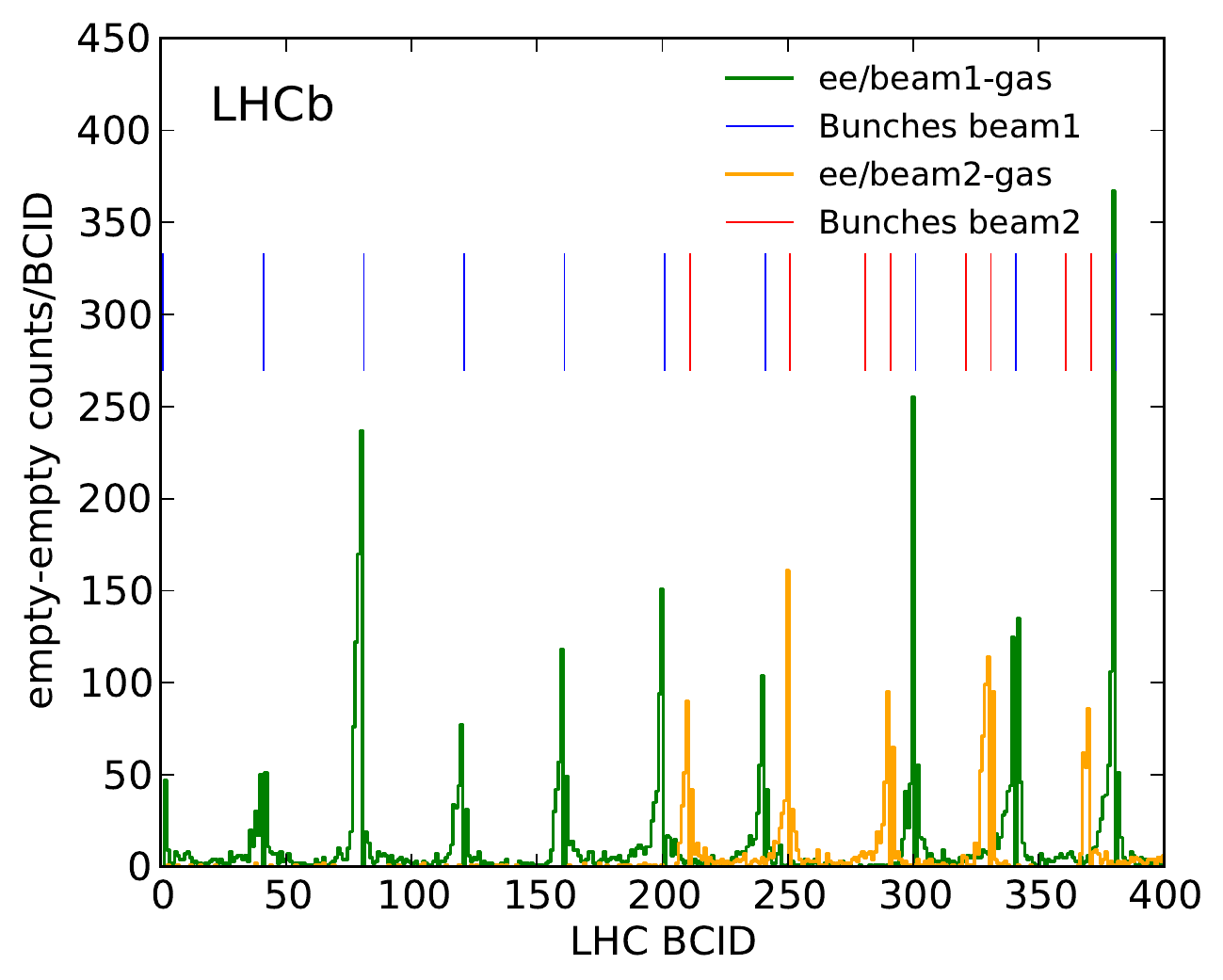}
  \end{minipage}\hfill
  \begin{minipage}[c]{0.5\textwidth}
    \includegraphics[width=\textwidth]{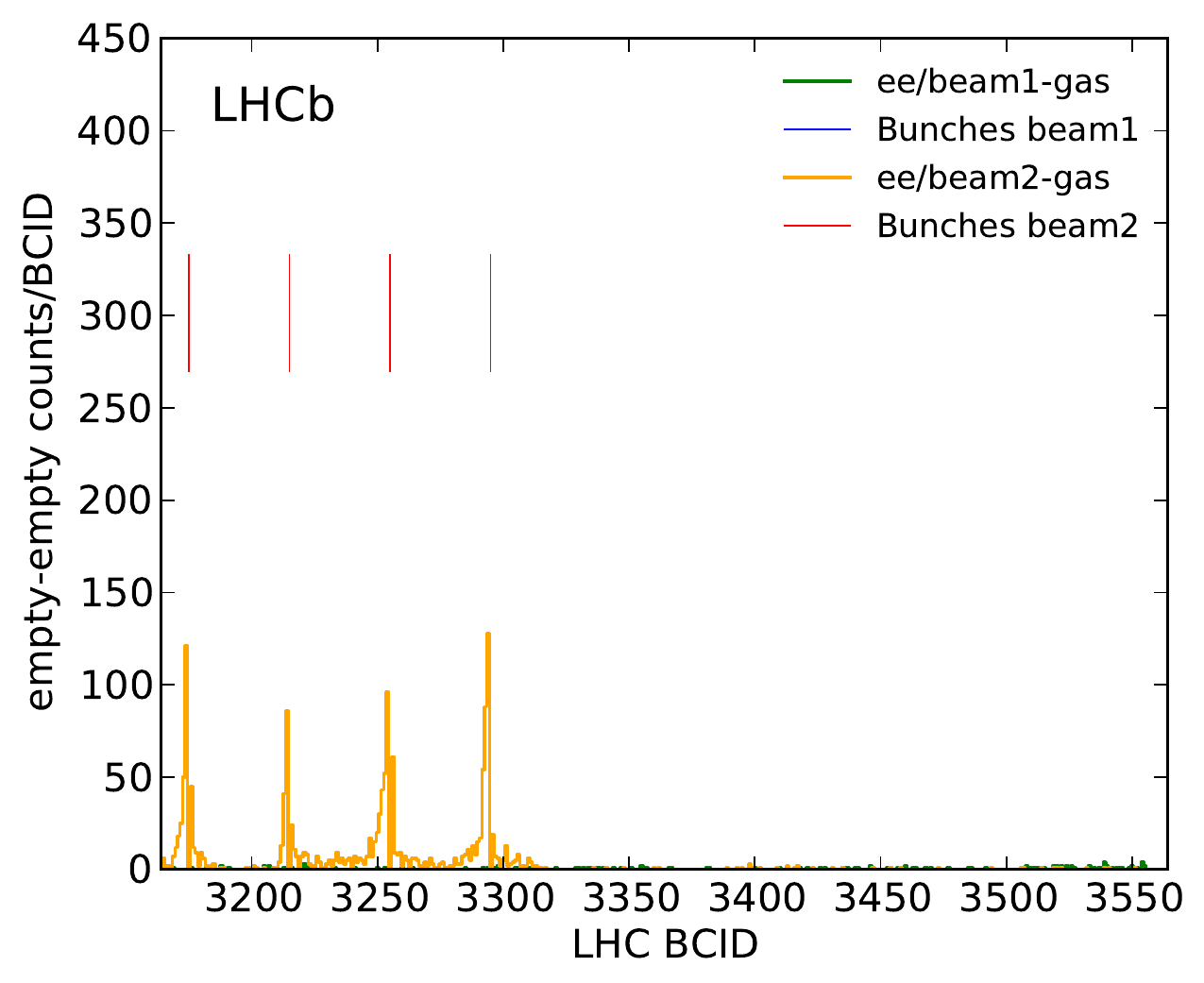}
  \end{minipage}\hfill
  \begin{minipage}[c]{1\linewidth}
    \caption{\small Histogram of ghost charge distribution as a function of LHC 
      bunch slot number (BCID) in fill 2520 for {\beamone} (green) and {\beamtwo} (yellow). 
      The BCID position of nominally filled bunches is indicated as small
      vertical blue and red lines for {\beamone} and {\beamtwo}, respectively. 
      The ghost charge distribution is shown for the (top) ring circumference and (bottom left) 
      first 400 and (bottom right) last 400 BCIDs. 
      Ghost charges are mostly absent in regions without nominally filled bunches.
      Note that only  \bx{ee} BCIDs are displayed.}
    \label{gc_2520}
  \end{minipage}
\end{figure}
The increased rate of beam-gas interaction data acquired with neon gas injection 
enables a measurement of the charge distribution over the ring circumference.
In Fig.~\ref{gc_2520} the ghost charge per 25\ns slot is shown as function of slot
number (BCID) using data from fill 2520 as an example.
Ghost charges are observed around the nominally filled bunches and are mostly absent 
further than about 20 slots away from filled bunches.

Ghost charge fractions during LHC luminosity calibration fills are measured in four-minute time bins.
For each time bin the ghost charge fraction is evaluated with both counting methods: 
including and excluding double-counted vertices and applying the corresponding 
average trigger efficiency of Table~\ref{tab:trigger_efficiency}.
If all charges are evenly spread within their bunch slot, each evaluation
would provide a different result before efficiency correction, but the same result
after efficiency correction.
After efficiency correction the differences between the two evaluations are small.
This observation is in agreement
with the LDM measurements~\cite{Boccardi:1556087}, which show that the ghost charge 
tends to be spread evenly over all RF buckets of a bunch slot.
The LDM information on the charge distribution within the nominally
empty bunch slots is not used in the results except for fill 3542 during which
the trigger was not configured to perform this measurement.
The average of the two efficiency-corrected evaluations is taken as final value 
for the ghost fraction, while their difference 
is taken as systematic uncertainty.
The trigger efficiency uncertainty taken from Table~\ref{tab:trigger_efficiency} is added in 
quadrature with the systematic uncertainty.
A summary of all ghost charge measurements performed for the special luminosity fills in 2011, 2012 and 2013 is provided in Table~\ref{tab:gc_results}.

\begin{figure}[tbp]
\centering
\begin{minipage}[c]{0.5\textwidth}
\includegraphics[width=\textwidth]{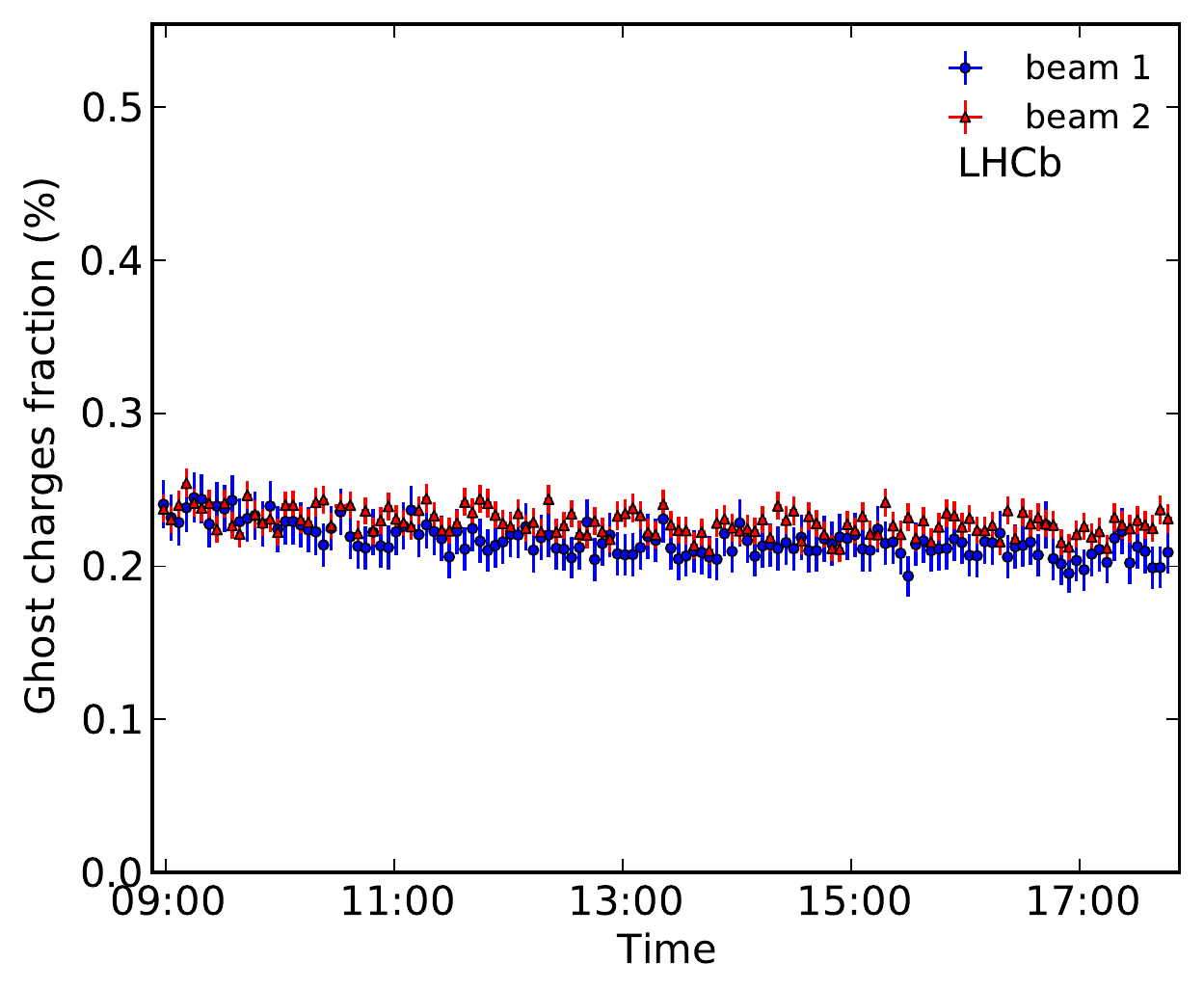}
\end{minipage}\hfill
\begin{minipage}[c]{0.5\textwidth}
\includegraphics[width=\textwidth]{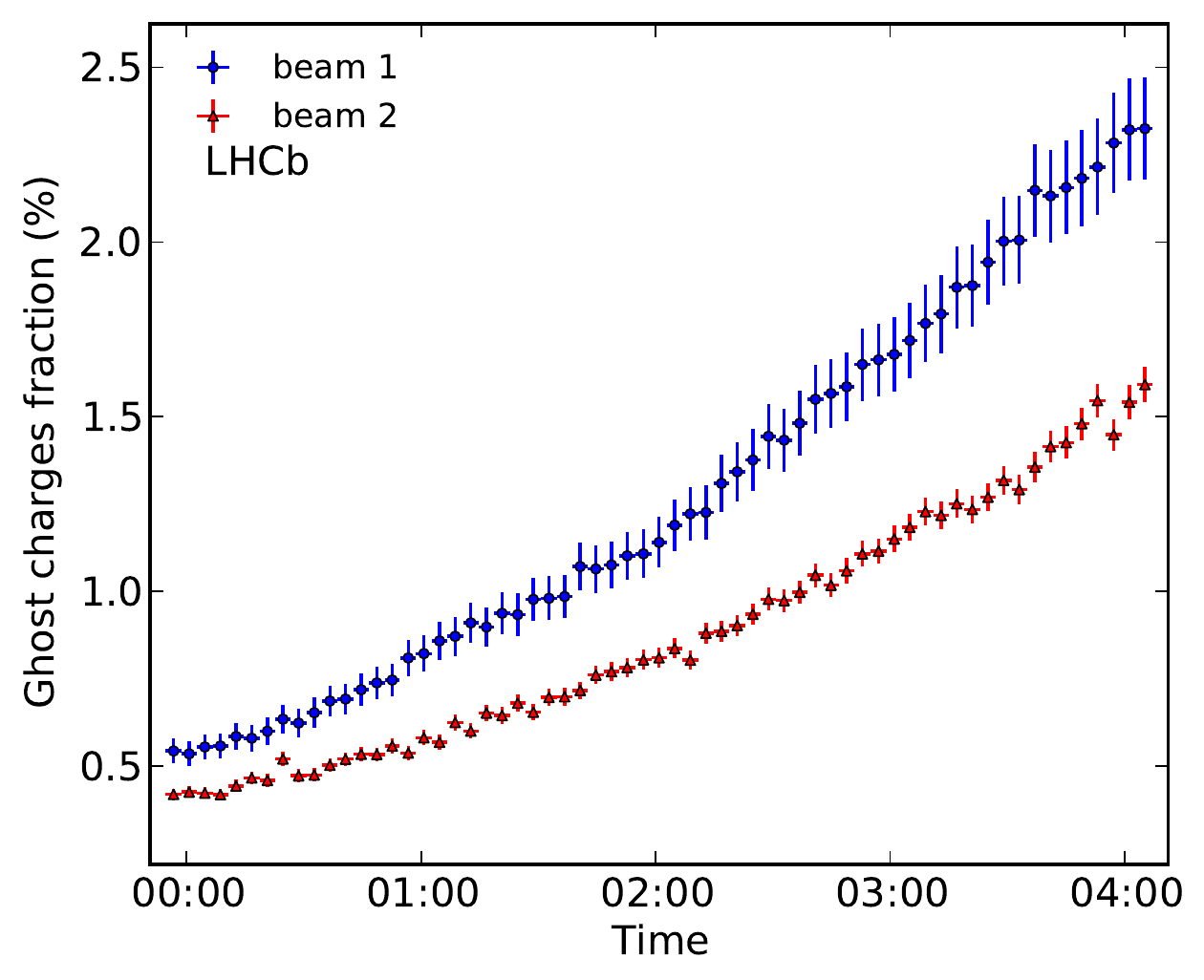}
\end{minipage}\hfill
\begin{minipage}[c]{1\linewidth}
\caption{\small Ghost charge fractions for (left) fill 2855 and (right) fill 3563. 
Fill 2855 with $\sqrt{s}=8\unit{TeV}$ shows a constant or slightly decreasing ghost charge fraction throughout the 
fill lasting about 9 hours. Fill 3563  ($\sqrt{s}=2.76\unit{TeV}$) shows an important 
increase of ghost charge over a period of 4 hours.}
\label{gc_2855}
\end{minipage}
\end{figure}

With the exception of intermediate energy fills at $\sqrt{s}=2.76\tev$, ghost charge fractions are 
stable within $\pm10$\% during a fill and the total beam intensity can be corrected with
good accuracy using an average value for a fill.
In this case the \rms over the fill, given in Table~\ref{tab:gc_results}, should be taken into account
in the uncertainty.
On the contrary, for the intermediate-energy fills, an increase in the ghost charge fraction over time 
warrants a time dependent correction to the total beam intensity.
As an example, the difference in ghost charge evolution seen between high- and intermediate-energy fills is shown 
in Fig.~\ref{gc_2855} comparing the long fill 2855 at $\sqrt{s}=8\tev$ and fill 
3563 at $\sqrt{s}=2.76\tev$.
\def\spc{\hspace*{40pt}}
\def\Spc{\hspace*{120pt}}
\def\mrw#1{#1}
\def\muluncert{\multicolumn{2}{l}{uncertainty}}
\begin{table}[tbp]
\centering
\centering
\caption[ghost charge results]{\small Measurements of ghost charge fractions for all luminosity
calibration fills in 2011, 2012 and 2013. 
The systematic uncertainty is assumed to be fully correlated between the two beams.
Therefore, the final systematic uncertainty on the beam intensity product due to the 
ghost charge correction is a linear sum of the ghost charge systematic uncertainty of each beam.
Proton-lead fills 
were acquired without neon gas injection and have 
a larger statistical uncertainty. 
For fill 3542  the ghost charge was only measured using the LHC LDMs.
}
\label{tab:gc_results}
  \begin{tabular}{lllllllll}\toprule
  Fill & \multicolumn{4}{l}{\Beamone\Spc} & \multicolumn{4}{l}{\Beamtwo}\\
       & \fghost{1}&\rms    & \muluncert     & \fghost{2}&\rms    & \muluncert    \\
  \spc & (\%)     & in fill & syst. &  stat. & (\%)     & in fill & syst. &  stat.\\
  \midrule
  \multicolumn{9}{l}{Fills with \PPP at $\sqrt{s}=8\tev$}\\
  \mrw{2520} &     0.30 & 0.01 & 0.02 & 0.001   &     0.35 & 0.01 & 0.01 & 0.002\\
  \mrw{2523} &     0.50 & 0.02 & 0.03 & 0.001   &     0.44 & 0.01 & 0.01 & 0.001\\
  \mrw{2852} &     0.62 & 0.01 & 0.04 & 0.002   &     0.53 & 0.01 & 0.01 & 0.002\\
  \mrw{2853} &     0.40 & 0.01 & 0.02 & 0.002   &     0.28 & 0.01 & 0.01 & 0.002\\
  \mrw{2855} &     0.22 & 0.01 & 0.01 & 0.001   &     0.23 & 0.01 & 0.01 & 0.001\\
  \mrw{2856} &     0.24 & 0.01 & 0.01 & 0.001   &     0.22 & 0.02 & 0.01 & 0.001\\
  \mrw{3311} &     0.15 & 0.03 & 0.01 & 0.001   &     0.06 & 0.01 & 0.01 & 0.001\\
  \mrw{3316} &     0.19 & 0.01 & 0.01 & 0.001   &     0.06 & 0.01 & 0.01 & 0.001\\
  \midrule
  \multicolumn{9}{l}{Fills with \PPP at $\sqrt{s}=7\tev$}\\
  \mrw{2234} &     0.84 & 0.10 & 0.05 & 0.012   &     0.76 & 0.13 & 0.02 & 0.015\\
  \midrule
  \multicolumn{9}{l}{Fills with \PPP at $\sqrt{s}=2.76\tev$}\\
  \mrw{3555} &     0.58 & 0.14 & 0.04 & 0.001   &     0.33 & 0.03 & 0.01 & 0.001\\
  \mrw{3562} &     0.78 & 0.30 & 0.05 & 0.003   &     0.52 & 0.22 & 0.01 & 0.003\\
  \mrw{3563} &     1.28 & 0.55 & 0.08 & 0.002   &     0.88 & 0.35 & 0.02 & 0.002\\
  \midrule
  \multicolumn{9}{l}{Fills with \PPB at $\sqrt{\sNN}=\Five\tev$}\\
  \mrw{3503} &     0.18 & 0.04 & 0.01 & 0.005   &     0.50 & 0.13 & 0.01 & 0.011\\
  \mrw{3505} &     0.29 & 0.05 & 0.02 & 0.007   &     0.66 & 0.12 & 0.02 & 0.015\\
  \midrule
  \multicolumn{9}{l}{Fills with \PBP at $\sqrt{\sNN}=\Five\tev$}\\
  \mrw{3537} &     0.50 & 0.12 & 0.03 & 0.010   &     0.88 & 0.11 & 0.02 & 0.015\\
  \mrw{3540} &     0.73 & 0.09 & 0.05 & 0.019   &     0.17 & 0.05 & 0.01 & 0.014\\
  \mrw{3542} &     \mul{4}{c}{{n.a.}}&\mul{4}{c}{{n.a.}}        \\
  \bottomrule
  \end{tabular}
\end{table}

\subsection{Total uncertainty}
\label{sec:beam_product_uncert}

A summary of the bunch population product uncertainties is given
in Table~\ref{tab:intensity_sys}  for each luminosity calibration fill.
The systematic uncertainties for the ghost charge corrections of the two beams 
described in the previous section are assumed to be fully correlated with each other,
\ie the final ghost charge uncertainty on the bunch population product is the 
linear sum of the ghost charge systematic uncertainty of each beam.

The satellite fractions provided by the LDM~\cite{Boccardi:1556087} are measured at 
the beginning and at the end of the fill.
Here, the average of these two measurements is used.
The average satellite fractions for all colliding bunches %
and fills with $\beta^*=10$~m 
at $\sqrt{s}=8\tev$ are 0.25\% and 0.18\% for {\beamone} and {\beamtwo}, respectively.
The uncertainty on the satellite fraction correction is taken as the full difference between the
fractions measured at the beginning and end of fill.
Assuming the uncertainties are fully correlated between the two beams, the uncertainty on the population
product due to the satellite fraction correction is taken as the linear sum of the average uncertainties per beam,
and is given as the average per fill in Table~\ref{tab:intensity_sys}. 

The beam population product normalization uncertainty is dominated by the DCCT measurement.
All fills listed in Table~\ref{tab:intensity_sys} are subject to the same procedure to evaluate 
the beam population product uncertainty. %
For fills with $\beta^*=10\m$ and $\sqrt{s}=8\tev$, the average uncertainty on the bunch population 
product weighted with the number of measurements amounts to
0.22\% at 68\% confidence level. %

\begin{table}[tbp]
\centering
\caption{Relative uncertainties (in percent) on colliding-bunch population products
for all relevant fills. %
}
\label{tab:intensity_sys}
\def\spz{\hspace*{80pt}}
\begin{tabular}{llll}\toprule
 Fill   & \mul{3}{c}{Sources of uncertainty on bunch population product}  \\%
 \spz   & DCCTs\hspace*{40pt}     & Ghost fractions           & Satellite fractions \\%
 \midrule
 \mul{4}{l}{Fills with \PPP at $\sqrt{s}=8\tev$}\\
 \mrw{2520} & 0.26 & 0.029 & n.a.  \\%
 \mrw{2523} & 0.22 & 0.043 & n.a.  \\%
 \mrw{2852} & 0.19 & 0.049 & 0.097 \\%
 \mrw{2853} & 0.24 & 0.032 & 0.019 \\%
 \mrw{2855} & 0.21 & 0.019 & 0.021 \\%
 \mrw{2856} & 0.21 & 0.020 & 0.031 \\%
 \mrw{3311} & 0.22 & 0.011 & 0.011 \\%
 \mrw{3316} & 0.23 & 0.013 & 0.011 \\%
 \midrule
 \mul{4}{l}{Fills with \PPP at $\sqrt{s}=7\tev$}\\
 \mrw{2234} & 0.24 & 0.064 & 0.25 \cite{BCNWG4} \\%
 \midrule
 \mul{4}{l}{Fills with \PPP at $\sqrt{s}=2.76\tev$}\\
 \mrw{3555} & 0.51 & 0.047 & 0.230 \\%
 \mrw{3562} & 0.22 & 0.062 & 0.020 \\%
 \mrw{3563} & 0.23 & 0.101 & 0.024 \\%
 \midrule
 \mul{4}{l}{Fills with \PPB at $\sqrt{\sNN}=\Five\tev$}\\
 \mrw{3505} & 0.31 & 0.137 & 0.070   \\%
 \midrule
 \mul{4}{l}{Fills with \PBP at $\sqrt{\sNN}=\Five\tev$}\\
 \mrw{3542} & 0.34 & 0.192 & 0.092   \\%
 \bottomrule
\end{tabular}
\end{table}

%% file: interactionrate.tex
\section{Relative luminosity calibration}
\label{sec:InteractionRate}

Absolute luminosity calibrations are performed during short periods of
data-taking.  To be able to determine the integrated luminosity for
any data sample obtained during long periods, 
the interaction rate of standard processes is measured continuously.
The effective cross-section corresponding to these 
standard processes is determined by counting the visible interaction rates during the
specific periods when the absolute luminosity is calibrated.  

\subsection{Interaction rate determination}
\label{subsec:InteractionRateDetermination}

The luminosity is proportional to the average number of
visible proton-proton interactions per beam-beam crossing, \mueff{\eff}. 
The subscript ``\eff'' is used to indicate that this particular definition of 
interaction rate does not need to have a simple physics interpretation.
Any interaction rate that can be measured under stable conditions can 
be used as such a relative luminosity monitor.
The interaction rates are acquired and stored together with the
physics data as ``luminosity data''. 
During further processing of the data the relevant
luminosity information is kept in the same storage entity.
Thus, it remains possible to select only part of the full
data set for analysis and still keep the capability to determine
the corresponding integrated luminosity.

Triggers, which initiate the full readout of the LHCb detector, are created 
for a random choice of beam crossings at a fixed average frequency.  
These are called ``luminosity triggers''.
During normal physics data-taking, the overall rate is chosen to be 1000\Hz.
Of this rate, 70\% is assigned to slots where two bunches cross (\bx{bb}), 
15\% to slots with only a beam-1 bunch (\bx{be}), 
10\% to those with only a beam-2 bunch (\bx{eb}) 
and the remaining 5\% to slots that are empty (\bx{ee}).
The events taken for crossing types other than \bx{bb} are used for
background subtraction and beam monitoring.

Interaction rates are measured by processing the random luminosity triggers 
and these rates are stored in a small number of ``luminosity observables''.
The set of luminosity observables comprises the number of vertices and tracks 
reconstructed in the VELO, the number of muons reconstructed in the muon system,
the number of hits in the PU and in the SPD in front of the calorimeters, and the
transverse energy deposition in the calorimeters.
The number of vertices in the VELO that fall within a limited region 
around the nominal interaction point and VELO tracks crossing this region 
are counted separately.
Some of these observables are directly obtained from the hardware trigger 
decision unit, others are the result of partial event reconstruction 
in the software trigger or in the off-line software.
Observables used in this analysis are summarized in Table~\ref{tab:lumiobservables}.

\begin{table}[bt]
 \centering
 \small
 \caption{\small Definition of luminosity observables used in the analysis. 
          The fiducial volume used here is a cylinder of radius $< 4\mm$ around the $z$ axis
          and bound by $|z|<300\mm$. It is used to cut either on the point of closest
          approach of a track relative to the $z$ axis or on the position of a vertex.
 \label{tab:lumiobservables}
 }  
 \vskip 1mm
 \begin{tabular}{lll}
  {\bf Name} & {\bf Description} & {\bf Origin}\\
  \midrule
  \emph{Track}     & Number of tracks reconstructed in the VELO in a
  & software reconstruction \\
  & fiducial volume  & \\
  \emph{TrackNR}   & Number of tracks reconstructed in the VELO not
  & software reconstruction \\
  & restricted to a fiducial volume & \\
  \emph{Vertex}    & Number of vertices reconstructed in the VELO in a
  & software reconstruction \\
  & fiducial volume & \\
  \emph{VertexNR} & Number of vertices reconstructed in the VELO not
  & software reconstruction \\
  & restricted to a fiducial volume & \\
  \emph{Muon}      & Number of muon tracks reconstructed in the muon &
  hardware trigger unit \\
  & system & \\
  \emph{PU}        & Number of hits counted in the PU   & hardware trigger unit \\
  \emph{SPD}       & Number of hits counted in the SPD  & hardware trigger unit \\
  \emph{ECalo}     & Energy deposition in the calorimeters  & hardware trigger unit \\
  \emph{Calo}      & Both \emph{SPD} and \emph{ECalo} over threshold & software reconstruction \\
  \bottomrule
 \end{tabular}
\end{table}

The luminosity for a given data set can be determined by integrating 
the values of observables that are proportional to the instantaneous 
luminosity and by applying the corresponding absolute calibration 
constant.  
However, this procedure sets stringent requirements on the stability 
of the observable and on its linearity in the presence of multiple interactions.
Alternatively, one may determine the relative luminosity from the
fraction of ``empty'' or invisible events in \bx{bb} crossings which we
denote by \PZero.  
An invisible event is defined by applying an  observable-specific threshold, 
below which it is considered that no \PPP interaction
is seen in the corresponding bunch crossing.
For a colliding bunch pair, the number of interactions per bunch 
crossing follows a Poisson distribution with mean value proportional 
to the luminosity, hence the luminosity is proportional to $-\ln \PZero$.  
In the absence of backgrounds, the average number of visible \PPP
interactions per crossing can be obtained from the fraction of empty
\bx{bb} crossings by $\mueff{\eff} = -\ln{\PZero[\bx{bb}]}$.
This ``zero-count'' method is both robust and easy to
implement~\cite{Zaitsev:473383}.
The choice of a low visibility threshold ensures a better behaviour
under gain or efficiency variations of the observable than the 
straightforward linear summing method.
In addition, any non-linearity encountered with multiple events does not 
play a role when counting empty slots.

Assuming equal particle populations in \bx{bb}, \bx{be}, and \bx{eb} bunches
and no particles in \bx{ee} slots, backgrounds are subtracted using 
\begin{equation} 
\label{eq:mu} 
\mueff{\eff} = -\left(\ln{\PZero[\bx{bb}]} - \ln{\PZero[\bx{be}]} -
\ln{\PZero[\bx{eb}]} + \ln{\PZero[\bx{ee}]}\right) \, ,
\end{equation}
where $\PZero[i] (i=\bx{bb},\bx{ee},\bx{be},\bx{eb})$ are the
probabilities to find an empty event in a bunch-crossing slot for the
four different bunch-crossing types.
In Eq.~\eqref{eq:mu} it is implicitly assumed that all bunches of the same type 
have the same properties.
The consequences of this approximation will be discussed in \sect\ref{sec:InteractionRate:syst}.  
The $\PZero[\bx{ee}]$ contribution is added because it is also contained in
the $\PZero[\bx{be}]$ and $\PZero[\bx{eb}]$ terms. 
The purpose of the background subtraction, Eq.~\eqref{eq:mu}, is to
correct the count-rate in the \bx{bb} crossings for the detector
response, which is due to beam-gas interactions and detector noise.  
In principle, the noise background is measured during \bx{ee} crossings.
In the presence of parasitic beam protons in \bx{ee} bunch positions 
(ghost charge), it is not correct to evaluate the noise 
from $\PZero[\bx{ee}]$.
In addition, the detector signals are not fully confined within one
25\ns bunch-crossing slot for some of the observables.
The empty (\bx{ee}) bunch-crossing slots immediately following a \bx{bb}, \bx{be}
or \bx{eb} crossing slot contain detector signals from interactions
occurring in the preceding slot (``spill-over'').
However, because the filling schemes used for the data-taking described here 
did not contain adjacent filled slots, 
the spill-over background is negligible in the \bx{bb}, \bx{be} and \bx{eb} crossings.
Since the detector noise for the selected observables is
small (see \sect~\ref{sec:InteractionRate:syst})
the term $\ln{\PZero[\bx{ee}]}$ in Eq.~\eqref{eq:mu} is neglected. 

The results of the zero-count method based on the number of tracks and 
vertices reconstructed in the VELO are found to be the most stable.
An empty event is defined to have $< 2$~tracks in the VELO.  A VELO track
is defined by at least three $R$ clusters and three $\phi$ clusters 
on a straight line in the VELO detector.
The number of tracks reconstructed in the VELO restricted to a fiducial 
region is chosen as the reference observable.  

\subsection{Systematic uncertainties}
\label{sec:InteractionRate:syst}

\begin{figure}[tb]
  \begin{center}
    \includegraphics[width=0.49\linewidth]{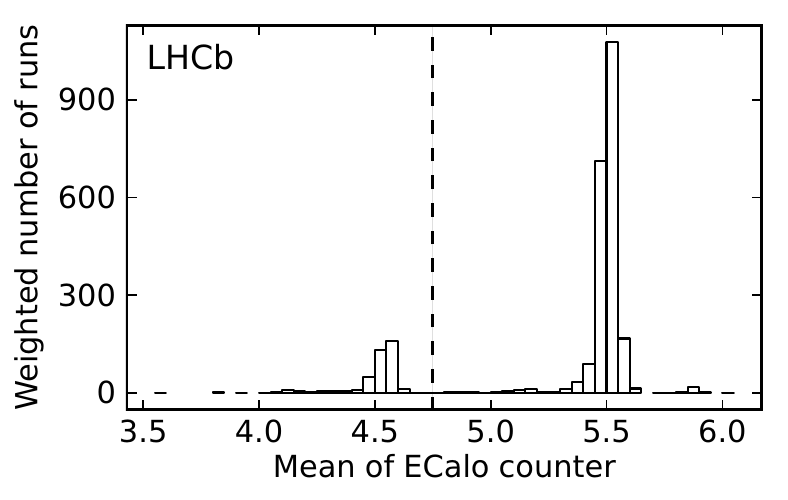}
    \includegraphics[width=0.49\linewidth]{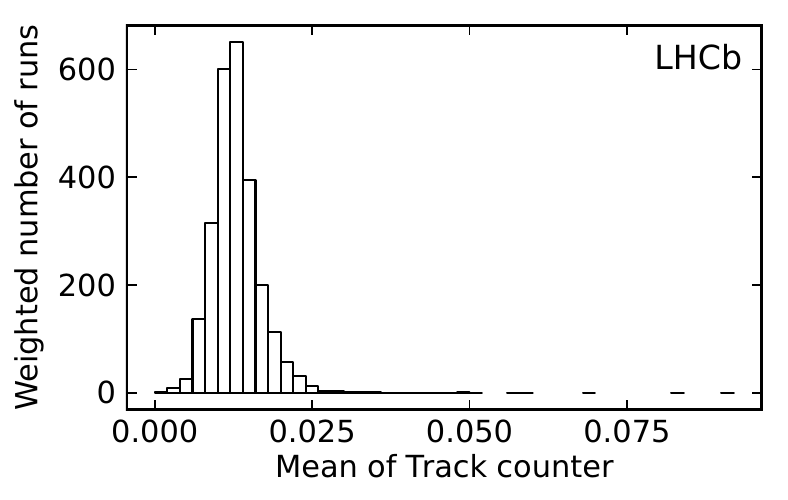}    
    \vspace*{-0.5cm}
  \end{center}
  \caption{\small
    Mean value of (left) \emph{ECalo} and (right) \emph{Track} observables in \bx{ee} crossings.
    Each histogram entry represents the average over a run\protect \footnotemark\ %
    in 2012 and is weighted with the 
    corresponding integrated luminosity.
    For measurements using the \emph{ECalo} and \emph{Calo} observables we discard the runs for 
    which the \emph{ECalo} pedestal mean is lower than $4.75$ (dashed vertical line).
  }
  \label{fig:rel_pedestals}
\end{figure}

The zero-count method is valid if an event is considered empty 
when the value of the observable is exactly zero.
However, if the observable is affected by noise such that its value is 
never zero, the threshold discriminating empty events has to be increased.
This is the case for the \emph{ECalo} and \emph{Calo} observables, used as a cross-check, for which a
positive threshold must be chosen.
The introduced bias depends on the noise distribution, the one-interaction 
spectrum and the average number of interactions per crossing.
While the latter was kept approximately constant during the 2012 data-taking period, the 
\emph{Calo} noise distribution was changing due to ageing of the hadron calorimeter.
In the second half of 2012, the HCAL gain was adjusted more frequently, thus 
keeping the noise distribution more stable.

\begin{figure}[tb]
  \begin{center}
    \includegraphics[width=0.98\linewidth]{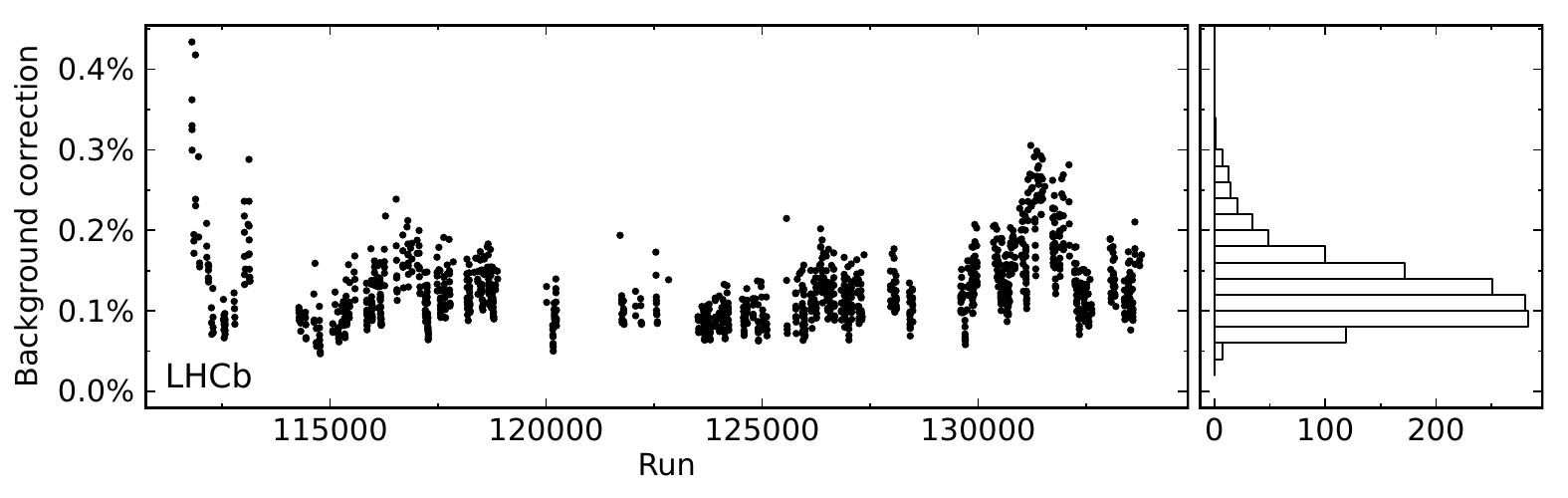}
    \vspace*{-0.5cm}
  \end{center}
  \caption{\small
    Beam-gas background fraction $\left(-\ln{\PZero[\bx{be}]} 
    -\ln{\PZero[\bx{eb}]}\right)/\mueff{\eff}$ for the \emph{Track} observable
    during the 2012 running period.
  }
  \label{fig:rel_bgr_frac}
\end{figure}

\footnotetext{
  A ``run'' is a consistent set of data,
  which usually spans about an hour of data taking, 
  mainly used as an administrative unit.}
The noise distribution is measured in \bx{ee} crossings.
Histograms of the mean value of the  noise are shown for the \emph{ECalo} and the \emph{Track} 
observable in Fig.~\ref{fig:rel_pedestals}.
For the \emph{ECalo} observable, two peaks are observed in the pedestal distribution.
This is attributed to a change of operating conditions, which is not easily
corrected for.
Therefore, for cross-checks using the \emph{ECalo} and \emph{Calo} observables we discard the 
runs for which the \emph{ECalo} pedestal mean is lower than $4.75$.
The remaining larger fraction of runs spans the full year and is subsequently 
used for assessment of systematic uncertainties.
The \emph{Track} observable has typically less than 2.5 tracks per 100 \bx{ee} 
crossings, which induces a negligible bias.
The systematic uncertainty due to noise is negligible.
Equation~\ref{eq:mu} assumes that the proton populations in the \bx{be}
and \bx{eb} crossings are the same as in the \bx{bb} crossings.
With a population spread of typically 10\% and a beam-gas background
fraction for the reference observable $< \, 0.3\%$ compared to the \PPP 
interactions (see Fig.~\ref{fig:rel_bgr_frac}) the effect of the spread is 
small and therefore neglected.

\begin{figure}[tb]
  \begin{center}
    \includegraphics[width=0.98\linewidth]{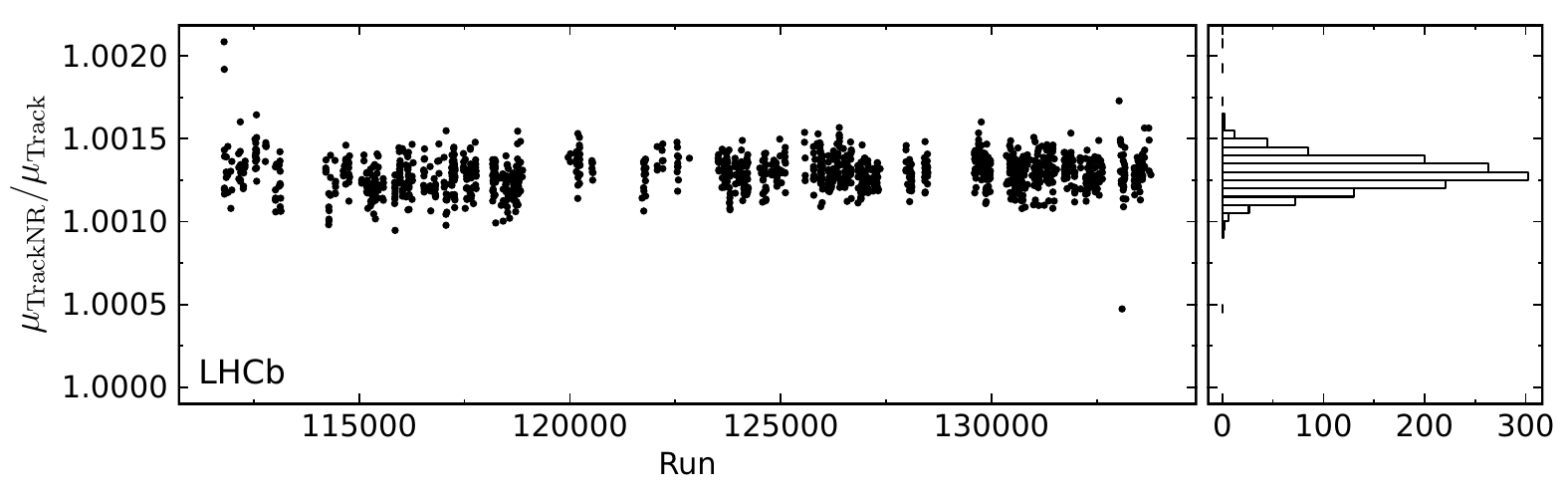}
    \vspace*{-0.5cm}
  \end{center}
  \caption{\small
    Ratio of the measured \mueff{\eff} values without (\emph{TrackNR}) and  with (\emph{Track})
    a fiducial volume cut during the 2012 running period.
    The observed deviation from unity is used as an estimate of the 
    systematic uncertainty due to potentially unaccounted background.
  }
  \label{fig:rel_ratio_unrestricted}
\end{figure}

The measured \mueff{\eff} values can be contaminated by 
other backgrounds than beam-gas interactions,
\eg collisions between satellite and main bunches, and interactions with material
in the \velo.
We reduce such effects by applying a fiducial volume cut to the \emph{Track} observable.
To assess the magnitude of potentially unaccounted background, a comparison 
is made between the \mueff{\eff} values measured with (\emph{Track}) and without (\emph{TrackNR}) the fiducial 
volume cut, see Fig.~\ref{fig:rel_ratio_unrestricted}.
The observed discrepancy is used as an estimate of the systematic uncertainty
due to beam-beam background.

\begin{figure}[tb]
  \begin{center}
    \includegraphics[width=0.98\linewidth]{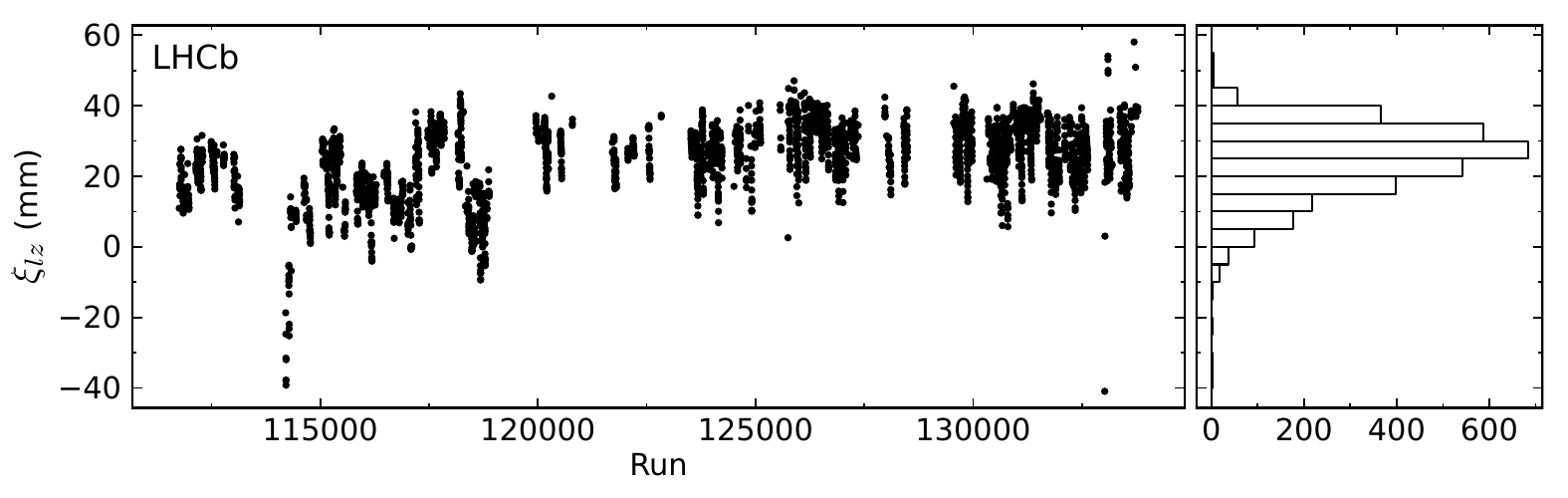}
    \includegraphics[width=0.98\linewidth]{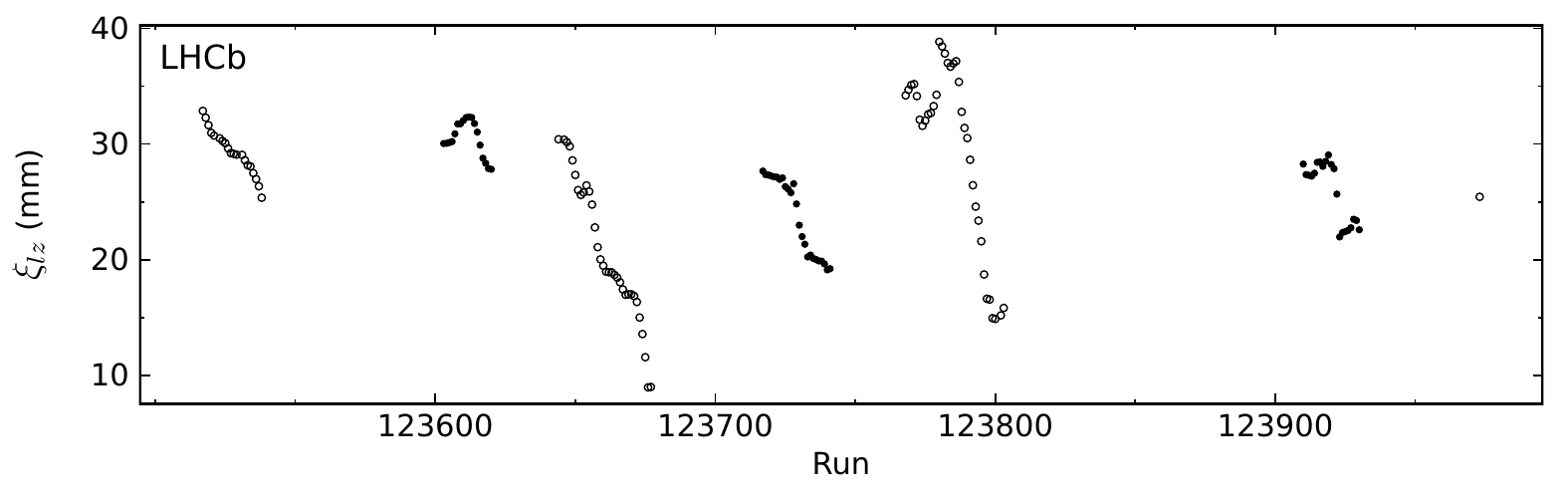}
    \vspace*{-0.5cm}
  \end{center}
  \caption{\small
    Longitudinal position of the luminous region $\xilz$ during the 2012 running period.
    The bottom plot shows a subset consisting of a few fills (distinguished 
    with alternating open and solid markers).
  }
  \label{fig:rel_zlr}
\end{figure}

The longitudinal position of the luminous region depends on the transverse beam 
separation in the crossing plane.
In 2012 the instantaneous luminosity at LHCb was kept approximately constant by
varying the separation of the two beams at the crossing point \cite{CERN-ACC-2013-0028} (so called ``luminosity levelling'').
Due to imperfections in the procedure, the beams were displaced in a direction
not exactly orthogonal to the crossing plane.
As a result, there was a significant variation in the 
longitudinal position of the luminous region $\xilz$,
as shown in Fig.~\ref{fig:rel_zlr}.
The reconstruction efficiency for tracks and vertices is not uniform along $z$ 
at the scale of the observed variations.
Therefore, a correction needs to be applied to the observed \mueff{\eff} values 
that are measured using VELO observables.
The \emph{Calo} observable is not affected.

\begin{figure}[tb]
  \begin{center}
    \includegraphics[width=0.49\linewidth]{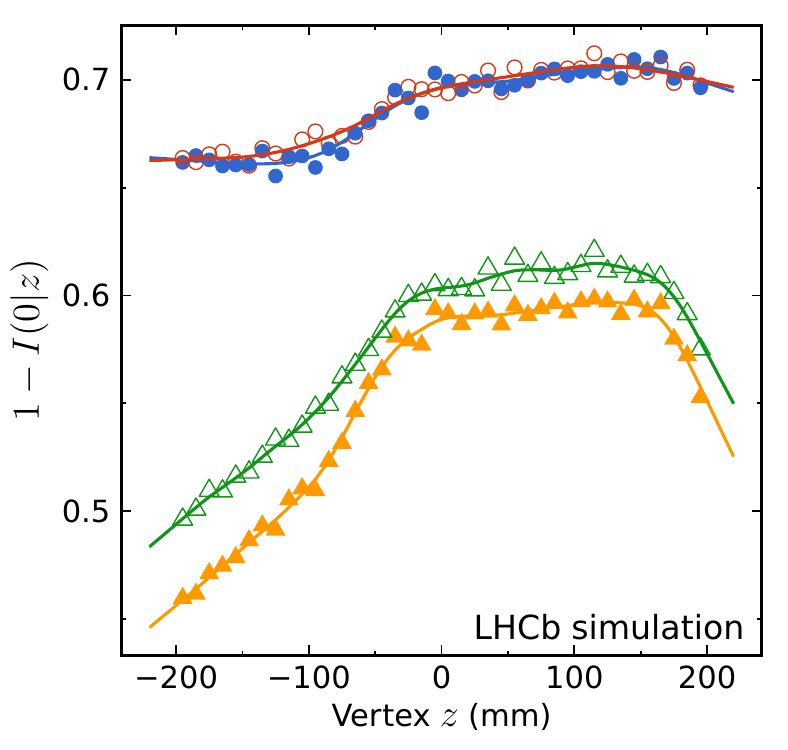}
    \includegraphics[width=0.49\linewidth]{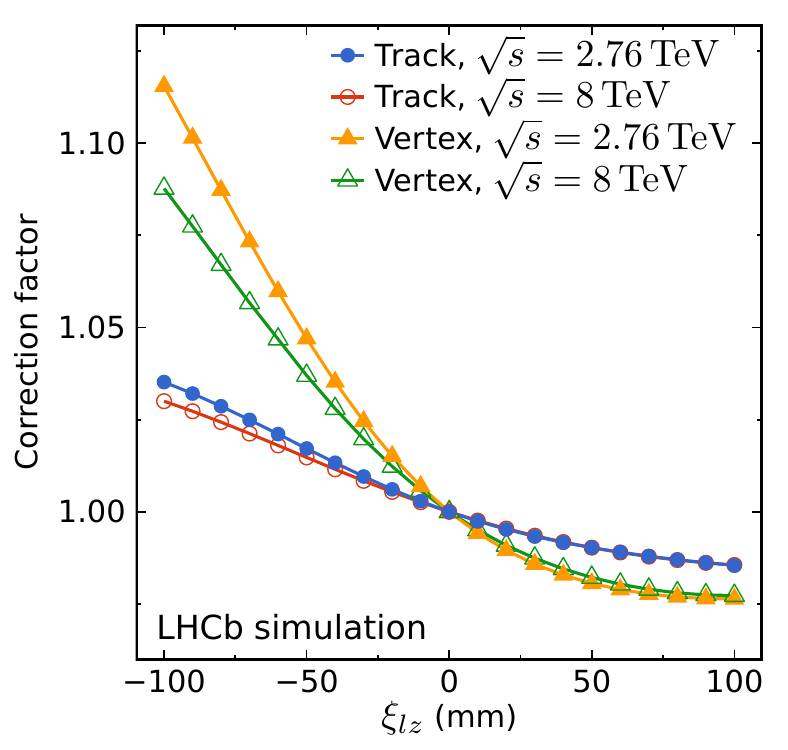}
    \vspace*{-0.5cm}
  \end{center}
  \caption{\small
    (Left) probability to see an event given an interaction at $z$ and (right) 
    correction factor as function of the longitudinal position $\xilz$ of the %
    luminous region with size $\siglz=50\mm$.
    Only data for one magnet polarity are shown as it is almost 
    identical to that for the other polarity.
  }
  \label{fig:sim_counter_efficiency}
\end{figure}

\def\fdistrib{f} %
From simulation we determine $I(0|z)$, the probability to obtain an empty event while
having one interaction at $z$,
\begin{equation}
I(0|z) \equiv P(\text{empty event}\,|\,\text{one interaction at } z) \, ,
\end{equation}
see Fig.~\ref{fig:sim_counter_efficiency}~(left).
Defining $\fdistrib(z)$ as the probability density of the longitudinal vertex 
distribution, the probability $\bar{I}(0|\fdistrib)$ to have an empty event while 
one interaction occurred is
\begin{equation}
\bar{I}(0|\fdistrib) = \int I(0|z)\, \fdistrib(z)\,dz \, .
\end{equation}
The absolute normalization of $I(0|z)$ and $\bar{I}(0|\fdistrib)$ depends on the underlying interaction 
generator. %
However, the normalization does not affect the luminosity measurement if used 
consistently. %
To avoid scaling \mueff{\eff} with factors largely different from unity, the 
correction is made with respect to a reference value $\bar{I}(0|\mathrm{ref})$
that corresponds to a Gaussian probability density $g(z)$ of the longitudinal
vertex distribution centred at zero and having an \rms of $\siglz = 50\mm$,
\begin{equation}
\bar{I}(0|\mathrm{ref}) = \int I(0|z) \, g(z) \, dz \, .
\end{equation}
The observed values, $\mueff[\mathrm{raw}]{\eff}$, are proportional to 
$1-\bar{I}(0|f)$. Thus, the corrected values are given by
\begin{equation}\label{eq:muvis_multiplicative_correction}
\mueff{\eff} =
  \frac{1 - \bar{I}(0|\mathrm{ref})}{1 - \bar{I}(0|\fdistrib)}
  \, \mueff[\mathrm{raw}]{\eff} \, .
\end{equation}
The $z$ distribution of the vertices is well approximated with a Gaussian function.
Examples of the correction 
factors for a Gaussian vertex distribution with $\siglz=50\mm$ are shown in
Fig.~\ref{fig:sim_counter_efficiency}~(right).

An efficiency correction is implemented as a factor, according to 
Eq.~\eqref{eq:muvis_multiplicative_correction},
evaluated as an average over about one-month running periods.
To take into account a possible inaccuracy of the efficiency obtained by simulation,
a comparison is made with an unaffected observable (\emph{Calo}).
The systematic uncertainty due to the residual dependence of the ratio 
$\mueff{Track}/\mueff{Calo}$ on the longitudinal position of the luminous region $\xilz$ is estimated as follows.
First, the data are divided in $5\mm$ bins in $\xilz$ and the median of the ratio in each bin is calculated.
The difference of the median with respect to that at $\xilz=0$ is assumed to be due to imperfect correction of the \emph{Track} observable.
The relative difference is then averaged over the full data set taking into account the luminosity content.
Finally, the resulting difference of 0.19\% is taken as the systematic uncertainty due to the imperfect correction for 
the efficiency of the observable.
\begin{figure}[tb]
  \begin{center}
    \includegraphics[width=0.49\linewidth]{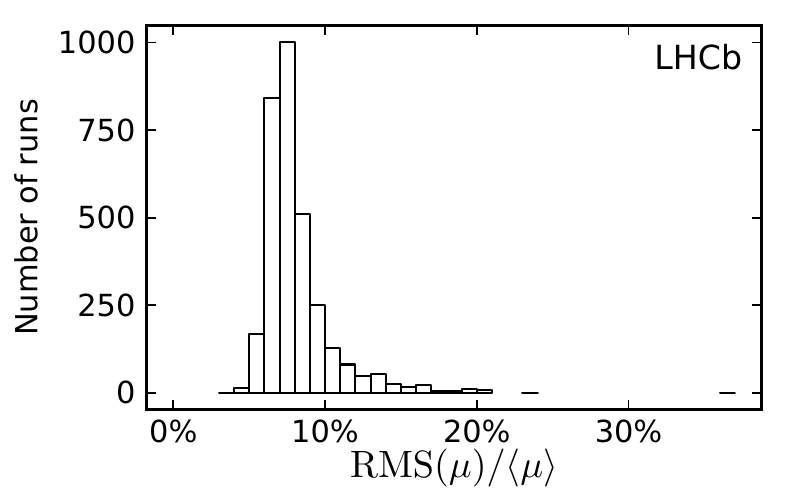}
    \vspace*{-0.5cm}
  \end{center}
  \caption{\small
    \rms of \mueff{\eff} across \bx{bb} bunch crossings relative to the mean 
    value for the 2012 running period.
  }
  \label{fig:rel_spread}
\end{figure}

The numbers of protons, beam sizes and transverse offsets at the
interaction point vary across bunches.
Thus, the \mueff{\eff} value varies across \bx{bb} crossings.
An estimate of the spread of \mueff{\eff} values is the \rms 
divided by the mean across bunch crossings, as shown in 
Fig.~\ref{fig:rel_spread}.
Due to the non-linearity of the logarithmic function, ideally one first needs to
compute \mueff{\eff} values for different bunch crossings and then 
take the average.
However, for short time intervals the number of samples is insufficient to make an 
unbiased measurement per bunch crossing using the zero-count method,
while \mueff{\eff} may not be constant when the intervals are too long due to 
\eg loss of bunch population and emittance growth.

During physics data-taking, the bandwidth reserved for luminosity triggers is limited.
Therefore, a statistically significant measurement of the luminosity cannot be obtained 
for each bunch crossing individually by integrating over periods shorter than about 30 minutes.
The bias and systematic uncertainty introduced by this limitation is evaluated with a 
simulation.
To reflect the luminosity integration for physics data, the 
number of visible interactions is counted in short time intervals ignoring the
spread of \mueff{\eff} values across bunch crossings.
A set of 30 consecutive short intervals is used to accumulate a sufficient number 
of events per BCID. %
Then, a correction for the spread is calculated and applied as described below.

For the following discussion backgrounds are not considered since their effect 
on the correction is negligible.
Let $n_{ti}$ and $k_{ti}$ denote the number of random triggers and the number 
of empty events, respectively, in \bx{bb} crossing slot $i$ and short time 
interval $t$.
Where the index $t$ is omitted, an implicit sum is assumed over the set $T$ of 
consecutive short intervals.
Similarly, in case the index $i$ is omitted, a summation over all \bx{bb} bunch 
crossings is assumed.
A correction factor is calculated for every set $T$ and is applied to each short period $t \in T$
\begin{equation}
\label{eq:rel_kappa}
\kappa_T = \frac{\langle -\ln \frac{k_i}{n_i} \rangle}
              {-\ln \frac{k}{n}} \, ,
\end{equation}
where the average in the numerator is taken over $i$.
The corrected estimate of the number of visible interactions is
\begin{equation}
\label{eq:rel_nvis}
N_{\eff,T} = \kappa_T f_\mathrm{trig}\sum_{t \in T}
\left( - n_t \ln\frac{k_t}{n_t} \right) \, ,
\end{equation}
where $f_\mathrm{trig}$ is the probability that a \bx{bb} crossing is randomly triggered.

\begin{figure}[tb]
  \begin{center}
    \includegraphics[width=0.49\linewidth]{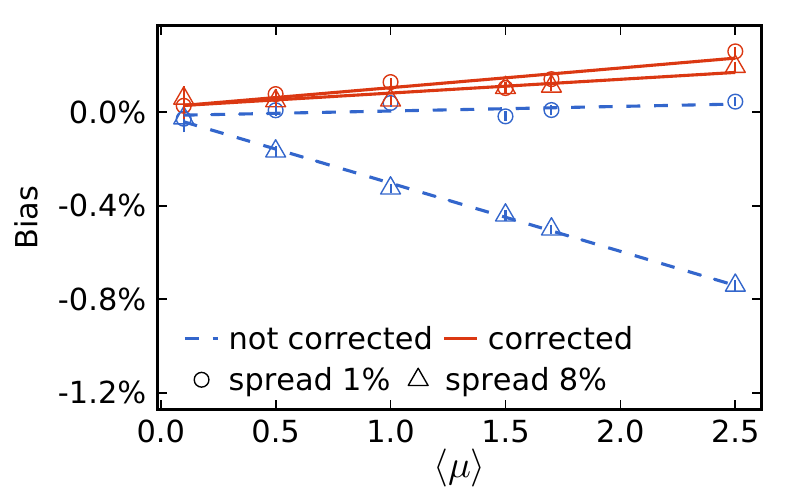}
    \includegraphics[width=0.49\linewidth]{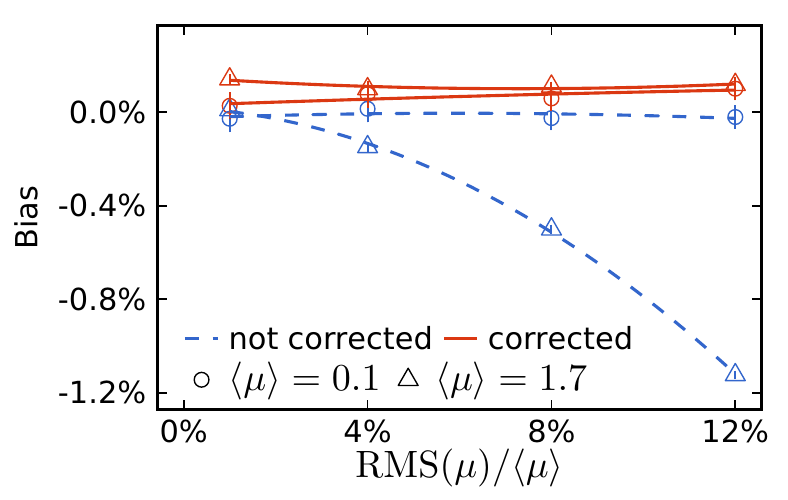}
    \vspace*{-0.5cm}
  \end{center}
  \caption{\small
    Bias of the estimated number of interactions as function of (left) mean 
    $\mu$ value and (right) relative \rms.
    The bias is fitted with a straight line and an even quadratic polynomial as function 
    of $\mean{\mu}$ and relative \rms, respectively.
    A quadratically increasing bias as function of relative \rms is present for 
    large $\mean{\mu}$ values before the correction (dashed blue line).
    For 2012 data taking conditions, the typical $\mean{\mu}$ for the \emph{Track} observable is 
    $1.7$ and the typical relative \rms is $8\%$.
  }
  \label{fig:rel_mc_bias}
\end{figure}
A simulation study is performed to compare the bias of the estimated 
number of interactions before and after the correction procedure.
The rate of triggers and the number of bunch crossings is chosen to reflect the 
typical running conditions.
The $\mu$ values across bunch crossings are sampled from a normal distribution.
A luminosity half-life of two hours is assumed. %
The bias is calculated as function of mean $\mu$ value and relative \rms, and is 
shown in Fig.~\ref{fig:rel_mc_bias}.

\begin{figure}[tb]
  \begin{center}
    \includegraphics[width=0.49\linewidth]{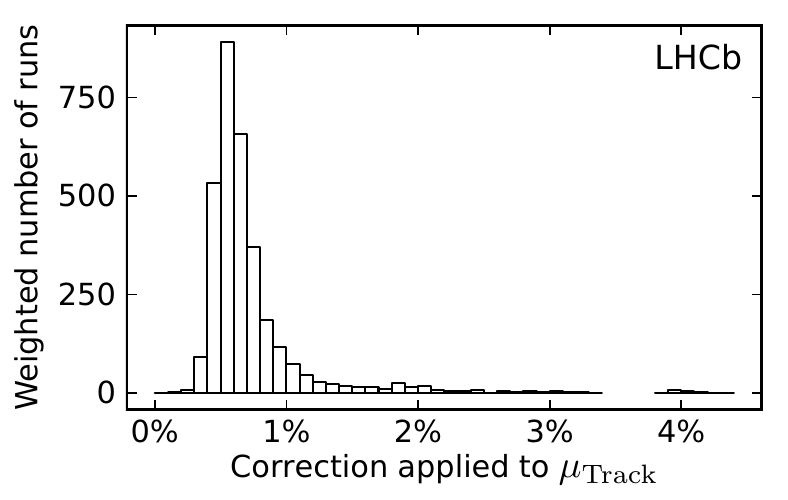}
    \includegraphics[width=0.49\linewidth]{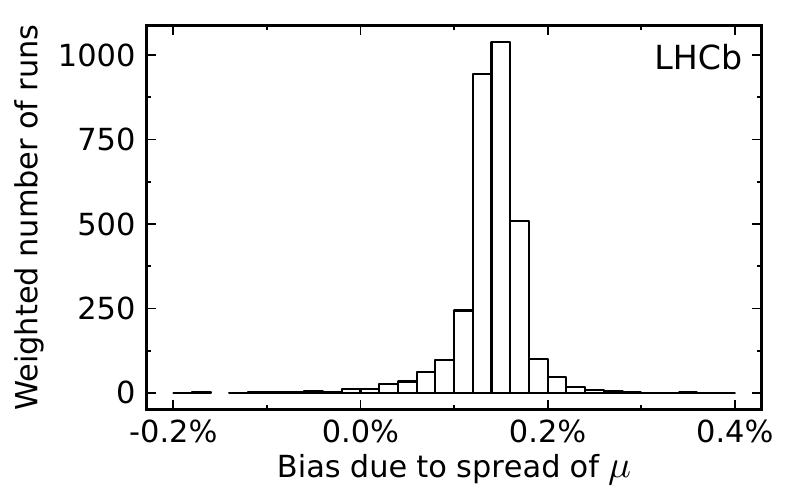}
    \vspace*{-0.5cm}
  \end{center}
  \caption{\small
    (Left) correction applied to the estimated values of $\mueff{Track}$
    for variations of its value across bunch crossings
    and (right) residual relative bias of the estimated number of visible 
    interactions after the correction.
    Each histogram entry represents a run in 2012 and is weighted with the 
    corresponding integrated luminosity.
  }
  \label{fig:rel_spread_cor}
\end{figure}

To estimate the residual bias of the correction technique on the data, we 
perform a simulation for each long period $T$.
First, the \mueff{\eff} value is estimated for each short period $t$ and each 
bunch crossing $i$ with
\begin{equation}
\mu_{ti} = - \frac{\ln{\frac{k_t}{n_t}} \, \ln{\frac{k_i}{n_i}}}{\ln \frac{k}{n}} \,  ,
\end{equation}
which has the desirable property that it coincides with the projection 
estimates ($\mu_{t}$ and $\mu_{i}$) when the true $\mu$ value does 
not change over time or across colliding bunches.
Then, for each $(t,i)$ pair, $k_{ti}$ is sampled from a binomial 
distribution with success probability $e^{-\mu_{i}}$ and number of trials 
equal to $n_{ti}$.
As for the actual data, Eqs.~\eqref{eq:rel_kappa} and~\eqref{eq:rel_nvis} are used 
to estimate the number of visible interactions.
Finally, the bias is obtained from the difference of the estimated and the true 
number of visible interactions, averaged over 25 independent repetitions of the 
simulation.
Histograms of the average values of $\kappa_T$ and the residual relative bias 
for each run
are shown in Fig.~\ref{fig:rel_spread_cor}.
The relative integrated bias over the full data set is assigned as a systematic 
uncertainty (0.14\%).

In addition, a cross-check is made using the \emph{Muon} observable, which is less sensitive to 
the spread owing to its low \mueff{\eff} values ranging from $0.07$ to $0.15$.\footnote{
  Due to a change of threshold mid-2012, the \emph{Muon} observable visible 
  cross-section changed significantly. Therefore, the periods before and after 
  the change are treated independently.
}
The ratio \mueff{Track}/\mueff{Muon} as a function of the relative \rms of 
\mueff{\eff} across bunch crossings is fitted with an even quadratic polynomial.
The $0.5\%$ of runs with extreme values of the ratio are excluded from the fit.
The maximum relative difference between the predicted value at the 
mean spread and at zero spread gives an estimate of the residual bias.
Since it has an opposite sign with respect to the residual bias obtained from 
the simulation, the result is taken as an additional systematic uncertainty 
(0.09\%).

\begin{figure}[tb]
  \begin{center}
    \includegraphics[width=0.98\linewidth]{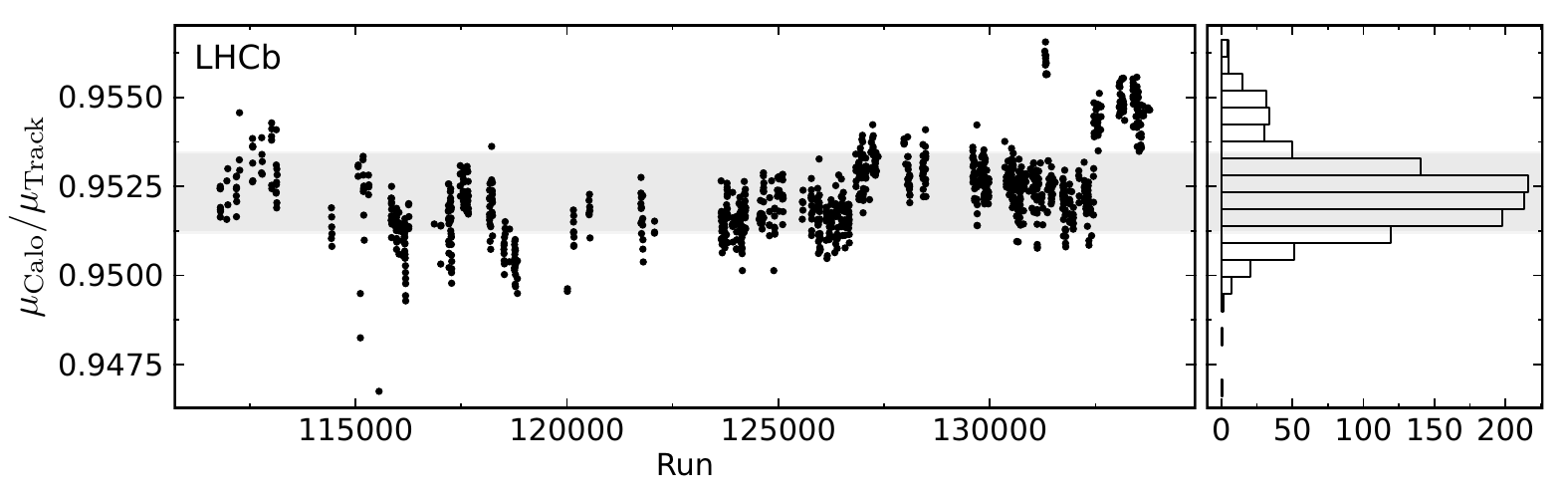}
    \vspace*{-0.5cm}
  \end{center}
  \caption{\small
    Ratio of the relative luminosities using the \emph{Track} and the \emph{Calo} observables 
    during the 2012 running period.
    Only data for runs that are longer than 30\unit{min} are plotted.
    The variation of the ratio after subtraction of the variation due to statistical fluctuations
    is shown with a shaded area spanning $\pm 1 \sigma$ around the mean.
  }
  \label{fig:rel_ratio_l0calo}
\end{figure}

The stability of the reference observable is demonstrated in 
Fig.~\ref{fig:rel_ratio_l0calo}, 
which shows the ratio of the relative luminosities determined with the 
zero-count method using the \emph{Track} and the \emph{Calo} observables.
These two observables use different sub-detectors and have different systematic uncertainties.
The variation of the ratio unexplained by statistical fluctuations is 
assigned as a systematic uncertainty to the relative luminosity 
measured using the \emph{Track} observable.
A similar cross-check with the ratio of the relative luminosities using the \emph{Track} and the \emph{Vertex} observables
shows negligible discrepancy.
\begin{table}[tb]
  \centering
  \caption{
    Top: systematic uncertainties of the relative luminosity measurement (in \%).
    Bottom: integrated effect of the applied corrections (in \%).
  }
  \begin{tabular}{lccccc}
     & \multicolumn{3}{c}{\pp} & \pA & \Ap \\
    Source & 8\tev & \ 7\tev & 2.76\tev & \Five\tev & \Five\tev \\
            
    \midrule
    Beam-beam background         & 0.13 & 0.24 & 0.13 & 0.95 & 0.73 \\
    Efficiency of the observable & 0.19 & 0.07 & 0.12 & 0.09 & 0.11 \\
    Bunch spread                 & 0.14 & 0.09 & 0.10 & 0.03 & 0.03 \\
    Bunch spread (cross-check)   & 0.09 & 0.44 &      &      &      \\
    Stability                    & 0.12 & 0.13 & 0.14 & 0.39 & 0.35 \\
    \midrule
    Total                        & 0.31 & 0.53 & 0.25 & 1.03 & 0.82 \\
    & & & & & \vspace*{-0.2cm} \\ 
    Correction & & & & & \\
    \midrule
    Efficiency of the observable & $-0.54\phantom{+}$ & $-0.11\phantom{+}$ & $-0.12\phantom{+}$ & $-0.09\phantom{+}$ & $-0.11\phantom{+}$ \\
    Bunch spread                 & $+0.72\phantom{+}$ & $+0.99\phantom{+}$ & $+0.10\phantom{+}$ & $+0.03\phantom{+}$ & $+0.03\phantom{+}$ \\
  \end{tabular}
  \label{tab:rel_syst}
\end{table}
The systematic uncertainties of the relative luminosity measurement are 
summarized in Table~\ref{tab:rel_syst}. By summing the effect of the
different sources in quadrature, we conclude that the relative luminosity 
measurement introduces a systematic uncertainty of 0.31\% for the \pp run at 8\tev.
The quoted uncertainty applies when the full dataset is used; for specific choices of partial datasets a different value may apply. 
In the case of the 2013 running conditions (proton-lead and \pp at 2.76\tev), the corrections due to 
the movement of the luminous region and the bunch spread are small.
Since the available data sample size is insufficient to reliably perform the corresponding cross-checks,
the full amount of each correction is assigned as an uncertainty.
The beam-beam background uncertainty is estimated to be up to 1\% for the proton-lead data taking, 
owing to the very low $\mu$ values (0.01--0.02) of these runs.
A higher uncertainty of about 0.5\% due to the bunch spread is estimated for the 2011 data taking.
This is explained by worse conditions in the beginning 
of the year, when the spread of $\mu$ across bunches reached 30\%, which leads to a correction of up to 7\% for some fills.
The 2011 data taking at 7\tev was affected by parasitic collisions due to a vanishing net crossing angle for one of the magnet polarity settings.
This background ranges between 0.2\% and 0.7\% and a correction is applied averaging over time intervals of a few weeks each, during which data were taken under similar conditions.
The average correction amounts to about 0.4\% and since only about half of the 2011 running period is affected, an uncertainty due to parasitic collisions of 0.2\% is assigned on the full period.
In addition, the estimated uncertainty due to beam-beam background from 2012 is added in quadrature to obtain 0.24\% uncertainty for 2011.
The stability of the effective process is estimated using only data that is not affected by parasitic collisions.

%% file: formalism.tex
\section{Formalism for the luminosity of colliding beams}
\label{sec:Luminosity formalism for colliding beams}

In a cyclical collider, such as the LHC, the average instantaneous
luminosity of one pair of colliding bunches can be expressed 
as~\cite{ref:moller}
\begin{equation}
 \label{eq:moller}
  L =  N_1 \, N_2 \, \frev \sqrt{|\Vec{v}_1-\Vec{v}_2|^2-\frac{|\Vec{v}_1\times \Vec{v}_2|^2}{c^2}}
       \, \displaystyle\int\!\rho_1(x,y,z,t) \, \rho_2(x,y,z,t) \, dx\, dy\, dz\, dt \, ,
\end{equation}
where we have introduced
the velocities $\Vec{v}_1$ and $\Vec{v}_2$ of the particles (in the 
approximation of zero emittance the velocities are the same within one bunch).
The particle densities $\rho_j(\Vec{r},t)$ ($j=1,2$) at position $\Vec{r}=(x,y,z)$ 
and time $t$ are normalized such that their 
individual integrals over all space are unity at all times. 
For highly relativistic beams colliding with a small
half crossing-angle \hca, the M{\o}ller factor 
$\sqrt{(\Vec{v}_1-\Vec{v}_2)^2-{(\Vec{v}_1\times \Vec{v}_2)^2}/{c^2}}$
reduces to $2c \cos^2{\hca} \simeq 2c$ and one recovers
Eqs.~\eqref{eq:luminosity} and \eqref{eq:overlapintegral}.
The LHCb system of coordinates, which is used here, is chosen as a right-handed cartesian 
coordinate system with its origin at the nominal interaction point IP8.
The $z$ axis points towards the LHCb dipole magnet 
along the nominal average beam-line, 
the $x$ axis lies in the horizontal plane, with $x>0$ pointing approximately toward the centre
of the LHC ring, and the $y$ axis completes the right-handed system. %
This system almost coincides with the LHC coordinate system.
Small angles due to the known LHC plane inclination and other magnetic lattice imperfections
have negligible influence on the measurement of the overlap integral as only the 
crossing angles are relevant, not the individual beam directions.

Up to a normalization factor, 
$\rho_{bb}(x,y,z,t) = \rho_1(x,y,z,t) \, \rho_2(x,y,z,t)$ is 
the distribution of interactions from the luminous region in the laboratory frame.
If both $\rho_1$ and $\rho_2$ factorize as a product of a longitudinal and a transverse density
(relative to the direction of motion of the bunch), the spatial distribution integrated over time\footnote{%
  When the time dependence is dropped, an integration over time is implied: $\rho(x,y,z) = \int\rho(x,y,z,t) dt \,$.} 
can be expressed as 
\begin{equation}\label{eq:bb_density}
  \rho_{bb}(x,y,z) = n(z) \, \rho_1(x,y,z)\, \rho_2(x,y,z) \, 
\end{equation}
where $n(z)$ is a shape factor which depends on $z$ only. 
This relation between the distributions of beam-beam and beam-gas interactions is
used in the BGI analysis.

Determining the luminosity or the reference cross-section %
requires measuring the bunch population products $N_1\,N_2$, 
as discussed in \sect~\ref{sec:Bunch current normalization},
and evaluating the overlap integral $\ovint$. 
We briefly describe the principles of the two methods that are used in this paper to determine the latter.

\subsection{Beam overlap measurement methods}
\label{subsec:Beam overlap measurement methods}

The first method was introduced by van der Meer to measure the 
luminosity of the coasting beams at the Intersecting Storage Rings (ISR) \cite{vanderMeer:296752}.
The method was further extended to measure the luminosity of a collider with bunched beams \cite{Rubbia:1025746} 
and is the main method used to determine the luminosity at the other LHC experiments. 
The key principle of the VDM scan method is to express the overlap integral in terms of rates that 
are experimental observables as opposed to measuring the bunch density functions.
Experimentally, the method consists in moving the beams across each other in two orthogonal directions.
The overlap integral can be inferred from the rates measured at different beam separations,
provided the beam displacements are calibrated as absolute distances.

A reaction rate $\rate$ per bunch crossing is measured that is proportional to the luminosity and depends
on the two orthogonal transverse separations of the two beams $\Dx$ and $\Dy$.
Measuring this rate relative to the revolution frequency $\frev$ (approximately 11245~Hz at the LHC)
defines the parameter $\mu$, which is the average number of reactions per bunch crossing.
In the case where the spatial distributions of the beams can be factorized in the two coordinates $x$ and $y$,
it is sufficient to measure $\mu$ (and thus $\rate$) as a function of $\Dx$  (at a fixed $\Dyz$)
and as a function of $\Dy$  (at a fixed $\Dxz$).
One can show that the interaction cross-section is then given by
\begin{equation}
\label{eq:vdm_formula}
\xsec =
   \frac{       \int\!\mu(\Dx,\Dyz)\,d\Dx 
          \cdot \int\!\mu(\Dxz,\Dy)\,d\Dy
        }{ N_1\,N_2 \, \mu(\Dxz,\Dyz) } \, .
\end{equation}
The pair of separation values ($\Dxz$,$\Dyz$) is called the working point and is typically chosen 
to be as close as possible to the point where the luminosity is at its maximum. %
However, Eq.~\eqref{eq:vdm_formula} is valid for any values of $\Dxz$ and $\Dyz$.
It can be shown that it is also valid in the presence of non-zero crossing angles \cite{2011NIMPA.654..634B}.

The VDM method has the advantage of using a measured rate as its only observable, which is experimentally simple.
The experimental difficulties of the VDM method arise mostly from the fact that the beams must be moved 
to perform the measurement.
The exact displacements $\Dx$ and $\Dy$ in Eq.~\eqref{eq:vdm_formula} steered by the LHC magnets 
are calibrated at each interaction point in a so-called length scale calibration (LSC). %
While the resulting corrections are typically of the order of 1\%, some non-reproducibilities 
have been observed between two consecutive scans without being able to identify the cause.
Another difficulty originates from beam-beam effects.
When the beams are displaced, a change in $\beta^*$ (dynamic beta effect) and a beam deflection may be produced, 
which both influence the observed rate. %
The resulting corrections to the visible cross-section depend on the LHC optics, the beam parameters 
and filling scheme, and must be evaluated at each interaction point (see \sect~\ref{sec:VDM:Beam-beam effects}). %

In addition, when performed with one vertical and one horizontal scan, 
the VDM method is valid only under
the assumption that the distributions along the transverse variables $x$ and $y$ are independent, 
\ie that the $x$ ($y$) shape measured at a working point $\Dyz$ ($\Dxz$) does not depend on
the working point position.
As will be shown in the analysis described here, this assumption is not valid at the required precision.

An alternative  to the  VDM scan method for measuring the luminosity is provided by
the BGI method~\cite{FerroLuzzi:2005em}, which was first applied at the LHCb experiment 
in 2009~\cite{LHCb-PAPER-2010-001} and 2010~\cite{LHCb-PAPER-2011-015}.
The principle of this method is to evaluate the overlap integral by measuring all required observables
in Eq.~\eqref{eq:overlapintegral} using the spatial distribution of beam-gas and beam-beam 
interaction vertices.
The details of the measurement are discussed in \sect~\ref{sec:Beam-gas imaging method}.
Measuring the shapes of stationary beams avoids changes due to beam-beam effects and other, non
reproducible, effects due to beam steering.
Furthermore, at the LHC the BGI measurements at a given IP (here at LHCb in IP8) can be made in parallalel
to the VDM scans of other LHC experiments and can therefore be made more frequently.

On the other hand, while the $\beta^*$ and crossing angles used at the LHC do not impact the VDM method
to first order, the BGI measurement relies on the vertex measurement to determine the bunch shape.
Therefore, an increased $\beta^*$ is preferable to avoid limitations introduced by the detector resolution.
At LHCb, in 2012, \PPP physics data were acquired at $\beta^*=3\m$, 
while the most precise BGI luminosity calibrations fills were carried out with $\beta^*=10\m$.
The knowledge of the crossing angle is also important since the luminosity reduction due to the crossing 
angle has been as large as 20\%.
A non-vanishing crossing angle is necessary to avoid interactions between the main bunch and out-of-time 
charges captured in the next RF bucket, which occur near $z=\pm37.5\cm$. %
Such displaced collisions, if present, must be disentangled from beam-gas interactions.  
They can be completely avoided by introducing a sufficiently large crossing angle.
The VDM measurement can exclude interactions occurring away from the interaction point 
and is therefore less affected by these satellite collisions.

The VDM and BGI methods are complementary, in the sense that %
their systematic uncertainties on the overlap integral are highly uncorrelated, and a luminosity calibration 
performed with both methods in the same fill permits their systematic uncertainties to be constrained further.
At present this can only be done at the LHCb experiment.

The analyses of the VDM and BGI luminosity calibration measurements presented here indicate that
the observed luminosity profiles and vertex distributions are not consistent with Gaussian bunch distributions.
It is found that a sum of two Gaussian functions (``double Gaussian'' shape model) is sufficient
to describe the $x$ and $y$ shapes of each bunch as well as the resulting luminous region.
However, the joint two-dimensional transverse distribution of the bunches is found to be non-factorizable
in the transverse coordinates.
Therefore, as explained in \sect~\ref{sec:double gaussian model}, the transverse shape of the bunches 
is modelled with a sum of four two-dimensional Gaussian functions, which is in general non-factorizable.

In order to explain the full analysis of the present work, which 
involves a  detailed fit model with a sum of Gaussian terms, it is useful to 
consider first the formalism for the ideal case of pure Gaussian beams and then
describe the two-dimensional (non-factorizable) Gaussian model used in this work.

\subsection{Luminosity in the case of purely Gaussian beams}
\label{sec:Luminosity in the case of pure Gaussian beams}

The overlap integral in Eq.~\eqref{eq:moller} can be calculated analytically when the single beam distributions
$\rho_j$ ($j=1,2$) are the product of three Gaussian functions, each one depending on a single spatial coordinate $m=\hx_j,~\hy_j$, or $\hz_j$.
The beam reference frames $\hat{x_j},\hat{y_j},\hat{z_j}$ are right-handed systems and
the longitudinal axis $\hz_j$ is assumed to be parallel to the velocity vector of the bunch $\Vec{v}_j$.
It is also assumed that the $\hy_j$ axes of the two colliding bunches are parallel to the $y$ axis of the laboratory frame.
The beam crossing plane, defined by the velocity vectors $\Vec{v}_1$ and $\Vec{v}_2$, is here assumed to coincide
with the $xz$ plane. This condition was not respected only for the April 2012 fills.
The relevant modifications of the formulae below are discussed 
in Sec.~\ref{sec:Overlap integral model}.
We assume the bunches are centred at $\Vec{r}_j=(\xi_{xj},\xi_{yj},\xi_{zj})$ at time $t=0$, 
with a particle density function described by a normalized Gaussian function
\begin{equation}
\label{eq:bunch_density}
\rho_{mj}(m)=\frac{1}{\sqrt{2\,\pi }\,\sigma_{mj}}\,
   {e}^{-\frac{1}{2}{\left( \frac{m-\xi_{mj}}{\sigma_{mj}}\right) }^{2}}
   \quad\text{for beam }j=1,2\;\text{and coordinate }m=\hx,\hy,\hz,
\end{equation}
where $\sigma_{mj}$ denotes the \rms of the corresponding Gaussian function. 

Assuming that $\rho_j(\Vec{r},t) = \rho_j(\Vec{r} - \Vec{v}_jt,0)$,
one can show that
the overlap integral becomes
\begin{equation}
\label{eq:ovrlap_compact}
\ovint = \frac{e^{-\frac{\Dx^2}{2\Sigma_x^2}-\frac{\Dy^2}{2\Sigma_y^2}}}{2\pi\,\Sigma_x\,\Sigma_y} \, ,
\end{equation}
where the following quantities have been introduced
\begin{equation}\label{eq:lumi_variable}
\begin{array}{lll}
    \Sigma^2_x = 2\sigma^2_z\sin^2\!\hca + 2\sigma^2_x\cos^2\!\hca &\quad\quad\text{with}\quad\quad& 2\sigma^2_x = \sigma^2_{\hx1} + \sigma^2_{\hx2}\\[2mm]
    \Sigma^2_y = 2\sigma^2_y && 2\sigma^2_y = \sigma^2_{\hy1} + \sigma^2_{\hy2}\\[2mm]
    &&2\sigma^2_z = \sigma^2_{\hz1} + \sigma^2_{\hz2} \, 
\end{array}
\end{equation}
and $\Dm=\xi_{m1}-\xi_{m2}$ (with $m=x,\,y$) are the transverse beam separations evaluated
at the moment $t=0$ when the colliding bunches are at the same $z$ position.
In the LHCb experiment, this $z$ position (called $\zrf$) is defined by the LHC RF timing and needs not coincide 
with the location $z=0$ 
of the LHCb laboratory frame nor with the geometrical crossing point of the two beam trajectories.

The longitudinal position $\xi_{lz}$ of the luminous region is related to the beam separation $\Dx$ 
and longitudinal bunch crossing point $\zrf$ with %
\begin{equation}
 \label{eq:zrf_muzl_deltazl}
  \xi_{lz} - \zrf = \frac{\sina\cosa \, (\sigma_x^2 - \sigma_z^2)}{ \Sigma_x^2 } \, \Dx \, .
\end{equation}
The index $l$ indicates here a property of the luminous region, as opposed to a single beam property. 

One can also show that the longitudinal size $\sigma_{lz}$ of the luminous region is related to the 
convolved bunch length $\sigma_z$ by %
\begin{equation}
\label{eq:sigma_zl}
  \frac{1}{\sigma_{lz}^2}= \frac{2 \ssqa}{\sigma_x^2}+\frac{2 \csqa}{\sigma_z^2} \, .
\end{equation}
Therefore, if one has a measurement of the transverse bunch size of the individual beams,
of the crossing angle and of the longitudinal size of the luminous region, one can evaluate the 
longitudinal bunch convolution.

\subsection{Double Gaussian shape model}
\label{sec:double gaussian model}

A factorizable transverse beam distribution with double Gaussian projections has the density
\begin{align}\label{eq:factorizable_density}
\rho(\hx,\hy) &= \rho(\hx)\,\rho(\hy) = \prod_{m=\hx,\hy}
  \left[ w_m   \, g(m;\xi_m,\sigma_{m\rmn}) +
        (1-w_m)\, g(m;\xi_m,\sigma_{m\rmw}) \right] \notag\\
&= \sum_{i_x i_y} w_{i_x i_y}\, g(\hx;\xi_{\hx},\sigma_{\hx i_x})\, g(\hy;\xi_{\hy},\sigma_{\hy i_y}) \ ,
\end{align}
where $g(m;\mu,\sigma)$ indicates a normalized Gaussian function of the variable $m$ with parameters $\mu$ and $\sigma$.
By convention, the narrow (\rmn) and wide (\rmw) components in each projection
have widths $\sigma_{m\rmn}$ and $\sigma_{m\rmw}$, and weights $w_m$ and $1-w_m$.
The weights $w_{i_x i_y}$ in the sum representation are defined as 
\begin{equation}\label{eq:factorizable_weights}
w_{i_xi_y} =
 \begin{bmatrix}
   w_{\rmn\rmn} & w_{\rmn\rmw} \\ w_{\rmw\rmn} & w_{\rmw\rmw}
 \end{bmatrix} =
 \begin{bmatrix*}[r]
   w_x w_y & w_x (1-w_y) \\ (1-w_x) w_y & (1-w_x)(1-w_y)
 \end{bmatrix*} \ .
\end{equation}
The wide and narrow components are assumed to have the same mean,
as supported by the data.
Moreover, it is assumed that the 3-dimensional bunch distribution factorizes in a transverse 
($\rho(\hx,\hy)$) and a longitudinal component, where the latter is modelled with a Gaussian function.

Non-factorizability can be introduced into the model in Eq.~\eqref{eq:factorizable_density} by modifying 
the weights $w_{i_xi_y}$ from Eq.~\eqref{eq:factorizable_weights}.
For instance, in an extreme case, one can have $w_{\rmn\rmw}=w_{\rmw\rmn}=0$ and $w_{\rmn\rmn}+w_{\rmw\rmw}=1$,
which corresponds to a sum of two 2-dimensional Gaussian functions.
To allow for a gradual transition between this extreme case and the case of factorizable beams, 
it is useful to define the weights as a linear combination
\begin{equation}\label{eq:non_factorizable_weights}
\begin{bmatrix}
w_{\rmn\rmn} & w_{\rmn\rmw} \\ w_{\rmw\rmn} & w_{\rmw\rmw}
\end{bmatrix} =
f \begin{bmatrix*}[r]
w_x w_y & w_x (1-w_y) \\ (1-w_x) w_y & (1-w_x)(1-w_y)
\end{bmatrix*} +
(1-f) \begin{bmatrix}
\frac{w_x+w_y}{2} & 0 \\ 0 & 1-\frac{w_x+w_y}{2}
\end{bmatrix} \ ,
\end{equation}
where the coefficient $f$ parametrizes the factorizability.
In the fully non-factorizable case ($f=0$) there is no distinction between the $x$ and $y$ weights, thus 
the parameters $w_x$ and $w_y$ are (arbitrarily) combined in a single weight.

As a result of the single beam model from Eq.~\eqref{eq:factorizable_density}, the shape of the luminous region and the overlap integral are described by a weighted sum of 16 components.
Explicitly, the beam overlap integral %
is given by
\begin{equation}\label{eq:bb_overlap_general}
\ovint = \sum_I w_I \, \ovint_I = \sum_I w_{i_x i_y,1} \, w_{j_x j_y,2} \, \ovint_I \ ,
\end{equation}
where $I$ denotes the set of indices $i_x,i_y,j_x,j_y$,
while  $w_{i_x i_y,1}$ and $w_{j_x j_y,2}$ are the weights from Eq.~\eqref{eq:non_factorizable_weights}
for \beamone and \beamtwo.
Each partial overlap integral $\ovint_I$ is evaluated with Eq.~\eqref{eq:ovrlap_compact}.

%% file: bgi.tex
\section{Beam-gas imaging method}
\label{sec:Beam-gas imaging method}
In this section, the BGI methodology and calibration results are presented in detail.
A description of the data taking conditions (trigger settings, vacuum conditions) and of the event selection are given.
Studies of the vertex position resolution and the unfolding method are presented.
The resolution is determined from data, separately for beam-beam and beam-gas
interactions.
An analysis of the resolution-corrected vertex position distribution is then performed, which uses
both beam-gas vertices and beam-beam vertices to perform a global fit of the
beam parameters (angles, luminous region length, longitudinal crossing position,
transverse beam shapes). 
A double Gaussian model is used which also allows for a non-factorizability of the
$x$ and $y$ distributions.
Simulation is used to verify the soundness of the fit procedure.
Several checks are made, based on data, to quantify systematic uncertainties.
The list of dedicated luminosity calibration fills discussed in this paper for
2011, 2012 and 2013 can be found in Table~\ref{tab:fills}. 
We focus on the 8\tev \PPP data set taken in 2012, because
it gives the most precise results.
For the other data sets the analysis is similar and only the differences with the former are discussed 
in \sect~\ref{sec:resolution}.

\subsection{Data-taking conditions and event selection}
\label{sec:data_taking}

For dedicated luminosity calibrations the LHC is filled with a low number of bunches,
of the order of 50 per beam or less, and a large gap between bunches ($\sim 1\mus$) is maintained. 
Under these conditions the vacuum pressure at the interaction point is  $\sim 10^{-9}$~mbar, 
producing a beam-gas trigger rate of about 0.5\Hz per $10^{11}$~protons.\footnote{During fills dedicated 
to luminosity the bunch population ranges between $0.7\cdot10^{11}$ and $1.1\cdot10^{11}$.}
Performing a BGI measurement with such low rates necessitates integration of a measurement 
over a period of up to 8 hours.
Significant limitations in the precision are caused by the low event rate, 
beam drift and by emittance growth over the integration time.
In order to mitigate this, in 2011 the VELO vacuum pumps located close to the interaction point were switched off,
thus increasing the beam-gas rate by about a factor of four.
To increase the rate further and to take full advantage of the BGI capabilities, the use of a gas injection system 
was proposed~\cite{FerroLuzzi:2005em}, developed and commissioned in the LHCb experiment~\cite{Colin-thesis}.

\begin{figure}[tbp]
\centering
\begin{minipage}[c]{0.5\textwidth}
\includegraphics[width=\textwidth]{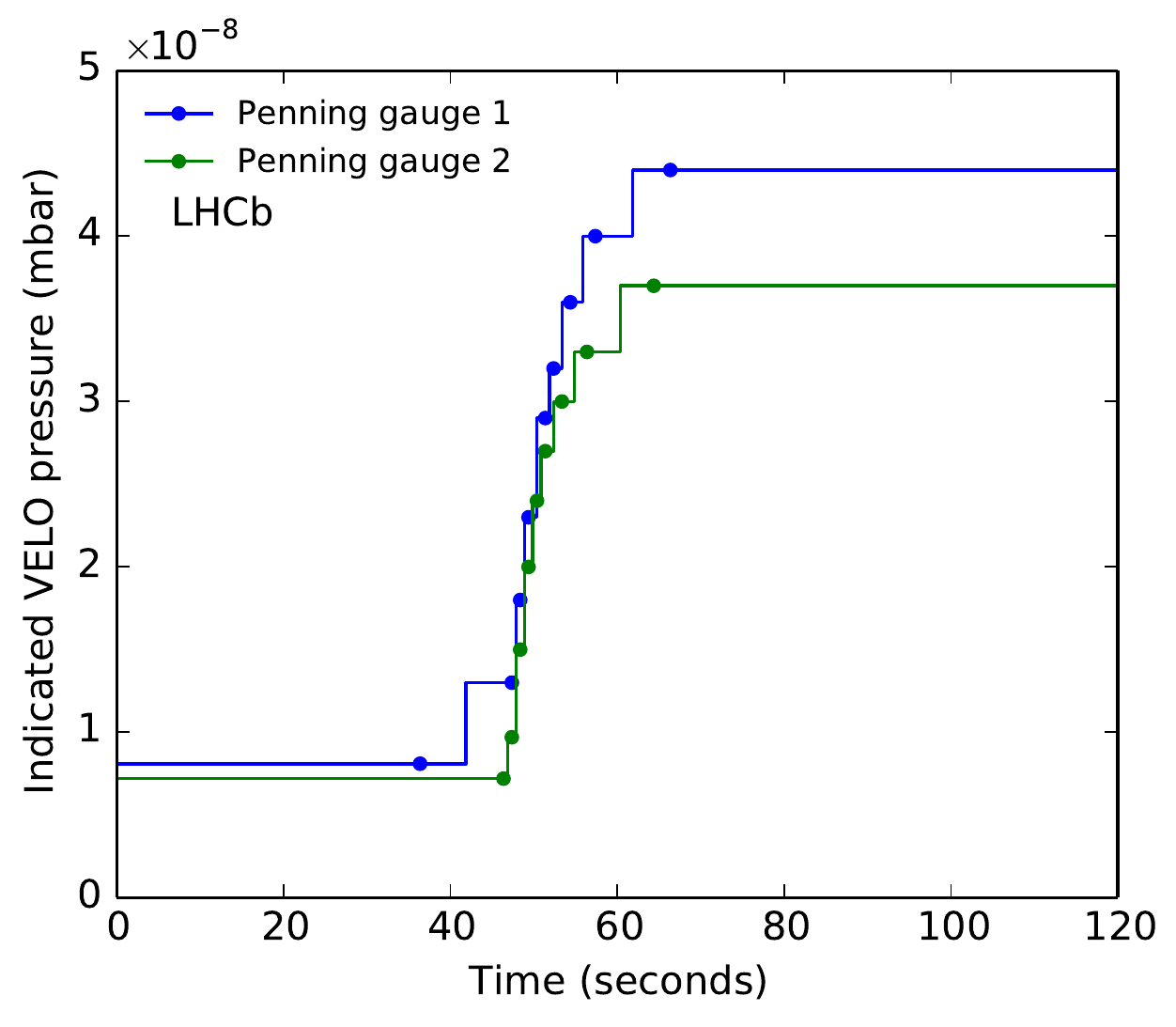}
\end{minipage}\hfill
\begin{minipage}[c]{0.5\textwidth}
\includegraphics[width=\textwidth]{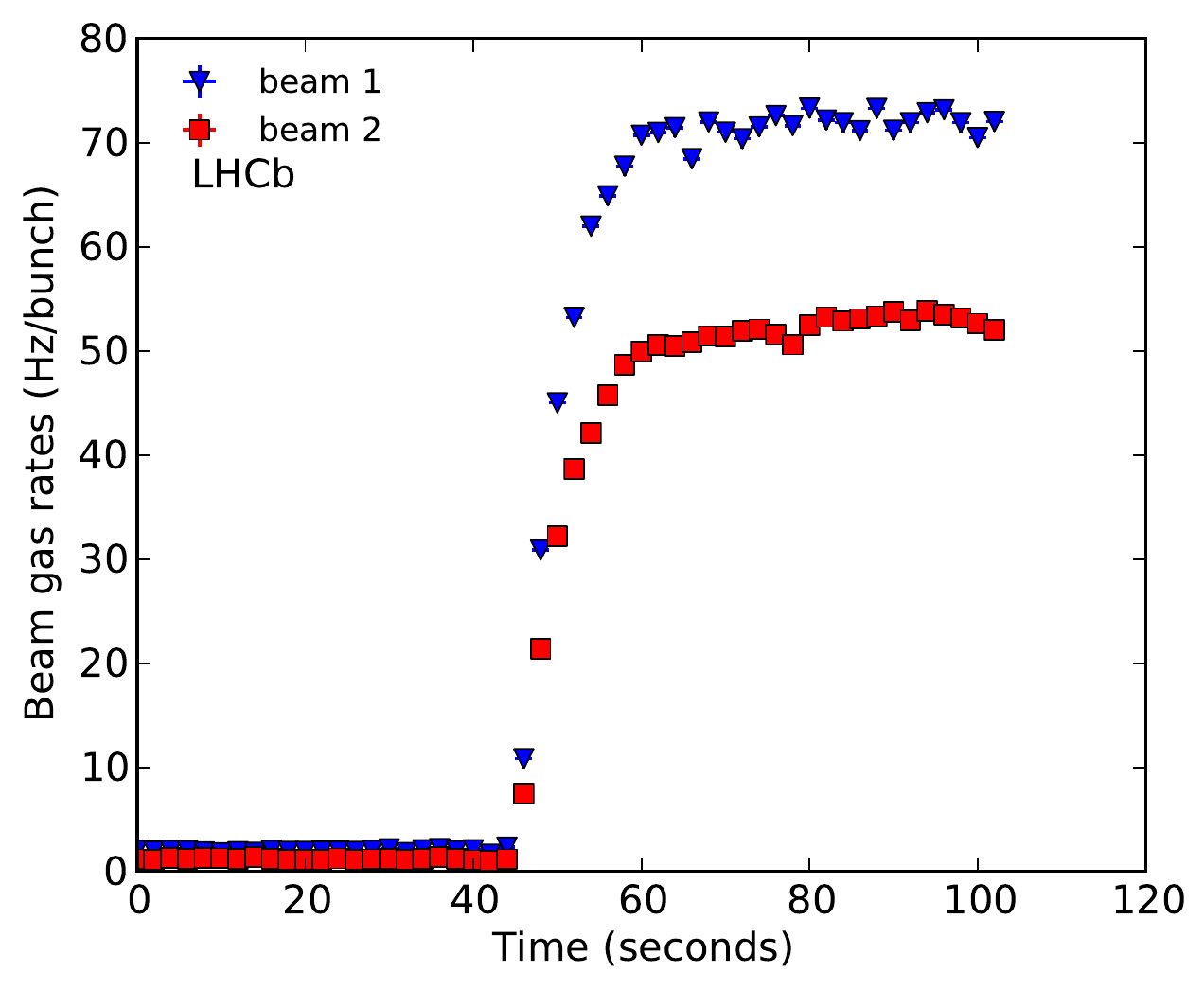}
\end{minipage}\hfill
\begin{minipage}[c]{1\linewidth}
\caption[Beam-gas event rate increase with gas injection]{\small Effect of the  gas injection (fill 2520), with 
(left) the pressure increase in the VELO and (right) the trigger rate increase. 
The VELO ion pumps are off.
The pressure is measured with Penning gauges in the vacuum vessel within $50\cm$ of the interaction point. 
The indicated value when neon is injected is to be multiplied by approximately a factor four to account for the lesser gauge sensitivity to neon gas.
The absolute pressure increase cannot be accurately determined as the gas composition in unknown before the
start of injection.
The beam-gas rate for triggered events increases from about 2 to 72 Hz/bunch for {\beamone} 
and 1.3 to 51 Hz/bunch for {\beamtwo}.}
\label{smog_rate_incr}
\end{minipage}
\end{figure}

A first gas injection test with circulating beams was performed in November 2011.
When activating the %
system, neon gas is injected in the VELO, thus raising the pressure from about $10^{-9}\unit{mbar}$
to slightly above $10^{-7}\unit{mbar}$.
Once the injection is stopped, the nominal pressure of $\sim 10^{-9}\unit{mbar}$ is recovered within 20 minutes.
The pumps are switched off during the gas injection.
The effect of %
gas injection on the pressure and beam-gas interaction rate is shown in Fig.~\ref{smog_rate_incr}.
For \PPP collisions at $\sqrt{s}=8\tev$, 
the recorded beam-gas event rates per $10^{11}$~protons were $98\Hz$ for {\beamone} and $82\Hz$
for {\beamtwo}.
The corresponding rates of the hardware trigger 
were about $2.1\kHz$ and $1.3\kHz$ for \beamone and \beamtwo,  respectively.

All luminosity calibrations acquired in 2012 and 2013 took advantage of a simplified activity trigger 
in all bunch crossing types.
For these fills the hardware trigger used information from the SPD and PU sensors and from the calorimeters.
The software trigger dedicated to the BGI measurement accepted events with a multiplicity larger than ten tracks.
The vertex position $z$ was required to lie within a range of $-2000<z<400$\mm for \beamone-gas events
and $0<z<2000$\mm \beamtwo-gas events.
Of the triggered events in \bx{bb} crossings with a vertex in the range $-300<z<300$\mm 
only a fixed fraction was accepted (``prescaled'') to keep their total rate below 15\kHz,
close to the maximum rate that can be recorded.
Interactions with a transverse vertex position more than 4\mm from the nominal beam line were not accepted in order to reject interactions with the material of the RF foil of the VELO \cite{LHCb-DP-2014-001}. 

\includefig{0.8}{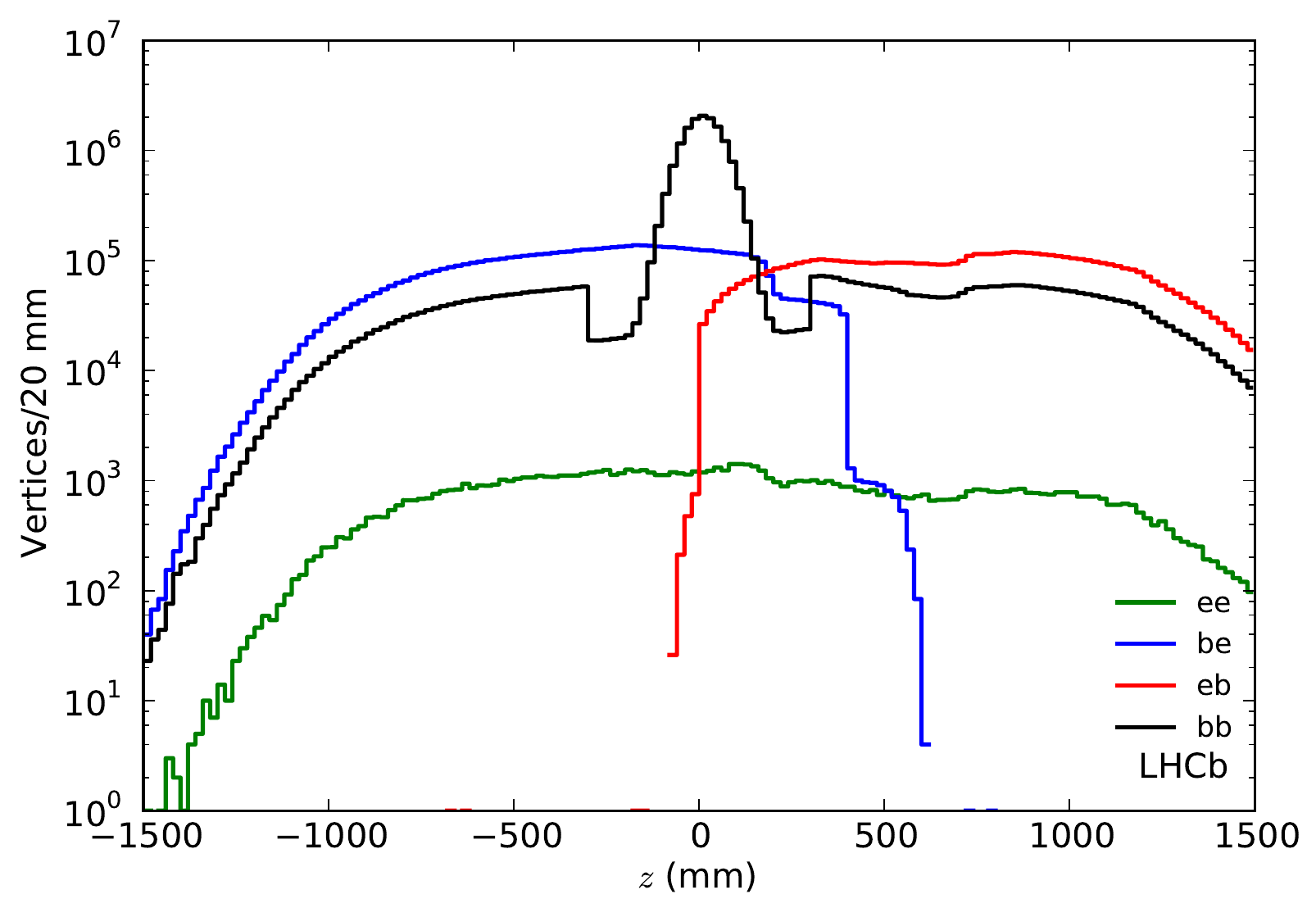}{\small Distribution of longitudinal $z$ position 
vertices for the various bunch crossing types}{Longitudinal distribution of vertices 
for the various bunch crossing types acquired in 40 minutes during fill 2852.
Crossing types \bx{ee} (green), \bx{be} (blue) and \bx{eb} (red) contain only beam-gas events while the \bx{bb} crossing 
type (black) contains beam-beam vertices in the central region and beam-gas vertices over the whole range.
The effect of the prescale factor on the beam-beam interaction rate is visible.
The exclusion region of $\pm300\mm$ for beam-gas vertices in \bx{bb} crossings is visible as 
the step given by the prescale factor applied to these vertices inside this range.
}

\begin{figure}[tbp]
\centering
\begin{minipage}[c]{0.5\textwidth}
\includegraphics[width=\textwidth]{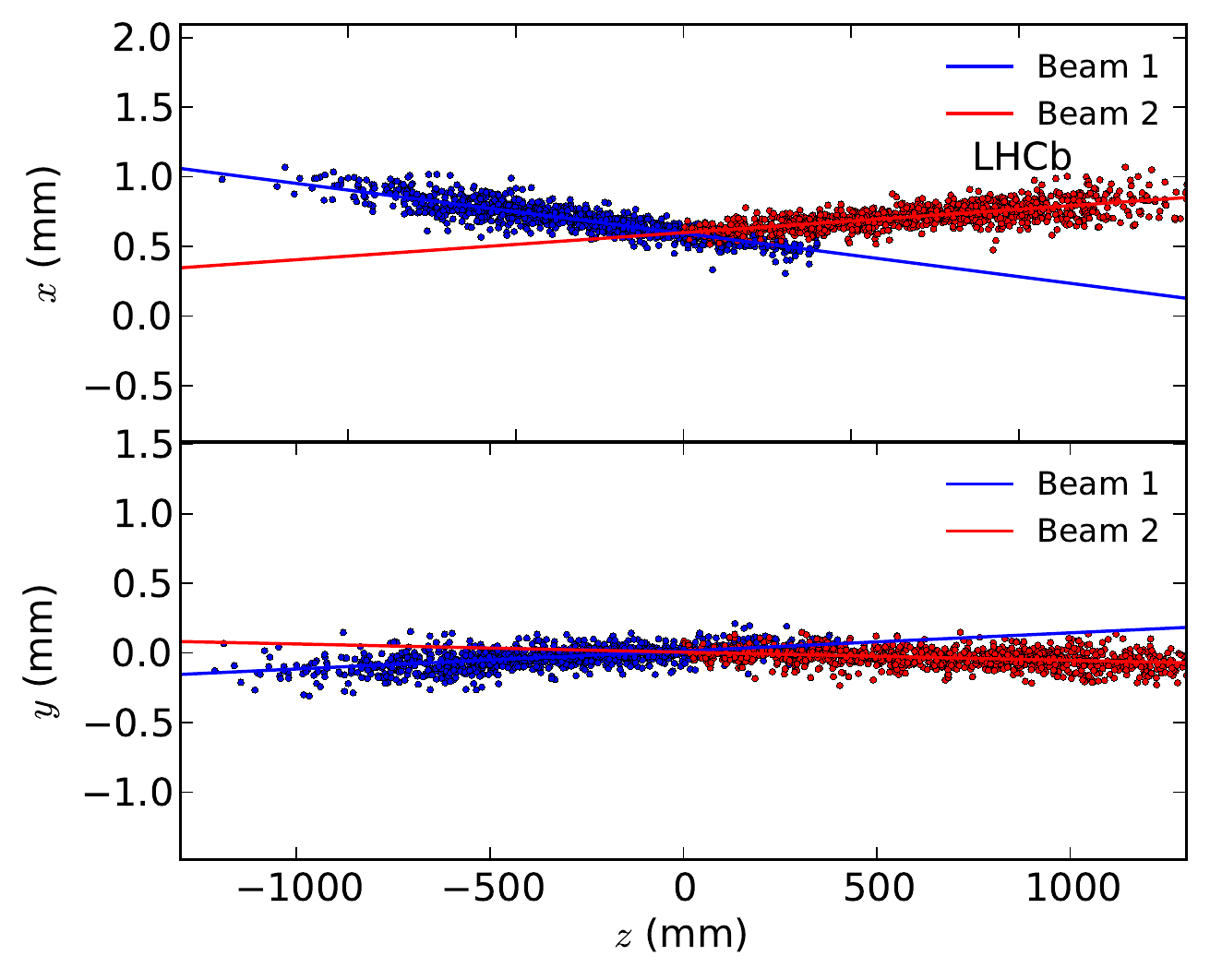}
\end{minipage}\hfill
\begin{minipage}[c]{0.5\textwidth}
\includegraphics[width=\textwidth]{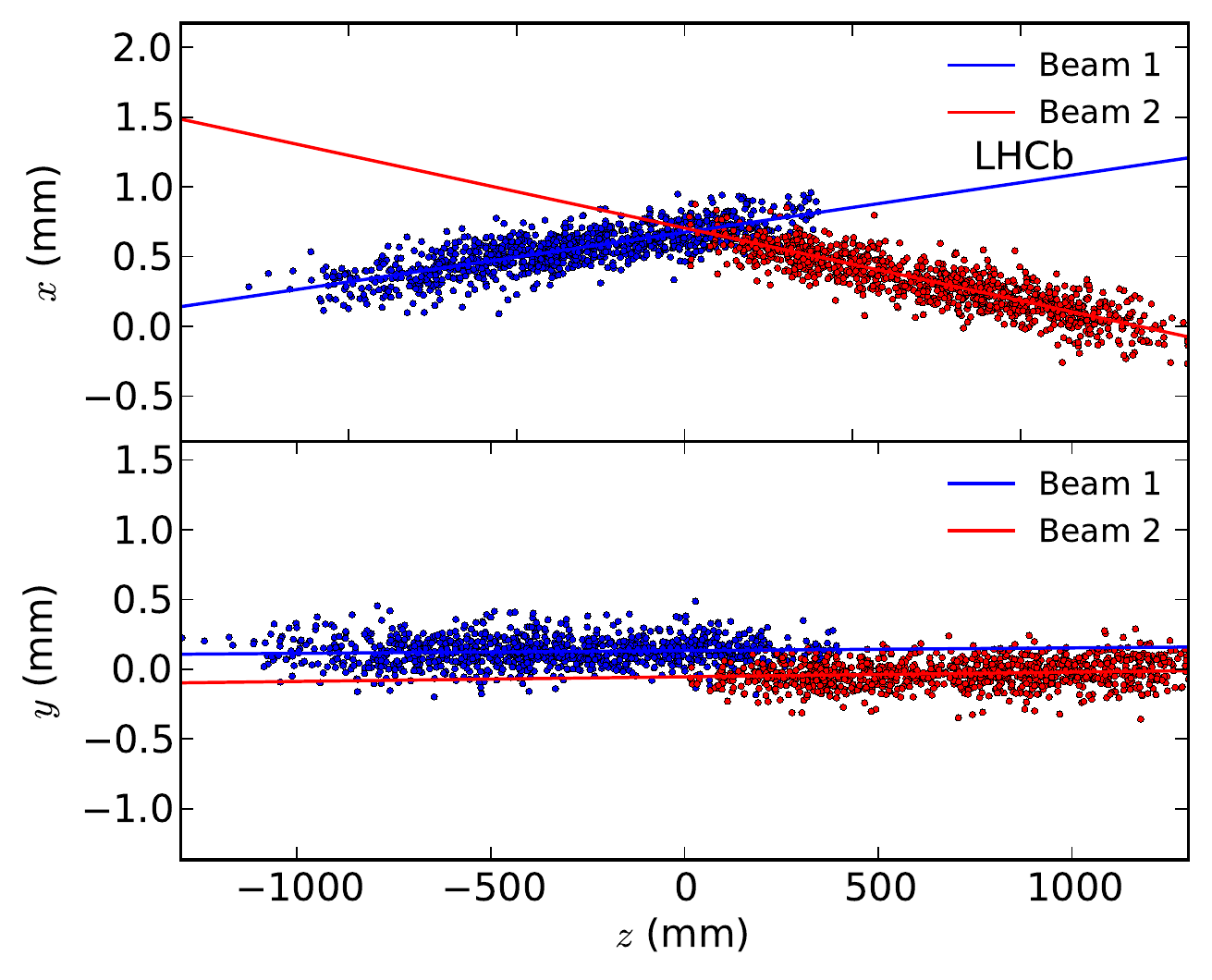}
\end{minipage}\hfill
\begin{minipage}[c]{1\linewidth}
\caption[Position of beam-gas vertices projected in the $xz$ and $yz$ planes]{\small Position of beam-gas 
vertices projected in the $xz$ and $yz$ planes.
The first 1000 vertices per beam are shown.
The crossing angles are visible in both planes (left) for fill 2520 and in the $xz$ plane only (right) 
for fill 2852.
The angles are measured by fitting a straight line through all vertices and indicated as solid lines.
The beams were offset in the $yz$ plane during this period of fill 2852; {\beamone} is slightly above {\beamtwo} 
in the $y$ plane (right plot, bottom).}
\label{scatter_xz_yz}
\end{minipage}
\end{figure}

All events acquired are reconstructed offline to determine their vertex position.
A vertex has a number of tracks $\Ntrks$ associated with it, each track having either a 
forward or backward direction with respect to {\beamone}.
Their multiplicities are defined as $\Nfwd$ and $\Nbwd$, respectively.
Forward tracks correspond to particles moving towards the LHCb spectrometer and benefit
from additional information such as energy and transverse momentum, which can improve the vertex resolution.
The VELO acceptance for forward tracks vanishes when they originate from $z\gtrsim 600\mm$.
Backward tracks are only recorded by the VELO and can be detected when they are produced 
from a vertex with longitudinal position $z\gtrsim -95\mm$.
The longitudinal vertex distribution for all bunch-crossing types before applying further selection criteria is shown in Fig.~\ref{BgStd_BG_2852_122364_0_z_distribution}.
The acceptance limits for beam-gas events from {\beamone} and {\beamtwo} are visible as a sharp drop in rate.
It can be seen that the acceptance drops in two stages for {\beamone} at about 200\mm and 400\mm; 
this effect is related to the positions of VELO sensors in the forward region.
The distribution of beam-beam interactions in the luminous region is reduced due to the prescale factor, 
which also affects beam-gas interactions in \bx{bb} crossings located in $|z|<300\mm$.
The $xz$ and $yz$ vertex distributions for non-colliding bunches are shown
in Fig.~\ref{scatter_xz_yz} for the first 1000 vertices per beam for two fills.

The BGI method relies on differentiation of beam-gas and beam-beam vertices.
The reconstructed vertex is required to contain at least 10 tracks, $\Ntrks\geq 10$.
To exclude interactions with material in the VELO RF foil, the vertex position must be within a 
radial distance of $2$\mm from the beam line.  %
The longitudinal position of a beam-gas vertex must satisfy
$-1000$\mm$\leq z\leq 500$\mm
for \beamone-gas vertices and $0$\mm$\leq z \leq 1000$\mm for \beamtwo-gas vertices.
In \bx{bb} crossings, these ranges are reduced to $-1000$\mm$\leq z\leq -250$\mm
and $250$\mm$\leq z \leq 1000$\mm, respectively.
For \beamone(\beamtwo)-gas vertices all tracks are required to be in the forward (backward) direction.
Beam-beam interactions are selected requiring at least two tracks in both directions in addition to a 
minimum requirement of 25 on the vertex track multiplicity.

The measurement of the beam angles combines beam-gas interaction vertices from colliding and non-colliding bunches.
For the measurements of the overlap integral, beam-gas and beam-beam events from colliding bunch pairs are used.
Beam-gas interactions in non-colliding and empty bunch crossings determine the ghost charge fractions 
and are used in the beam-gas vertex resolution determination.

\subsection{Vertex position resolution}
\label{sec:resolution}
The knowledge of the vertex position resolution is a central ingredient for the measurement of the absolute 
luminosity as the observed vertex distribution is a convolution of the physical beam or luminous region 
with the detector resolution.
To reduce the impact of the resolution on the measurement of the overlap integral, the beam optics for dedicated 
luminosity calibration fills had a $\beta^{*}$ value of $10\m$ (compared to $3\m$ used for physics production runs) and an 
increased transverse emittance.
This resulted in a beam width about twice as large as the transverse vertex resolution.

The vertex resolution used for the BGI analysis is understood as the standard deviation of
the distribution of the residual distance in one coordinate 
between the true vertex position and its measured position. 
The resolution depends on the number of tracks associated with a vertex, the longitudinal position and
whether the vertex originates from a beam-gas collision with only forward or only backward tracks, or from 
a beam-beam collision with both forward and backward tracks. 
Although the value of the resolution in $z$ is about ten times worse than that in the transverse directions, 
its effect can be neglected owing to the much larger luminous region length ($\sim 60\mm$).
Therefore, only the resolution in the transverse $x$ and $y$ directions is considered here.
The resolution measurement method (described below) has been verified with simulated events.
Our studies show that a better resolution is predicted by simulation compared to the measurement from data.
The difference could be explained by the imperfect alignment of the VELO sensors.

\subsubsection{Resolution for beam-beam interaction vertices}
\begin{figure}[tbp]
\centering
\begin{minipage}[c]{0.5\textwidth}
\includegraphics[width=\textwidth]{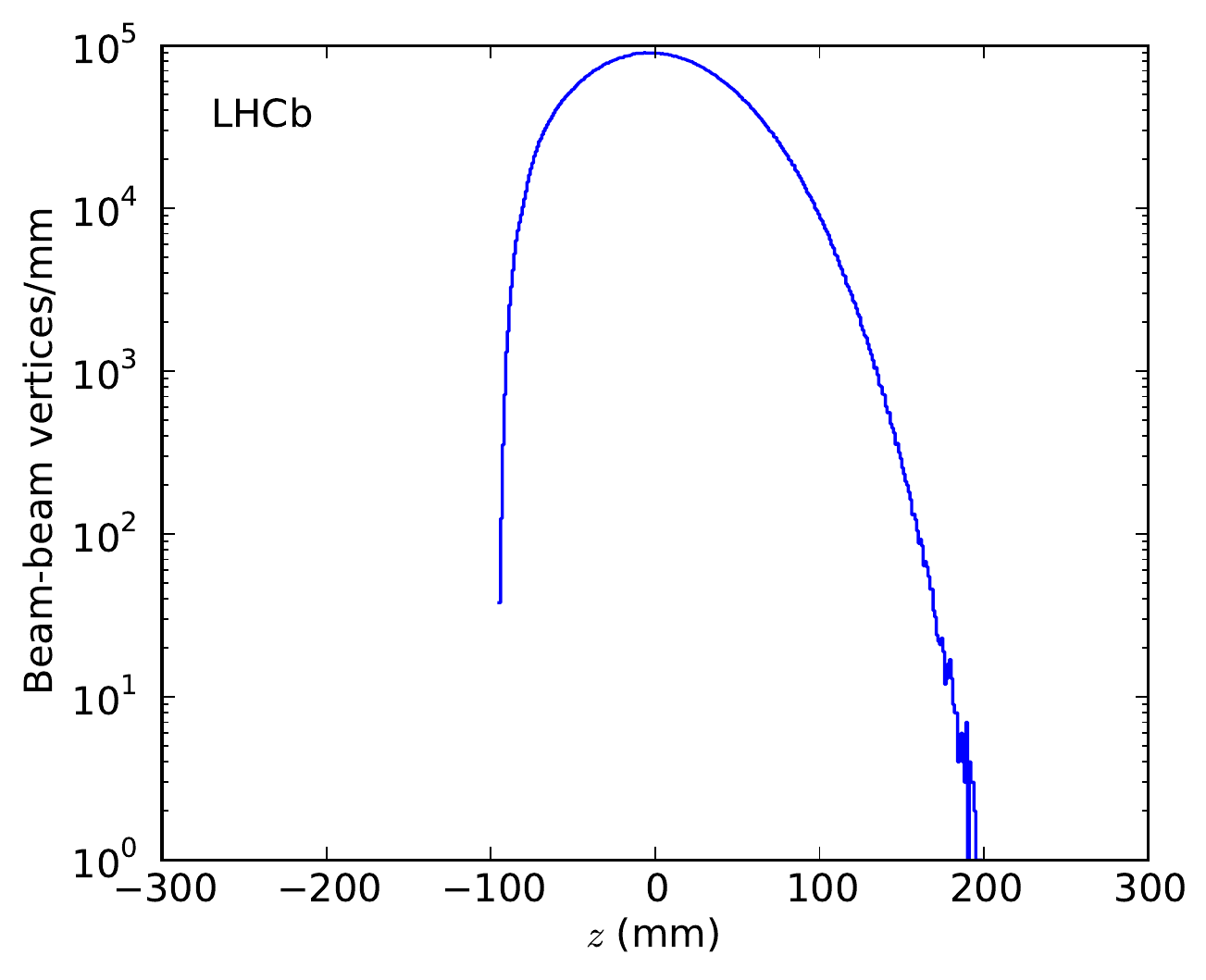}
\end{minipage}\hfill
\begin{minipage}[c]{0.5\textwidth}
\includegraphics[width=\textwidth]{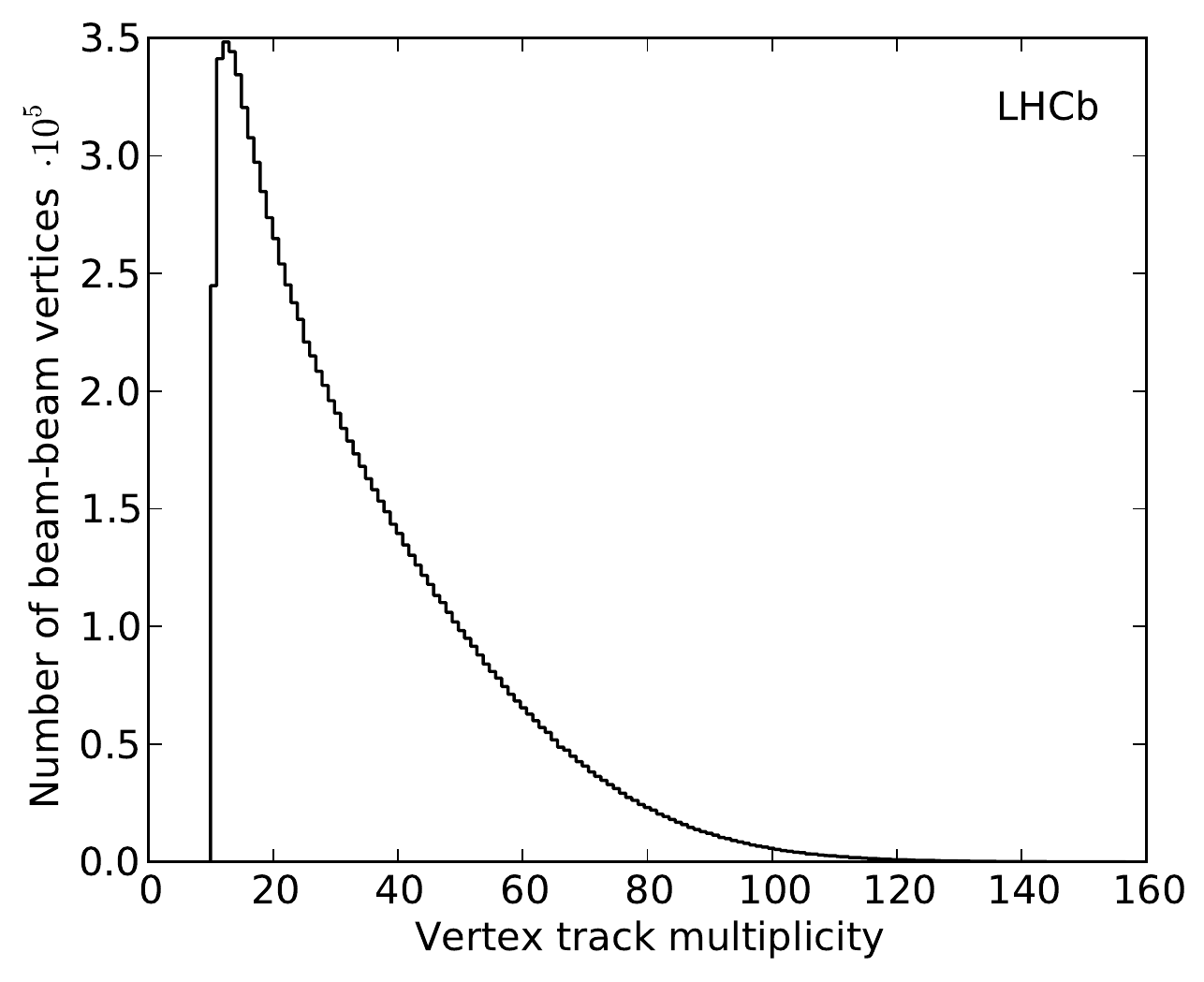}
\end{minipage}\hfill
\begin{minipage}[c]{1\linewidth}
\caption{\small Distributions for beam-beam vertices selected for the resolution measurement with
(left) the longitudinal position distribution and (right) the track multiplicity distribution.
The requirement to have at least two forward and two backward tracks 
in each vertex causes the sharp efficiency drop close to $z=-100$\mm. 
Fill 2520 is selected as example for this figure. 
}
\label{res_bb_sel}
\end{minipage}
\end{figure}

The longitudinal distribution of selected beam-beam interaction vertices %
and the corresponding distribution 
of the number of tracks per vertex are shown in Fig.~\ref{res_bb_sel}.

Without external knowledge of the true position of primary vertices, the residual distance to the 
true position and therefore the resolution cannot be measured directly.
Instead, one can measure the residual distance between two reconstructed vertices originating from 
the same collision by dividing the tracks forming one vertex into two randomly chosen samples (``split vertex method'').
Defining the absolute vertex positions $v_1$ and $v_2$ ($v=x$ or $y$) resulting from the 
primary vertex splitting and $\Delta v = v_{1}-v_{2}$ the distance between the two measurements,
the Gaussian width $\sigma_{\Delta v}$ of all $\Delta v$ distances is a convolution of both 
vertex resolutions and depends on the vertex track multiplicity $N_{\rm Tr,1}$ and $N_{\rm Tr,2}$
of each vertex
\begin{align}
\label{eq:vertex_resol}
\sigma_{\Delta v}^2 &= \sigma_{\text{res},v1}^2(N_{\rm Tr,1})+\sigma_{\text{res},v2}^2(N_{\rm Tr,2}) \, ,
\end{align}
where $\sigma_{\text{res},v}(N_{\rm Tr})$ is the resolution for the vertex track multiplicity $N_{\rm Tr}$.
Indices 1, 2 denote here the two vertices resulting from the splitting.
It is observed that 
the position of the 
original primary vertex is not identical to the average position of the two sub-vertices. 
To minimize any systematic bias due to the resolution measurement procedure, the analysis 
of the beam parameters is performed using the average position 
of the two sub-vertices rather than the position of the original primary vertex.
\begin{figure}[tbp]
\centering
\begin{minipage}[c]{0.47\textwidth}
\includegraphics[width=\textwidth]{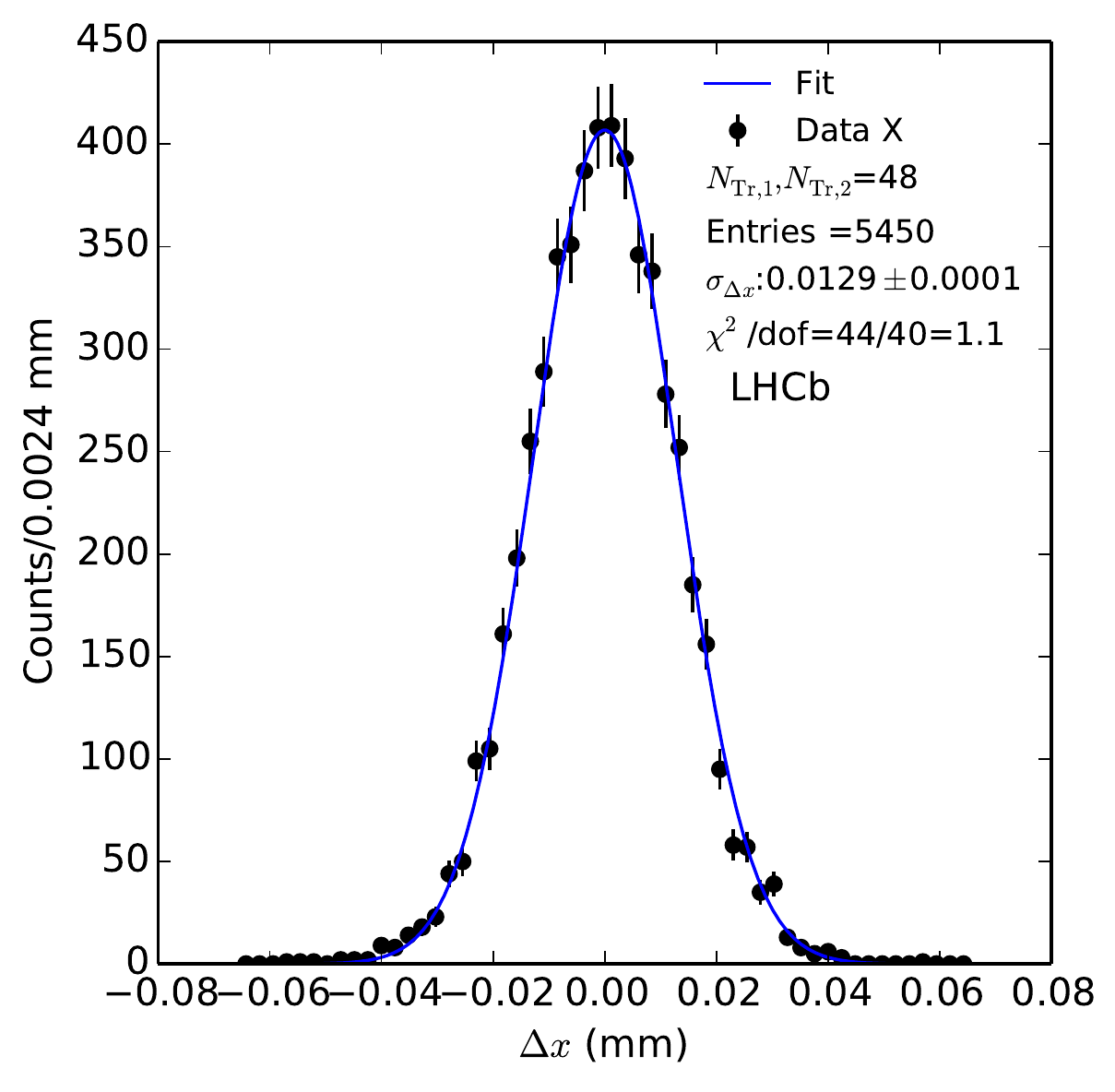}
\end{minipage}\hfill
\begin{minipage}[c]{0.52\textwidth}
\includegraphics[width=\textwidth]{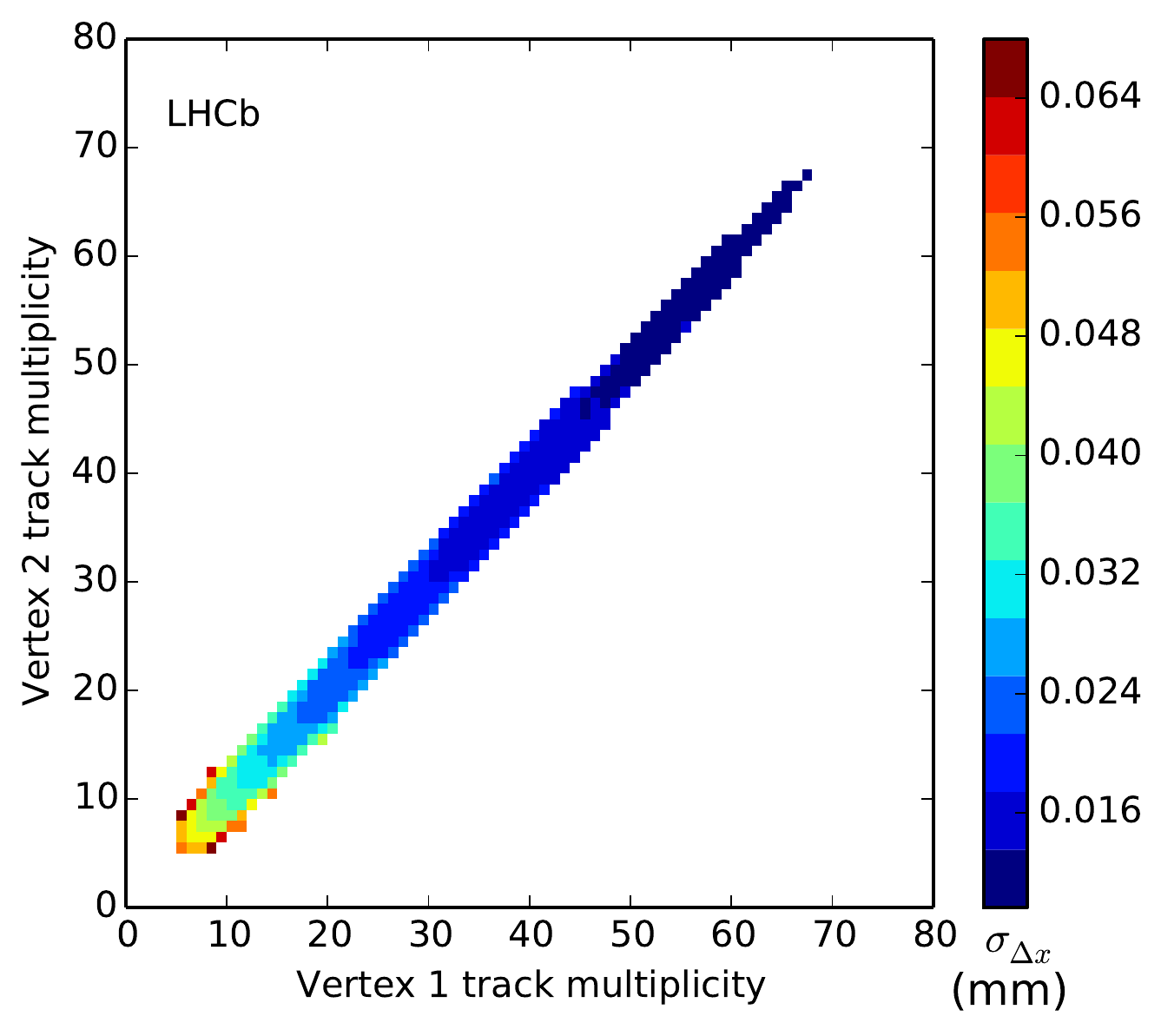}
\end{minipage}\hfill
\begin{minipage}[c]{1\linewidth}
\caption{\small
Measurement of the position difference between two sub-vertices from beam-beam interactions (fill 2520). 
The left-hand panel shows the distribution of distances in $x$ between the two measurements resulting 
from primary vertex splitting, selecting the case with exactly 48 tracks in both sub-vertices.  
The data are fitted with a Gaussian function.
The right-hand panel shows as a colour code the standard deviation $\sigma_{\Delta v_x}$ of measured distances in $x$ between the two vertices 
as a function of the vertex track multiplicities.
The measured \rms of $0.0129\mm$ for two vertices with 48 tracks as shown on the left plot 
corresponds to the blue square at a multiplicity of 48 for both vertex 1 and vertex 2.
}
\label{res_bb_n1n2}
\end{minipage}
\end{figure}

The dependence on  $N_{\rm Tr}$ is determined using the full $z$ range of the luminous region.
In a second step, the variation of the resolution as function of $z$ is addressed by introducing $z$-dependent factors.
Measured distances $\Delta v$ are sorted according to the possible combinations 
$i=(N_{{\rm Tr},1}, N_{{\rm Tr},2})$ of the two vertex track multiplicities.
For each combination $i$ %
the distribution of all $\Delta v_i$ measurements are fitted with a Gaussian function to determine the width of the distribution 
$\sigma_{\Delta v_i}$. 
An example of such a distribution is shown in Fig.~\ref{res_bb_n1n2} (left) together with the result of the fit.
In this particular case, the number of tracks in both sets being equal, the measured distribution 
width $\sigma_{\Delta v_i}$ is directly related to the vertex resolution %
$\sigma_{\text{res},v}(N_{\rm Tr})$ using Eq.~\eqref{eq:vertex_resol}.
All measured distribution widths $\sigma_{\Delta x_i}$ as function of ($N_{{\rm Tr},1}$, $N_{{\rm Tr},2}$) combinations
are shown in Fig.~\ref{res_bb_n1n2} (right).
\begin{figure}[tbp]
\centering
\includegraphics[width=0.535\textwidth]{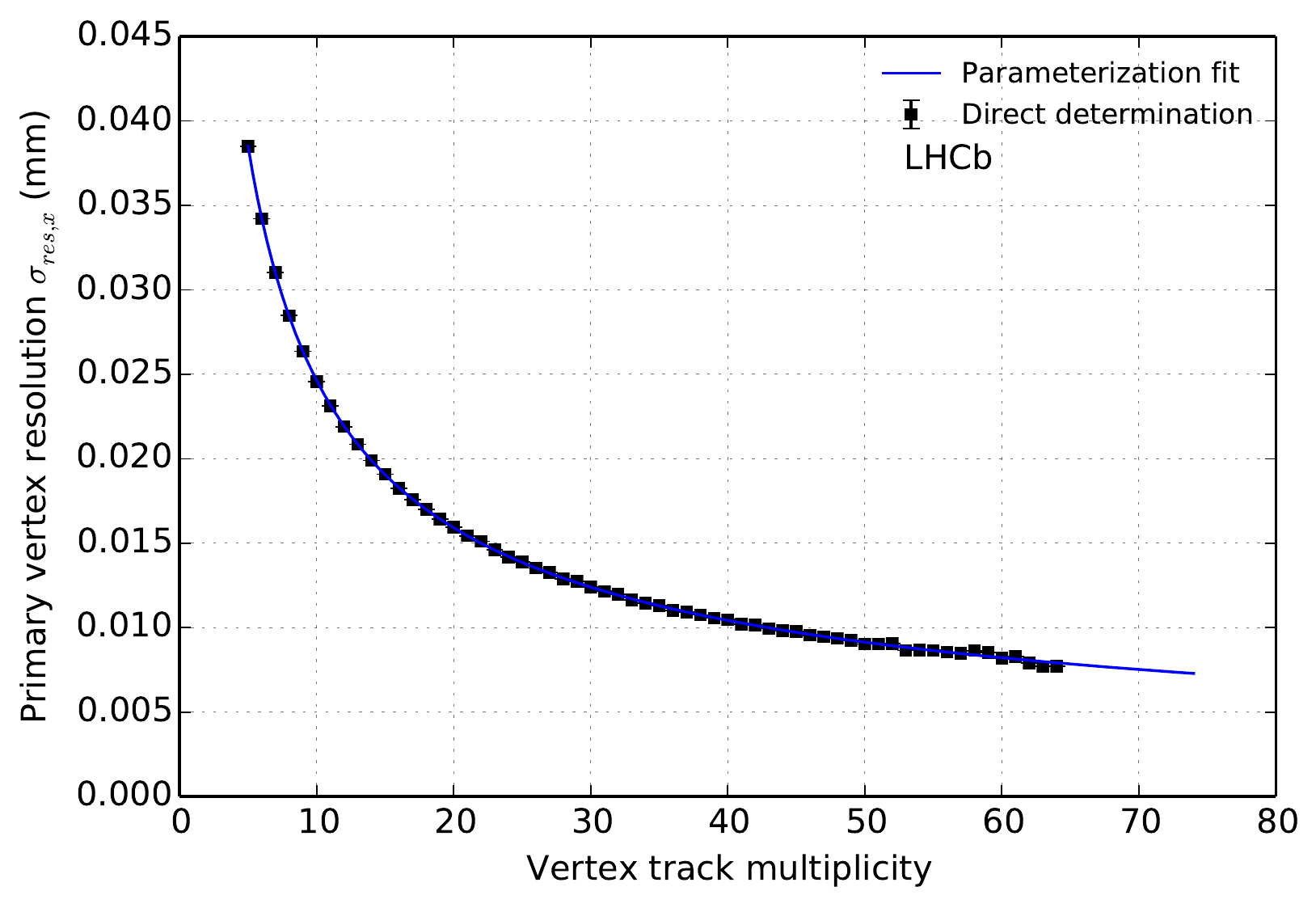}
\hfill
\includegraphics[width=0.455\textwidth]{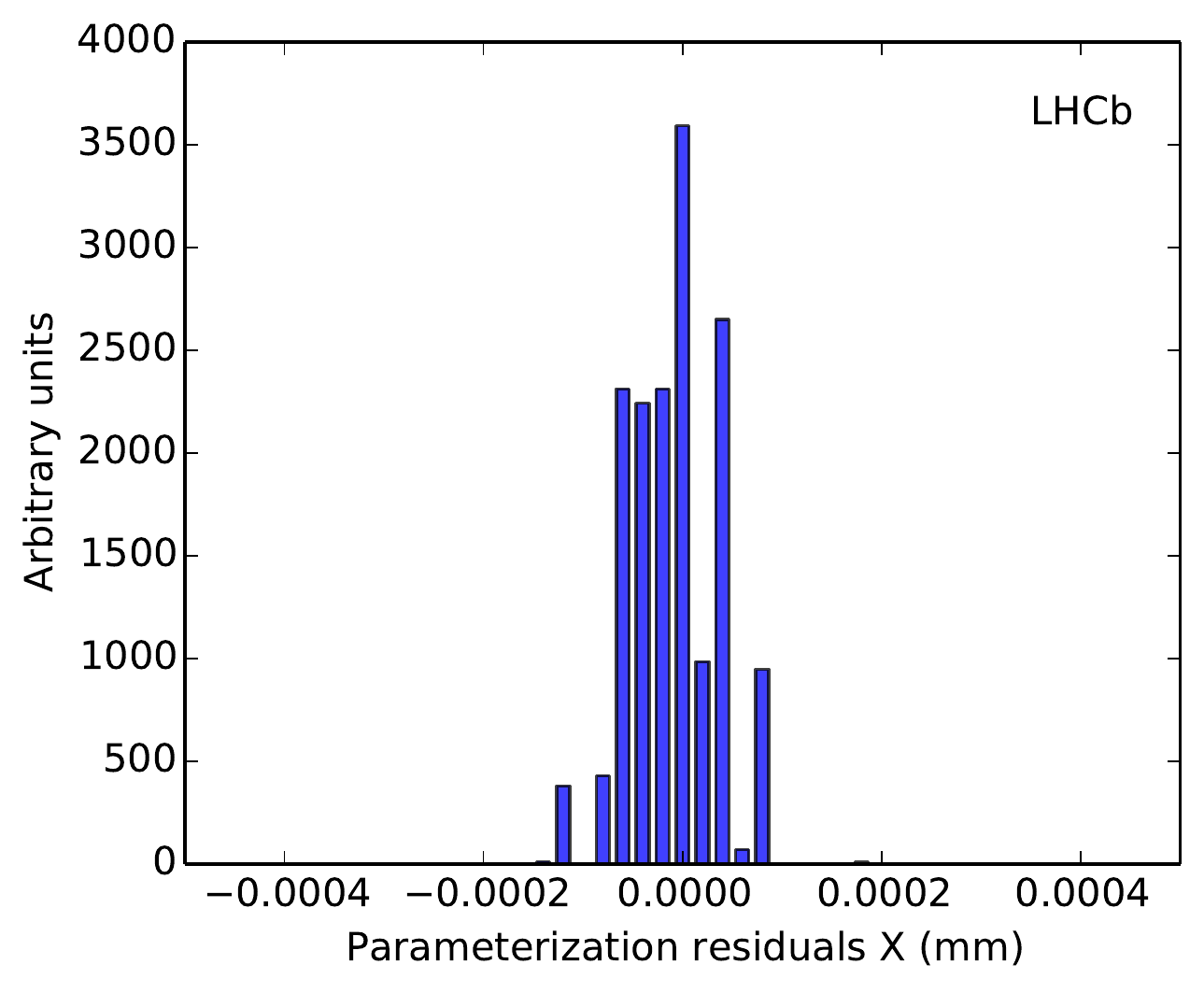}
\caption{\small 
Results of the beam-beam vertex position resolution measurement in fill 2520 with (left)
the resolution as a function of the track multiplicity and (right) the distribution of parametrization residuals.
The markers in the left plot show the direct determination of the individual resolution for a given vertex track multiplicity for the $x$ coordinate. Results for the $y$ coordinate are similar.
The curve shows the results of the resolution parametrization fitted using Eq.~\eqref{eq:vertex_param}.
In this fill the fit gives the values 
$A_x =  0.110\mm$, $B_x= 0.669$, $C_x= 0.0011\mm$, $A_y = 0.101\mm$, $B_y = 0.640$, $C_y = 0.0006\mm$
for the parameters defined in Eq.~\eqref{eq:vertex_param}.
The entries in the right plot are weighted by the number of vertices
observed with a given vertex track multiplicity.
}
\label{BgStd_BG_2520_resolution_BB_dXY_resolution_tracks}
\end{figure}

The resolution for a given vertex track multiplicity $\sigma_{\text{res},v}(N_{\rm Tr})$ is obtained 
by fitting all distribution widths $\sigma_{\Delta v_i}$ 
with a least squares minimization constrained by Eq.~\eqref{eq:vertex_resol}.
Results for the resolution as a function of vertex track multiplicity are shown in 
Fig.~\ref{BgStd_BG_2520_resolution_BB_dXY_resolution_tracks}.
As can be seen in Fig.~\ref{res_bb_sel} (right), the number of vertices with more than 120 tracks
vanishes, limiting the resolution measurement up to about 60 tracks per vertex.
The resolution for vertices with a larger number of tracks is obtained by extrapolation of a parametrization,
\begin{equation}
\label{eq:vertex_param}
\sigma_{\text{res},v}(N_{\rm Tr}) = \frac{A}{N_{\rm Tr}^B}+C.
\end{equation}
The factor $A$, the power $B$ and constant term $C$ are measured by fitting all $\sigma_{\Delta v_i}$ measurements.
Typical values are given for fill 2520 in the caption of Fig.~\ref{BgStd_BG_2520_resolution_BB_dXY_resolution_tracks}.
Results for the resolution parametrization functions are shown in 
Fig.~\ref{BgStd_BG_2520_resolution_BB_dXY_resolution_tracks}  and are in good 
agreement with results from the direct determination of individual resolutions for each vertex track multiplicity.
No bias is observed and the statistical fluctuations are less than $0.1\mum$.
Beam-beam vertex resolution parametrization results are evaluated for each fill independently.

The variation of the vertex resolution for beam-beam interactions as a function of $z$ is determined by
comparing the residual distribution with the parametrization.
A correction factor $F_z$ is introduced for each $z$ bin such that the residuals are minimized.
The resulting values of $F_z$ range between 0.98 and 1.04~\cite{Colin-thesis}. 

\subsubsection{Resolution for beam-gas interaction vertices}
\label{sec:resolution_beam_gas}

In previous measurements~\cite{LHCb-PAPER-2010-001}, the beam-gas vertex resolution had to be
extrapolated from beam-beam resolution measurements due to an insufficient number of beam-gas events.
With neon gas injection, the increased rate allows a direct measurement of the beam-gas vertex resolution 
to be made for both beams on a fill-by-fill basis.
The measurement principle is similar to that used for the beam-beam vertex resolution.
The main differences reside in the vertex selection and in the $z$ dependence 
of the resolution measurement.

\begin{figure}[tbp]
\centering
\begin{minipage}[c]{0.5\textwidth}
\includegraphics[width=\textwidth]{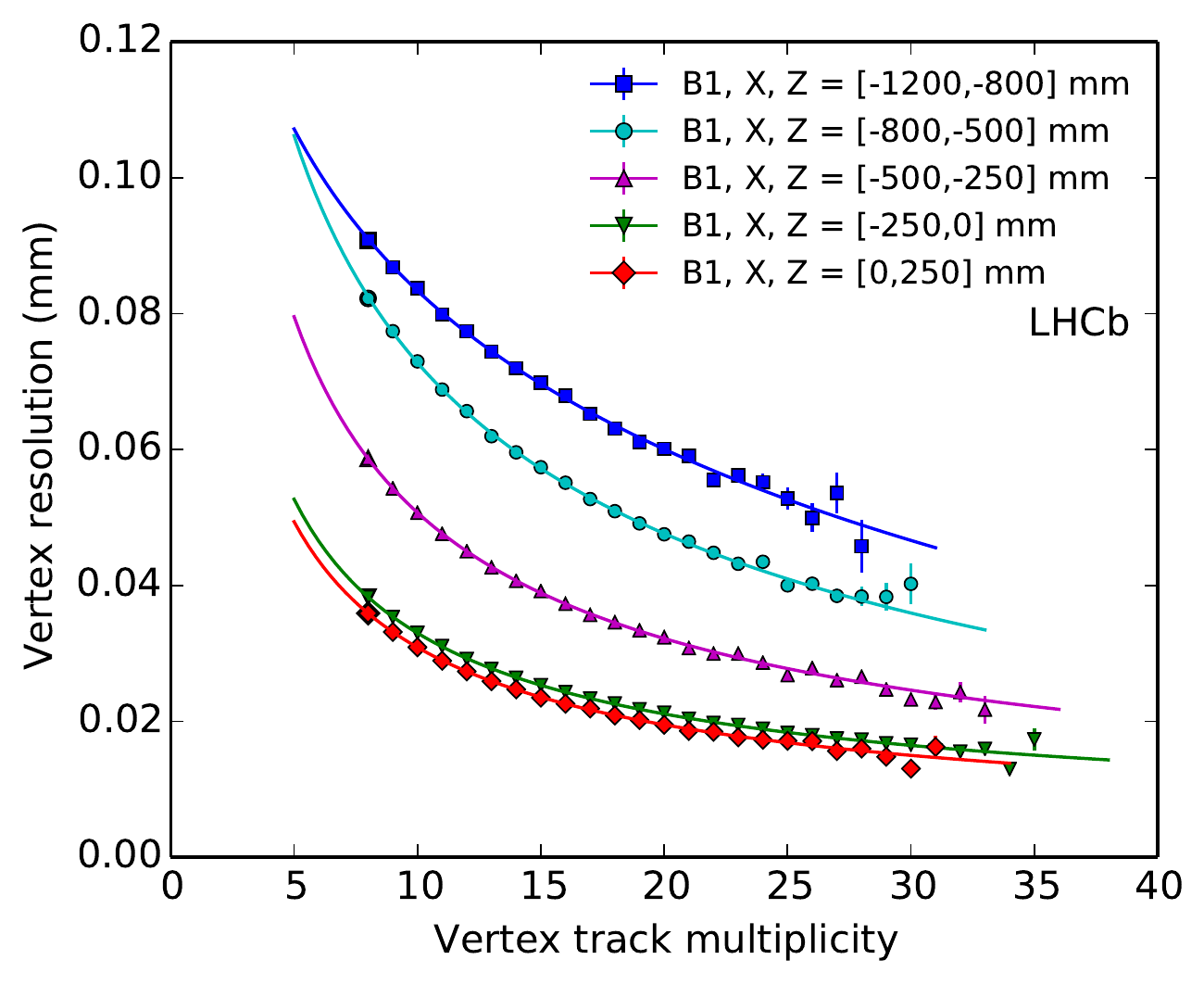}
\end{minipage}\hfill
\begin{minipage}[c]{0.5\textwidth}
\includegraphics[width=\textwidth]{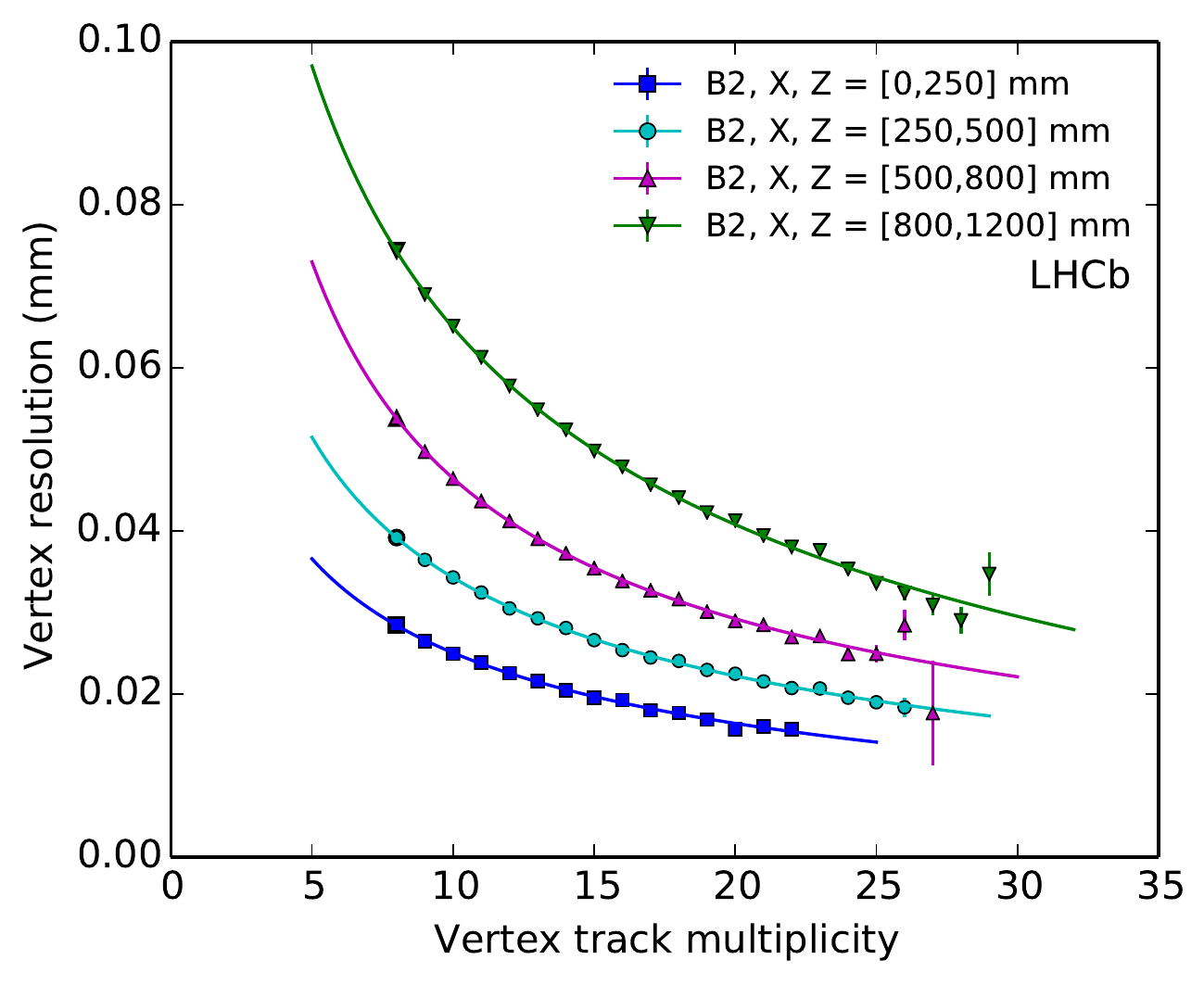}
\end{minipage}\hfill
\begin{minipage}[c]{1\linewidth}
\caption{\small
Beam-gas vertex resolution in $x$ as a function of vertex track multiplicity
for (left) \beamone and (right) \beamtwo.
Fill 2855 is displayed here as an example.
Single markers indicate one resolution measurement for a given vertex track multiplicity.
The continuous curves indicate the results of the fits using the parametrization in the same $z$ ranges.
The statistical uncertainties are shown by error bars. %
The \beamone (\beamtwo) resolution is determined separately in five (four) $z$ regions.
The results for the $y$ resolution are similar.
}
\label{res_bg_param}
\end{minipage}
\end{figure}
 
\begin{figure}[tbp]
\centering
\begin{minipage}[c]{0.5\textwidth}
\includegraphics[width=\textwidth]{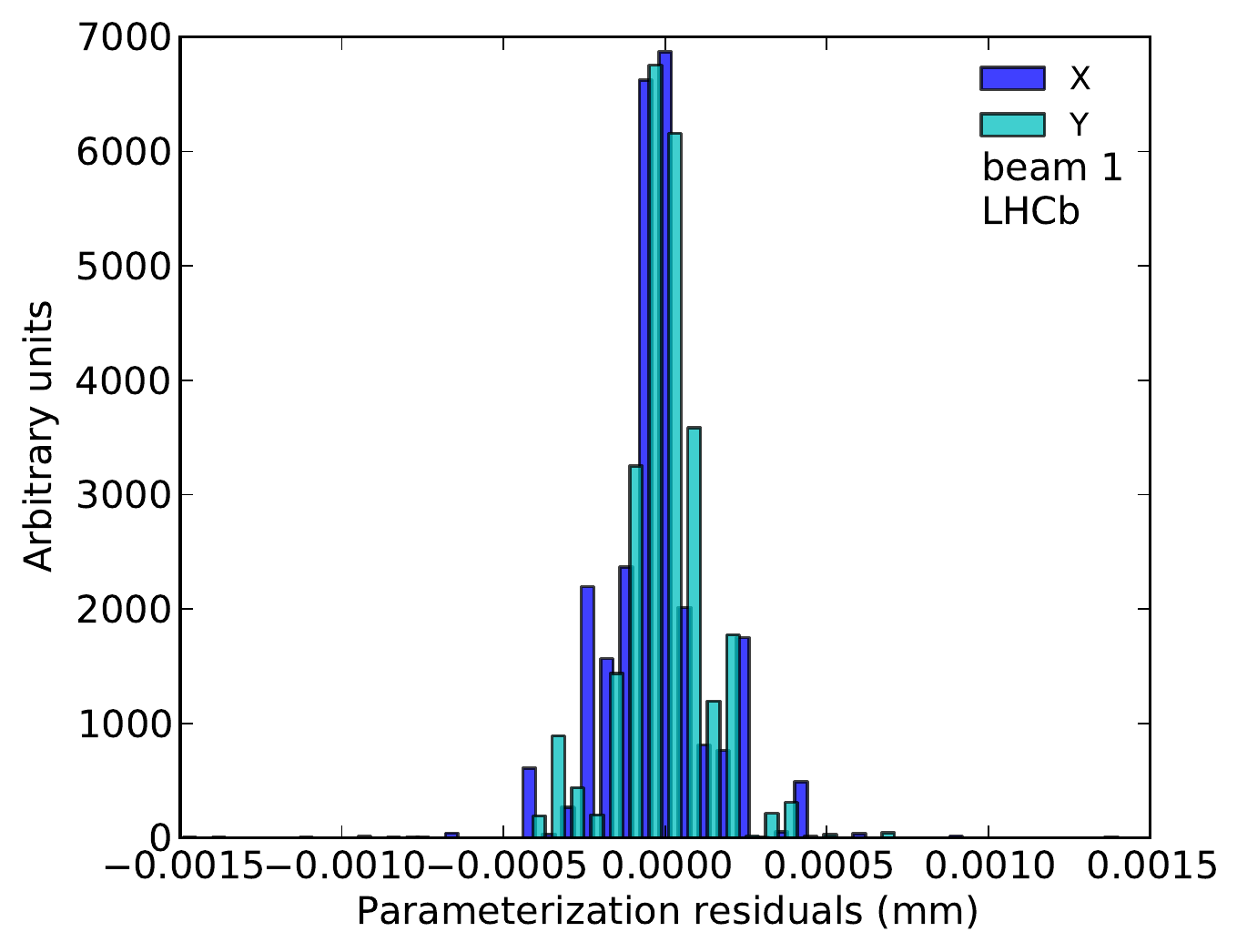}
\end{minipage}\hfill
\begin{minipage}[c]{0.5\textwidth}
\includegraphics[width=\textwidth]{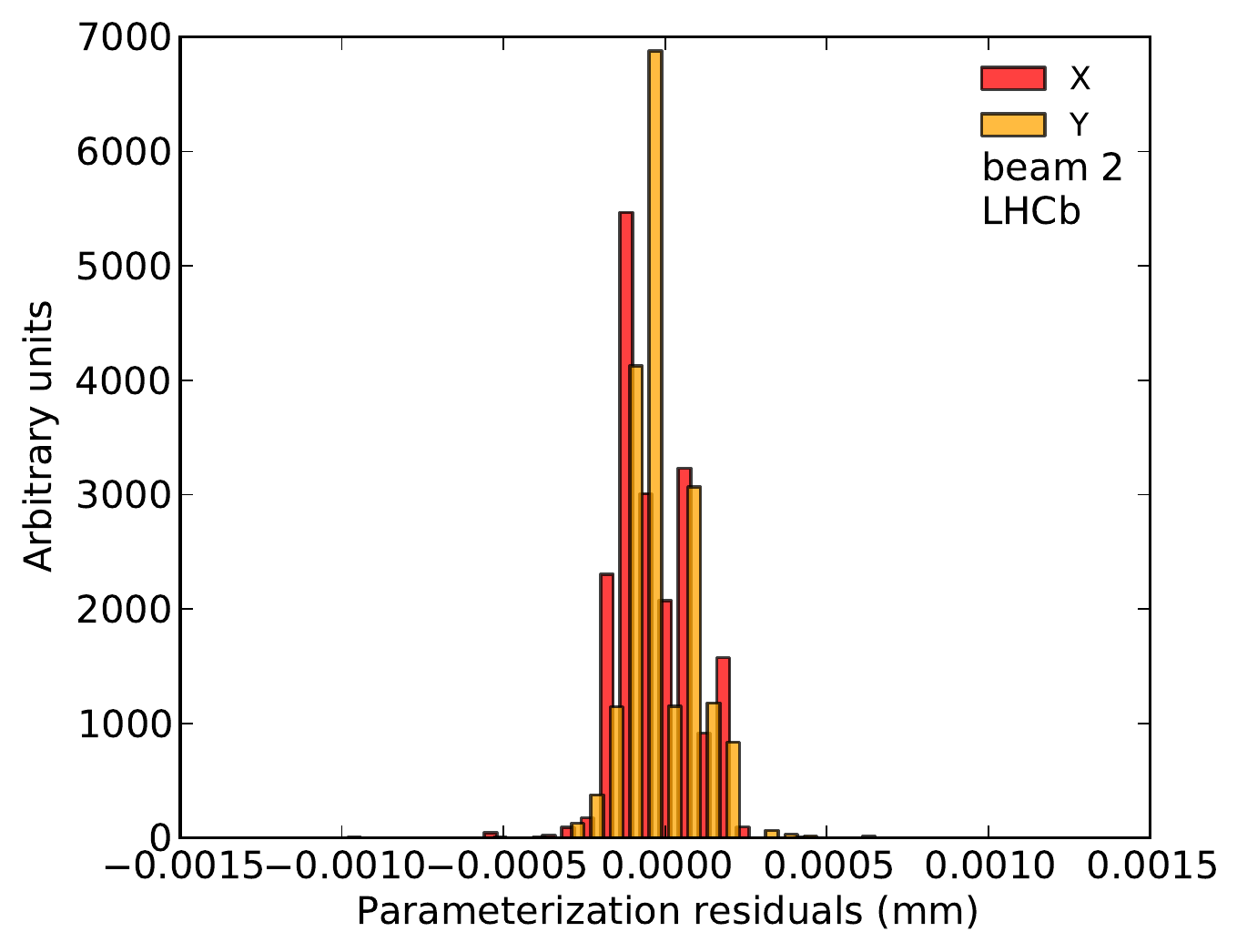}
\end{minipage}\hfill
\begin{minipage}[c]{1\linewidth}
\caption[Weighted residuals between the direct beam-gas resolution measurement and the parametrization.]{%
\small Distributions of residuals between the parametrization and the direct fits for (left) \beamone and (right) \beamtwo.
Entries are weighted by the number of vertices observed with a given vertex track multiplicity.
Fill 2855 is shown here as example.
}
\label{res_bg_resid}
\end{minipage}
\end{figure}

The detector acceptance and extrapolation distance vary considerably within the $\pm1\m$ $z$ range 
used for the BGI analysis, leading to different distributions of vertex track multiplicity
and vertex resolution. 
Therefore, beam-gas interaction vertex resolutions were measured separately in five $z$ ranges
for {\beamone} ([$-$1200,~$-$800], [$-$800,~$-$500], [$-$500,~$-$250], [$-$250,~0], [0,~250]\mm)
and four for {\beamtwo} ([0,~250], [250,~500], [500,~800], [800,~1200]\mm).
Results are shown in Fig.~\ref{res_bg_param}.
The distributions of the residuals between the direct resolution measurements and the 
resolution parametrization are shown in Fig.~\ref{res_bg_resid}. %
There is no significant bias in the parametrization and the statistical spread 
of about $0.2\mum$ can be neglected.
The $z$ dependence is evaluated with a procedure similar to that used for beam-beam interactions.

\includefig{1}{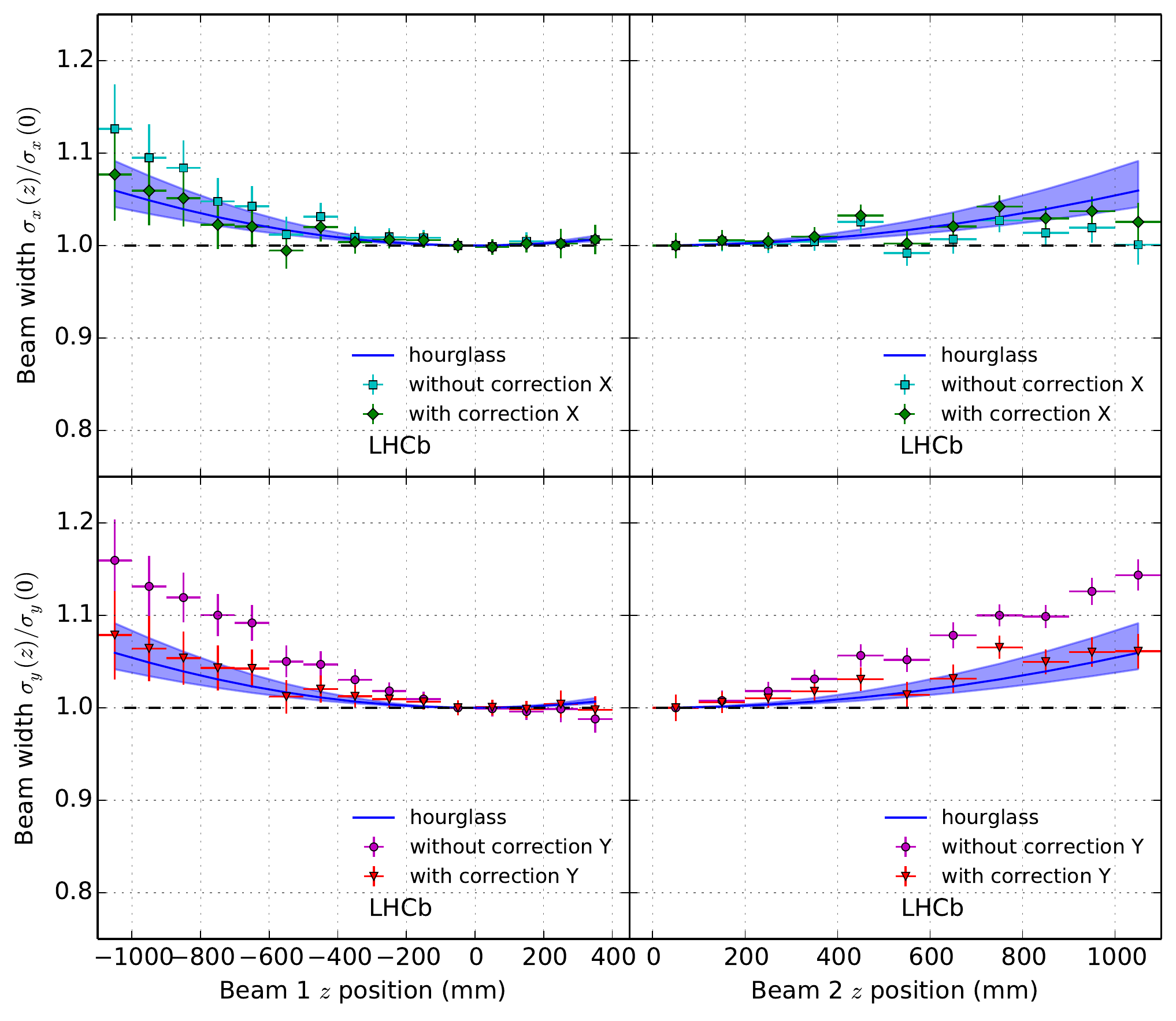}{%
\small Measurement of single beam width $\sigma_v$ as a function of $z$ normalized to the width at $z=0$ including 
resolution corrections.}{%
Resolution-deconvolved single beam widths $\sigma_x$ and $\sigma_y$ as a function of $z$ for fill 2520 ($\beta^*=3\m$).
Values with and without correction factors are shown.
Each data point for a given $z$ position is an average of all normalized widths from non-colliding bunches.
The curved blue line shows the expected evolution of the beam width due the $\beta^*$ hourglass effect, 
the shaded blue surface indicates the boundaries corresponding to a 10\% uncertainty on $\beta^*$.
All points are normalized to the values nearest to $z=0$.
}

The beam-gas vertex resolution can be tested by measuring the single beam width at different $z$ positions.
While the physical beam width is unknown, its relative change as a function of $z$ can be predicted 
from the machine optics and is expected to behave as
\begin{equation}
\sigma_{\text{beam}}(z) = \sigma_{\text{beam}}(0)~\sqrt{1+\Big(\frac{z}{\beta^{*}}\Big)^2}~,
\end{equation}
assuming the waist position is at $z=0$.
This is called the ``hourglass'' effect. %
In fill 2520,  measurements were performed  with $\beta^*=3\m$ optics, providing smaller
beam sizes and a stronger hourglass effect than in other luminosity calibration fills.
Therefore, the beam width measurements for this fill are more sensitive to the resolution.
A deviation from the expected $z$ dependence of the beam width was observed and
additional resolution factors $f_z$ are introduced based on this deviation.

These corrections factors are measured for each fill separately.
With a $\beta^*$ value of $10\m$ the hourglass effect is negligible (less than 1\% at $1\m$).
Similarly the beam width to resolution ratio is also larger and the correction factors have a smaller impact on the measured width.
A final verification of single beam width measurements as function of $z$ is shown in 
Fig.~\ref{widhtZbinsNormed_beam_combined3_BgStd_BG_2520_112867_0dbv76CorrTrue_sigmas} %
after applying the correction factors.
The resolution has been measured independently for all dedicated fills.

\subsubsection{Resolution function for a sample of vertices}

\def\Ng{N_{\rm g}}
The average position of the two sub-vertices from a split primary vertex is used to measure the beam shapes.
The observed vertex distribution is a convolution of the density function 
such as Eq.~\eqref{eq:factorizable_density} with the detector resolution.
The resolution function of a sample of vertices for a given vertex type (\beamone, \beamtwo, beam-beam) 
is the superposition of the resolution parametrizations based on the number of tracks  and the $z$ position
of each of the vertices in the sample.
This resolution function of the sample, $R_m$, can be parametrized in each coordinate, $m=x,\,y$, by 
a sum of a limited number $\Ng$ of normalized Gaussian functions $g_{k_m}$ 
with weights $c_{k_m}$ and widths $\sigma_{\text{res},{k_m}}$
\begin{equation}
R_m = \sum\limits^{\Ng}_{k_m=1}\, c_{k_m}\, g_{k_m}(\sigma_{\text{res},{k_m}})\, .
\end{equation}
The resolution is unbiased, \ie the functions are centred at the origin.
By dividing the full range of the distribution of the resolution estimates of 
all vertices in the sample into $\Ng$ equal-sized bins, the widths and weights are 
determined by taking the centre and population of the bins, respectively.
The weights $c_{k_x}$ and $c_{k_y}$ are the relative resolution weights of the effective 
resolution functions for $x$ and $y$.
Choosing $\Ng = 3$ gives a good description of the sample resolution-function;
a larger number of Gaussian functions does not change the results.

Since the transverse distributions of single beams and of the luminous region are expressed 
as a superposition of Gaussian functions, describing the resolution function also as a sum of Gaussian functions 
results in an analytical expression for their convolution.
In this approximation, the observed transverse distribution for one of the beam components 
is written as a superposition of convolved Gaussian density functions. %
Each intervening Gaussian width ($\sigma_{m\text{n}}$, $\sigma_{m\text{w}}$) %
is replaced by a resolution-convolved width  in the coordinate $m=x,\,y$:
\begin{equation}
\label{eq:convolved_width}
\sigma^*_{k_m\text{n}} = \sqrt{\sigma_{\text{res},{k_m}}^2+\sigma^2_{m\text{n}}}\, 
\quad\mbox{and}\quad
\sigma^*_{k_m\text{w}} = \sqrt{\sigma_{\text{res},{k_m}}^2+\sigma^2_{m\text{w}}}\, .
\end{equation}
A similar treatment can be applied to the transverse distribution of the luminous region, 
using in this case the products of single beam density functions as given by 
Eq.~\eqref{eq:bb_density}~\cite{Colin-thesis}.

\subsection{Measurement of the overlap integral} %
\label{sec:bunch_shape_model}

The knowledge of the three-dimensional bunch shapes $\rho_j(x,y,z)$ is required to evaluate 
the overlap integral defined in Eq.~\eqref{eq:overlapintegral}.
To determine the value of the parameters described in Eq.~\eqref{eq:non_factorizable_weights},
a fitting procedure is performed that proceeds in several steps.
First, the directions of the single beams are obtained from the positions of beam-gas 
interaction vertices using all bunches in the filling scheme simultaneously.
These beam directions determine the crossing angles and are used to project vertex positions in 
relatively large $z$ ranges onto a reference plane to ease the fitting procedure.
The rest of the procedure is applied only to colliding bunches, taking each bunch-pair individually.
To avoid problems with beam drifts and emittance growth, the data are grouped  
into data-taking periods of about 20 minutes.
In a second step, transverse properties of single beams are obtained using the 
assumption that the shapes in the two transverse coordinates are factorizable.
The parameter values obtained are used as initial values for the following step.
The third step consists of a fit to beam-gas vertices of bunch pairs of both 
beams simultaneously, together with beam-beam interaction vertices in their luminous region.
This fit is performed in two transverse coordinates separately, still assuming factorizability.
In the following, fourth, step the two beams and their luminous region are fitted 
in both transverse coordinates simultaneously with the full two-dimensional model.
The initial parameters of the latter fit are provided by the fit performed in the previous step.
Finally, the $z$ positions of the beam-beam vertices are used to determine 
the longitudinal properties of the bunches.
At this point, all parameters needed to evaluate the overlap integral are determined.
The individual steps of the procedure are described in more detail below.

The beam angles $\alpha_{m,j}$ for beam $j=1,2$ and axis $m=x,\,y$ in the laboratory reference frame 
are measured using beam-gas vertex positions.
While the luminosity measurement is based on vertices in the colliding bunch pairs, vertices originating 
from non-colliding bunch crossings are valuable to measure the crossing angles as they cover the 
full $z$ range owing to the absence of beam-beam background.
In a first pass, a straight-line least-square fit is performed to all beam-gas interaction vertices 
weighting the positions according to their resolution and to an initial estimate of the beam width.
An example of a crossing angle measurement using vertex positions directly is shown in Fig.~\ref{scatter_xz_yz}.
In a second pass, events are binned in 50\mm $z$ intervals with centre $z_c$.
Their transverse vertex position $v_{j,m}$ is projected to $z_c$ 
using as initial estimates values of $\alpha_{m,j}$ obtained in the first pass.
A weighted straight-line fit is then performed through the transverse positions 
obtained by Gaussian fits to the distributions in the $z$ intervals. 
The statistical uncertainty in the angles is about $10^{-2}\unit{{\micro}rad}$.
The half crossing angle, which is the angle of interest to measure the overlap integral, 
is defined as $\phi_{m} \equiv (\alpha_{m,1} - \alpha_{m,2})/2$.

In the second step of the fitting procedure, the transverse shapes of each individual bunch are analysed using beam-gas interaction data from both beams.
The data are divided into different $z$ ranges to combine only data of similar resolution.
Three slices for the beam-gas samples in the ranges $-1000$ mm $<z<-250$ mm  ({\beamone})
and $250$ mm $<z<1000$ mm ({\beamtwo}) are chosen.
The vertices within a slice are projected along the beam direction onto a plane perpendicular to the $z$ axis.
This coordinate translation neglects the hourglass effect by assuming a constant beam shape along the beam axis.
This is justified since, with $\beta^*=10\m$, the beam is broadened by 0.5\% over 1\m. %
The transverse beam-shape model is fitted to the data in the three slices simultaneously using the 
double-Gaussian density model of Eq.~\eqref{eq:factorizable_density} and
convolution with the resolution model.%

For the third and fourth steps, the beam-beam data are divided into $z$ slices and 
a new fit is performed using beam-gas and beam-beam data simultaneously. 
The transverse distribution  of the luminous region changes as function of $z$.
Since the number of events is sufficient, the analysis is simplified 
by taking many slices so that in each of these slices the fit model can be approximated 
by the function value at the centre of the $z$ range.
Thus, for the beam-beam vertices, the range  $-100$ mm $<z<110$ mm is divided into 18 slices.
The properties of the luminous region follow from the single beam properties, thus no new shape parameters are 
introduced, while the values obtained from the single beam fits are used as initial values.
Free amplitude parameters are introduced for each single beam and luminous region 
$z$ slice to take into account the combined effect of trigger and reconstruction efficiency and absolute rates.
The third step consists of a \oneD fit assuming a factorizable description in the $x$ and $y$ coordinate.
As is shown below, a two-dimensional model is needed for the transverse shapes of the beams
that can accommodate non-factorizable distributions in the $x$ and $y$ coordinates.
Such a fit is performed in the fourth step, where a \twoD fit is performed to both coordinates simultaneously.

\begin{figure}[tbp]
\centering
\includegraphics[width=0.93\textwidth]{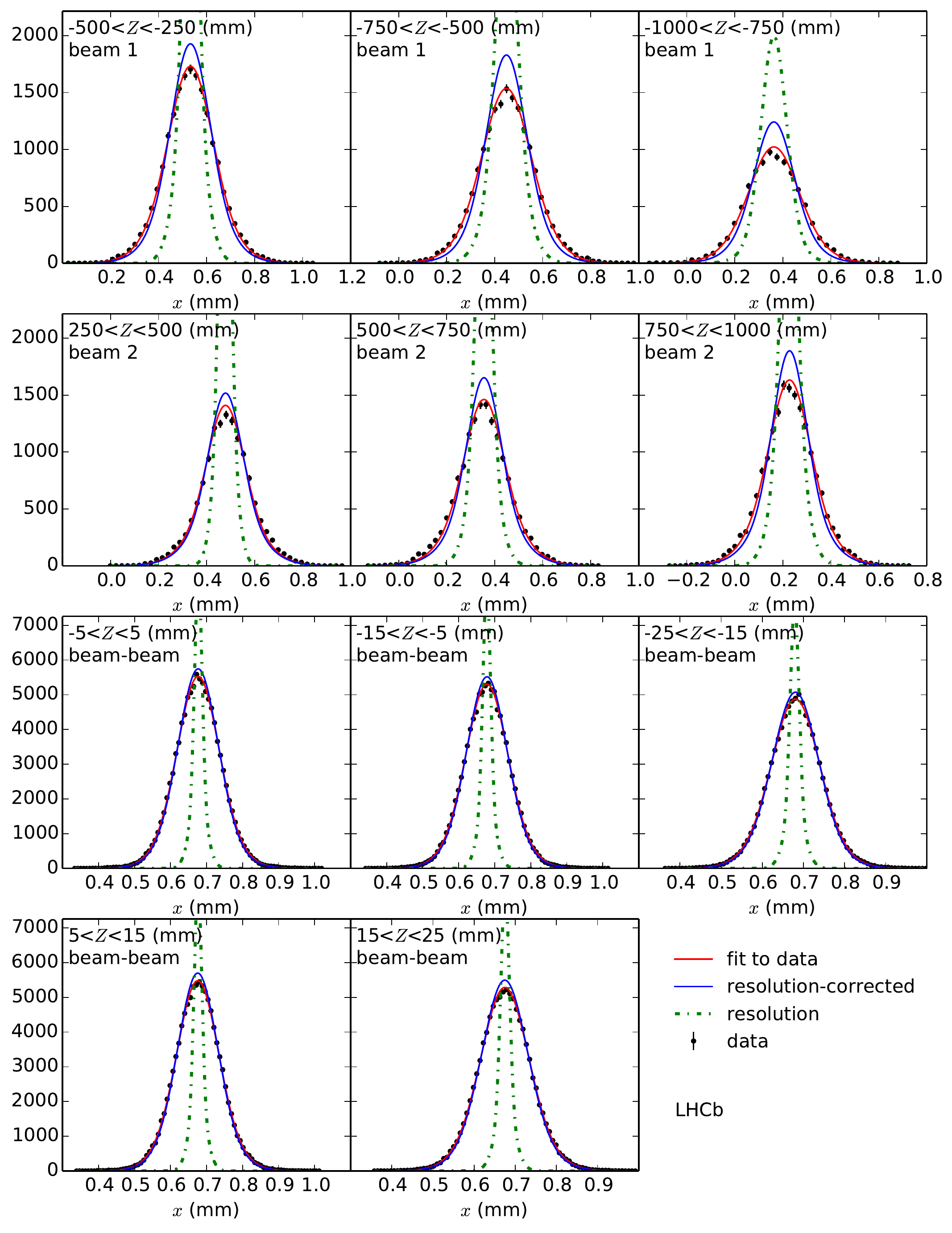}
\caption[Global fit results of a colliding bunch pair in 1-dimension.]{\small Results of a \oneD global fit of one colliding bunch
pair and for the $x$ coordinate (fill 2852).
The first (second) row shows the results of the three {\beamone} ({\beamtwo}) $z$ slices.
The third and fourth row show the results of the central five beam-beam $z$ slices
(the remaining 13 slices used in the fit are not shown for better readability). 
The double Gaussian fit results are shown as solid red lines.
The dashed green lines show the effective $x$ resolution functions.
The resolution-corrected distributions are shown as solid blue lines.}
\label{GlobalFit_doubleGaussianSel_BgStdrun122467_bunch1909_X_1}
\end{figure}

The model for the transverse distributions in the two fit passes is similar.
The single-beam density function is defined in terms of two Gaussian functions 
for each coordinate.
A factorizability parameter $f_j$ (for beam $j$) is used as defined in Eq.~\eqref{eq:non_factorizable_weights}. 
For the \oneD fits the values of $f_j$ ($j=1,\,2$) are fixed to 1,
decoupling the two transverse coordinates. 
Following this model, described in \sect~\ref{sec:double gaussian model}, the observed 
transverse vertex distribution per beam for a given $z$ range is fitted with the resolution-convolved width defined in Eq.~\eqref{eq:convolved_width}.
The fit parameters are, per beam $j$, the Gaussian parameters
$w_{m,j}$, $\xi_{m,j}$, $\sigma_{m\text{n},j}$ and $\sigma_{m\text{w},j}$ 
for both axes $m=x,\,y$, the factorizability parameter $f_j$, and a free amplitude $A_{j,k}$ per $z$ slice $k$.
Figure~\ref{GlobalFit_doubleGaussianSel_BgStdrun122467_bunch1909_X_1}  
shows \oneD global fit results for the first measurement of the first bunch pair of fill 2852 as example.

The next step introduces values of the factorizability parameters $f_j$ different from unity, and therefore
a combined fit coupled in the two transverse coordinates is mandatory.
The large number of parameters (18 beam parameters + 24 amplitudes) for the \twoD global fit 
requires good starting values.
The results of the \oneD global fits are used as starting values for the final global fits
except for the starting values for the factorizability parameters $f_j$, which are set to 0.5.
An example of a global fit result displaying only one $z$ slice per beam and one luminous region slice
is shown in Fig.~\ref{GlobalFit_3D_BgStdrun122476_bunch1949_0} for the first bunch and 
first measurement performed with gas injection and a $\beta^*=10\m$ lattice (fill 2853).
Evidence for a significant non-factorizability of the beam shape is discussed further in 
\sect~\ref{sec:non_fac_beams}.
The $\chi^2/$ndf, with ndf the number of degrees of freedom of the fit, is typically between 1 and 1.1 for the \twoD fit. 
The non-factorizability of the beams in the transverse coordinates can affect the overlap integral 
by up to 3\%.

\includefig{1}{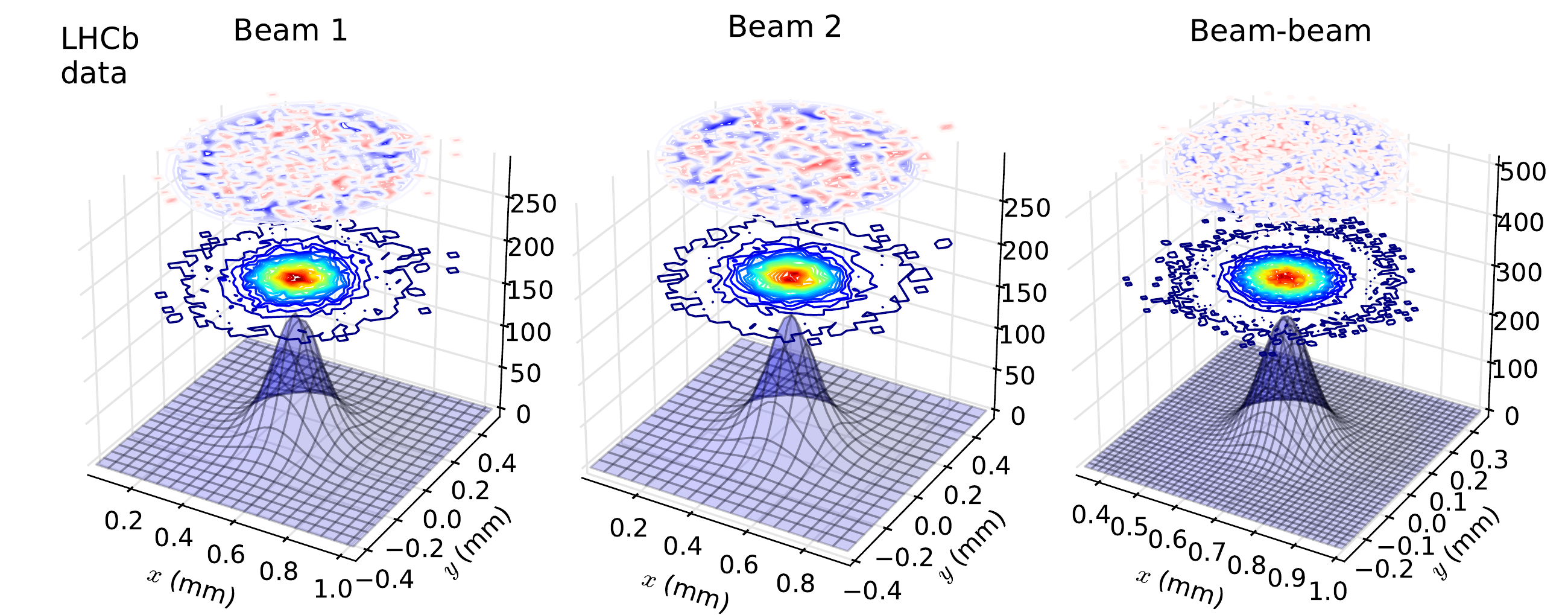}{Global fit in 2-dimensions.}{\small 
Results of the global \twoD fit for one bunch pair in fill 2853 as an example, with
(left) the central $z$ slice of {\beamone},  
(middle) the central $z$ slice (out of 21) of {\beamtwo} and
(right) the central $z$ slice of the luminous region. 
The fit result of the resolution-corrected beam shape is shown as a three-dimensional shape with amplitude indicated on the vertical scale. 
The data are displayed as a contour plot above the fit result with an arbitrary colour scale with red indicating the maximum. 
The pulls of the fit are given on the top with a colour scale ranging from $-3$ (dark red) to $+3$ (dark blue). 
}

Finally, to be able to calculate the overlap integral from Eq.~\eqref{eq:ovrlap_compact},
the convolved bunch length $\sigma_z$ and the longitudinal position $\zrf$, %
where the bunch centres coincide longitudinally, must be known.
The transverse offsets $\Dm$ of Eq.~\eqref{eq:ovrlap_compact} have to be evaluated at
$\zrf$, which is defined by the LHC RF phase.
The values of the parameters $\sigma_z$ and $\zrf$ can be obtained from an analysis of the  
longitudinal vertex distribution of beam-beam interactions using the relations given in 
Eqs.~\eqref{eq:zrf_muzl_deltazl} and~\eqref{eq:sigma_zl}.
A fit is performed to the luminous region distribution of beam-beam interaction vertices 
for the same data sets as the transverse fits.
Following Eqs.~\eqref{eq:bb_density} and~\eqref{eq:factorizable_density},
the luminous region distribution $\rho_{bb}(z)$ is represented by the sum of sixteen Gaussian contributions. 
Complete factorization of the $z$ dependence of the bunch distribution is assumed.
The uncertainty introduced by this assumption is discussed further in \sect~\ref{sec:bgi_results}.
Under this assumption each contribution to the luminous region has a length, 
amplitude and longitudinal offset, which depend only on the transverse single beam parameters and the quantities 
$\zrf$, the combination $\sigma_z^2 = \sigma_{z1}^2+\sigma_{z2}^2$ in Eq.~\eqref{eq:lumi_variable}
and an arbitrary overall amplitude parameter $A_l$.  %
The quantities $\zrf$ and $\sigma_{z1}^2+\sigma_{z2}^2$ are common to all Gaussian contributions.
The relative weights of the sixteen contributions follow from Eq.~\eqref{eq:bb_overlap_general} 
according to the fraction of luminosity they carry and do not introduce new parameters.
Thus, the longitudinal distribution of the luminous region is fitted with only three free parameters
and makes use of transverse fit parameters fixed by the global transverse fit.
Because the reconstruction efficiency in the VELO is not constant over the 
full luminous region, the $z$ distribution is corrected for the efficiency obtained from simulation.

At this point, all parameters are measured and the overlap integral
can be calculated following Eqs.~\eqref{eq:ovrlap_compact} and \eqref{eq:lumi_variable}.
The statistical uncertainty is evaluated by sampling the multivariate normal distributions of 
the parameters using the fit results as mean values and the covariance matrix provided by the last fitting step.
The resulting statistical uncertainty on the overlap integral evaluated per bunch pair and in 20 minute periods
is typically less than 0.5\%. %

\subsection{Generic simulation}
\label{sec:generic_mc}

The BGI method relies on an accurate beam shape description.
To test whether the fitting procedure described in the previous section gives unbiased 
results, simulated data sets are created with a Monte Carlo method.
The development of the more complex \twoD fit model described in \sect~\ref{sec:double gaussian model} 
was motivated by the possible non-factorizability of the beams in the $x$ and $y$ directions and the
observation that the \twoD properties of the beams could, in principle, be measured with beam-gas interactions.
This capability was first tested with simulated data.
Results showing evidence for beam non-factorizability are presented in \sect~\ref{sec:non_fac_beams}.

Datasets of simulated vertices are generated for single beams and the luminous region as follows.
Single beam vertex positions $v_m$ (in the axis $m=x,\,y$) are generated by sampling 
Eq.~\eqref{eq:factorizable_density} at a fixed $z=0$ position for both beams.
A random $z$ position $v_{z,j}$ is then assigned to each vertex in the range $-1000<z<-250\mm$ 
for {\beamone} and $250<z<1000\mm$ for {\beamtwo}.
The transverse coordinates of the vertices are translated to $v_{z,j}$ according to the beam direction.
The $z$ dependence of the reconstruction efficiency is implemented with a linear reduction of
vertices as function of $z$.
Per simulated dataset, about $5\times10^4$ vertices are generated per beam, similar to the number of events 
acquired per bunch pair during 20 minutes of data taking.

\begin{figure}[tbp]
\centering
\begin{minipage}[c]{0.5\textwidth}
\includegraphics[width=\textwidth]{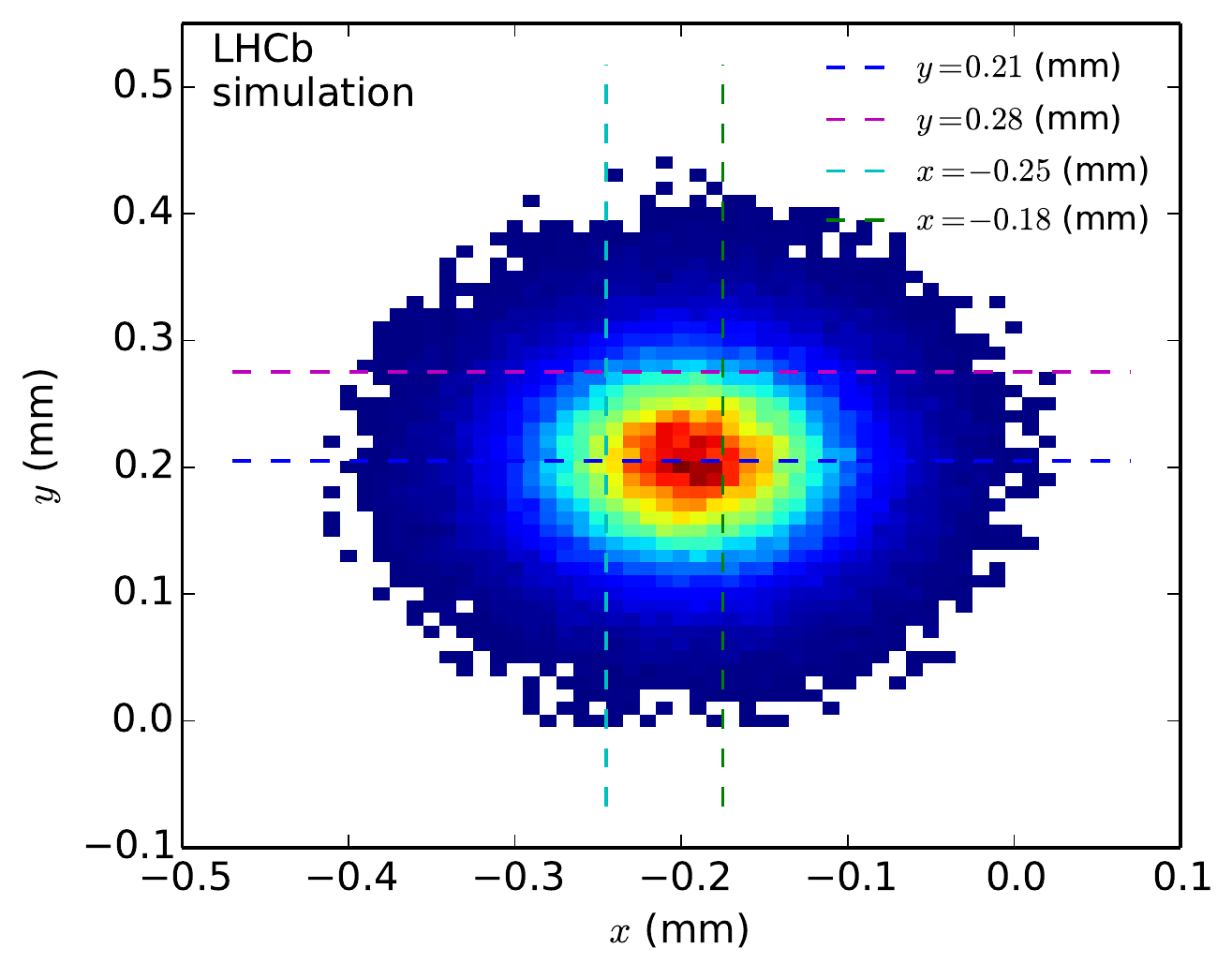}
\end{minipage}\hfill
\begin{minipage}[c]{0.5\textwidth}
\includegraphics[width=\textwidth]{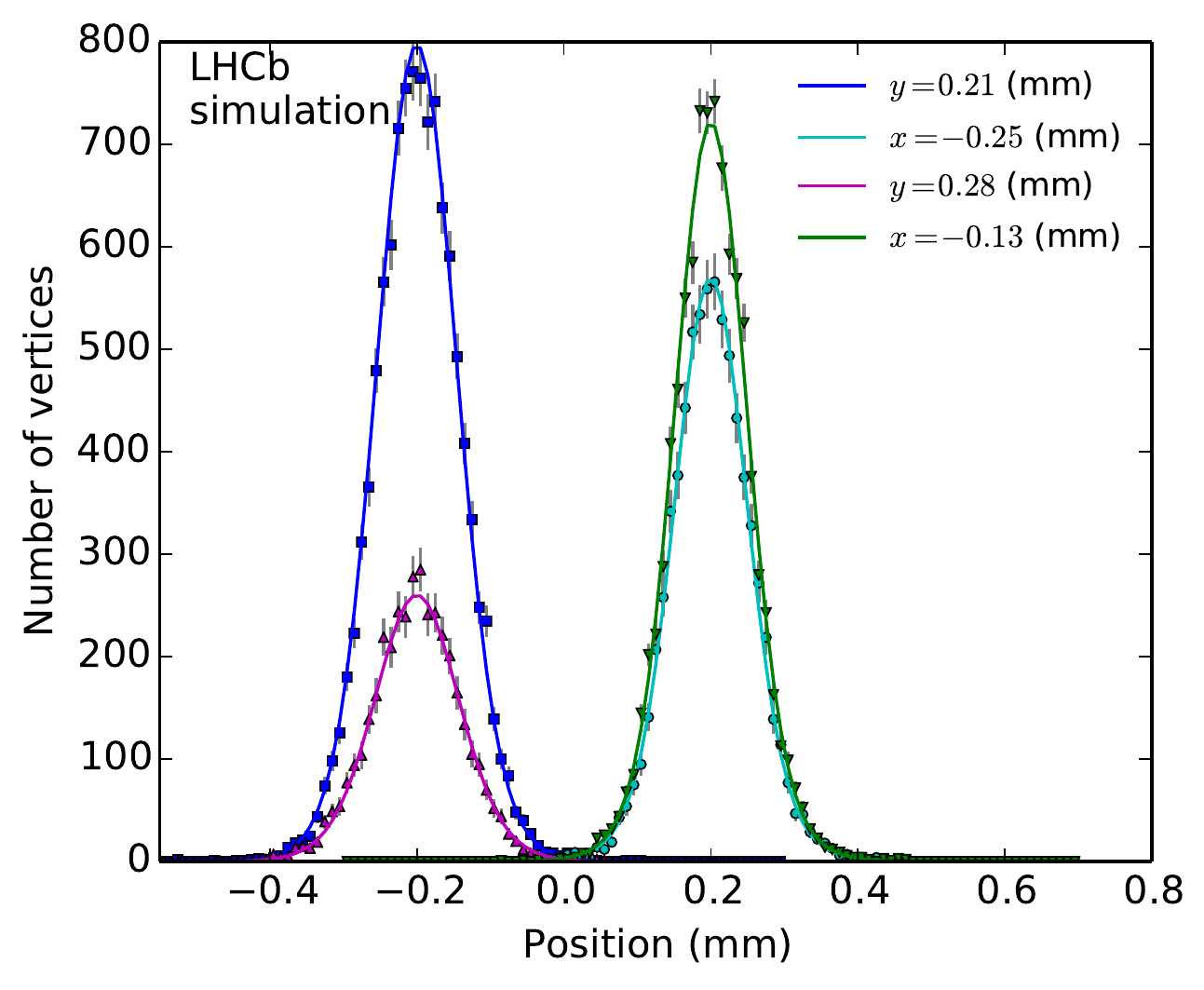}
\end{minipage}\hfill
\begin{minipage}[c]{1\linewidth}
\caption[Fit model test on simulated data with realistic beam parameters]{\small 
Test of the fit model on simulated data with 
(left) the transverse view of vertex density of the central $z$ slice at $z=0$ and  
(right) the comparison of simulated data with predictions for different $x$ and $y$ slices of the luminous 
region central $z$ region. 
The colour scale is relative with red indicating the highest number of vertices per bin. 
The dashed lines indicate the $x$ and $y$ slices in the distribution shown on the right. 
The markers indicate the simulated data while solid lines show the prediction.
Both simulated data and predictions are convolved with the resolution.
The beam parameters are simulated with values similar to those observed in the data.
}
\label{mc_example2}
\end{minipage}
\end{figure}

The shape of the luminous region and the overlap integral are sampled over space and time using
the single beam transverse distributions $\rho_j$ ($j=1,2$) 
defined in \sect~\ref{sec:Luminosity formalism for colliding beams}
and the longitudinal bunch shape $\rho_{zj}$, which is assumed to be a single Gaussian function. 
A sufficiently large sampling volume $S_{\text{vol}}=\Delta x\,\Delta y\,\Delta z\,\Delta t$ is chosen
($\pm0.8\mm$ in $x$ and $y$, $\pm250\mm$ in $z$ and $\pm1.2\ns$ in $t$).
\def\Abb{A_{bb}}  
A number $N_s$ of random samples of $(x$, $y$, $z$, $t)\,\in S_{\text{vol}}$ are generated uniformly over the volume.
Vertices are retained according to the probability density
\begin{equation}
\label{eq:2d_rhobb}
\rho_{bb}^s(x,y,z,t) = \rho_{1}(x,y,z,t)\rho_{z1}(z-ct)\,\rho_{2}(x,y,z,t)\rho_{z2}(z+ct)\, ,
\end{equation}
where the $z$ and $t$ dependence in $\rho_{1,2}$ just expresses a translation along the beam direction. 
Following the usual rejection sampling method,
a randomly sampled vertex is retained if a uniform random number $u$ assigned to it in a 
range $[0,\Abb]$ satisfies $u \le \rho_{bb}^s(x,y,z,t)$, with the arbitrary constant $\Abb\ge\text{max}(\rho_{bb})$.
The numerical value of a generated overlap integral is calculated using the fraction $N_{bb}$ 
of vertices retained compared to the total number of samples $N_s$ generated in the volume $S_{\text{vol}}$ with
\begin{equation}
\label{eq:2d_overlap_num}
\ovint  = 2c\, \frac{N_{bb}}{N_{s}} S_{\text{vol}}\, \Abb\, .
\end{equation}
Each simulated primary vertex generated with the above method is assigned a track multiplicity 
according to the distributions found in data.
Using the resolution parametrization measured with data as described in \sect~\ref{sec:resolution}, 
each vertex is assigned a measurement deviation in $x$ and $y$ by sampling a normal Gaussian distribution.

The generated datasets are stored with the same format and are processed with the same algorithms as used for the data.
The fitting algorithms are tested with different beam parameters and 
are validated with simulated datasets before being applied to the data.
An example generated with beam parameters similar to those found in the data is shown in Fig.~\ref{mc_example2}.
Detailed studies validate that the simulation input parameters can consistently be recovered
with the fitting procedure.

\subsection{Evidence of non-factorizable beam shapes}
\label{sec:non_fac_beams}

Discrepancies of the order of 3\% are observed in visible cross-section measurements
performed with the BGI method in the four July 2012 fills when fitted with a model factorizable 
in the $x$ and $y$ coordinates.
Since the beam-gas interaction vertices provide a complete transverse view of the beams,
the factorizability hypothesis can be tested.

\begin{figure}[p]
\centering
\includegraphics[width=\textwidth]{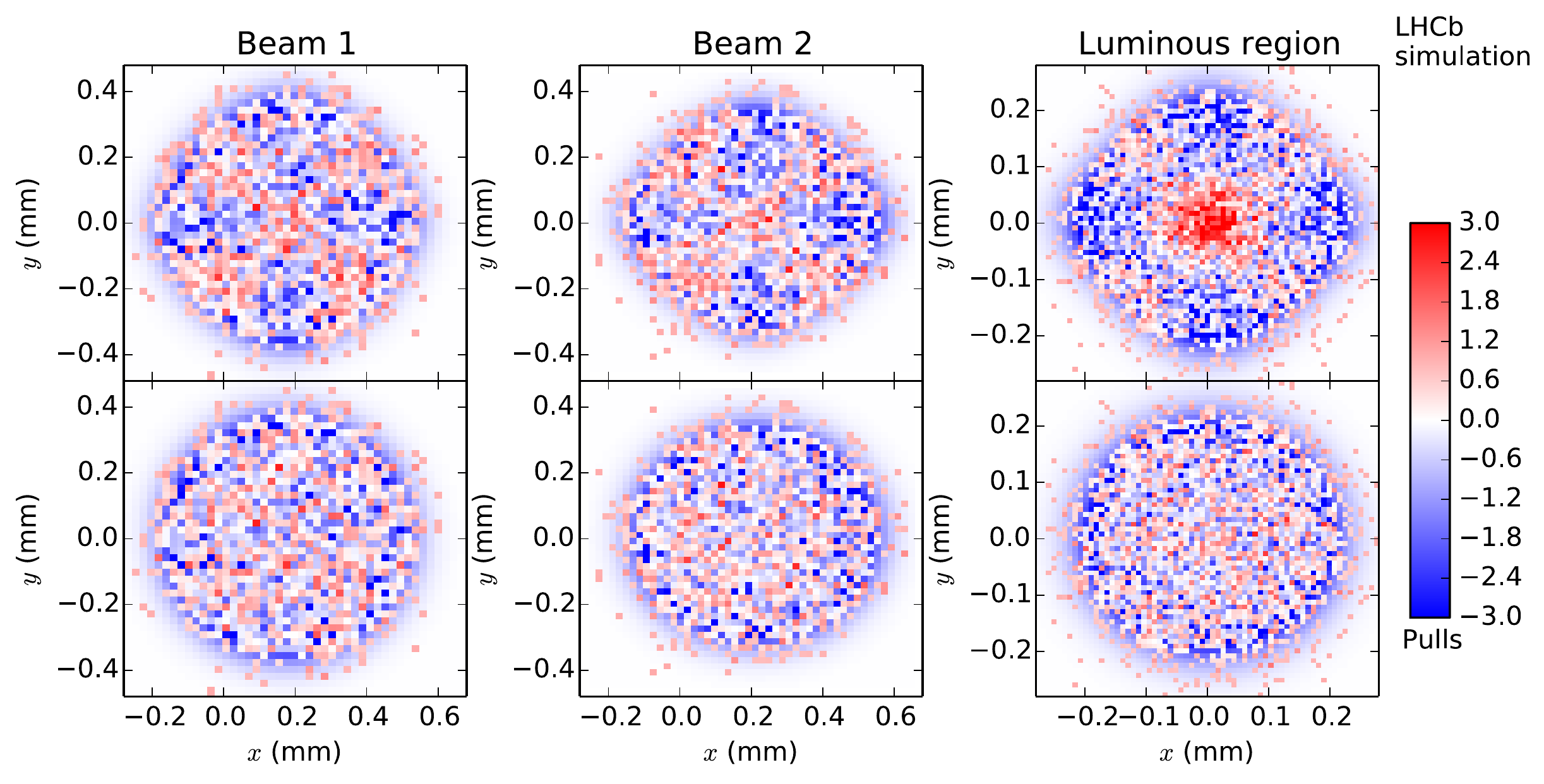}
\includegraphics[width=\textwidth]{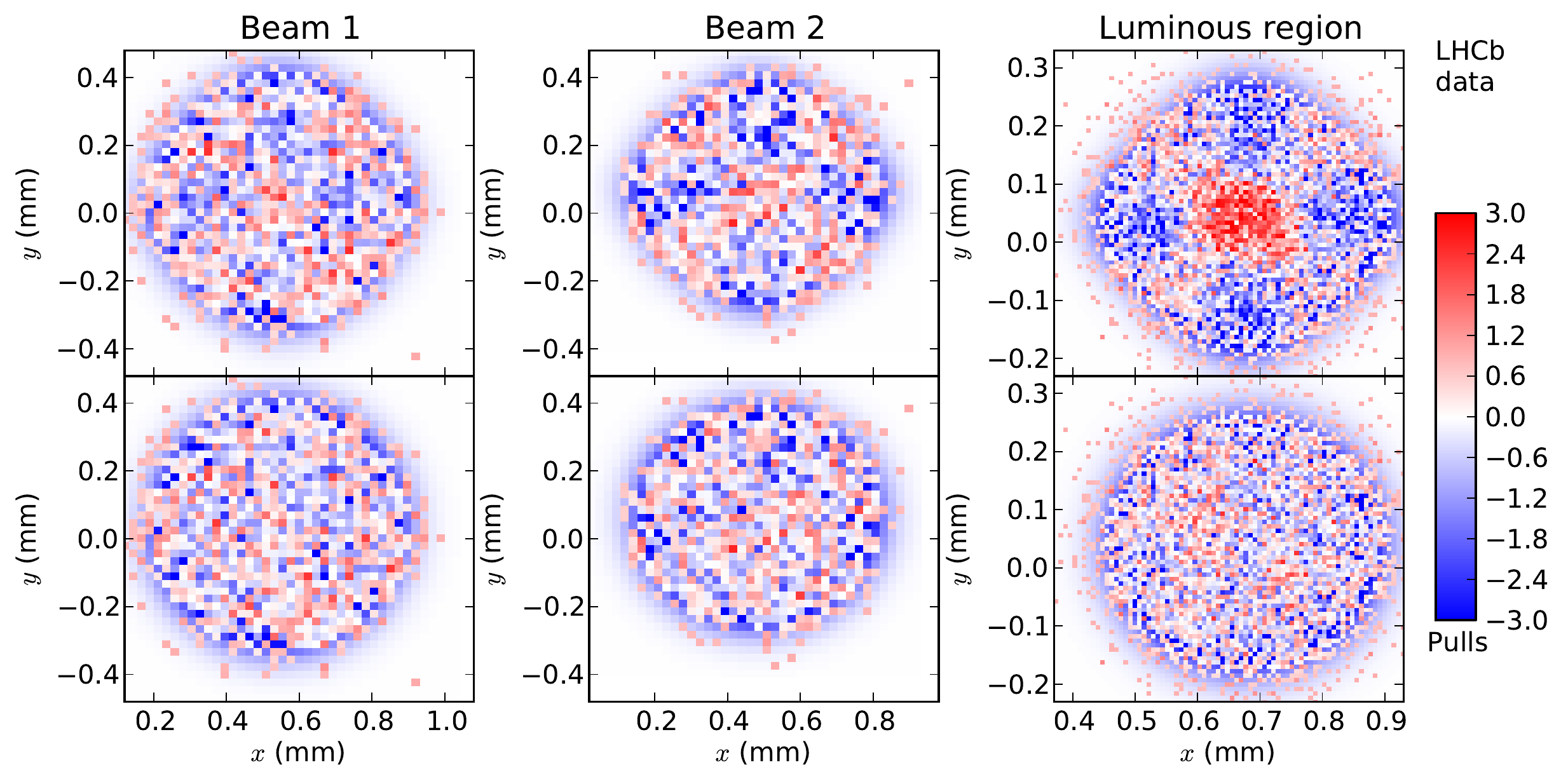} 
\caption{\small %
Measurement of transverse beam shape factorizability.
The upper six graphs are the results of a simulation.
From left to right: fit pulls of a beam $z$ slice for {\beamone}, {\beamtwo} and luminous region.
The $z$ slices [$-$500, $-$250]\mm for \mbox{{\beamone}}, [250, 500]\mm 
for \mbox{{\beamtwo}} and [$-5$, 5]\mm for beam-beam are shown here out of the 24 $z$ slices.
The beams are generated assuming non-factorizability, $f_{1,2}=0$; the same dataset is used in 
both rows but the data are fitted with two different models.
Top row: the fit assumes factorizable beams ($f_{1,2}=1$), which is equivalent to the \oneD model.
Bottom row: fit with the additional beam factorizability parameters $f_{1,2}$.
The fit converges to the correct parameter valuess $f_{1,2}=0$.
The lower six graphs are the results of a fit to data (fill 2855, BCID 1335), which converges 
to non-factorizable beam shapes in this example.
\\
\\
}
\label{GlobalFit_2D_pulls_beams_and_lumi}
\end{figure}

A set of simulated data samples with non-factorizable beams ($f_{1,2}=0$) is fitted with the \twoD global fit model. 
A factorizable version of the model is obtained by fixing the factorizability parameters at unity.
A fit with these parameters left free can describe also non-factorizable distributions.
The effect of the pulls showing the difference between a factorizable and non-factorizable model is
shown in Fig.~\ref{GlobalFit_2D_pulls_beams_and_lumi}.
The fit converges towards the correct value of $f_j=0$ when left free.
In addition, the ($x,y$) distribution of pulls reveals a clearly visible cross-like structure
when fitted with a factorizable model, which cannot fully describe the beam shape.
The measurement performed on one data set acquired in July 2012 is also shown 
in Fig.~\ref{GlobalFit_2D_pulls_beams_and_lumi} as an example.
The fit correctly describes the beam shapes and the fit pulls are more uniformly distributed.
The \twoD fit model converges towards non-factorizable beams and the pulls of the fit assuming a 
factorizable beam display the same structure as in the distributions simulated with non-factorizable beams.

\begin{figure}[tbp]
\centering
\includegraphics[width=0.49\textwidth]{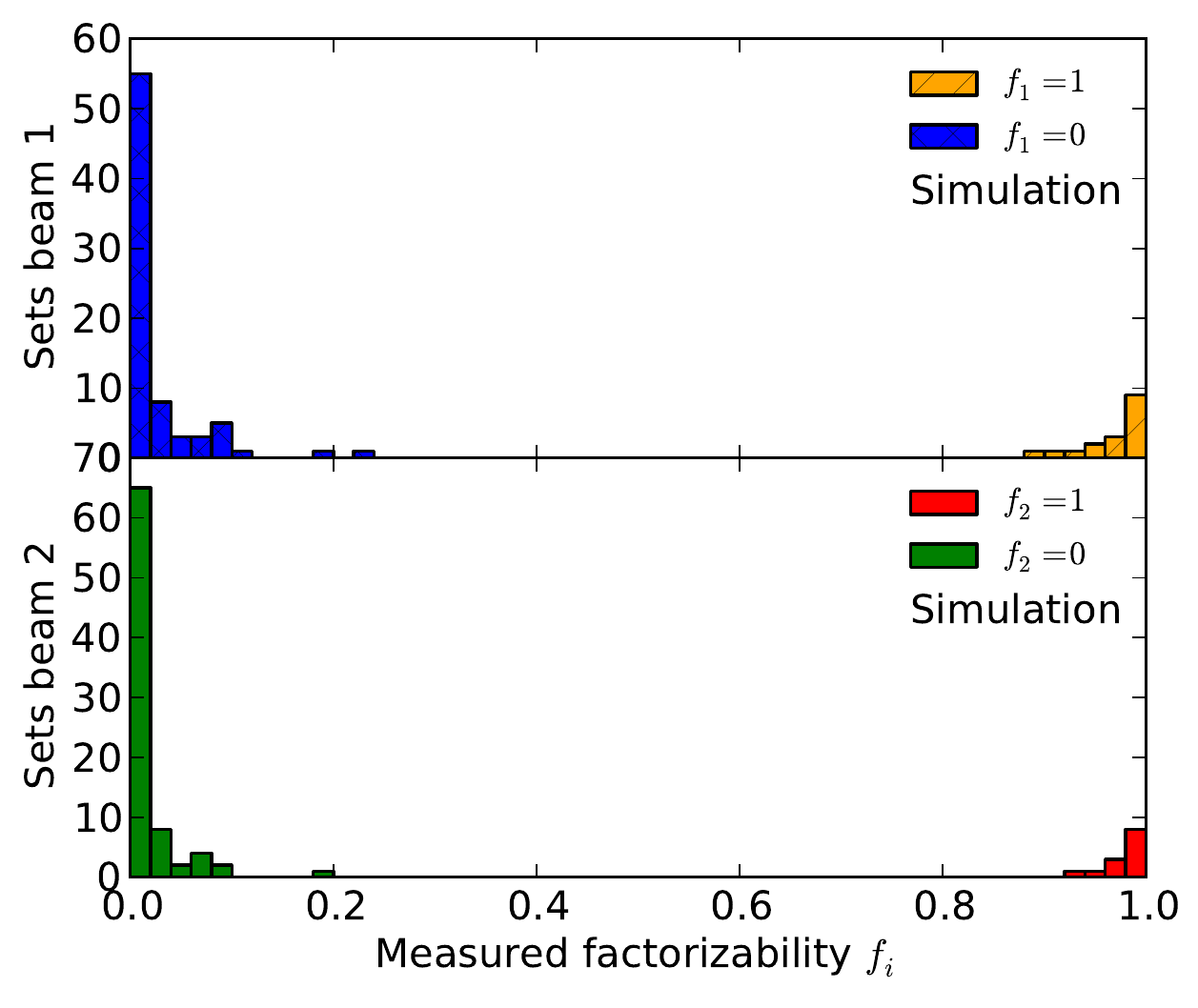}
\includegraphics[width=0.49\textwidth]{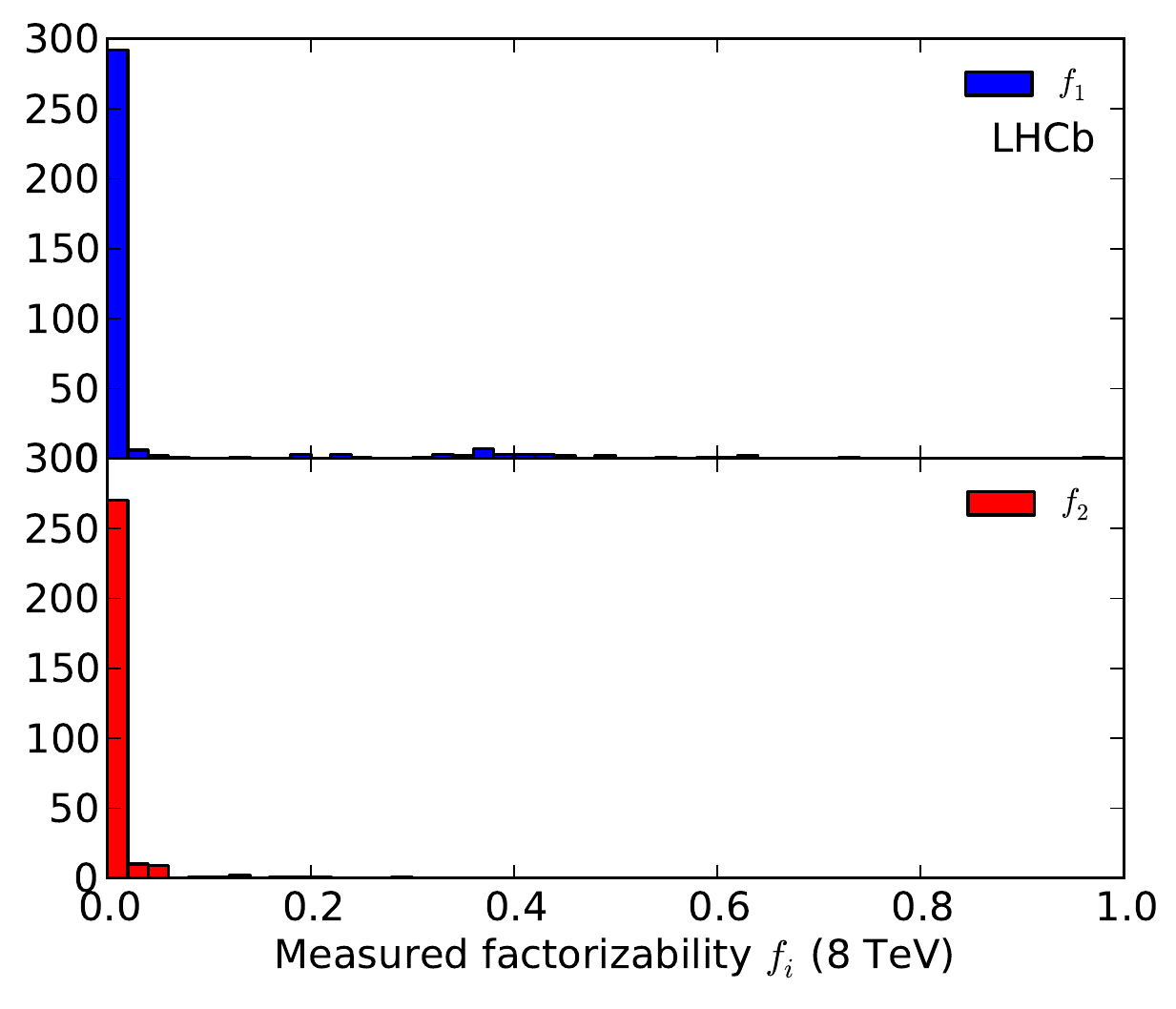}
\caption{\small Fit results for factorizability parameters $f_{1,2}$ for
(left) the simulation and 
(right) all measurements performed in 2012 at $\sqrt{s}=8\unit{TeV}$.
In the simulation, different double-Gaussian beam parameters were used and the values of $f_{1,2}$ were set to 0 or 1.
For the data, only bunch pairs with a double Gaussian strength $S_{j,m}>0.02$ and with a fit uncertainty on the 
factorizability parameter of $\delta(f_j)<1$ are displayed.
The top (bottom) panels refers to {\beamone} ({\beamtwo}). 
}
\label{fitted_fn_results}
\end{figure}

Data sets have been generated with the parameter $f_j=0$ and another set of samples with $f_j=1$.
Results for the measured factorizability parameter $f_j$ are shown in Fig.~\ref{fitted_fn_results} 
both for simulated samples and for data.
In general, the fit algorithm can reliably measure the value of $f_{1,2}$ except when the beams 
are close to a single Gaussian shape, where the parameters $f_{1,2}$ have little or no meaning.
The distribution of values of $f_j$ in the fits to the data shows a clear dominance of non-factorizable bunch shapes.
The factorizability parameter is only meaningful if the beam has a double Gaussian shape in both transverse coordinates.
If the beam shape is single Gaussian in one plane only, the beam is by definition factorizable in the model used 
and $f_j$ can not be measured.
The ability to measure $f_j$ depends thus on the ``strength'' of the double Gaussian shape of the beams defined here as
\begin{equation}
\label{eq:sjm}
S_{j,m} = 1-\frac{\sigma_{{\rm main},m,j}}{\sigma_{{\rm rms},m,j}}\, ,
\end{equation}
with 
\begin{equation}
\sigma_{{\rm rms},m,j}^2 = w_{m,j}\sigma_{m\text{n},j}^2+(1-w_{m,j})\sigma_{m\text{w},j}^2
\end{equation}
for beam $j=1,2$ and plane $m=x,\,y$.
The indices n and w denote the narrow and wide width Gaussian component, respectively,
while the width $\sigma_{{\rm main}}$ is the width of the Gaussian (n or w) which carries the largest weight.
A single Gaussian shape has therefore a vanishing strength parameter.

With few exceptions, all bunch pairs measured in April and July 2012 have a double Gaussian shape 
and $S_{j,m}$ is significantly larger than zero for both beams and planes.
The November fills 3311 and 3316 are clearly different from the earlier fills,
as all bunches have $S_{j,m}$ values smaller than 0.04.

\subsection{Results and systematic uncertainties}
\label{sec:bgi_results}

\begin{figure}[btp]
\vspace{-3mm}
\centering
\begin{minipage}[c]{0.5\textwidth}
\includegraphics[width=\textwidth]{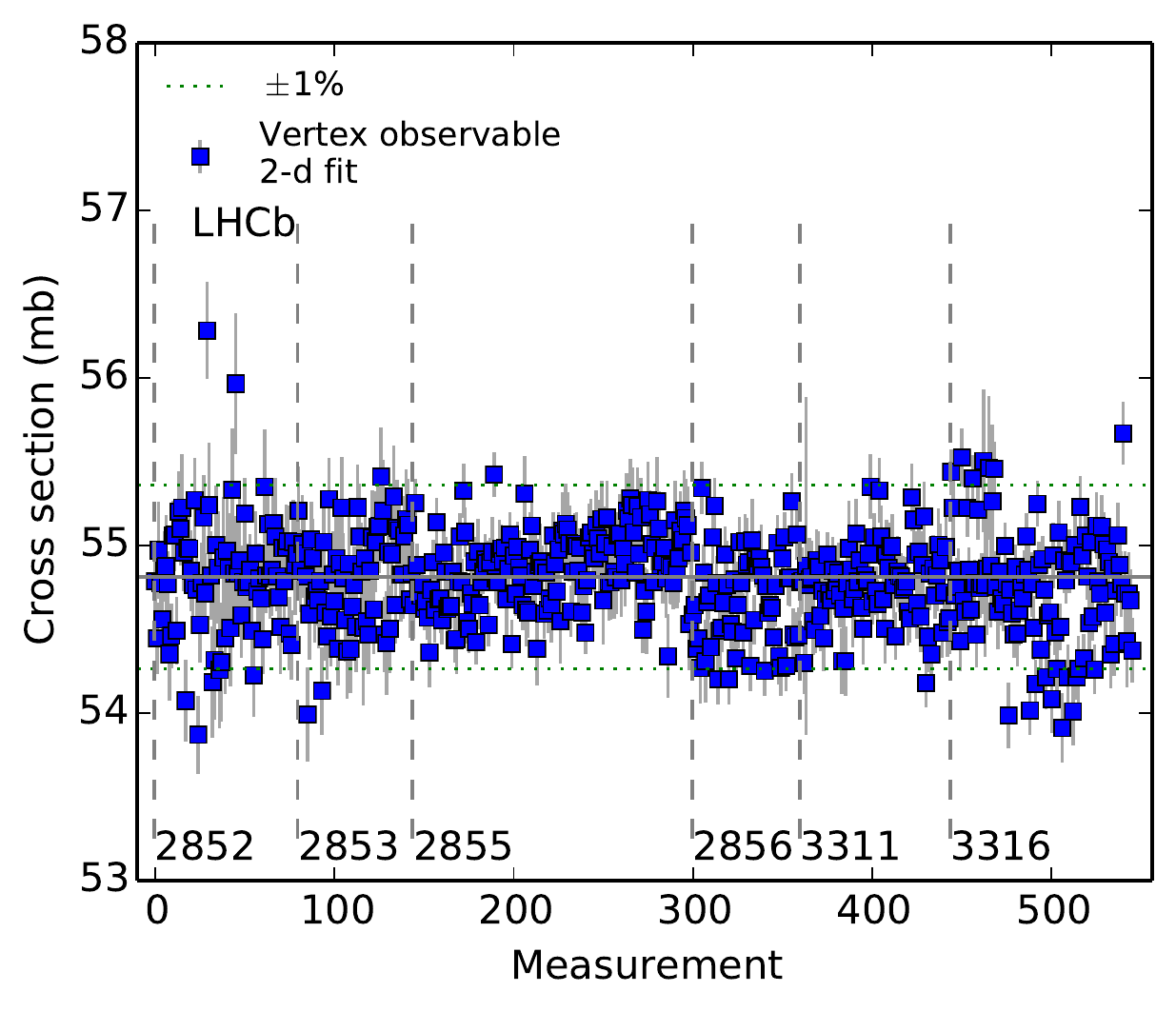}
\end{minipage}\hfill
\begin{minipage}[c]{0.5\textwidth}
\includegraphics[width=\textwidth]{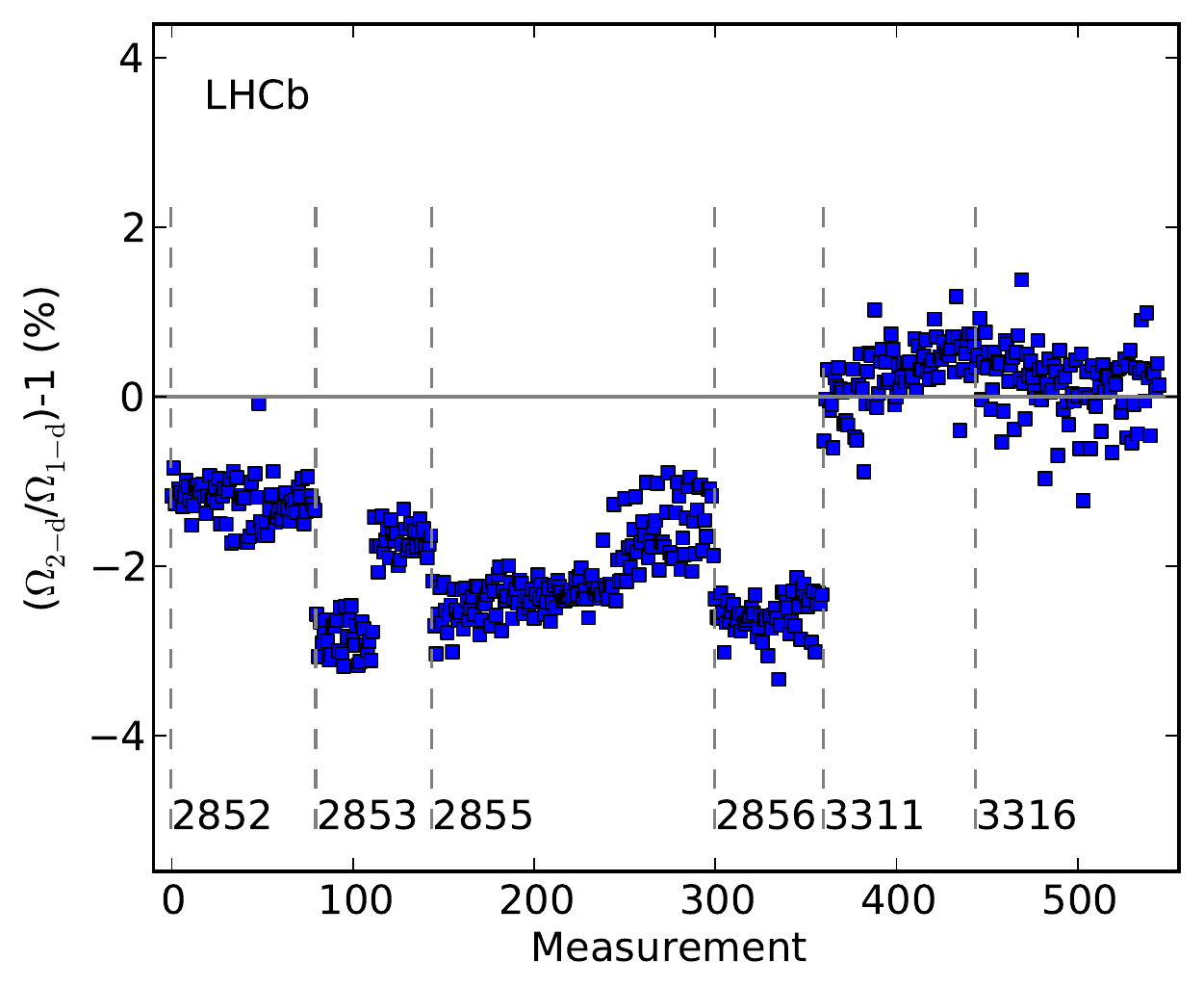}
\end{minipage}\hfill
\begin{minipage}[c]{1\linewidth}
\caption[Cross-section results for head-on beams with $\beta^*=10$ m at $\sqrt{s}=8$~TeV.]{\small Cross-section results 
for head-on beams with $\beta^*=10$~m at $\sqrt{s}=8$~TeV using the \emph{Vertex} observable,
(left) the \twoD fit and (right) the relative difference between the \twoD and \oneD fit.
Each data point is a cross-section measurement from a colliding bunch pair using integrated data over about 20 minutes.
The measurements are sorted by time and BCID.
The fills are indicated in the figure and are separated by dashed vertical lines.
Two dotted horizontal lines indicate a $\pm 1\%$ deviation from the central value.
The error bars show the statistical uncertainties of the overlap integral.
}
\label{bgi_results_perbunch}
\end{minipage}
\end{figure}

Instantaneous luminosity values for each colliding bunch pair are evaluated using
Eq.~\eqref{eq:cross-section} and \eqref{eq:luminosity}
with the overlap integral $\ovint$ measured with data integrated over intervals of about 20 minutes.
The luminosity measurements per colliding bunch pair are used to evaluate the visible cross-section 
for specific observables (\sect~\ref{sec:InteractionRate}) with
\begin{equation}
\sigeff{\eff} =\frac{\mueff{\eff}\,\frev}{\instL}=\frac{\mueff{\eff}}{N_1 N_2\,\ovint}\, ,
\end{equation}
where $\mueff{\eff}$ is the visible average interaction rate for the reference observable. 
In the BGI measurement, the \emph{Vertex} observable is chosen as reference owing to its time stability and low background.
Cross-section results for the \emph{Vertex} observable for all dedicated luminosity calibration fills in 2012 
at $\sqrt{s}=8\tev$ with $\beta^*=10\m$ and nominally head-on beams are shown in Fig.~\ref{bgi_results_perbunch}, 
together with a comparison between \oneD and \twoD fits of the transverse bunch properties.
One observes that the statistical uncertainties are small and that the results using the \twoD fits are consistent between fills.
Comparing results from both fit methods shows the importance of measuring the shapes in two dimensions
to take the non-factorizability of the description between the two transverse coordinates into account;
the \oneD fits do not display the same consistency.
One notes also that the difference between the \oneD and \twoD fit results ranges
between 1 and 3\% in the July fills, for which the bunches clearly had double 
Gaussian shapes, while it is $<1\%$ in the November fills, for which the bunches 
were almost single Gaussian (see \sect~\ref{sec:non_fac_beams}).

\begin{figure}[tbp]
\vspace{-1mm}
\centering
\begin{minipage}[c]{0.5\textwidth}
\includegraphics[width=\textwidth]{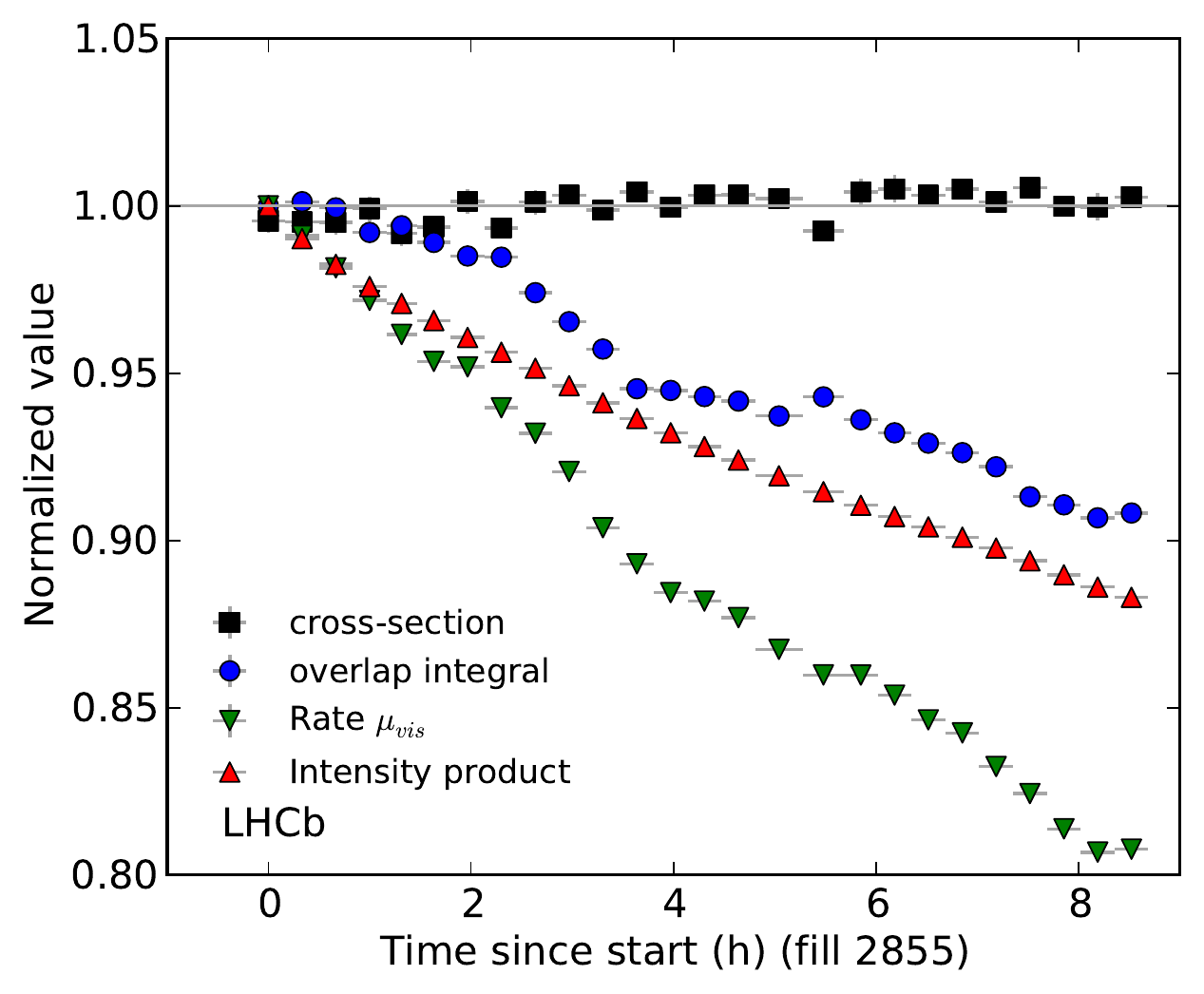}
\end{minipage}\hfill
\begin{minipage}[c]{0.5\textwidth}
\includegraphics[width=\textwidth]{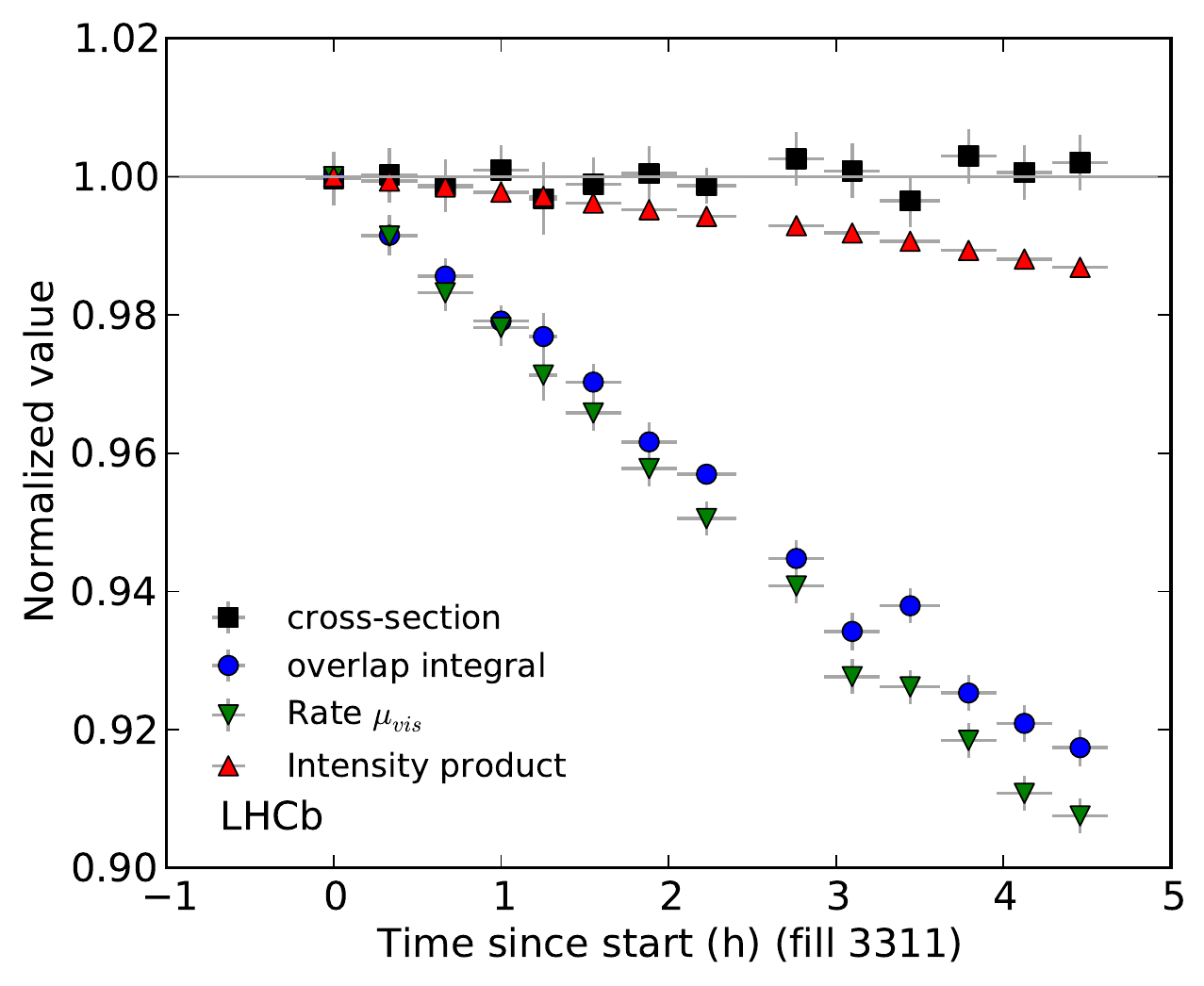}
\end{minipage}\hfill
\begin{minipage}[c]{1\linewidth}
\caption[Evolution of normalized cross-section and related values for a colliding bunch pair]{\small Evolution 
of normalized cross-section and related values for a colliding bunch pair for  
BCID 1335 of (left) fill 2855 and (right) of fill 3311.
The cross-section values are normalized to their average while other parameters are normalized to 
the first data point. Fill 2855 lasted more than eight hours with a beam intensity product decrease of about 10\%.
Fill 3311 lasted about 4h30 with a beam intensity product decrease of less than 2\%; in this 
fill the luminosity reduction is mostly caused by emittance growth.}
\label{fig_bgi_evolution}
\end{minipage}
\end{figure}

The bunch shapes change over the course of a fill due to emittance growth and other factors, 
such as beam-beam effects or beam position drifts. %
Any change in the beam shape influences the overlap integral.
Furthermore, the bunch population product decays over time, reducing the luminosity.
In contrast, the cross-section is a physical observable and must be stable over time.
The consistency of the measured cross-section, together with the corresponding overlap integral, 
rate and bunch population product is shown in Fig.~\ref{fig_bgi_evolution} for one colliding bunch pair 
for two different fills.
While the intensity product decay is typically smooth, the rate fluctuations follow the variations 
of the overlap integral.
The figure also shows that the variation of the overlap integral in adjacent 20 minute intervals 
is very small (much less than a percent), indicating that the effect of \eg emittance growth 
and beam drifts during the short intervals is negligible.
In the following, sources of systematic uncertainties and their effect on the measurement precision will be described.

\label{sec:uncert_bb_res}
The impact of the beam-beam interaction vertex resolution on the cross-section depends on the transverse size of the luminous region.
Comparing the results obtained with different ratios of the resolution to this transverse size 
permits evaluating the corresponding systematic uncertainty.
The measurement of the width of the luminous region can be biased by \eg a misalignment in the 
VELO sensors, which would correlate the position of the sub-vertices 
and broaden the beam without being detected by the resolution measurement.
Considering head-on beams by setting $\Dx=\Dy=0$ in Eq.~\eqref{eq:ovrlap_compact}, one sees that
$\ovint\propto ({\Sigma_x\Sigma_y})^{-1}$.
Assuming similar size for both beams in each transverse coordinate, one can derive 
from the formalism for pure Gaussian beams that the quantity
\begin{equation}
\label{eq:beambeamR}
R= \frac{2\sigma^2_{\text{res},l x}}{4\sigma'^2_{l x}+\sigma^2_{z}\sin^2\phi_x}
  +\frac{\sigma^2_{\text{res},l y}}{2\sigma'^2_{l y}}
\end{equation}
is representative of the importance of the beam-beam resolution in the cross-section measurement.
Here, the $\sigma'_{l m}$ represent the measured values of $\sigma_{l m}$
and $\sigma_{\text{res},l m}$ the beam-beam resolutions ($m=x,\,y$).
A value of $R=0$ means that the resolution is negligible compared to the beam size.
Cross-section measurements with different $R$ values are obtained by using different cuts on 
the vertex track multiplicity and by using data acquired with a $\beta^*$ value of 3 and 10\m.
This procedure gives four sets of results with different values of $R$.

\includefig{0.75}{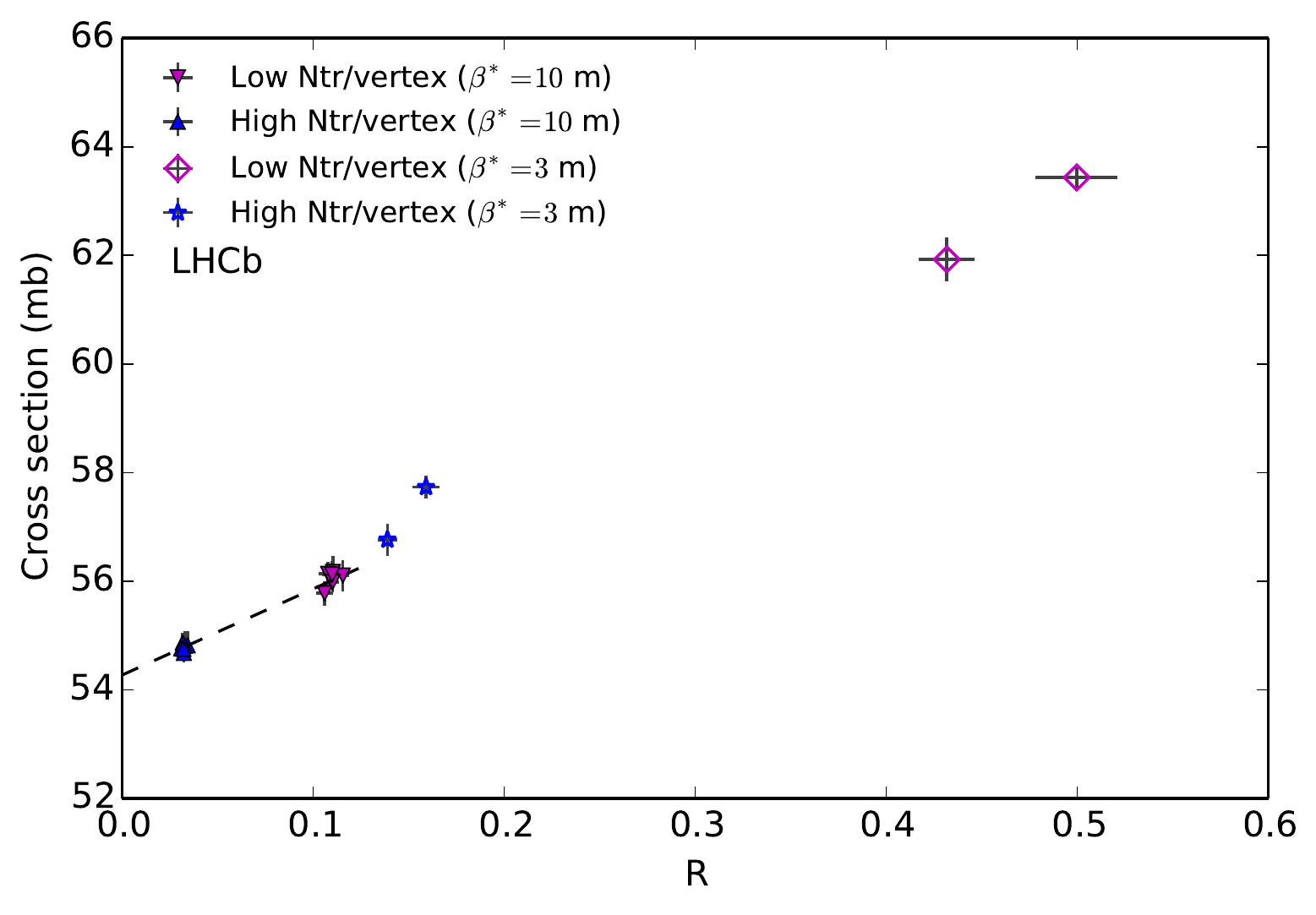}{\small Dependence of the cross-section 
measurement on beam-beam resolution compared to beam width.}{Illustration of the evaluation of the 
systematic uncertainty due to the beam-beam resolution.
Measured visible cross-section values for different event samples are shown as a function of $R$.
The $R$ parameter is calculated with Eq.~\eqref{eq:beambeamR}.
Each data point is an average of all measurements from a fill.
The error bars are the \rms of the cross-section and $R$ values per fill.
Plain markers are measurements performed with $\beta^*=10\m$ (6 fills) and open markers are performed 
with $\beta^*=3\m$ (2 fills).
The samples with $R \approx 0.5$ are selected to enhance the effect of the resolution.
The dashed line visualizes the determination of the systematic uncertainty as described in the text.
}

The results for the six fills with $\beta^*=10\m$ and two fills with $\beta^*=3\m$ are combined 
in Fig.~\ref{cross_sections_8TeV_cs_bbresol_lowhigh_fill_lc14cBgStdRF}.
Measurements performed with the larger $\beta^*$ and the best resolution (high track multiplicity cut) 
provide the smallest $R$ value, while a worse resolution combined with the smaller $\beta^*$ results 
in a larger $R$ value.
The clear correlation between the cross-section and the $R$ value visible in 
Fig.~\ref{cross_sections_8TeV_cs_bbresol_lowhigh_fill_lc14cBgStdRF} is an indication that the resolution is not perfectly understood.
The cross-section obtained for similar $R$ values and different $\beta^*$ are similar.
This shows that the difference between the measurements at $\beta^*=3\m$ and $\beta^*=10\m$ can be attributed to the resolution description.
The leftmost group of data points are the results presented in Fig.~\ref{bgi_results_perbunch}.
Those results are obtained with the high track multiplicity cut and with a $\beta^*=10\m$ beam optics.
This combination provides the best measurement conditions and those measurements, called here $\sigma_c$ 
are used as central value for the final results.
The cross-section $\sigma_e$ obtained by extrapolating the cross-section to $R=0$ based on the
$\beta^*=10\m$ fills with the low and high track multiplicity cuts (blue and magenta leftmost
 measurement groups) is used to evaluate the uncertainty due to the beam-beam resolution.
The difference of $\Delta\sigma=\sigma_c-\sigma_e = 0.93\%$ between the central value $\sigma_c$ and
the extrapolated value $\sigma_e$ is taken as systematic uncertainty. 
This is the largest single contribution to the uncertainty of the cross-section result. 
\label{sec:sys_beam_gas_resol}
As discussed in \sect~\ref{sec:resolution}, a set of correction factors to the beam-gas
interaction vertex resolution have been determined to reach consistent beam width measurements at
different $z$ positions.
The necessity to include correction factors is an indication of additional systematic uncertainties.
The overlap integrals have also been evaluated without the resolution correction factors.
The full difference between the results with and without correction factors amounts to 0.55\%,
which is used as the corresponding systematic uncertainty.

\label{sec:sys_angle_align}

A misalignment of the VELO detector can correlate the positions of the sub-vertices and degrade the vertex resolution.
This effect is included in the systematic uncertainties assigned to the resolution as described above. 
However, detector misalignment can also affect the crossing-angle measurement.
Various versions of the detector alignment are produced, all of which provide acceptable and comparable 
results in the overall alignment quality.
The different alignment versions arise from introducing different sets of constraints 
to satisfy internal consistency checks.
The same data set is reconstructed with all alignment versions and measurements of the half crossing angle are performed.
In the $x$ direction, differences of the order of 10 \textmu rad are found, an order of magnitude
larger than the statistical uncertainty.
The crossing angle uncertainty in the $y$ axis is about 3 \textmu rad and has a negligible impact on the luminosity.
The luminosity uncertainty from the crossing angle correction depends on the bunch width and length 
and is different for each bunch pair. 
The full range of 0.45\% in the various measured cross-sections for different alignment versions is taken as systematic 
uncertainty related to this source.
The different alignment versions have negligible impact on the bunch shape measurements.

\label{sec:sys_fit_model}
The bunch shape model defined in \sect~\ref{sec:double gaussian model} can describe all observed bunch shapes with a $\chi^2/$ndf close to one.
Nevertheless, the accuracy of the fit procedure as well as the capacity to describe different shapes is verified.
The simulation (\sect~\ref{sec:generic_mc}) is used to generate datasets of vertex distributions with different input parameters in order to test the capability to reproduce the input values by the fit procedure.
In addition, the systematic uncertainty arising from fitting a double Gaussian model to different simulated shapes is estimated.
Single, double or triple Gaussian shapes, or a ``Super Gaussian shape''\footnote{A generalized 
Gaussian shape with a power parameter different from 2.}~\cite{decker_1994}, are tried.

A given set of input parameters is used to generate a sample of statistically independent datasets, 
each of which is analysed using the same algorithms as for real data.
All shapes give a $\chi^2/$ndf close to unity and can be considered to be well described by the fit model.
The results of the measurements of the simulated data are compared to the input parameters on the 
basis of the value of the overlap integral. 
The difference in the results indicates a 0.5\%  systematic uncertainty due to the fit model and accuracy of the fitting procedure.
The double Gaussian fit model used to describe the transverse bunch shapes does not allow a description of a possible third component.
For example, a fraction of the protons measured in a bunch could be present in a wider 
Gaussian shape that would not be measured with a double Gaussian function.
In this case, the tails of the measured distributions would have a larger population fraction than expected.
The fraction of vertices in the tails beyond the double-sided 99 percentile predicted by the fit,
is checked for all measurements with $\beta^*=10$ m and for the simulated datasets.
The tail population is about 2\% for the single beams and about 1.5\% for the luminous region (while 1\% is expected).
This tail population, however, is also observed in the simulated datasets,
indicating that the higher tail population is a result of the fitting procedure. 
Therefore, the corresponding bias is already included in the fit model uncertainty given above, 
and no additional systematic uncertainty is assigned.

\begin{figure}[tbp]
\centering
\begin{minipage}[c]{0.48\textwidth}
\includegraphics[width=\textwidth]{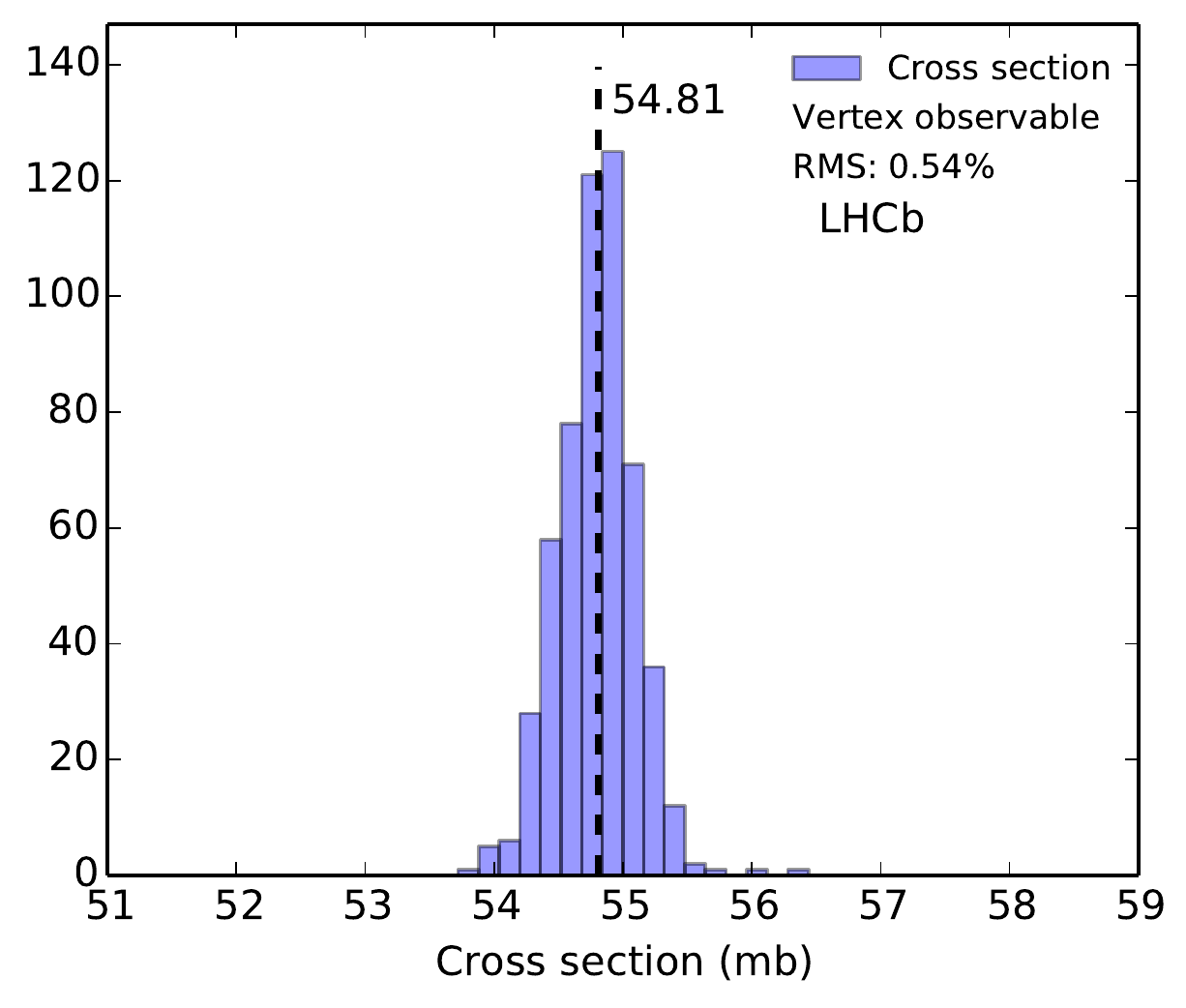}
\end{minipage}\hfill
\begin{minipage}[c]{0.52\textwidth}
\includegraphics[width=\textwidth]{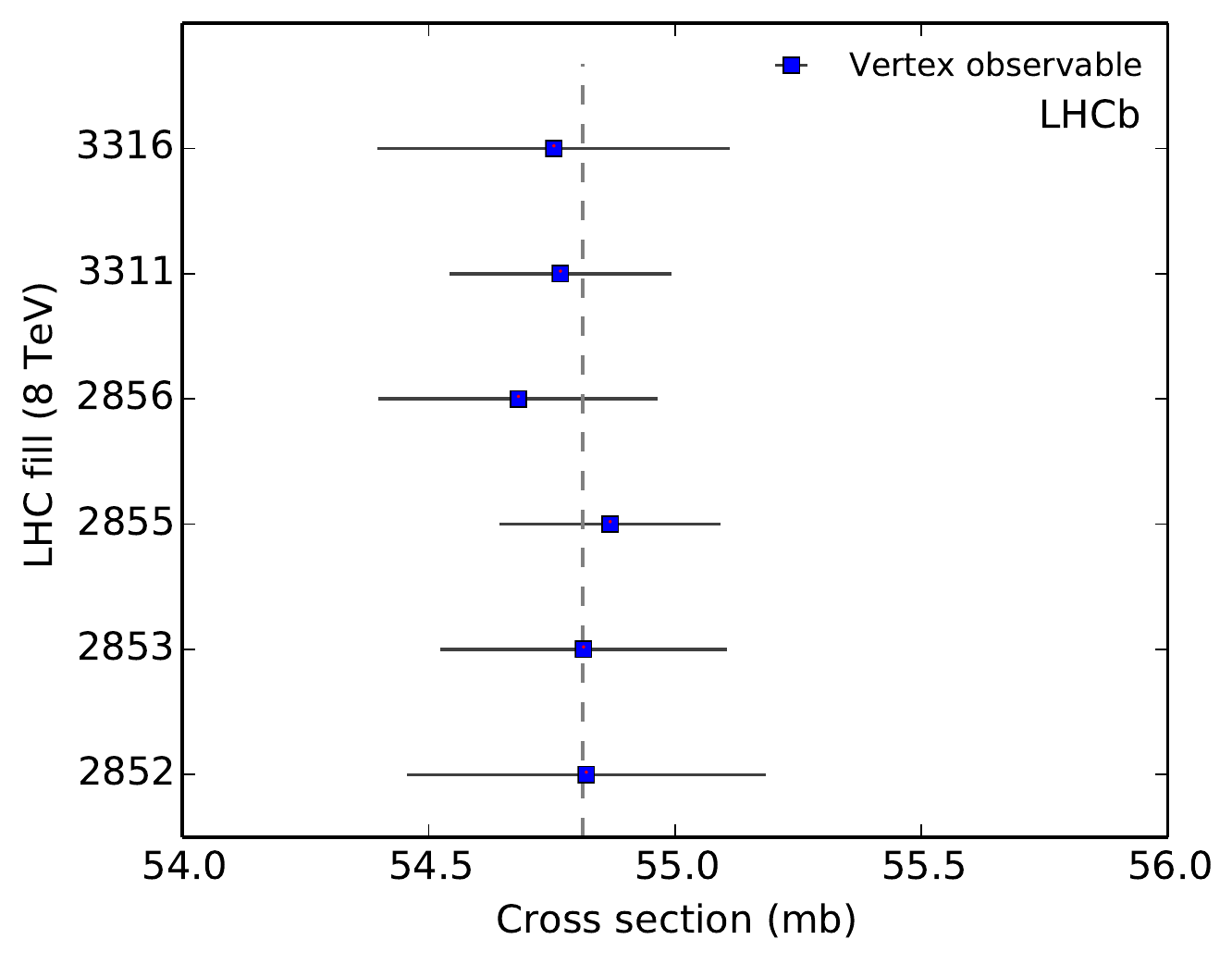}
\end{minipage}\hfill
\begin{minipage}[c]{1\linewidth}
\caption[Cross-section overview for head-on periods with $\beta^*=10\m$ fills at $\sqrt{s}=8$~TeV.]{\small Overview of 
cross-section measurements for head-on periods in fills with $\beta^*=10\m$ at $\sqrt{s}=8\unit{TeV}$ with
(left) the histogram combining all data from Fig.~\ref{bgi_results_perbunch} and
(right) the central value of cross-sections averaged per fill.
The dashed vertical line indicates the median value of all measurements.
The error bars indicate the standard deviation of all measurements from a fill.
The statistical uncertainty of each average is smaller than the marker size.}
\label{bgi_results}
\end{minipage}
\end{figure}

\label{sec:sys_spread}
Measurements of cross-sections for all colliding bunch pairs in the 
fills of 2012 with $\beta^*=10$~m are shown in Fig.~\ref{bgi_results}.
The measurements use the \emph{Vertex} observable as reference and have an \rms spread of 0.54\%.
Cross-section results are also shown on a fill-by-fill basis in Fig.~\ref{bgi_results}; 
the indicated error bar is the \rms of all measurements of the corresponding fill.
Since the statistical uncertainty per measurement is significantly less than 0.5\%, the 
observed spread of 0.54\% is due to the combination of statistical fluctuations and additional 
systematic effects.  %
The full \rms is taken as systematic uncertainty on the cross-section.
This uncertainty covers uncorrelated bunch-by-bunch uncertainties such as the shape description, 
which is influenced by the fit model and detector resolutions, or uncertainties in the bunch population measurements.

Some structure remains visible in the measurement shown in Fig.~\ref{bgi_results_perbunch},
pointing towards additional systematic uncertainties or a less than perfect bunch shape description.
Correlations between the cross-sections and the major variables entering into the cross-section measurement
(interaction rate,  bunch population product, overlap integral, luminous region $z$ position, crossing angle corrections)
have been checked and are found to be negligible.

\begin{figure}[tbp]
\centering
\begin{minipage}[c]{0.5\textwidth}
\includegraphics[width=\textwidth]{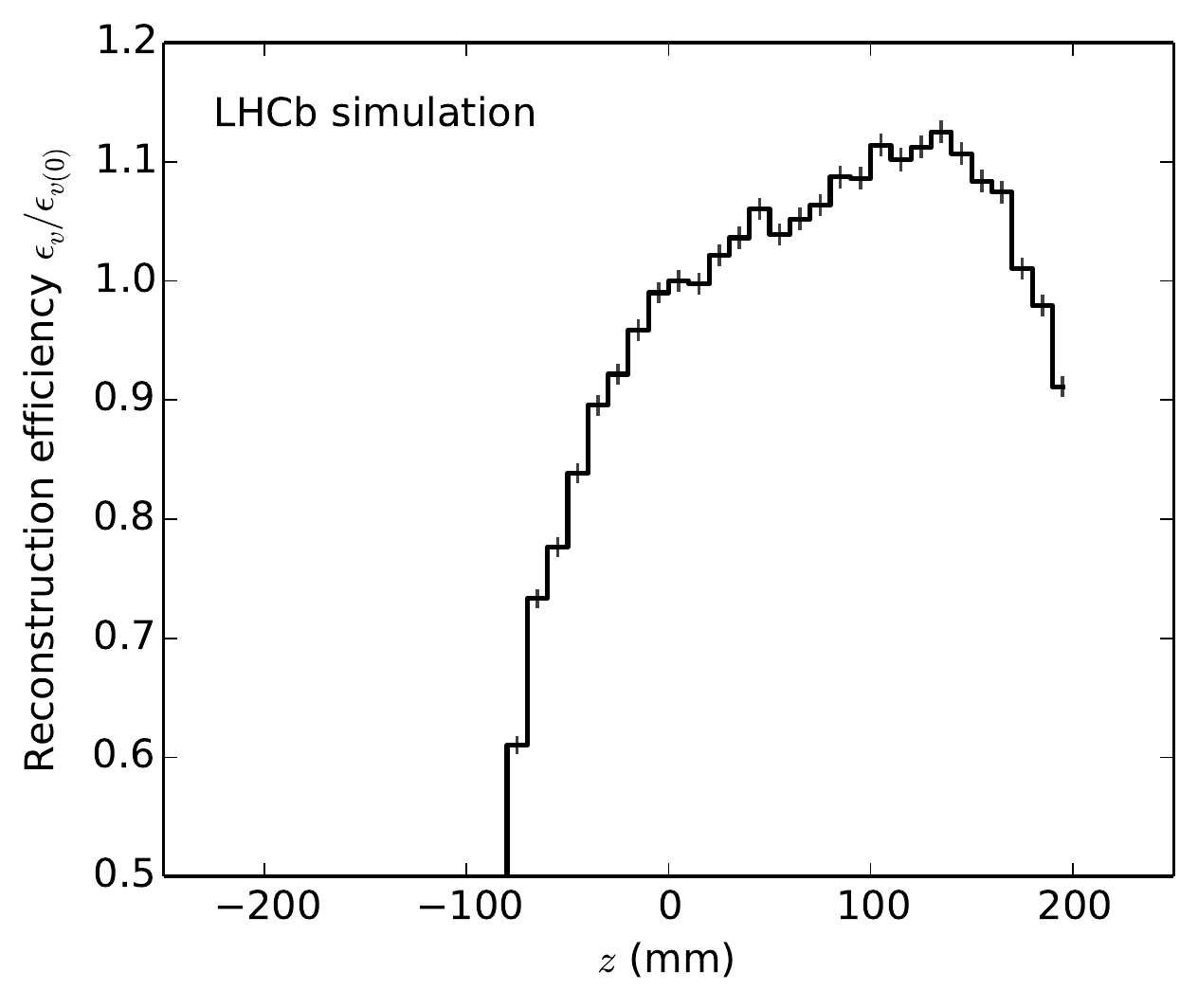}
\end{minipage}\hfill
\begin{minipage}[c]{0.5\textwidth}
\includegraphics[width=\textwidth]{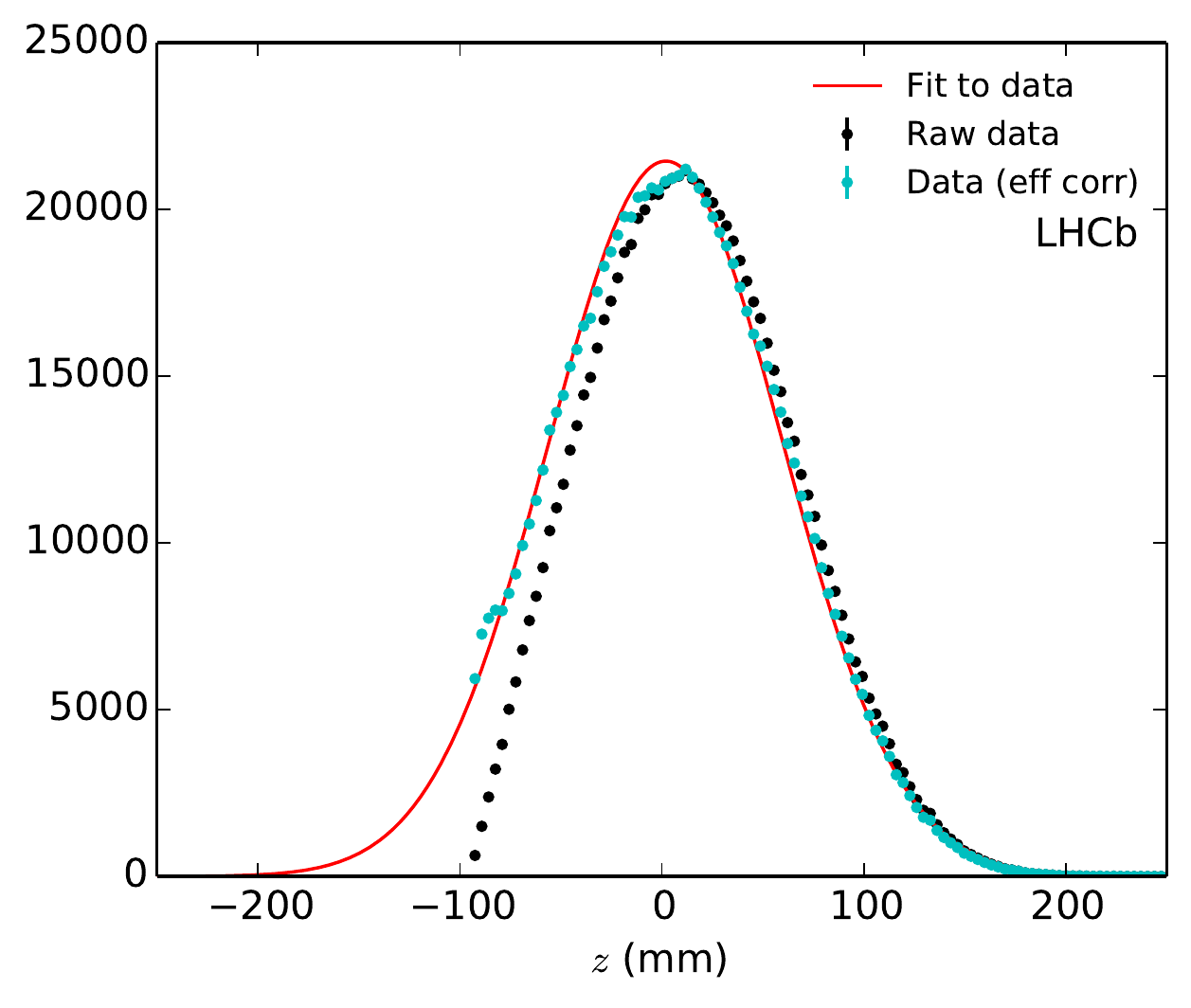}
\end{minipage}\hfill
\begin{minipage}[c]{1\linewidth}
\caption[Bunch length and $\zrf$ measurement with asymmetric tracks cut]{\small 
Determination of the bunch length of a colliding bunch pair with (left) the vertex 
reconstruction efficiency as function of $z$ and (right) the fit to the longitudinal vertex distribution after efficiency correction.
The statistical uncertainty on the corrected points is dominated by the limited amount of simulated data.
This component of the uncertainty is not shown in the right-hand plot.
The data are selected requiring vertices with at least two forward and two backward tracks.
The efficiency is normalized to the value at $z=0$ to keep similar amplitudes between the raw and corrected data; 
the absolute scale of the correction does not change the fit results.
This example displays data for BCID 1335 in fill 2855.
The raw data (black dots) show the distortion resulting from the requirement of having at least two tracks in both directions.}
\label{fit_zzrf2}
\end{minipage}
\end{figure}

\label{sec:sys_bunch_length}
As described in \sect~\ref{sec:bunch_shape_model},
the convolved bunch length $\sigma_{z1}^2+\sigma_{z2}^2$ and bunch crossing position $\zrf$ 
are measured with a fit to the longitudinal distribution of the luminous region.
For each colliding bunch pair measurement, all vertices in the range $|z|<250\mm$ are selected, regardless of the track directions.
This selection reduces the effect of the $z$ dependence in the VELO region and limits the distortion of the luminous region shape.
On the other hand, some beam-gas interaction vertices can be included in the distribution.
An additional measurement of $\sigma_{z1}^2+\sigma_{z2}^2$ and $\zrf$ is performed by 
requiring that vertices contain at least two forward and two backward tracks to exclude beam-gas interactions.
This requirement distorts its longitudinal shape as backward tracks with an origin $z\gtrsim -95\mm$ are not detected by the VELO.
The reconstruction efficiency $\epsilon_{\rm vtx}$ obtained from simulation and corresponding 
to this requirement is shown in Fig.~\ref{fit_zzrf2} and is used to correct the raw distribution.
An example of a fit to the longitudinal distribution of the luminous region with this
requirement is also shown in the figure for one bunch-pair measurement.
All cross-section measurements have been analysed with both track requirements.
The difference between the results obtained with the two methods is 0.04\% and is taken as 
systematic uncertainty related to the reconstruction efficiency.

\includefig{0.65}{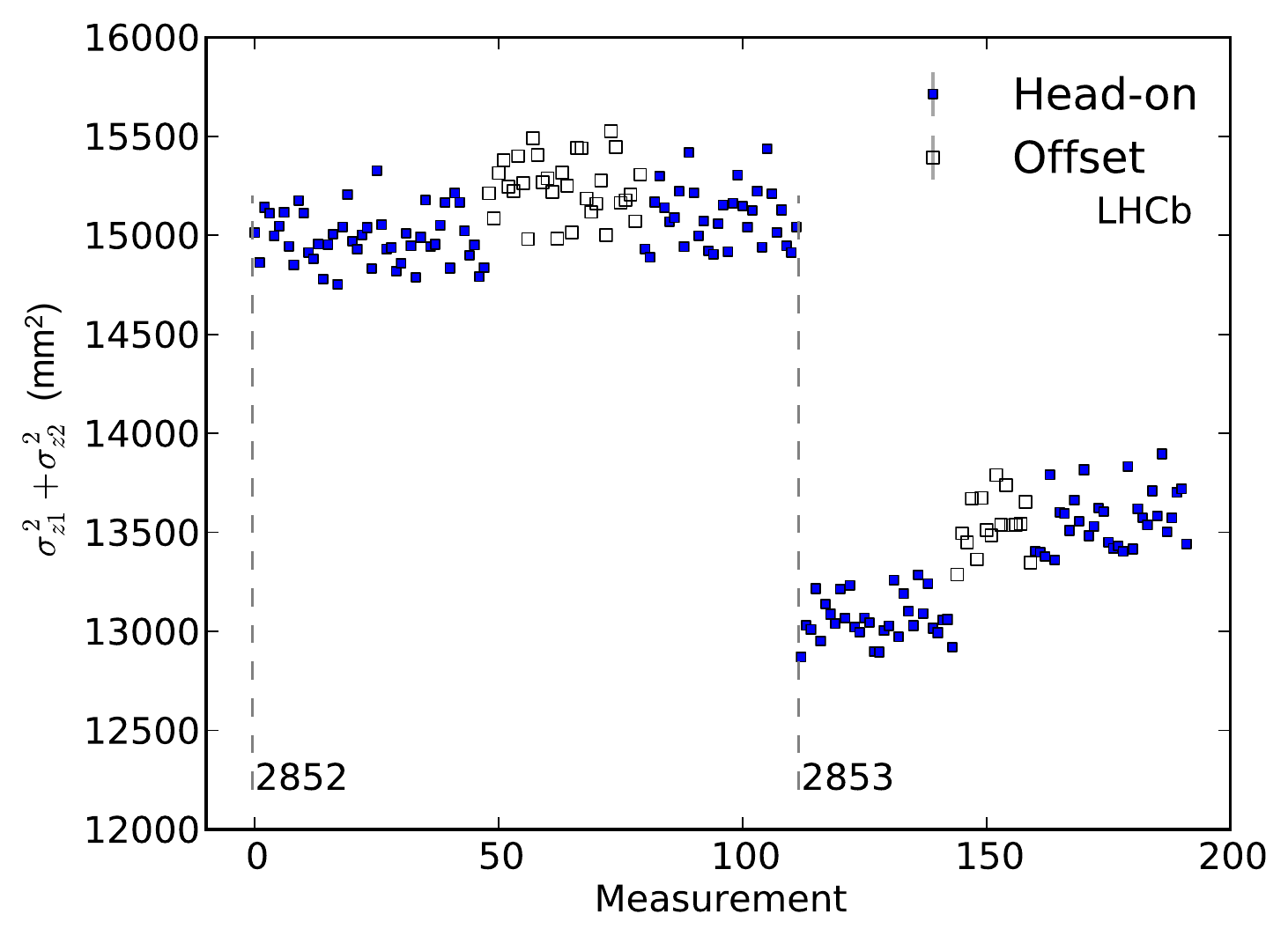}{\small Measurement of convolved bunch 
length $\sigma_{z1}^2+\sigma_{z2}^2$ with displaced beam.}{Measurement of convolved bunch
length $\sigma_{z1}^2+\sigma_{z2}^2$ with displaced beam during fills 2852 and 2853.
The fills are indicated in the plot and separated by a dashed vertical line.
The measurements are sorted by time and BCID.
About 3.5 hours separate the first head-on period in fill 2853 from the  
period with offset beams explaining the higher values of the bunch length observed during that period.
The next head-on period follows the period with offset beams without time interval.}
Two dedicated BGI measurements were performed with beams displaced vertically 
with repect to each other by 180\mum in fills 2852 and 2853
with a $\beta^*=10\m$ optics and were preceded and followed by periods with no displacement.
The beam offset reduced the overlap integral by a factor of about four.
The convolved bunch length $\sigma_{z1}^2+\sigma_{z2}^2$ measured during those fills is shown
in Fig.~\ref{cross_sections_8TeV_zzoffset_BgStd}.
Periods where the beams were offset show the same $\sigma_{z1}^2+\sigma_{z2}^2$ values as when the beams were head-on.
With a relative displacement of 180\mum, the luminosity is dominated by the interaction of the wide beam components of both beams.
On the other hand, the luminosity is dominated by the interaction of the narrow beam components when the beams collide head-on.
The equality of the convolved bunch length values for head-on and offset collisions is an indication 
that the wide and narrow bunch components share the same length,
justifying the assumption that the $z$ coordinate is factorizable.
The observed difference of 0.05\% in the cross-section for the periods with offset beams compared 
to the head-on beams is taken as systematic uncertainty 
related to the convolved bunch length $\sigma_{z1}^2+\sigma_{z2}^2$ measurement.

The beam-gas interaction rate is proportional to the residual gas pressure at the interaction point.
The BGI method measures the beam shapes with beam-gas interaction vertices, and therefore assumes that 
the pressure is uniform in the transverse directions.
The relative error induced by a pressure gradient is estimated by evaluating the effect of 
a distortion on the overlap integral considering a constant pressure along the $y$ axis and a pressure gradient along the $x$ axis
at the experimentally determined limit.
A measurement was performed in 2010 during fill 1422 to verify the homogeneity of the pressure in 
the $x$ direction by displacing the beams by 0.3\mm and has been used in the luminosity measurement 
for 2010~\cite{LHCb-PAPER-2011-015}. %
The relative uncertainty on $\ovint$ introduced by the pressure gradient is at most 0.03\%. %

The VELO transverse dimensions, which fix the absolute scale of the vertex transverse distribution measurements,
were checked in the laboratory at different temperatures on individual silicon sensors.
The relative uncertainty on the cross-section determination due to the transverse scale uncertainty is 
estimated to be at most 0.05\%.
The FBCT response is approximatively linear with respect to bunch intensities.
However, small deviations from linearity lead to a non zero offset when extrapolating to zero bunch intensity.
This offset can be inferred from the combination of all cross-sections measured during a fill,
taking advantage of the fact that the cross-section is independent of the choice of the bunch pair.
An analysis of the FBCT non-linearity is performed as outlined in Ref.~\cite{BCNWG3}.
A fit is performed with the offsets and the improved cross-section value as free parameters.
As expected, the $\chi^2/$ndf values are improved by this procedure.
All cross-section results obtained after applying the FBCT offset corrections for fills with $\beta^*=10\m$ and $\sqrt{s}=8\tev$
give a distribution with an \rms of 0.48\% instead of 0.54\% without the correction. The central value is changed by 0.04\%.
This small deviation is not applied to the final result, but is taken as systematic uncertainty related to a potential FBCT offset.
The systematic uncertainties introduced by the bunch population determination are described in 
\sect~\ref{sec:Bunch current normalization}.

A difference of 0.2\% is observed between the background-subtracted interaction rate measurement 
with the restricted \emph{Vertex} observable and the corresponding unrestricted observable.
This difference is attributed to the background subtraction and assigned as systematic uncertainty.
The reference cross-section used for physics data taking is based on the \emph{Track} observable, which 
is affected by beam-gas interaction background when the neon gas injection system is used for
the BGI calibration data taking.
To transport the visible cross-section based on the \emph{Vertex} observable to that based on the \emph{Track} observable,
their ratio is measured in periods without gas injection.
A variation of 0.2\% is observed, which is taken as systematic uncertainty.
Using the relation between the \emph{Vertex} and \emph{Track} observable of $\mu_{\text{Track}}/\mu_{\text{Vertex}}=1.106$, 
the final calibration result is
$\sigma_{\text{Track}} = 60.62\pm0.87\unit{mb}$.
A summary of all uncertainties is provided in Table~\ref{tab:bgi_uncert}.
The values shown for the other energies will be discussed below.

\def\sigvtx{\sigma_{\text{Vertex}}}
\def\sigtrk{\sigma_{\text{Track}}}
\def\dsigtrk{\delta\sigtrk}
\begin{table}[!htbp]
\centering
\caption[Systematic and statistical uncertainties on the luminosity calibration using the beam-gas imaging method.]{
\small Relative systematic uncertainties on the reference cross-section for the BGI calibrations at 
$\sqrt{s}$ values of 8, 7 and 2.76\tev (in \%).
The uncertainties are divided into groups affecting the measurement of the 
overlap integral (\sect~\ref{sec:Beam-gas imaging method}), the bunch population product 
(\sect~\ref{sec:Bunch current normalization}) and the rate measurement 
(\sect~\ref{sec:InteractionRate}).   %
}
\label{tab:bgi_uncert}
\begin{tabular}{@{}lcccc@{}}
Source of uncertainty       & 8\tev & 7\tev & 2.76\tev \\ %
\midrule
\mul{4}{c}{Overlap integral measurement}\\ 
Beam-beam resolution        & 0.93 & 1.00  & 0.40  \\
Beam-gas resolution         & 0.55 & 2.80  & 1.31  \\
Detector alignment          & 0.45 & 0.30  & 0.90  \\
Fit model                   & 0.50 & 1.50  & 0.50  \\
Measurement spread          & 0.54 & 1.00  & 1.30  \\
Reconstruction efficiency   & 0.04 & 0.20  & 0.17  \\
Bunch length and $\zrf$     & 0.05 & 0.10  & 0.10  \\
Pressure gradient           & 0.03 & 0.03  & 0.03  \\
VELO transverse scale       & 0.05 & 0.05  & 0.05  \\
Statistical uncertainty     & 0.01 & 0.25  & 0.03  \\
\midrule
\mul{4}{c}{Bunch population uncertainties} \\ %
FBCT non-linearity          & 0.04 & 0.05 & 0.05  \\
DCCT population product     & 0.22 & 0.24 & 0.28  \\
Ghost charge                & 0.02 & 0.10 & 0.07  \\
Satellite charge            & 0.06 & 0.25 & 0.08  \\
\midrule
\mul{4}{c}{Rate measurement} \\ %
Background subtraction      & 0.20 & 0.05 & 0.20  \\
Ratio of observable \roc    & 0.20 & n.a. & 0.20  \\       %
\midrule                                          
Final $\dsigtrk/\sigtrk$    & 1.43 & 3.50  & 2.20  \\       %
\end{tabular}
\end{table}

\subsection{Beam-gas imaging results at lower energies}
\label{sec:bgi_final_results}
\label{sec:intermediate_energy}

The analysis presented in this paper focuses on the calibration of the reference cross-section at 
$\sqrt{s}=8\tev$ (\PPP) since it gives the most precise results.
Comparable results are obtained with the BGI method for the 2013 \PPP calibration at
$\sqrt{s}=2.76\tev$ 
and the 2011 \PPP calibration at $\sqrt{s}=7\tev$. 

\begin{figure}[bt]
\centering
\begin{minipage}[c]{0.51\textwidth}
\includegraphics[width=\textwidth]{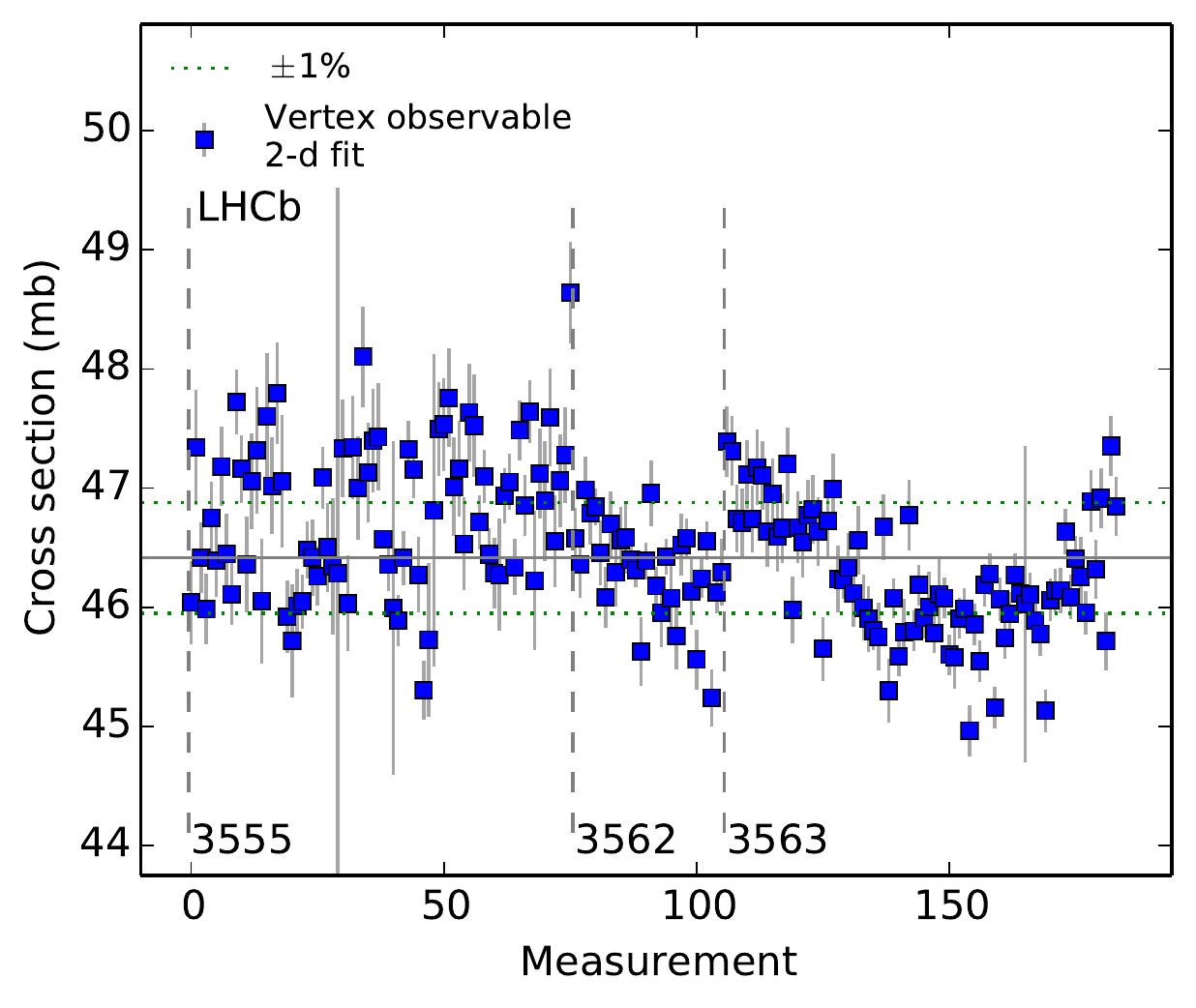}
\end{minipage}\hfill
\begin{minipage}[c]{0.49\textwidth}
\includegraphics[width=\textwidth]{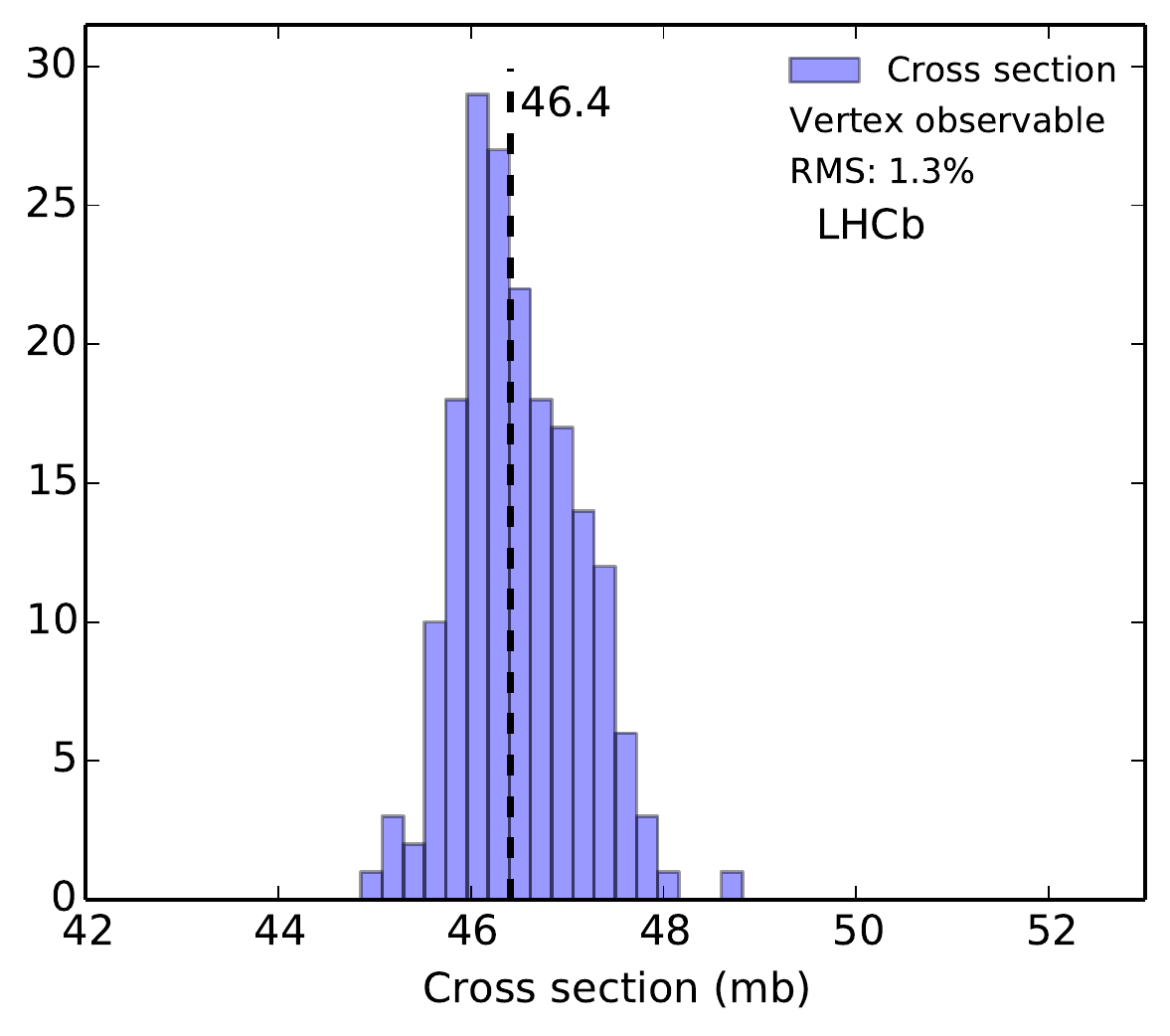}
\end{minipage}\hfill
\begin{minipage}[c]{1\linewidth}
\caption[Cross-section results for head-on beams and 10 m $\beta^*$ fills at $\sqrt{s}=2.76$~TeV.]{\small Cross-section 
results with head-on beams with $\beta^*=10$ m at $\sqrt{s}=2.76$~TeV for the \emph{Vertex} observable, with
(left) the individual measurements and (right) the histogram of the values.
The individual measurements are made per colliding bunch pair and 20 minutes time integration and 
are sorted by time and BCID.
The different fills are separated by a dashed vertical line in the figure.
Two dotted horizontal lines show the $\pm1\%$ deviation from the central value.
The measurement spread has a 1.3\% \rms.
The median value is indicated by a dashed vertical line in the right-hand plot.}
\label{fig_inten_results}
\end{minipage}
\end{figure}

The  $\sqrt{s}=2.76\tev$ calibration was performed in three dedicated fills (3555, 3562 and 3563). %
Trigger conditions were identical to those used in July and November 2012 and the neon gas injection system was active.
The luminosity measurement and evaluation of systematic uncertainties follows the same procedure as with $\sqrt{s}=8\tev$ data.
Cross-section results for $\sqrt{s}=2.76\tev$ are shown in Fig.~\ref{fig_inten_results}.
The \rms of the measurements is 1.3\%.

A half crossing angle of 885~\textmu rad was chosen to avoid collisions between satellite and main bunches.
Differences observed in the data as compared to 2012 data are discussed below.
The beams have a double Gaussian shape and are significantly non-factorizable ($f_j$ is close to zero).
However, the factorizability is more difficult to measure than with $\sqrt{s}=8\tev$ data and the uncertainties on the $f_j$ parameter are larger.
The beam-beam resolution has a small impact on the cross-section compared to 2012 (uncertainty of 0.40\%), 
owing to the comparatively large transverse bunch size. %
On the other hand, the uncertainty related to the beam-gas vertex position resolution has significantly larger impact than in 2012, 
due to the lower number of vertices in the luminous region.
The cross-section difference measured with and without all beam-gas resolution correction factors amounts to 1.31\%.
The uncertainty related to the detector alignment of 0.9\% is estimated at twice the value obtained in 2012 because 
the crossing angle correction is about twice as large.
The convolved bunch length $\sigma_{z1}^2+\sigma_{z2}^2$ measurement plays a more important role compared 
to the $\sqrt{s}=8\tev$ data due to the larger crossing angle correction.
An uncertainty of 0.1\% is assigned to the determination of 
$\sigma_{z1}^2+\sigma_{z2}^2$ and $\zrf$ and 0.17\% to the reconstruction efficiency. 
The FBCT offset fit changes the cross-section by 0.05\%
and reduces the overall \rms to 1.1\%.
The systematic uncertainty related to the ghost charge amounts to 0.07\%.
Uncertainties for the observable background subtraction and fit model are taken from the 2012 measurements.
The reference cross-section for the \emph{Vertex} observable is $46.4\pm1.0\unit{mb}$.
A summary of all uncertainties is provided in Table~\ref{tab:bgi_uncert}.
As for the $\sqrt{s}=8\tev$ data, the reference cross-section used for physics data taking is based on the \emph{Track} observable.
Using the relation between the \emph{Vertex} and \emph{Track} observable of $\mu_{\text{Track}}/\mu_{\text{Vertex}}=1.135$, 
the final calibration result is
$\sigma_{\text{Track}} = 52.7\pm1.2\unit{mb}$.

\begin{figure}[bt]
\centering
\begin{minipage}[c]{0.51\textwidth}
\includegraphics[width=\textwidth]{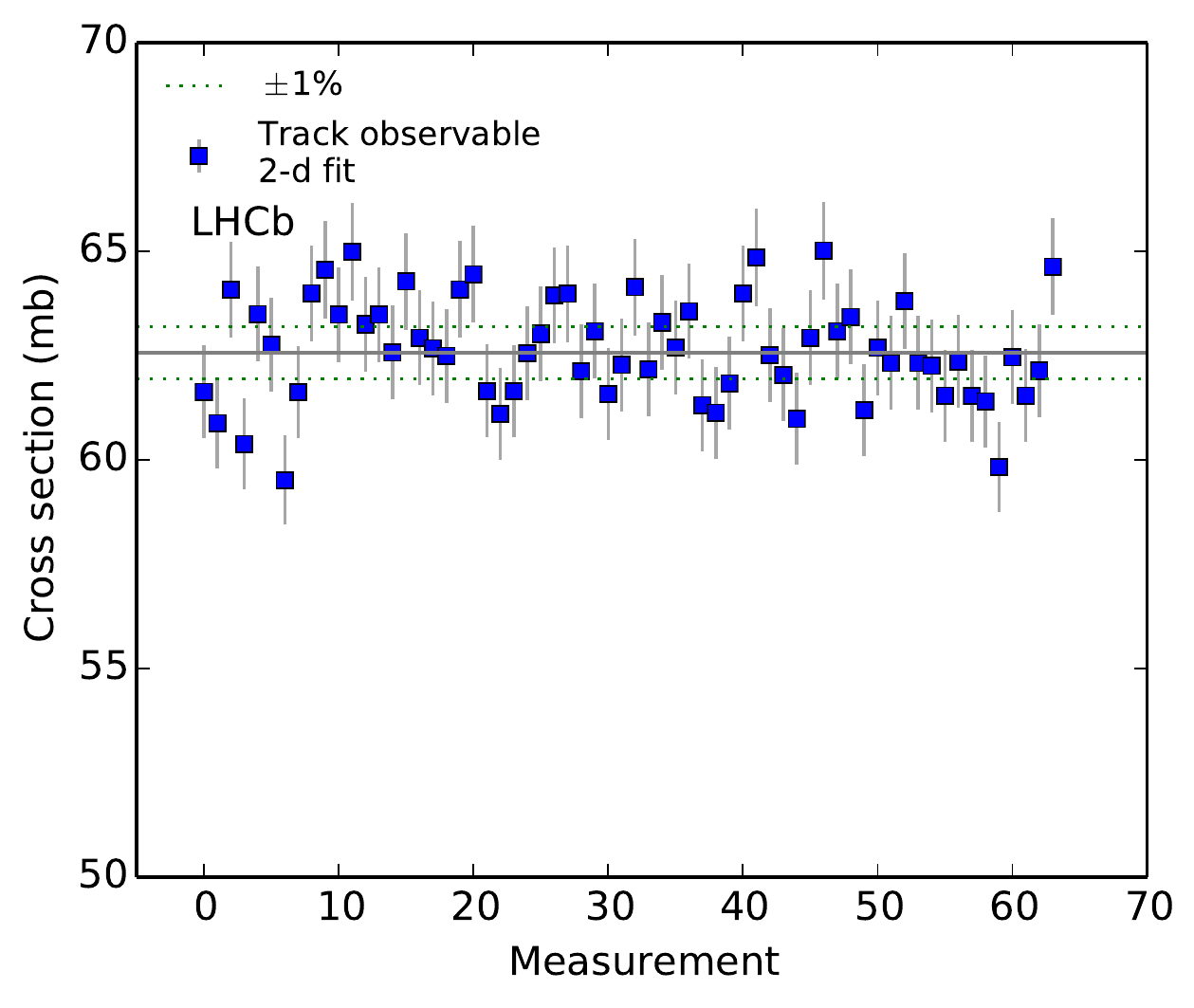}
\end{minipage}\hfill
\begin{minipage}[c]{0.49\textwidth}
\includegraphics[width=\textwidth]{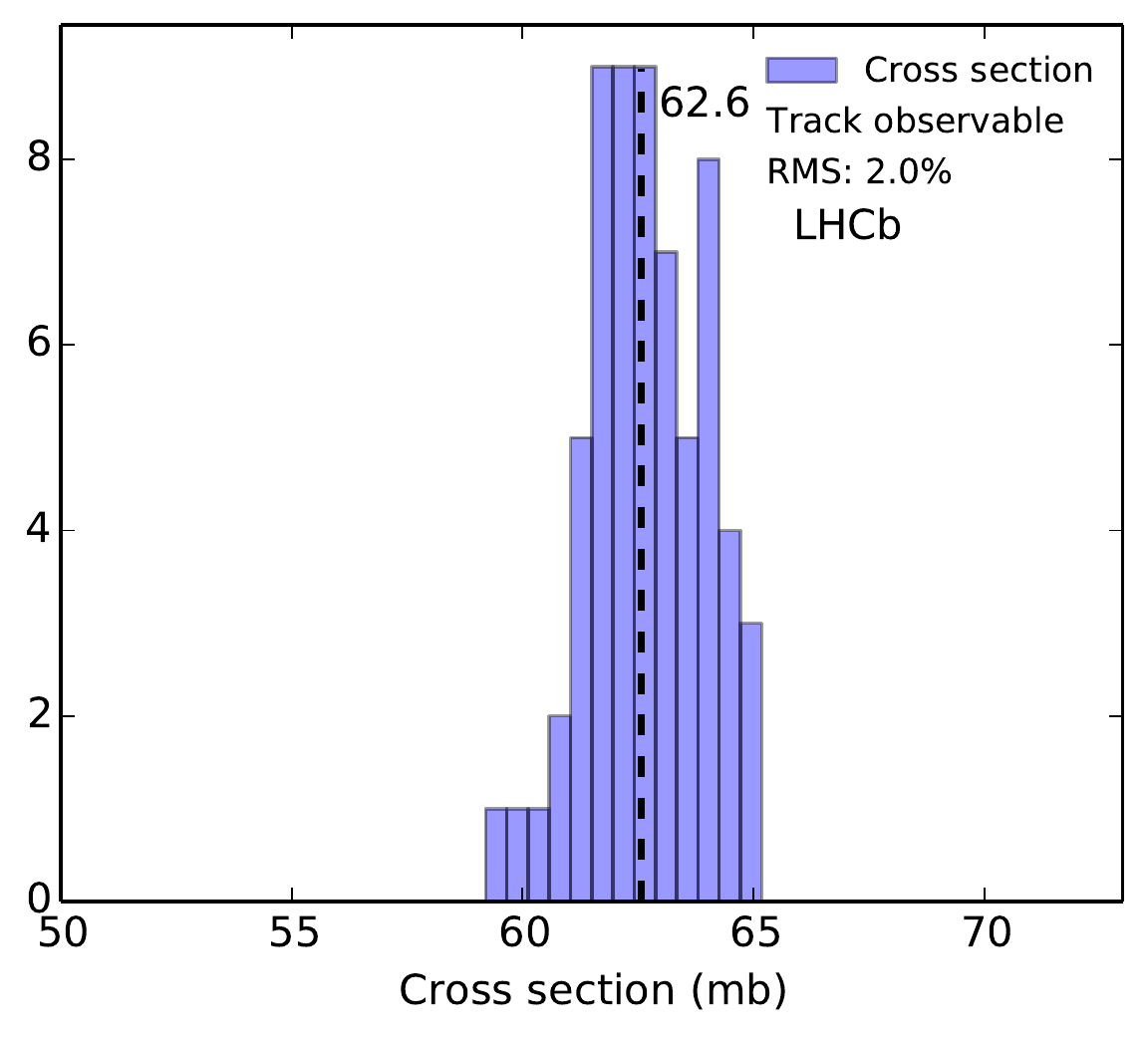}
\end{minipage}\hfill
\begin{minipage}[c]{1\linewidth}
\caption[Cross-section results for at $\sqrt{s}=7$~TeV in 2011.]{\small Cross-section 
results at $\sqrt{s}=7$~TeV in 2011 for the \emph{Track} observable, with
(left) the individual measurements and (right) the histogram of the values.
The individual measurements are made per colliding bunch pair, for a one hour integration time, and 
are sorted by time and BCID.
The different fills are separated by a dashed vertical line in the figure.
Two dotted horizontal lines show the $\pm1\%$ deviation from the central value.
The measurement spread has a 2.0\% \rms.
The median value is indicated by a dashed vertical line in the right-hand plot.}
\label{fig_7tev_results}
\end{minipage}
\end{figure}

In the 2011 \PPP calibration, the main differences with the situation in 2012 are the
absence of the neon gas-injection system and a different trigger configuration.
Therefore, the beam-gas interaction rate is a factor 20 lower than in 2012.
To partly compensate for the lower rate, measurements are performed in one-hour periods,
potentially introducing effects of emittance growth and beam drifts.
In addition, data are available in one dedicated calibration fill only.
Cross-sections results for $\sqrt{s}=7\tev$ are shown in Fig.~\ref{fig_7tev_results}.
The \rms of the measurements is 2.0\%.

Collisions between satellite and main bunches were observed at $z=\pm375$~mm due to the low half crossing angle of 270~\textmu rad.
Therefore, beam-gas vertices in the region $-550$~mm $<z<550$~mm are discarded to exclude beam-beam vertices in the single beam selection.
This requirement, however, reduces the number of vertices available for the single beam measurement.
Furthermore, the remaining vertices are measured with a worse resolution leading to a bigger impact on the beam size measurement.
The limited amount of beam-gas vertices reduces the accuracy of the resolution determination and the correction factors described in \sect~\ref{sec:resolution_beam_gas} cannot be determined.
An uncertainty of 10\% is assumed for the beam-gas vertex resolution leading to an uncertainty of 2.8\% on the cross-section calibration.

The ratio of the resolution width to the beam size is smaller in 2011, potentially reducing the uncertainty resulting from the beam-beam resolution.
However, the same uncertainty as for $\sqrt{s}=8\tev$ is conservatively assumed for the impact of the beam-beam vertex resolution.
Given the smaller crossing angle, the uncertainty related to the detector alignment effects is estimated at 0.3\%.
The beam factorizablility is more difficult to measure in 2011 due to the lower number of vertices.
The beams are mostly non-factorizable ($f_j$ is typically close to zero).

Owing to the low beam-gas induced background, the reference cross-section can be based on the \emph{Track} observable directly without using the ratio \emph{Vertex} to \emph{Track}.
A global satellite fraction correction of 0.78\% is applied to the result shown in Fig.~\ref{fig_7tev_results}.
The correction cannot be applied on a per bunch basis as the LDM instrument was not operational yet for this fill.
A summary of all uncertainties is provided in Table~\ref{tab:bgi_uncert}.
The final calibration result for $\sqrt{s}=7\tev$ based on the \emph{Track} observable is
$\sigma_{\text{Track}} = 63.0\pm2.2\unit{mb}$.
The difference with respect to the 8\tev calibration result is not significant given the uncertainties, which are mostly uncorrelated.

This value is consistent with the result obtained with the BGI method described in a previous LHCb publication ~\cite{LHCb-PAPER-2011-015}.
The improvement in the present result is due to the better bunch population measurement, while the uncertainties in the overlap integral are similar.

%% file: vdm.tex
\section{Van der Meer scan method}
\label{sec:Van der Meer scan method}

\subsection{Experimental conditions}
\label{sec:VDM:Experimental conditions}
\input{vdm_conditions}

\subsection{Overlap integral model}
\label{sec:Overlap integral model}
\input{vdm_model}

\subsection{Cross-section determination}
\label{sec:Cross-section determination}
\input{vdm_xsection}

\input{vdm_corrections}

\input{vdm_results}

%% file: vdm_conditions.tex
\newcommand{\seprng}{$\pm 6 \sign$/$\pm 6 \sign$}
\begin{table}[pb]
  \centering
    \caption{
      Parameters of \vdm scans.
      The scan pairs marked with an asterisk are not used in the determination of the central value of the cross-section (as explained in the text).
      The step duration indicates the period of stable conditions available for the measurement.
    }
    \label{tab:vdm_scans}
    \begin{tabular}{lccccccc}
      \toprule
      Pair & Scans & Axes & Offset & Number & Separation & \multicolumn{2}{c}{Duration} \\
      & & & & of steps & range & step (s) & total (min) \\
      \midrule
      \multicolumn{8}{l}{\PPP at $\sqrt{s}=8\tev$, Apr 2012, Fill 2523} \\
      1  & 2/3 & $x/y$     & $\phantom{\sim}$0/0 & 31/31 & \seprng & 15 & 12/12 \\
      2* & 1/4 & $x''/y''$ & $\phantom{\sim}$0/0 & 33/31 & \seprng & 15 & 13/12 \\
      3  & 5/6 & $x/y$     & $\phantom{\sim}$0/0 & 17/17 & \seprng & 15 & 7/7 \\
      4* & 7/8 & $y/x$     & $\sim 2\sign$/$2\sign$ & 17/17 & \seprng & 15 & 7/7 \\
      \midrule
      \multicolumn{8}{l}{\PPP at $\sqrt{s}=8\tev$, Jul 2012, Fill 2853} \\            
      1  & 1/2   & $x/y$ & $\phantom{\sim}$0/0 & 27/27 & \seprng & 15 & 12/12 \\
      2  & 3/4   & $x/y$ & $\phantom{\sim}$0/0 & 27/27 & \seprng & 15 & 12/12 \\
      3  & 5/6   & $x/y$ & $\phantom{\sim}$0/0 & 27/27 & \seprng & 15 & 12/12 \\
      4  & 7/8   & $x/y$ & $\phantom{\sim}$0/0 & 27/27 & \seprng & 15 & 12/12 \\
      5* & \phantom{1}9/10  & $x/y$ & $\phantom{\sim}$0/0 & 51/51 & \seprng & 15 & 21/21 \\
      6* & 11/12 & $y/x$ & $\sim 2\sign$/$2\sign$ & 27/27 & \seprng & 15 & 12/12 \\
      \midrule
      \multicolumn{8}{l}{\PPP at $\sqrt{s}=7\tev$, Oct 2011, Fill 2234} \\      
      1 & 1/2 & $x/y$ & $\phantom{\sim}$0/0 & 31/31 & \seprng & 15 & 13/13 \\
      2 & 3/4 & $x/y$ & $\phantom{\sim}$0/0 & 17/17 & \seprng & 15 & 8/8 \\
      3 & 5/6 & $x/y$ & $\phantom{\sim}$0/0 & 17/17 & \seprng & 15 & 8/8 \\
      \midrule
      \multicolumn{8}{l}{\PPB at $\sqrt{\sNN}=\Five\tev$, Jan 2013, Fill 3505}  \\
      1  & 1/2   & $x/y$ & $\phantom{\sim}$0/0 & 27/27 & $\pm6\sign$/$\pm6\sign$ & 30 & 17/17 \\
      2  & 3/4   & $x/y$ & $\phantom{\sim}$0/0 & 27/27 & $\pm6\sign$/$\pm5\sign$ & 30 & 17/17 \\
      \midrule
      \multicolumn{8}{l}{\PBP at $\sqrt{\sNN}=\Five\tev$, Feb 2013, Fill 3542} \\   
      1  & 1/2   & $x/y$ & $\phantom{\sim}$0/0 & 27/27 & $\pm6\sign$/$\pm5\sign$ & 30 & 17/17 \\
      2  & 3/4   & $x/y$ & $\phantom{\sim}$0/0 & 27/27 & $\pm6\sign$/$\pm5\sign$ & 30 & 17/17 \\
      \bottomrule
    \end{tabular}
\end{table}

\vdm scans at $\sqrt{s}=8\tev$ (\PPP) were performed in \lhcb during dedicated \lhc fills in the 2012 running period, one in April and one in July.
The \vdm scan method is presented in detail for these calibrations.
The calibrations at $\sqrt{s}=7\tev$ (\PPP) and at $\sqrt{\sNN}=\Five\tev$ (\PPB and \PBP) are summarized in Sec.~\ref{sec:VDM:Results 2}.
The list of dedicated luminosity calibration fills can be found in Table~\ref{tab:fills} and
the scan parameters are listed in Table~\ref{tab:vdm_scans}.
Four $x$-$y$ scan pairs were performed in April and six in July.
In all scans the two beams were moved symmetrically, typically covering a $\pm6\sign$ range of beam separations,
where $\sign$ is the nominal beam width assuming nominal values of \betastar and transverse normalized emittance $\epsilon_\mathrm{n}=3.75\mum\rad$.
The last scan pair in each fill had a nominal working point at a relatively large offset $\Dxz,\Dyz \approx +2\sign$.
These offset scan pairs are only used for cross-checks and are not considered in the cross-section determination because of their high sensitivity to systematic effects (\eg beam orbit drift, factorizability, linear correlations).
The first and fourth scans in April were performed along $x''$ and $y''$ axes that are rotated with respect to the principal axes of the \lhc optics.
This particular scan pair is analysed, but not used in the final result.

\begin{figure}[ptb]
  \begin{center}
    \includegraphics[width=0.49\linewidth]{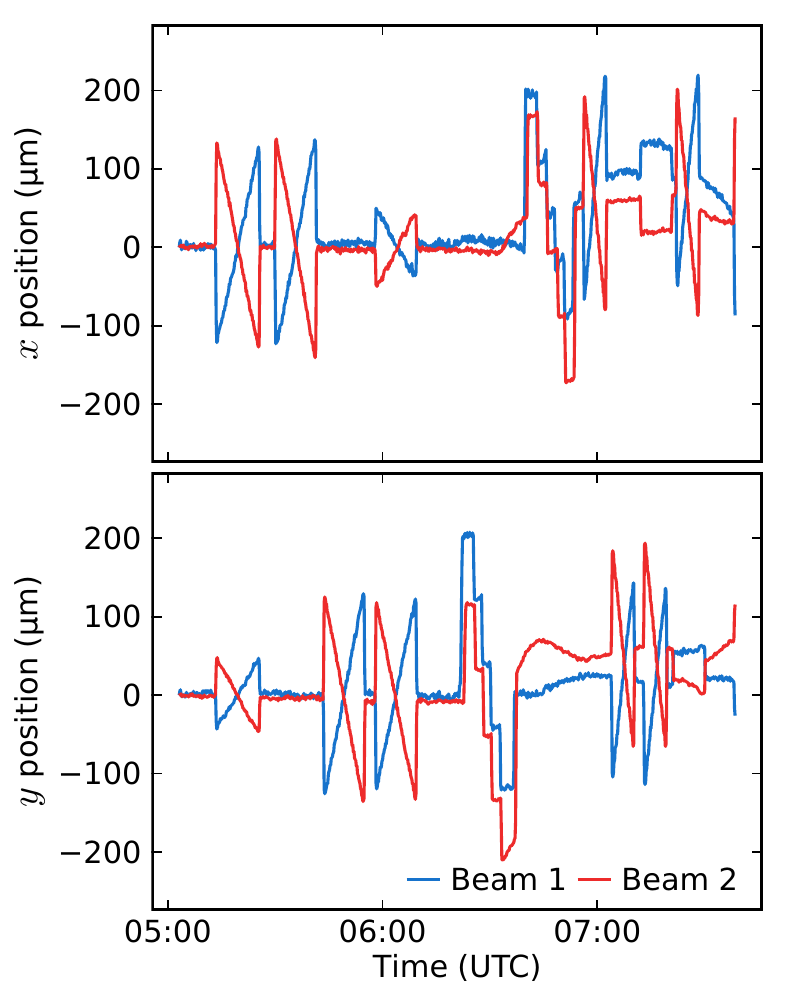}
    \includegraphics[width=0.49\linewidth]{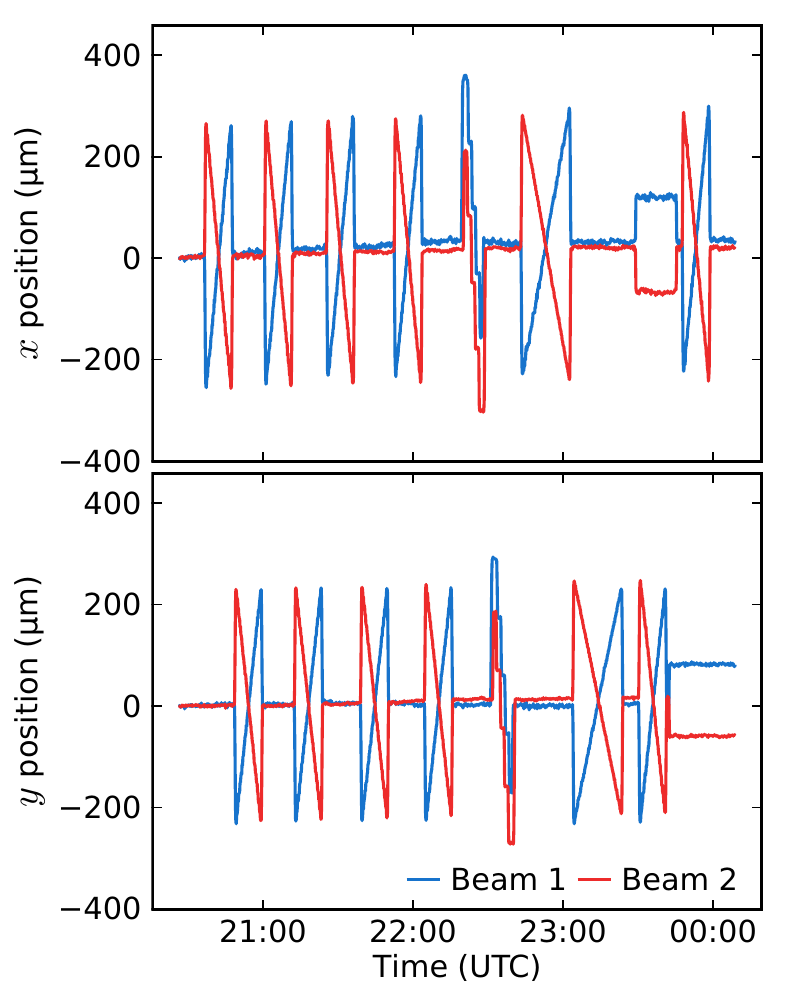}
    \vspace*{-0.5cm}
  \end{center}
  \caption{
    Beam positions recorded with the LHC beam position monitors around 
    the \lhcb interaction point in the (left) April and (right) July \vdm sessions.
    The initial values are set to zero. The top (bottom) panel shows horizontal 
    (vertical) positions.
    In all scans the beams are moved symmetrically.
    The beam movements around 6:30 (22:40) in April (July) correspond to the length scale calibration by the constant separation method.
  }
  \label{fig:cond_bpm}
\end{figure}

Beam movements recorded with \lhc beam position monitors (BPMs) upstream and downstream of \lhcb are shown in Fig.~\ref{fig:cond_bpm}.
The BPM measurements are not used in the analysis as their time stability is insufficiently accurate. %
During a scan the beam separation values are calculated on the basis of a detailed model of the \lhc magnets and are set by the accelerator control system.
Dedicated length scale calibration scans are performed in each \vdm calibration fill to experimentally verify and calibrate the beam displacements using the precisely known geometry of the \velo (see Sec.~\ref{sec:VDM:Length scale}).

During \vdm scan sessions, a large fraction of the available trigger bandwidth was allocated to randomly triggered bunch crossings; 20\kHz were devoted to the crossings with collisions, 2\kHz to the crossing slots where only one of two beams was present, and 0.5\kHz to the empty crossing slots.
In addition, starting from the July 2012 session, beam-gas events were recorded simultaneously.
Due to the small beam-gas interaction rate of a few hundred $\Hz$ in total, the collected events are only used for a cross-check of the beam positions.

The average decay time of the bunch population product ${N_1N_2}$ was 36 (70) hours in the April (July) session.
The value of ${N_1N_2}$ changed by about 1\% during a scan pair.
Therefore, the rates are normalized by ${N_1N_2}$ of each colliding bunch pair at every scan point by defining a specific average number of interactions per crossing
\begin{equation}
\mueff{sp} = \frac{\mueff{\eff}}{N_1N_2} \, .
\end{equation}
To reduce the noise associated with the $N_{1,2}$ measurements, the data  for each beam are approximated by a smoothing spline \cite{hastie1990generalized}.

\begin{figure}[ptb]
  \begin{center}
    \includegraphics[width=0.49\linewidth]{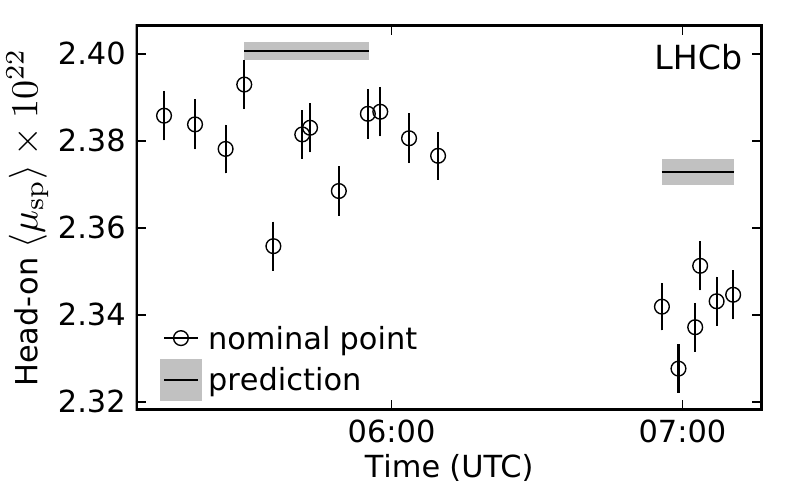}
    \includegraphics[width=0.49\linewidth]{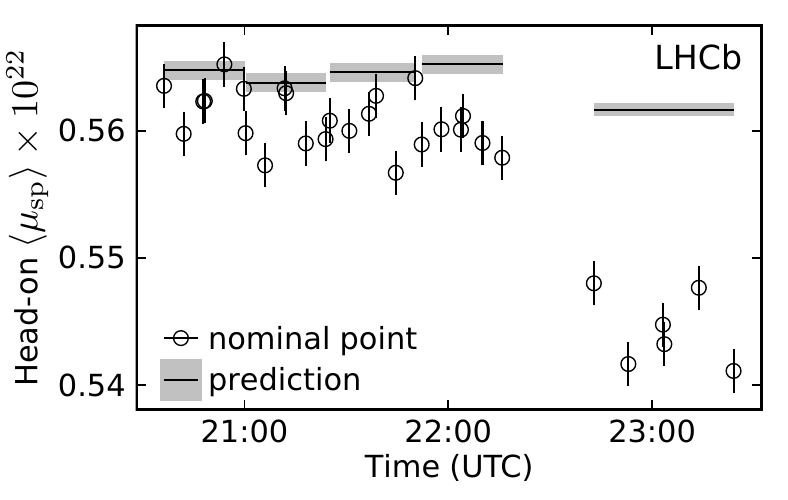}
    \vspace*{-0.5cm}
  \end{center}
  \caption{
    Evolution of the specific average number of interactions per crossing ($\mueff{sp}$) at the nominal head-on beam positions during the (left) April and (right) July \vdm sessions.
    In each scan the nominal point was measured three times and the average 
    over bunches is plotted with open markers. %
    The value at the actual zero beam separation is predicted for each scan pair (excluding offset and tilted scans) and shown with a horizontal band.
    The difference with respect to the values at zero nominal separation is due to the working point not being exactly at zero beam separation.
    The beam positions were not re-optimized during the sessions.
  }
  \label{fig:cond_headonmu}
\end{figure}
In addition to the bunch population changes, the luminosity stability may be limited by changes in the bunch profiles, \eg by emittance growth.
The luminosity stability is checked several times during the scans when the beams were brought back to their nominal position.
The evolution of $\mueff{sp}$, averaged over bunch pairs, is shown in Fig.~\ref{fig:cond_headonmu}.
The average luminosity decay time is measured to be 29 (58) hours in the April (July) session, which is largely due to the decay in the bunch population product.
The average luminosity drop caused by emittance growth (drop of $\mueff{sp}$) amounts to 1.2\% (0.5\%) during the entire calibration session in April (July).
The scan points have been taken from lower to higher \Dx, \Dy values, therefore, the luminosity drop due to emittance growth effectively enhances the %
rate at negative values and reduces the rate at positive values,
so that the net effect on the integrals in Eq.~\eqref{eq:vdm_formula} cancels to first order since the curves are symmetric.
Therefore, the systematic error due to the emittance growth is considered to be negligible.
As will be shown in Sec.~\ref{sec:VDM:Reproducibility}, the measurement using the fifth scan pair in July is more sensitive to beam orbit drifts, because it was performed at a slightly offset working point and took longer than usual (42 instead of 24 minutes).
Since the orbit drifts cannot be precisely corrected for, this scan pair is not used for the central value of the result but is accounted for in the systematic uncertainty.
In total, for the cross-section measurement, two \emph{reference} scan pairs are used from the April session and four from the July session.

%% file: vdm_model.tex
The rate measurements from the orthogonal \vdm scans performed provide no information on the factorizability of the single beams.
However, the cross-section measurement is sensitive to the latter.
In order to impose constraints on the factorizability, the BGI measurements performed in the same fills on the same bunch pairs can be used.
To facilitate such approach, it is advantageous to use an identical model of the single beam densities for both BGI and \vdm analyses.
In this section, a model for the overlap integral as function of beam separation is developed based on the double Gaussian beam shape description from Sec.~\ref{sec:double gaussian model}.
It should be noted that 
the application of the classic \vdm method in the case of factorizable beam distributions (see Eq.~\eqref{eq:vdm_formula}) only requires a good empirical description of the \vdm scan data.

The formalism discussed in Sec.~\ref{sec:Luminosity in the case of pure Gaussian beams} is only valid in the case of a horizontal beam crossing plane.
However, in the April calibration the crossing plane was rotated along $z$, which necessitates the following extended treatment.
In this more general case, the exponent in the overlap integral for Gaussian beams from Eq.~\eqref{eq:ovrlap_compact} gains a dependence on the product of the transverse beam separations $\Dx \Dy$ and the normalization factor is modified.
It is useful to extend the definition of the effective convolved widths of the luminous region $\Sigma_m$ ($m=x,y$) from Eq.~\eqref{eq:lumi_variable} (taking $\cos\phi_m \approx 1$) as
\begin{equation}
\Sigma_{m}^2 = \sigma_{m1}^2 + \sigma_{m2}^2 + (\sigma_{z1}^2 + \sigma_{z2}^2)\tan^2\phi_m \, ,
\end{equation}
so that $\Sigma_y$ also has a term that depends on the corresponding half crossing angle in the $yz$ plane $\phi_y$.
The general expression for the overlap integral (in the case of pure Gaussian beams) is then given by
\begin{align}\label{eq:overlap_with_rho}
&\ovint(\Dx,\Dy\,|\,\Sigma_x,\Sigma_y,\corcoef) = \notag\\ 
&~~~\sqrt{1-\corcoef^2}
\exp\left(\frac{\corcoef\Dx\Dy}{\Sigma_x\Sigma_y(1-\corcoef^2)}\right)
g(\Dx\,|\,\sigma=\Sigma_x \sqrt{1-\corcoef^2})
g(\Dy\,|\,\sigma=\Sigma_y \sqrt{1-\corcoef^2}) \, ,
\end{align}
where $g$ is the normalized Gaussian probability density function and $\corcoef$ is the correlation coefficient, 
which is related to the bunch lengths and the half crossing angles by
\begin{equation}\label{eq:overlap_correlation}
\corcoef(\Sigma_x,\Sigma_y) = \frac{C_{xy}}
{\Sigma_x\Sigma_y} + \bar{\corcoef},\quad 
C_{xy} = (\sigma_{z1}^2 + \sigma_{z2}^2) \tan\phi_x \tan\phi_y,\quad
\bar{\corcoef} = \frac{\corcoef_1 + \corcoef_2}{2} \, .
\end{equation}
The variables $\corcoef_{1,2}$ are the individual bunch correlation coefficients, which take into account a possible rotation of the principal axes of the bunches (around the direction of motion).
It is seen from Eq.~\eqref{eq:overlap_correlation} that $C_{xy}$ is only non-zero when both $\phi_x$ and $\phi_y$ are non-zero, which occurs when the beam crossing plane is tilted (\ie neither strictly horizontal nor vertical).

The overlap integral in the double Gaussian model, which is discussed in Sec.~\ref{sec:double gaussian model}, is given by Eq.~\eqref{eq:bb_overlap_general} with each partial overlap integral expressed as
\begin{equation}
\ovint_{i_x,i_y,j_x,j_y} = \ovint(\Dx,\Dy\,|\,
  \Sigma_{x,i_x j_x},\Sigma_{y,i_y j_y},
  \corcoef(\Sigma_{x,i_x j_x},\Sigma_{y,i_y j_y})) \, ,
\end{equation}
where $i_x$, $i_y$, $j_x$, $j_y$ take the values \rmn and \rmw.
Here, it is implicitly assumed that all Gaussian components have the same centres, same bunch lengths and same correlation coefficients (\ie $C_{xy}$ and $\bar{\corcoef}$).
The weights in Eq.~\eqref{eq:bb_overlap_general} depend on the factorizability parameter $f_j$ and two weight parameters ($w_{x,j}$ and $w_{y,j}$) for each beam $j=1,2$.

The rate that is measured in \vdm scans only provides direct information on the beam overlap integral and not on the single beam densities.
Therefore, it is convenient to parametrize the model of the overlap integral using the effective convolved widths rather than the underlying beam width parameters.
Only three of the four effective convolved widths per plane are linearly independent, thus two parameters are sufficient to quantify their relative magnitudes, with the choice
\begin{align}
\label{eq:vdm_params_Rm}
R_m &= \frac{\Sigma_{m,\rmw\rmw}}{\Sigma_{m,\rmn\rmn}} \, ,\quad R_m \geq 1 \, , \\
\label{eq:vdm_params_Am}
A_m &= \frac{\Sigma_{m,\rmn\rmw}^2 - \Sigma_{m,\rmw\rmn}^2}
  {\Sigma_{m,\rmw\rmw}^2 - \Sigma_{m,\rmn\rmn}^2} \, ,\quad A_m \in [-1,1] \, ,
\end{align}
for $m = x,y$.
In terms of beam width parameters we have
\begin{equation}
A_m = -\frac{(\sigma_{m1,\rmw}^2 - \sigma_{m1,\rmn}^2) - (\sigma_{m2,\rmw}^2 - \sigma_{m2,\rmn}^2)}
{(\sigma_{m1,\rmw}^2 - \sigma_{m1,\rmn}^2) + (\sigma_{m2,\rmw}^2 - \sigma_{m2,\rmn}^2)} \, ,
\end{equation}
thus $A_m$ describes the asymmetry between the differences in the widths of the wide and narrow components of the two beams.
The special case $A_m = 1$ corresponds to
$\sigma_{m1,\rmn} = \sigma_{m1,\rmw}$,
\ie the distribution of \beamone is Gaussian in the $m$ axis.
In the parametrization of Eqs.~\eqref{eq:vdm_params_Rm}~and~\eqref{eq:vdm_params_Am},
the effective convolved widths are expressed as
\begin{align}\label{eq:dgsb_capsigmas}
\begin{bmatrix}
  \Sigma_{m,\rmn\rmn}^2 & \Sigma_{m,\rmn\rmw}^2\\ \Sigma_{m,\rmw\rmn}^2 & \Sigma_{m,\rmw\rmw}^2
\end{bmatrix}
&= \Sigma_{m,\rmn\rmn}^2
\begin{bmatrix}
r_{m,\rmn\rmn}^2 & r_{m,\rmn\rmw}^2\\ r_{m,\rmw\rmn}^2 & r_{m,\rmw\rmw}^2
\end{bmatrix}  \notag\\
&= \Sigma_{m,\rmn\rmn}^2 \left(
\begin{bmatrix}
  1 & \frac{R_m^2+1}{2}\\ \frac{R_m^2+1}{2} & R_m^2
\end{bmatrix}
+ A_m \frac{R_m^2-1}{2}\begin{bmatrix} 0 & 1 \\ -1 & 0 \end{bmatrix}
\right) \, ,
\end{align}
where the ratios $r_{m,ij}$ are defined as $r_{m,ij} \equiv \Sigma_{m,ij}/\Sigma_{m,\rmn\rmn}$.

For the actual fitting procedure, it is advantageous to use scale (or width) parameters that have a model-independent meaning.
For example, the \rms of the luminosity profiles at $\Dx=0$ and $\Dy=0$ are such parameters, which can be easily estimated from the scan data to obtain starting values for the fit.
Using the parameters defined above, the \rms $S_x$ of the luminosity profile at $\Dy=0$ is given by
\begin{align}
S_x^2 &= \frac{\int \Dx^2 \ovint(\Dx,0) d\Dx}
  {\int \ovint(\Dx,0) d\Dx} %
  = \Sigma_{x,\rmn\rmn}^2 \frac{\sum_I \frac{w_I}{r_{y,i_y j_y}} r_{x,i_x j_x}^2
     (1-\corcoef^2(\Sigma_{x,\rmn\rmn}r_{x,i_x j_x},\Sigma_{y,\rmn\rmn}r_{y,i_y j_y}))}
  {\sum_I \frac{w_I}{r_{y,i_y j_y}}}
\end{align}
and similarly for the other coordinate.
The values $\Sigma_{x,\rmn\rmn}$ and $\Sigma_{y,\rmn\rmn}$ are obtained by solving the system of equations defined by the above equation.
Finally, in the double Gaussian model, the shape of the overlap integral as function of beam separation is parametrized with the following 14 parameters
\begin{equation}\label{eq:overlap_params_vdm}
S_x,S_y,R_x,R_y,A_x,A_y,f_1,f_2,w_{x,1},w_{y,1},w_{x,2},w_{y,2},C_{xy},\bar{\corcoef} \, .
\end{equation}
Only some of those parameters remain free for the cross-section determination as explained in the following section.

%% file: vdm_xsection.tex
All colliding bunch pairs are analysed individually.
The data are fitted simultaneously for pairs of $x$ and $y$ scans.
The value of $\mu_\eff$ at each step $k$ is described with
\begin{equation}\label{eq:vdm_2d_fitfun}
\mu_{\eff}(\Dx_k, \Dy_k) = \sigma_\eff\,N_1 N_2\,\ovint (\Dx_k-\Dxz,\Dy_k-\Dyz) +
  N_1\mu_\bx{be}^\mathrm{sp}(t_k) + N_2\mu_\bx{eb}^\mathrm{sp}(t_k) \, ,
\end{equation}
where $N_{1,2}$ are the bunch intensities and $\mu_{\bx{be}(\bx{eb})}^\mathrm{sp}$ is the specific $\mu$ value of the beam-gas background for beam~1(2).
Two position parameters, \Dxz and \Dyz, were introduced to account for the fact that the luminosity may reach a maximum at a non-zero nominal separation (\Dx, \Dy) due to an imperfect alignment of the beams.

The last two terms in Eq.~\eqref{eq:vdm_2d_fitfun} are due to beam-gas interactions and are proportional to the beam intensity and the residual gas pressure.
The value of $\mu_{\bx{be}(\bx{eb})}^\mathrm{sp}$ is determined using events from data taken simultaneously for non-colliding bunches for each scan independently
and is typically $2\times10^{-14}$ ($1\times10^{-14}$).
In most cases the pressure was very stable during \vdm scans.
During the first pair of scans in the July session a drop of about 10\% was observed due to the neon gas injection system being used beforehand.
In order to take this into account, an exponential dependence on time is assumed for the first pair of scans in July, while in all other cases constant specific background is assumed.

As already discussed above, due to the pattern of two orthogonal movements used in the scans, the data provide no information on the factorizability parameters and the linear correlation parameters.
Therefore, in order to obtain the required additional information, we use the BGI measurements performed in the same fills as the \vdm scans.
The BGI analysis gives $f_{1,2}$ values compatible with zero.
In the \vdm analysis the $f_{1,2}$ parameters are fixed to zero.
In this fully non-factorizable case, the two weight parameters $w_{x}$ and $w_{y}$ for each beam are perfectly anticorrelated as seen from Eq.~\eqref{eq:non_factorizable_weights}.
Therefore, only one weight parameter per beam is used for the \vdm analysis.
The value of $C_{xy}$ is computed using the BGI measurements of the convolved bunch length and crossing angles.
No significant linear correlation in the beam distributions is observed, thus the value of $\bar{\corcoef}$ is set to zero.
The available number of counts per scan point and the relatively coarse scanning grid do not allow measuring both $R_m$ and $A_m$.
The parameter $R_m$ describes the main features of $\Sigma_{m,ij}$, Eq.~\eqref{eq:dgsb_capsigmas}, and is determined in the fit, while $A_m$ is fixed to zero for the determination of the central value of $\sigma_\eff$.
The systematic uncertainties arising from the assumptions on the fit parameters are discussed in Sec.~\ref{sec:VDM:Fit model uncertainty}.
In total nine parameters remain free, including the visible cross-section.

Initially, the fit is performed using the uncertainty estimates obtained from the data ${\delta \mu_{\eff} = \sqrt{1/N_0 - 1/N}}$, where $N_0$ is the number of empty events in a total of $N$ events.
For small values of $\mu_{\eff}$, these uncertainty estimates are correlated with the $\mu_{\eff}$ values themselves, thus biasing the data weights and the fit result.
To mitigate this problem, the data are fitted a second time using
uncertainties $\delta\mu_{\eff} = \sqrt{\exp(\hat{\mu}_{\eff})/N - 1/N}$, which are based on the predictions $\hat{\mu}_{\eff}$ of the first fit.

\begin{figure}[ptb]
  \begin{center}
    \includegraphics[width=0.98\linewidth]{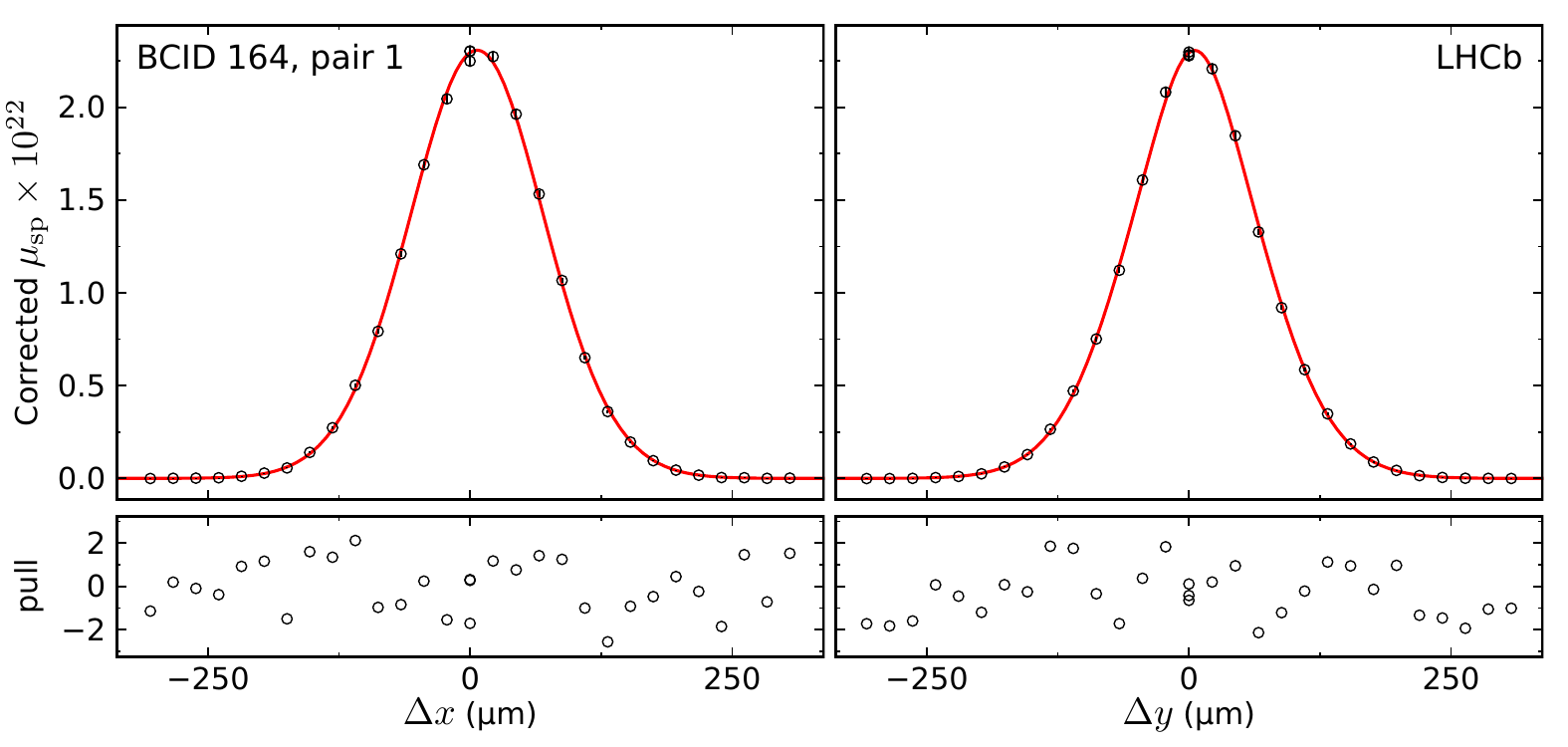}
    \includegraphics[width=0.98\linewidth]{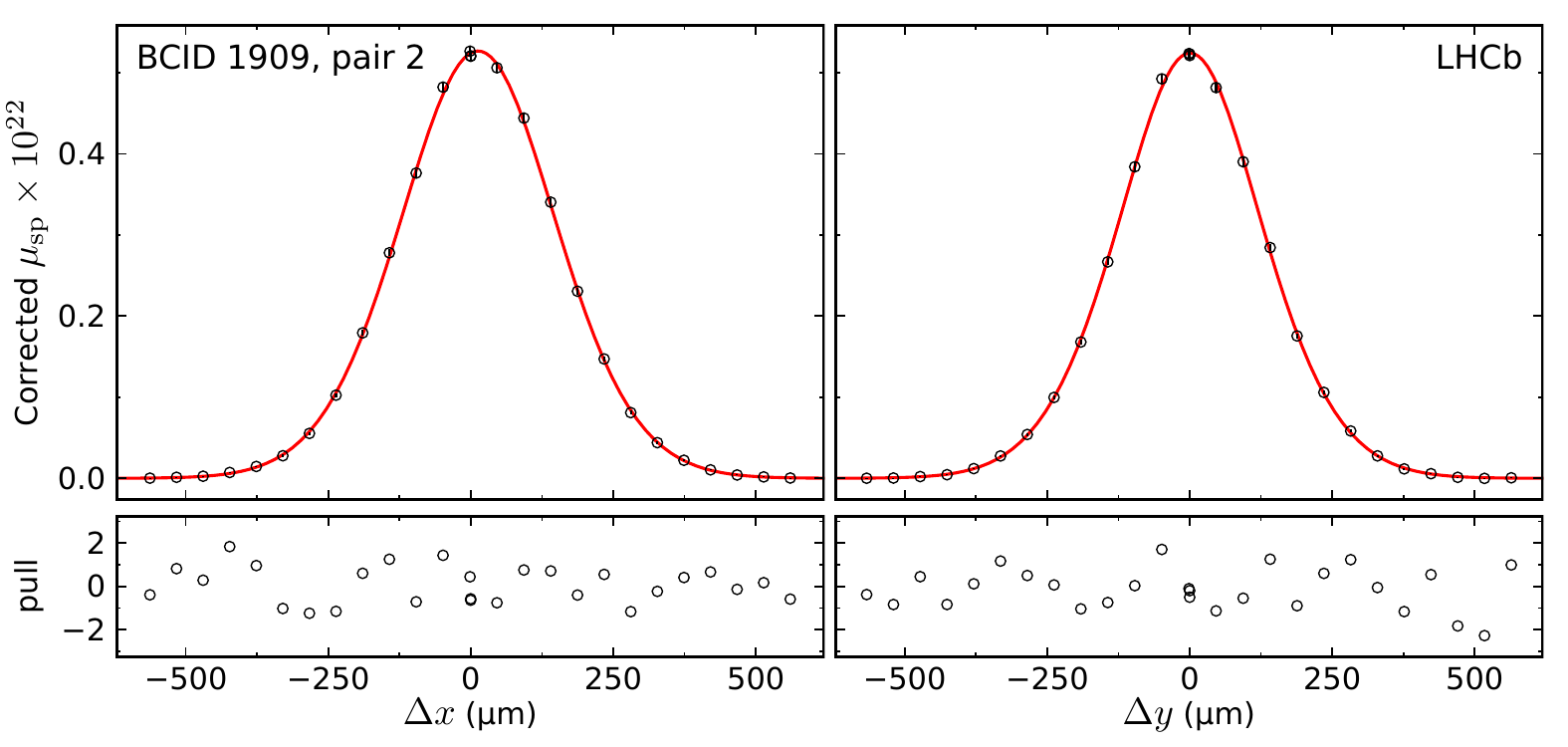}
    \vspace*{-0.5cm}
  \end{center}
  \caption{
    Fit of \vdm profiles for a single bunch and scan pair in the (top) April and (bottom) July \vdm sessions.
    The left (right) panels show data and fit predictions corresponding to the $x$ ($y$) scan.
    The data points are represented with circles, while the fitted curve is shown as a solid line.
    The error bars are smaller than the symbol size.
    The fit pulls are displayed below each fit projection and show no systematic structure.
    There are three data points at $\Delta m = 0$ as the nominal point was measured three times for each scan.
  }
  \label{fig:xsec_fit_example}
\end{figure}
\begin{figure}[ptb]
  \begin{center}
    \includegraphics[width=0.49\linewidth]{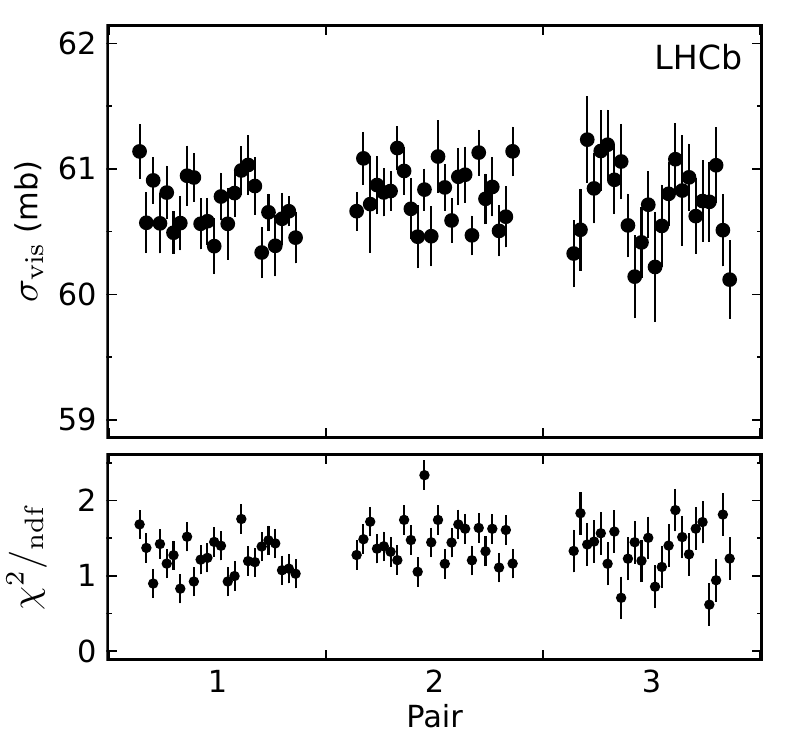}
    \includegraphics[width=0.49\linewidth]{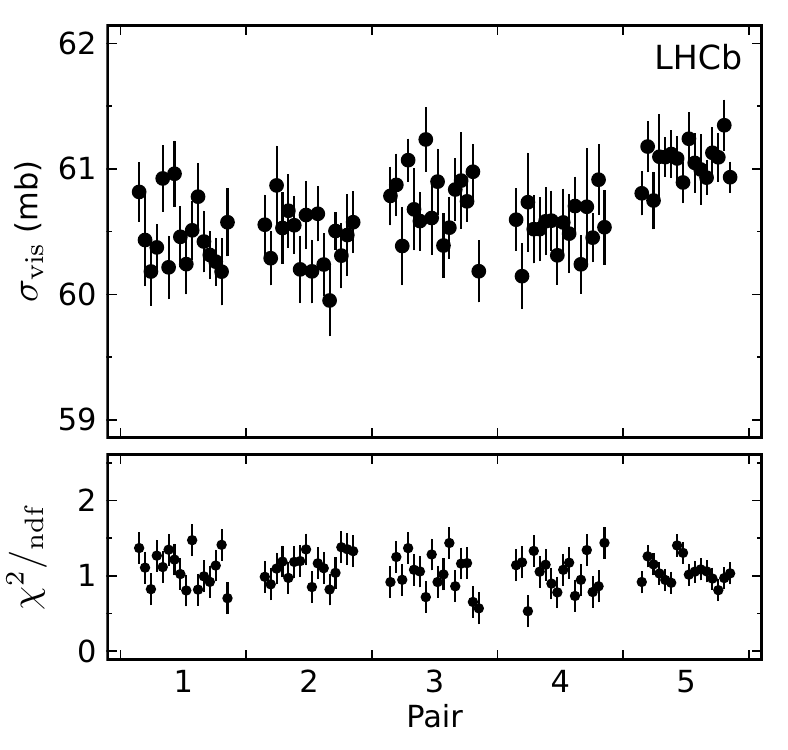}
    \vspace*{-0.5cm}
  \end{center}
  \caption{
    (Top) measured cross-section values and
    (bottom) the corresponding values of \chisqndf
    for all bunch pairs and for non-offset scan pairs in the (left) April and (right) July \vdm sessions.
    Measurements within one scan pair are separated from the rest by a larger distance.
  }
  \label{fig:xsec_fit_xsec}
\end{figure}
For presentation purposes we define a corrected specific interaction rate per crossing by rearranging Eq.~\eqref{eq:vdm_2d_fitfun}
\begin{equation}
\mu_{\eff,i}^\mathrm{sp} = \frac{\mu_{\eff}(\Dx_i, \Dy_i) 
  - N_1\mu_\bx{be}^\mathrm{sp}(t_i) - N_2\mu_\bx{eb}^\mathrm{sp}(t_i)}{N_1 N_2} \, .
\end{equation}
Then, the function $\sigma_\eff \ovint(\Dx,\Dy)$ represents the fit to the corrected data points $\mu_{\eff,i}^\mathrm{sp}$.
An example of a fit for a single bunch pair is shown in Fig.~\ref{fig:xsec_fit_example} for one scan pair.
The measured cross-section values and the values of $\chi^2$ per degree of freedom, \chisqndf, obtained for all bunch and scan pairs are shown in Fig.~\ref{fig:xsec_fit_xsec}.
Higher values of $\sigma_\eff$ are obtained for the fifth scan pair in July.
The associated uncertainty is discussed in detail in Sec.~\ref{sec:VDM:Reproducibility}.
For the April scan session, the values of \chisqndf are on average higher than one. This can be explained by the smaller beam size and the fact that the uncertainty of the beam separation is not taken into account.
Finally, the cross-section is obtained for each calibration session by calculating the weighted average of all measurements from reference scan pairs.

The sources of systematic effects that influence the result of the fit described with Eq.~\eqref{eq:vdm_2d_fitfun} can be grouped into two major categories.
First, there can be systematic effects related to the inputs of the fit, namely bunch intensities (described in Sec.~\ref{sec:Bunch current normalization}), rates (of signal and background) and beam positions (Secs.~\ref{sec:VDM:Rate measurement}--\ref{sec:VDM:Beam-beam effects}).
Second, the influence of the \vdm profile model, the technical aspects of the fitting procedure and discrepancies between repeated measurements are described in Secs.~\ref{sec:VDM:Fit model uncertainty} and \ref{sec:VDM:Reproducibility}.
The variation in the obtained cross-section among bunch pairs is used to estimate the uncertainty due to the relative bunch intensity measurement.
The uncertainty due to a potential non-linearity of the FBCT device is estimated to be about $0.05\%$ using the same method as described in Sec.~\ref{sec:bgi_results}.
Moreover, by using the independent measurements from the ATLAS BPTX system~\cite{ATLAS-BPTX}, a discrepancy in the final result of about $0.1\%$ is found, which is assigned as an additional systematic uncertainty.

%% file: vdm_corrections.tex
\subsection{Rate measurement}
\label{sec:VDM:Rate measurement}

For the absolute calibration, the beam-gas related backgrounds are subtracted taking into account the difference in bunch populations.
There is a statistical uncertainty associated with the measured specific background per proton used for subtraction.
The value of the specific background is shared among cross-section fits of bunch pairs and introduces a correlation,
which is taken into account when combining the measurements.
This uncertainty is estimated to be less than 0.1\%.

The beam-beam related background is estimated to be 0.1--0.2\% by taking the difference between the visible cross-sections of the restricted and the non-restricted \emph{Track} observables.
For the central value of the visible cross-section we use the less affected \emph{Track} observable, while the full difference is taken as systematic uncertainty.

In the presence of a non-zero beam crossing angle, the luminous region position varies with transverse beam separation.
The track reconstruction efficiency is not uniform as function of the primary vertex longitudinal position.
Therefore, a correction to the $\mu_\eff$ values is applied using the same principle as described in
Sec.~\ref{sec:InteractionRate}.

\begin{figure}[ptb]
  \begin{center}
    \includegraphics[width=0.49\linewidth]{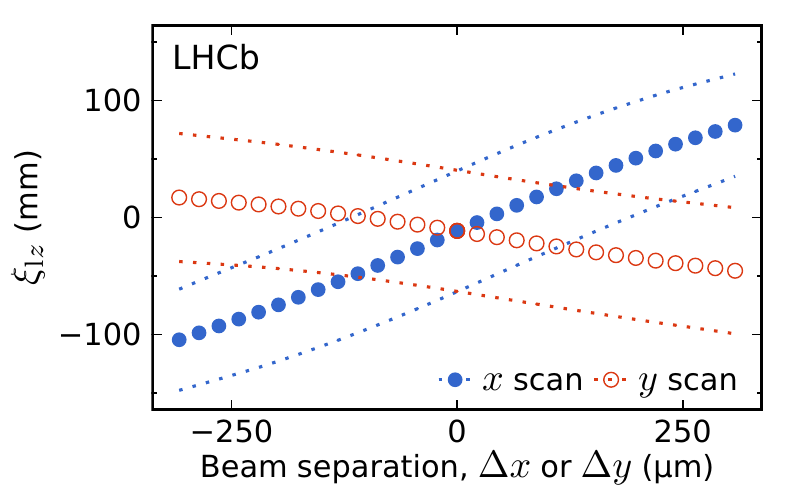}
    \includegraphics[width=0.49\linewidth]{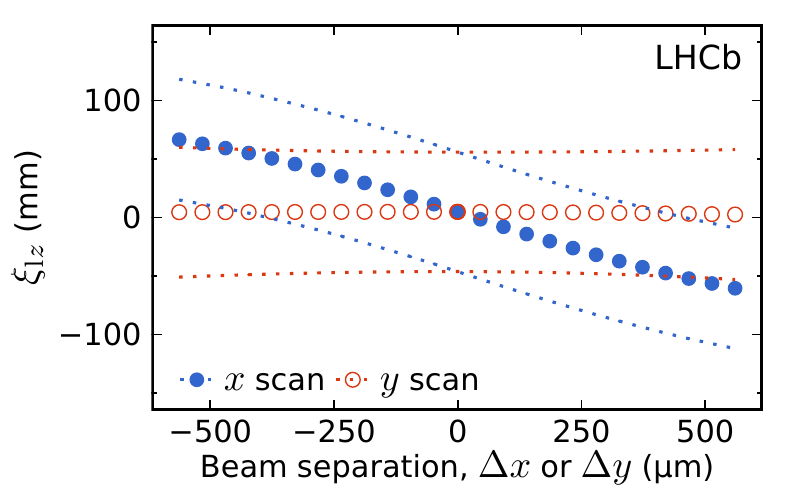}
    \vspace*{-0.5cm}
  \end{center}
  \caption{
    Luminous region longitudinal position and size as function of beam separation for a scan pair in the (left) April and (right) July \vdm sessions.
    The values for other scan pairs are similar.
    Markers indicate the mean while the bands between dotted lines contain 68\%
    of the vertices.
    In April, the luminous region moves in both horizontal and vertical scans 
    because of the crossing angle configuration.
    The inversion of the slope in the horizontal scan is due to
    the change of sign of the crossing angle between the two sessions.
  }
  \label{fig:cor_zeff_pos}
\end{figure}
\begin{figure}[ptb]
  \begin{center}
    \includegraphics[width=0.49\linewidth]{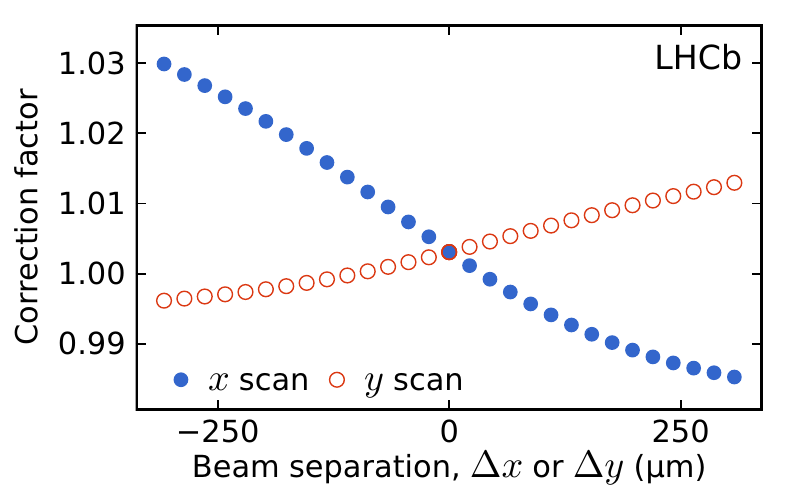}
    \includegraphics[width=0.49\linewidth]{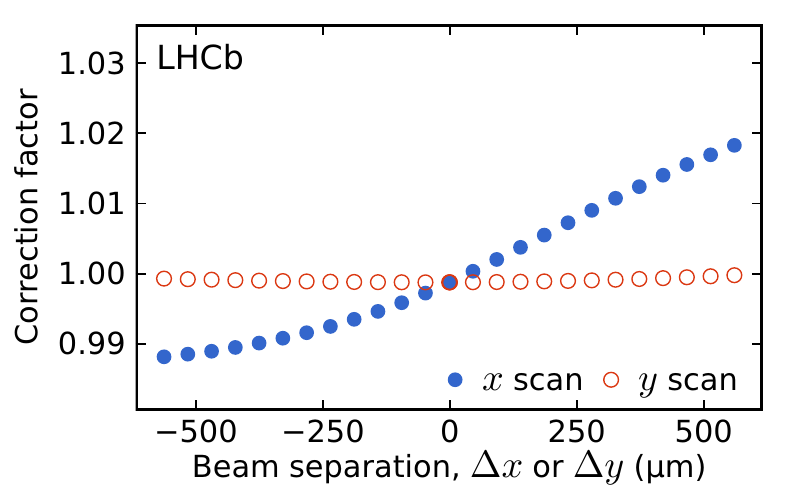}
    \vspace*{-0.5cm}
  \end{center}
  \caption{
    Efficiency correction factors for the \emph{Track} observable 
    for a scan pair in the (left) April and (right) July \vdm sessions.
    The values for other scan pairs are similar.
  }
  \label{fig:cor_zeff_factor}
\end{figure}
The longitudinal position and size of the luminous region are measured for each step using a single Gaussian fit to the selected beam-beam vertices, taking into account the vertex reconstruction efficiency as a function of $z$.
The quantities are linearly extrapolated to large separations, where the
luminosity vanishes and not enough vertices are available ($< 1000$).
The measured longitudinal parameters of the luminous region are shown in Fig.~\ref{fig:cor_zeff_pos} and
the calculated correction factors to the observed $\mu_\eff$ for each scan and step are shown in Fig.~\ref{fig:cor_zeff_factor}.
It can be noted that the correction factors are approximately linear as function of separation.
Therefore, no large effect on the VDM profile width is expected.
The effect of the correction amounts to $+0.32\%$ ($-0.03\%$) in the April (July) calibration.
An estimate of the associated systematic uncertainty is made using a comparison with an unaffected observable (\emph{Calo}).
In the approximation of pure Gaussian beams, the rate at zero beam separation is proportional to $\sigma_\eff/(\Sigma_x\Sigma_y)$, see Eqs.~\eqref{eq:cross-section}, \eqref{eq:luminosity} and \eqref{eq:ovrlap_compact}.
Equivalently, the measured cross-section is proportional to the rate at zero separation and the product of the widths of the \vdm profiles. 
It is useful to separate the correction factor dependence on $\xilz$ (Fig.~\ref{fig:sim_counter_efficiency}) in a linear and a higher order part.
The former only affects the rate at zero separation, while the latter mainly affects the widths. 
The residual (after correction) slope of the ratio $\mueff{Track}/\mueff{Calo}$ as function of $\xilz$ is measured and normalized to the ratio at $\xilz=0$.
The $\xilz$ value at the working point is multiplied by that slope to obtain the uncertainty on the cross-section due to the linear part of the correction.
The uncertainty due to the higher order part of the correction is estimated by comparing the product $\Sigma_x\Sigma_y$ measured using the \emph{Track} and the \emph{Calo} observable.
The two uncertainties are summed linearly to obtain the systematic uncertainty of 0.13\% (0.07\%) for the April (July) calibration.

\subsection{Length scale}
\label{sec:VDM:Length scale}
The nominal beam separation values \Dx and \Dy are calculated from the \lhc magnet currents at every scan step.
An absolute calibration of the beam separation is made using the more precise \velo scale.
In principle, each beam can have different calibration constants.
However, since for all scans the beams were moved symmetrically (as opposed to one beam at a time), only the average (or common) length scale matters to first order for calibrating the separation.
During the length scale calibration scans the beams were moved in several steps, with each measurement lasting from one to four minutes.
The time intervals during which the beams were stationary are determined using the recorded states of the deflection magnets in the \lhc logging data.
Two methods are used in order to obtain the absolute length scale calibration.

\paragraph*{Constant beam separation method:}
Both beams are moved in five equidistant steps in \Dx and \Dy keeping their nominal separation constant.
During a simultaneous parallel translation of both beams, the centre of the luminous region should follow the beam positions regardless of the bunch shapes.
The luminous region centre can be determined using beam-beam vertices measured with the \velo.

Beam-beam vertices are selected using randomly triggered events in \bx{bb} crossings.
In the 2013 calibrations the average number of interactions per crossing was very low.
Therefore, the standard trigger requirement for beam-beam events in \bx{bb} crossings was used in order to collect a sufficient amount of data.
To remove background from material interactions, only vertices that lie within $2\mm$ from the median in each axis are retained.
Potential beam-gas interaction background is reduced 
by applying a loose cut on the longitudinal position of the vertices $|z| < 350\mm$ %
and by requiring at least one track in each direction.
The vertex coordinates are binned in $x$ and $y$.
The bin width is chosen automatically based on the data according to the Freedman--Diaconis rule.\footnote{
  The bin width is determined as two times the interquartile range (the difference between the 75\textsuperscript{th} and the 25\textsuperscript{th} percentiles) divided by the number of observations in the sample to the power of $1/3$ \cite{Freedman:6360487}.
}
Data that deviates more than $5\sigma$ from the median is discarded.

An empirical model is fitted to the histogram for each coordinate and the mean of the model distribution is used as an estimator for the position of the luminous region.
A sum of two Gaussian functions, where the two means are not required to be equal, is found to fit the data well in all cases.
The distribution of the luminous region is not expected to change during the scans if the beam separation is kept constant.
Using this property as a constraint, an additional global fit per scan is performed, which has a single set of shape parameters with the exception of the mean, which is independent for each step.

A simultaneous drift of both beams in the same direction affects the length scale measurement.
The drift is estimated using the steps before and after the scans, when the beams are nominally centred.
Differences of the luminous region position before and after a scan are attributed entirely to simultaneous beam drift.
A correction is applied for each step during the scan by using a linear interpolation.
Since the beam separation is kept constant, beam-beam effects do not influence the measurement.

\begin{figure}[ptb]
\begin{center}
  \includegraphics[width=0.98\linewidth]{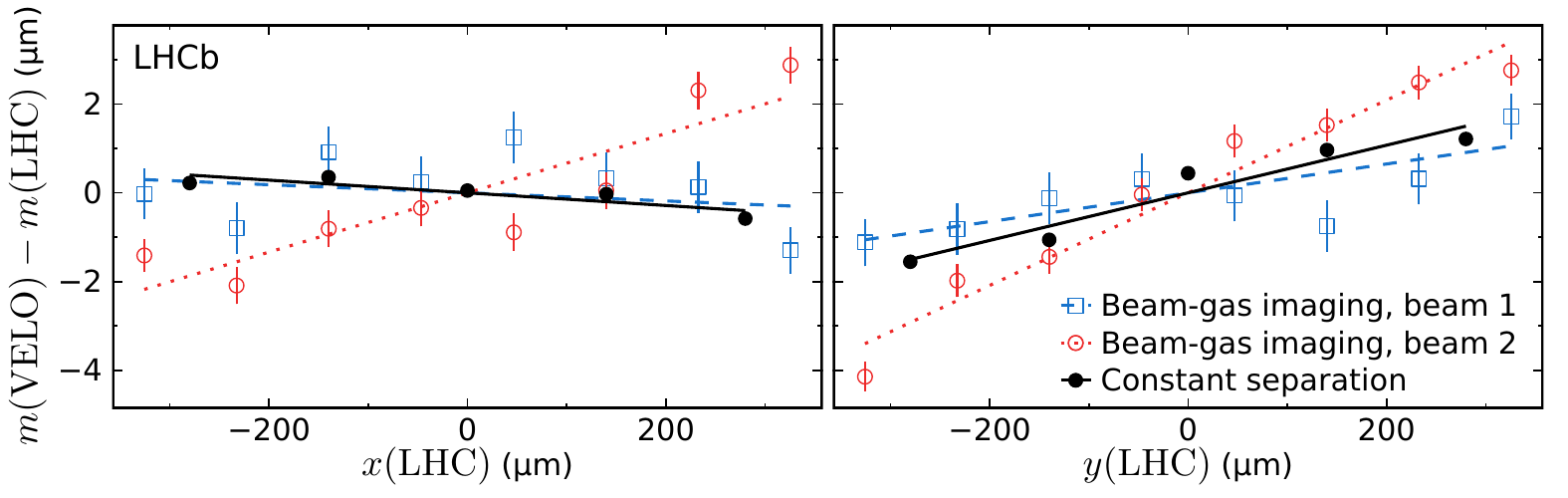}
  \vspace*{-0.5cm}
\end{center}
\caption{
  Length scale calibration fits in the (left) $x$ and (right) $y$ axis in the July \vdm session.
  The fitted intercepts are subtracted from both the fit curves and the data points as they have no significance.
  Each data point is an aggregate of the measurements from 16 colliding bunch pairs for the constant separation method and 35 individual bunches for the BGI method.
  The fit slope obtained using the constant separation method (solid line) should be compared with the average of the slopes obtained with the BGI method (dashed and dotted lines).
}
\label{fig:lsc-fit}
\end{figure}

The measured luminous region position for each step is fitted against the average of the nominal beam positions using a weighted linear fit.
The fit is performed simultaneously for all colliding bunch pairs using independent intercepts and a common slope.
The results of this fit are shown in Fig.~\ref{fig:lsc-fit}.
The slope parameter of the fit is an estimate of the length scale calibration factor.

\paragraph*{Beam-gas imaging method:}
Individual beam positions are measured using beam-gas interactions.
The neon gas injection system was used to enhance the beam-gas interaction rate.
The beams were moved similarly to \vdm scans, but in a few large steps.
Beam-gas vertices for each beam are selected using the corresponding standard trigger requirements for beam-empty and beam-beam crossings.
In order to remove background from material interactions, only vertices that lie within $2\mm$ from the median in each axis are retained.
Potential beam-beam interaction background is reduced by requiring no tracks in the opposite direction of the corresponding beam.
Consistent comparison between measurements from colliding and non-colliding bunches is ensured by imposing a cut on the longitudinal position of the vertices $|z| > 300\mm$.
To obtain the single-beam offsets, the $x$ and $y$ coordinates of the beam-gas vertices are fitted with a straight line as function of $z$.
The weight assigned to the individual vertices takes into account the beam width and the vertex resolution.

Individual beam positions are measured before and after the scans, when the beams are nominally centred.
No significant beam drifts are observed at the $0.5\mum$ level, thus no correction is made.
The beam-beam deflection effect influences beam positions in \bx{bb} bunch crossings.
We take this into account as described in Sec.~\ref{sec:VDM:Beam-beam effects}, where $\Sigma$ values are taken from the \vdm scan measurements.
Taking this effect into account improves the consistency between measurements from colliding and non-colliding bunches.

The measured beam position for each step is fitted against the nominal setting using a weighted linear fit.
The fit is performed simultaneously for all bunches in each beam using one slope and one intercept parameter.
The results of this fit are shown in Fig.~\ref{fig:lsc-fit}.
To obtain an effective common length scale, the average between both beams is taken.

\begin{table}[ptb]
  \centering
  \caption{ 
    Length scale calibration constants. All uncertainties are statistical.
  }
  \begin{tabular}{llcc}
    Calibration & Method & 
    $\left(\frac{\Dx(\mathrm{VELO})}{\Dx(\mathrm{LHC})}-1\right)\times 10^2$ &
    $\left(\frac{\Dy(\mathrm{VELO})}{\Dy(\mathrm{LHC})}-1\right)\times 10^2$ \\
    \midrule
    April 2012 & Constant separation
                          & $-1.10\pm0.02$ & $-0.30\pm0.02$ \\
    July 2012  & Constant separation 
                          & $-0.14\pm0.03$ & $\phantom{-}0.54\pm0.04$ \\    
    July 2012  & Beam-gas imaging    
                          & $\phantom{-}0.29\pm0.06$ & $\phantom{-}0.69\pm0.05$ \\
  \end{tabular}
  \label{tab:vdm_lsc_results}
\end{table}
The length scale calibration constants are summarized in Table~\ref{tab:vdm_lsc_results}.
A statistically significant discrepancy of about $0.5\%$ is observed between the calibration factors for the $x$ axis obtained using the two methods in July.
The origin of this difference is not understood.
Therefore, we assign $0.5\%$ uncertainty to all length scale calibrations.
It is assumed that this uncertainty is correlated between calibrations.
The statistical uncertainties are small compared to the discrepancy and are neglected.
The calibration constants used for July are those obtained by the constant separation method.
Thus, we maintain consistency of treatment with other calibration sessions, for which we have not performed a beam-gas imaging calibration.

\subsection{Beam-beam effects}
\label{sec:VDM:Beam-beam effects}
The electromagnetic interaction between charged particles of two colliding bunches is called \emph{beam-beam effect}.
There are two aspects of this interaction that affect the \vdm scan measurements.
The first, called \emph{dynamic $\beta$ effect}, is the result of the mutual (de)focusing of the two colliding bunches.
It leads to a change in \betastar (and thus beam size), which depends on the transverse beam separation.
Therefore, the transverse size of the bunches is not constant during the \vdm scans.
Secondly, the closed orbits of the bunches are distorted by the angular kick induced by their electromagnetic repulsion.
This \emph{beam-beam deflection} effect has a different magnitude depending on transverse separation, thus distorting the scan profiles.

\begin{figure}[ptb]
  \begin{center}
    \includegraphics[width=0.49\linewidth]{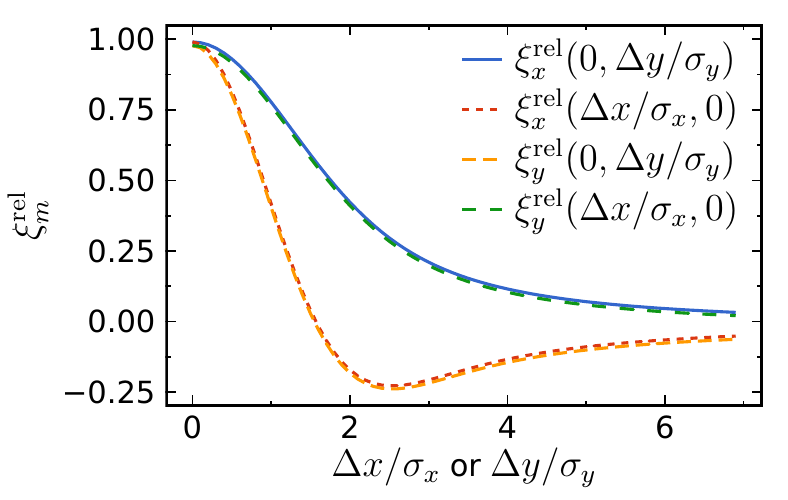}
    \vspace*{-0.5cm}
  \end{center}
  \caption{
    Relative change of the beam-beam parameter as function of beam separation.
    The values are obtained using a simulation of the machine optic elements.
  }
  \label{fig:cor_xi_rel}
\end{figure}

The so-called \emph{beam-beam parameter} that quantifies the strength of the dynamic $\beta$ effect in the axis $m$, for a normally distributed bunch in beam $j$ is given by \cite{Herr:CAS2003}
\begin{equation}\label{eq:beam_beam_param}
\xi_{m,j} = \frac{\alpha \hbar c}{2\pi}
  \frac{N_{j'}Z_{j'} \betastar_{m,j}}
  {E_{j} \sigma_{m,j'} (\sigma_{x,j'} + \sigma_{y,j'} )}
  \xi_m^\mathrm{rel}(\textstyle{\frac{\Delta x}{\sigma_{x,j}}},\textstyle{\frac{\Delta y}{\sigma_{y,j}}}) \, ,
\end{equation}
where $j'$ denotes the bunch in the opposite beam,
$N_{j'}Z_{j'}$ is the bunch charge in units of elementary charge,
$\betastar_{m,j}$ is the value of the $\beta$ function at the IP,
$E_{j}$ is the ring energy setting (particle energy divided by particle charge) and $\sigma_{m,j'}$ are the bunch sizes.
The function $\xi_m^\mathrm{rel}$ was modelled \cite{Herr:lumidays12} using 
the MAD-X optics software \cite{MADX} and is shown in Fig.~\ref{fig:cor_xi_rel}.
It is independent of the machine optics parameters and the bunch properties.
For small separations $\Dx$ and $\Dy$ the value of the $\xi_m^\mathrm{rel}$ is close to unity and approximately constant.
Therefore, the beam-beam force is approximately linear, resembling the force of a quadrupole field.
However, the force becomes non-linear for large separations.
The above equation is also valid in the case of ion beams.

The ratio between the perturbed \betastar and the unperturbed $\betastar_0$ as function of separation is given by 
\begin{equation}
\frac{\betastar_{m,j}}{\betastar_{0,m,j}} =
  (1 + 4\pi\xi_{m,j}\cot(2\pi Q_{m,j}) - 4\pi^2 \xi_{m,j}^2)^{-1/2} \, ,
\end{equation}
where $Q_{m,j}$ is the machine tune of beam $j$ in the axis $m$.
While there is a recursive dependence between $\xi$ and $\betastar$, this is only a second order effect and can be neglected by using the nominal $\betastar_0$ in Eq.~\eqref{eq:beam_beam_param}.
Possible collisions at other IPs also contribute to the dynamic $\beta$ independently from the local beam separation.
Such contributions, provided beams are not moving at other IPs, are constant and effectively modify the nominal value of \betastar.
Since these modifications are much smaller than the uncertainty in the value of the nominal \betastar itself, they are neglected.

\begin{figure}[ptb]
  \begin{center}
    \includegraphics[width=0.49\linewidth]{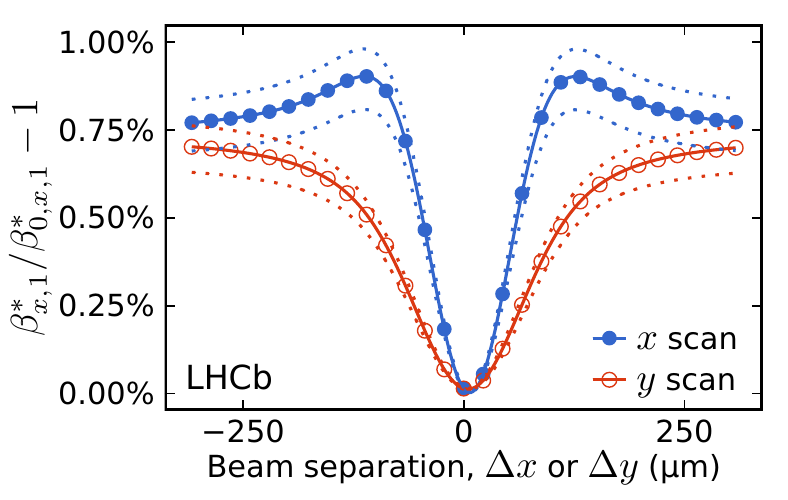}
    \includegraphics[width=0.49\linewidth]{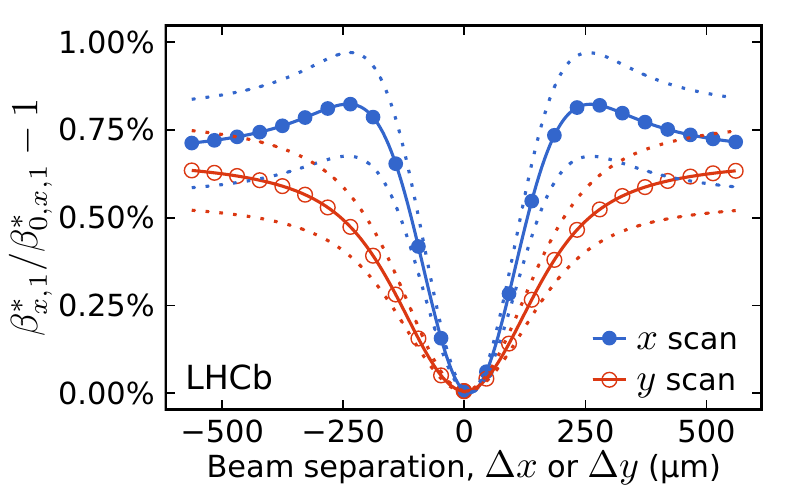}
    \vspace*{-0.5cm}
  \end{center}
  \caption{
    Relative \betastar change for \beamone in the $x$ axis as as function of beam separation
    for a scan pair in the (left) April and (right) July \vdm sessions.
    The values for other scan pairs are similar.
    The values for \beamtwo are similar owing to similar bunch intensities.
    The values for the $y$ axis are also similar 
    with $x$ and $y$ scans exchanged.
    The markers and the solid line represent the average, while the dotted 
    lines correspond to the extremes among colliding bunches.
    The reference \betastar is chosen such that the relative change is zero at the separation corresponding to maximum luminosity.
  }
  \label{fig:cor_dynbeta_rdb}
\end{figure}

\begin{figure}[ptb]
  \begin{center}
    \includegraphics[width=0.49\linewidth]{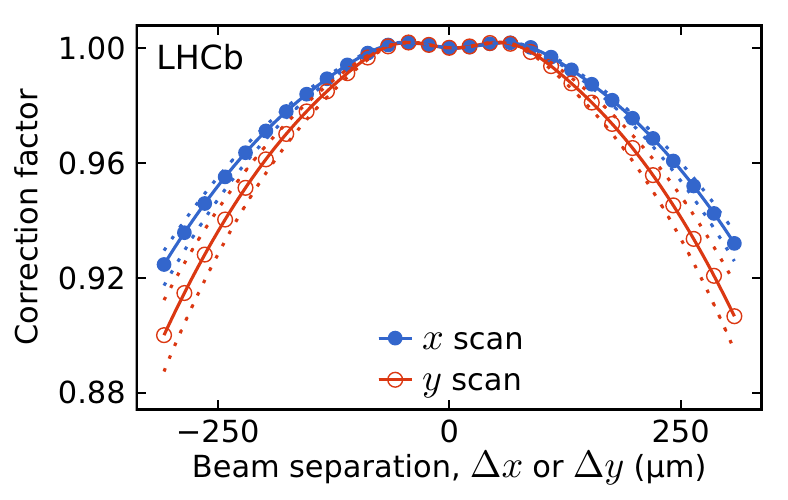}
    \includegraphics[width=0.49\linewidth]{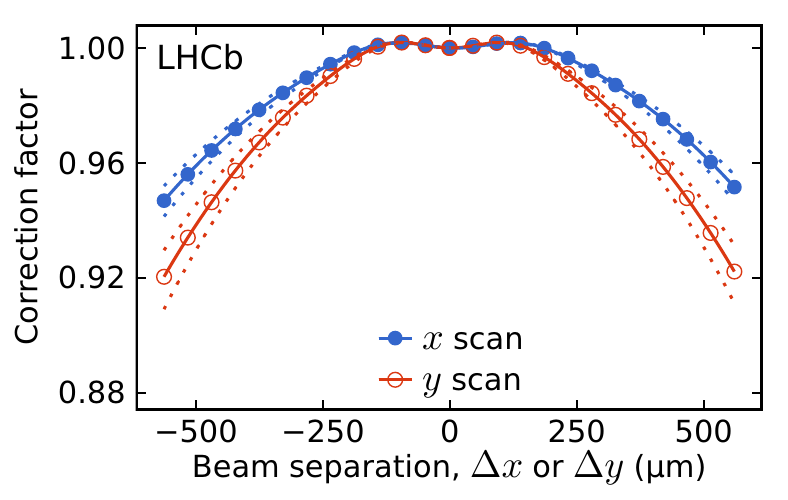}
    \vspace*{-0.5cm}
  \end{center}
  \caption{
    Rate correction factor for the dynamic beta effect
    for a scan pair in the (left) April and (right) July \vdm sessions.
    The values for other scan pairs are similar.
    The markers and the solid line represent the average, while the dotted 
    lines correspond to the extremes among colliding bunches.
    Owing to the dynamic $\beta$ effect, the $\beta$'s and thus the beam sizes 
    depend on the beam separation.
    To take this into account, the effect on the luminosity is compensated by multiplying the observed rates with the correction factors.
  }
  \label{fig:cor_dynbeta_factor}
\end{figure}

To calculate the effect on the luminosity, the relative change of \betastar is computed with respect to a reference value.
Applying a correction to the rate (or the overlap integral) effectively removes the dependence of bunch shapes on separation.
The choice of the reference \betastar value is arbitrary and determines the values of the effective separation-independent bunch parameters.
It can be shown that this choice has a negligible impact on the net correction.
It is beneficial to use $\betastar(0,0)$ as the reference, in order to enable consistent comparisons of \vdm profile parameters from different scans. 
The relative \betastar change (see Fig.~\ref{fig:cor_dynbeta_rdb}) is calculated for each colliding bunch pair using the parameters of the nominal optics, the measured bunch intensities and the convolved widths.
In this case, the shape of the bunches is approximated with a single Gaussian distribution and the size of the colliding bunches in each pair is assumed equal.
The relative change of the luminosity is obtained using the perturbed bunch widths.
Finally, the correction factor to the rate is obtained, which is shown in Fig.~\ref{fig:cor_dynbeta_factor}.

\begin{figure}[ptb]
  \begin{center}
    \includegraphics[width=0.49\linewidth]{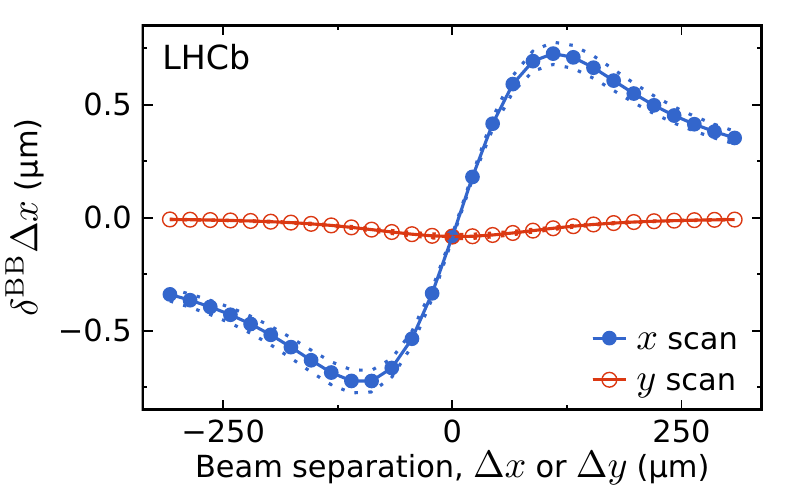}
    \includegraphics[width=0.49\linewidth]{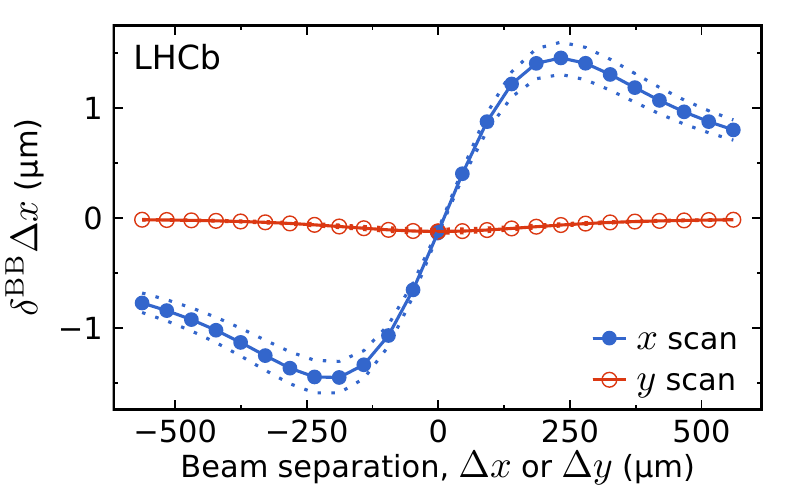}
    \vspace*{-0.5cm}
  \end{center}
  \caption{
    Correction to the nominal separation in the $x$ axis for beam-beam-induced orbit shift
    for a scan pair in the (left) April and (right) July \vdm sessions.
    The values for other scan pairs are similar.
    The values for the $y$ axis are similar with $x$ and $y$ scans exchanged.
    The markers and the solid line represent the average, while the dotted 
    lines correspond to the extremes among colliding bunches.
  }    
  \label{fig:cor_beambeam_shift}
\end{figure}

The beam-beam angular kick causes a deflection of beam $j$, which is calculated numerically using the formalism of Ref.~\cite{Ziemann:1991sb}
\begin{equation}
\theta_{y,j} + i\,\theta_{x,j} =
  -2\alpha \hbar c \frac{N_{j'}Z_{j'}}{E_j}
  F_0(\Dx, \Dy, \Sigma_x, \Sigma_y) \, ,
\end{equation}
where $N_{j'}Z_{j'}$ is the bunch charge in the opposite beam in units of elementary charge, $i$ is the imaginary unit, $F_0$ is a complex function
and Gaussian bunch profiles are assumed. 
The shift of the closed orbit in the $m$ coordinate for beam $j$ is given by
\begin{equation}
\delta_{m,j}^\mathrm{BB} = \theta_{m,j} \betastar_{m,j} \frac{1}{2 \tan(\pi Q_{m,j})} \, ,
\end{equation}
while the effect on the beam separation is the sum of the individual beam shifts
\begin{equation}
\delta^\mathrm{BB}\Delta m = \delta_{m,1}^\mathrm{BB} + \delta_{m,2}^\mathrm{BB} \, .
\end{equation}
The corrections that are added to the nominal beam separation are shown in
Fig.~\ref{fig:cor_beambeam_shift}.
The procedure outlined above was verified by the direct observation of beam-beam deflections with \lhc orbit data \cite{Kozanecki:1581723}, which were found to agree well with the expectations.

The systematic uncertainty of the correction for beam-beam effects is estimated by numerically propagating the uncertainties of the input parameters.
In addition to \betastar and machine tune, uncertainties are assigned to the ratio of bunch widths and the convolved bunch width to account for the assumption of identical bunch shapes and Gaussian bunches, respectively.
The assumed uncertainties on input parameters and their correlations are listed in Table~\ref{tab:beam_beam_input_unc}.
The uncertainty is similar for the two 8\tev \PPP calibrations and amounts to about $0.3\%$.
\begin{table}[ptb]
  \centering
  \caption{
    Uncertainties of input parameters to the correction for beam-beam effects.
    All quantities are correlated between axes. The quantities $\betastar_{m,j}$ and $Q_{m,j}$ are not correlated between the two beams.
  }
  \begin{tabular}{lcccc}
    Source & $\betastar_{m,j}$ & $Q_{m,j}$ & $\sigma_{m,1}/\sigma_{m,2}$ & $\Sigma_m$ \\
    \midrule
    Uncertainty & 20\% & 0.02 & 50\% & $\sim 10\%$ \\
  \end{tabular}
  \label{tab:beam_beam_input_unc}
\end{table}

\subsection{Fit model uncertainty}
\label{sec:VDM:Fit model uncertainty}
The choice of a particular model to describe the single beam shapes gives rise to a systematic uncertainty.
The bias in the case of the BGI method is studied by simulating various deviations of the single-beam distributions from the assumed model as described in Sec.~\ref{sec:sys_fit_model}.
We assume that the same uncertainty of $0.5\%$ applies to the \vdm calibration and that it is fully correlated between the methods.
In addition,
the fitting procedure applied to \vdm scan data might introduce a bias.
The potential fit bias is estimated to be less than $0.2\%$ by applying the analysis to simulated \vdm scans with experimental conditions and beam properties similar to those of data.

The linear correlations in \vdm scan profiles can be non-zero due to a tilt of the beam crossing plane or due to non-zero linear correlations with respect to the scanning axes of the transverse bunch distributions in Eq.~\eqref{eq:overlap_correlation}.
The latter are found to be small and are neglected.
The bunch crossing plane is nominally horizontal in all calibration fills, except in April, when the tilt was approximately $-21.5\degrees$.
Taking into account the correlation, the April calibration result changes by about $+0.3\%$.
The effect of the correlation is most pronounced if the scan pair crossing point is not centred at zero, as seen from the exponent factor in Eq.~\eqref{eq:overlap_with_rho}.
For the nominal scans, the correction from the exponent factor is negligible.
However, the correction to the fourth (offset) scan pair in April amounts to $-8\%$, improving significantly the consistency with the reference scans.
Rather than propagating the uncertainty of the fixed parameter $C_{xy}$ to the fit result, half of the effect of the correction is used as an estimate of the systematic uncertainty.

The major uncertainty of the \vdm calibration is due to the assumptions made on the parameter values as described in Sec.~\ref{sec:Cross-section determination}.
In order to estimate the corresponding bias, the fits are redone without fixing parameters but using constraints obtained from BGI measurements.
The information that is used from the BGI is limited to the factorizability of the beams, thus the \vdm calibration is as independent as possible.
In particular, constraints are not applied directly using dimensional parameters (widths) from BGI measurements, which are subject to resolution and alignment uncertainties.
Moreover, for each scan pair, all bunch pair fits are subject to the same constraints, with their width taking into account the spread of the parameter values among bunch pairs.

\begin{figure}[ptb]
  \begin{center}
    \includegraphics[width=0.49\linewidth]{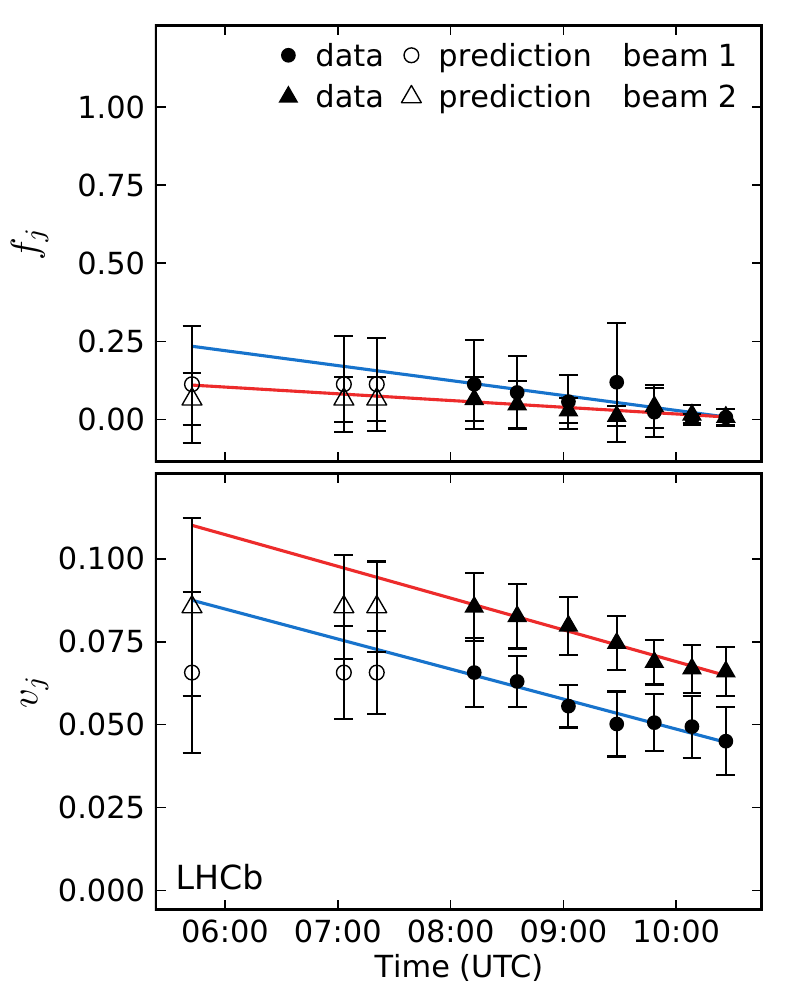}
    \includegraphics[width=0.49\linewidth]{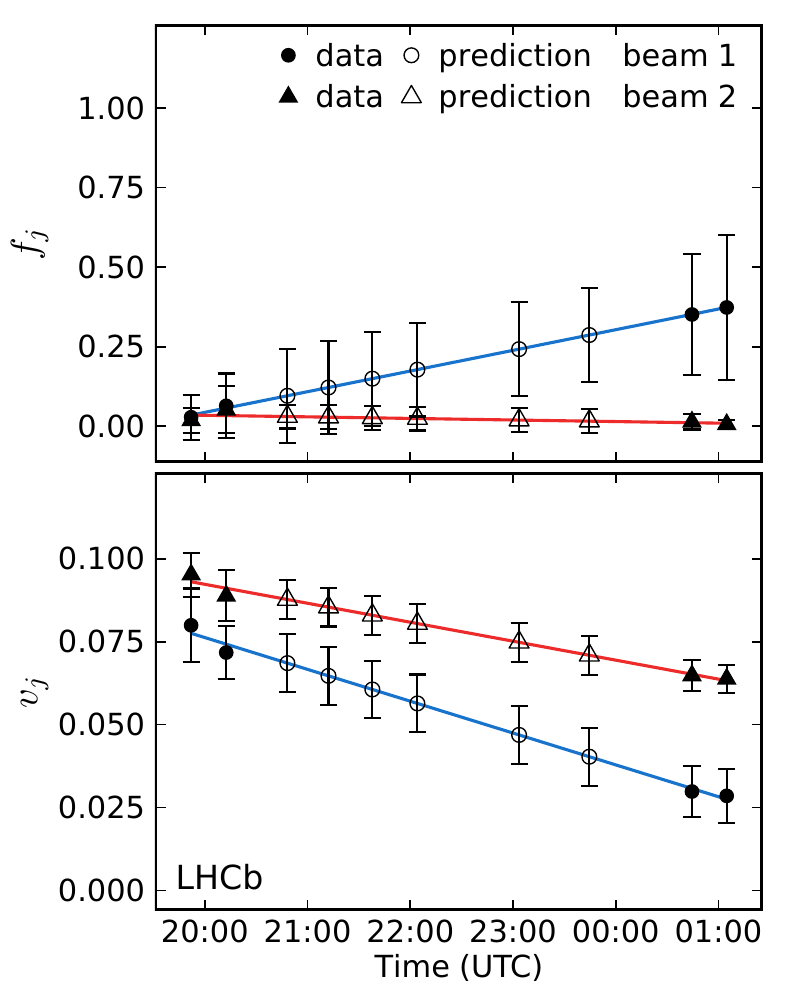}
    \vspace*{-0.5cm}
  \end{center}
  \caption{
    Values of the (top) factorizability parameter $f_j$ and (bottom) generic factorizability measure $v_j$
    in the (left) April and (right) July \vdm sessions.
    The filled markers and the corresponding error bars indicate the average and the RMS of BGI measurements for individual bunch pairs.
    In the case of the July scans, BGI measurements are available before and after the \vdm session, thus the values are linearly interpolated to obtain predictions at the time of \vdm scan pairs (open markers).
    For April, as BGI data are only available after the \vdm scans, the closest data points are used as a prediction, while the difference to the linear extrapolation is added in quadrature to the uncertainties.
    The trend of the generic factorizability measure shows that beam distributions become more factorizable as function of time.
  }
  \label{fig:factorizability}
\end{figure}

First of all, Gaussian constraints on the factorizability parameters $f_{1,2}$ are added to the fit.
However, these constraints are not sufficient to ensure that the beams are non-factorizable.
This is easily seen by considering sets of degenerate parameters for which the beam distribution is Gaussian in at least one of the axes.
Therefore, a generic measure of the transverse factorizability is introduced as
\begin{equation}
v_j = 1 - \frac{\int \rho_j(x,y=0) dx \int \rho_j(x=0,y) dy}
               {\rho_j(x=0, y=0)} \, ,
\end{equation}
where $\rho_j$ is the transverse beam distribution of beam $j=1,2$.
The measure $v_j$ is zero if %
$\rho_j$ is factorizable.
The values of $f_{1,2}$ and $v_{1,2}$ measured with BGI, as well as the predicted values at the time of the \vdm scans are shown in Fig.~\ref{fig:factorizability}.
The values of $v_{1,2}$, being functions of the individual beam distributions, cannot be computed using only the parameters describing the overlap integral from Eq.~\eqref{eq:overlap_params_vdm}.
Therefore, two additional parameters for each coordinate $m=x,y$ are added to the overlap fit model, which are sufficient to compute $v_{1,2}$:
\begin{align}
C_m &= (\sigma_{z1}^2 + \sigma_{z2}^2)\tan^2{\phi_m} \\
a_{m} &= \frac{\sigma_{m1,\rmn}^2}
              {\sigma_{m1,\rmn}^2 + \sigma_{m2,\rmn}^2} =
 \frac{\sigma_{m1,\rmn}^2}{\Sigma_{m,\rmn\rmn}^2 - C_m} \, ,
 \quad a_{m} \in [0,1] \, .
\end{align}
Gaussian constraints of 10\% around the measured value from BGI are applied to $C_{x,y}$.
Additionally, to avoid unphysical sets of beam parameters, beams are required to have similar transverse sizes in each coordinate.
This requirement is ensured by imposing a weak Gaussian constraint of $0\pm0.1$ on the asymmetry of the \rms of the two beam distributions.
Such constraints are justified by the equal design \betastar and emittance of the two beams.
The asymmetries estimated from the BGI measurements for all bunch pairs lie within $\pm 1\sigma$, indicating that the width of the constraints is sufficiently large.

\begin{figure}[ptb]
  \begin{center}
    \includegraphics[width=0.49\linewidth]{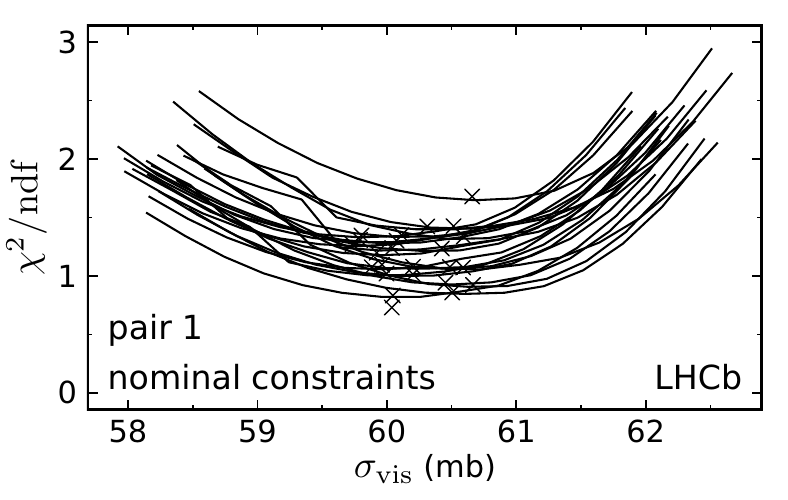}
    \includegraphics[width=0.49\linewidth]{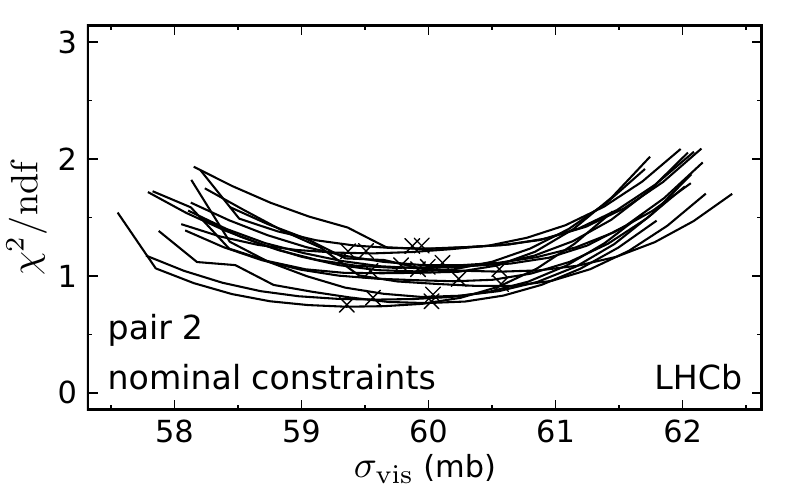}
    \vspace*{-0.5cm}
  \end{center}
  \caption{
    Examples of the constrained fit \chisqndf profiles as function of $\sigma_\eff$ for (left) scan pair 1 in April and (right) scan pair 2 in July.
    Each curve corresponds to one bunch pair.
    The values of $\sigma_\eff$ obtained with direct fitting are shown with cross markers.
  }
  \label{fig:vdm_syst_chi2_example}
\end{figure}

The \vdm data for each bunch and scan pair are fitted using the additional parameters and the constraints described above.
Moreover, profiles of \chisqndf are obtained
as function of $\sigma_\eff$.
The minima of the \chisqndf profiles are very close to the values of $\sigma_\eff$ obtained
with the fit where $\sigma_\eff$ is a free parameter
as seen in Fig.~\ref{fig:vdm_syst_chi2_example}.
The new values of $\sigma_\eff$ are averaged for each scan pair and the obtained value is compared to the baseline cross-section for that pair.
A difference of $0.6\%$ and $0.9\%$ is found for the April and the July calibration, respectively.
The difference is attributed to the fact that $f_{1,2}$ and $A_{x,y}$ are fixed for the baseline fits and is assigned as a systematic uncertainty.

\begin{figure}[ptb]
  \begin{center}
    \includegraphics[width=0.98\linewidth]{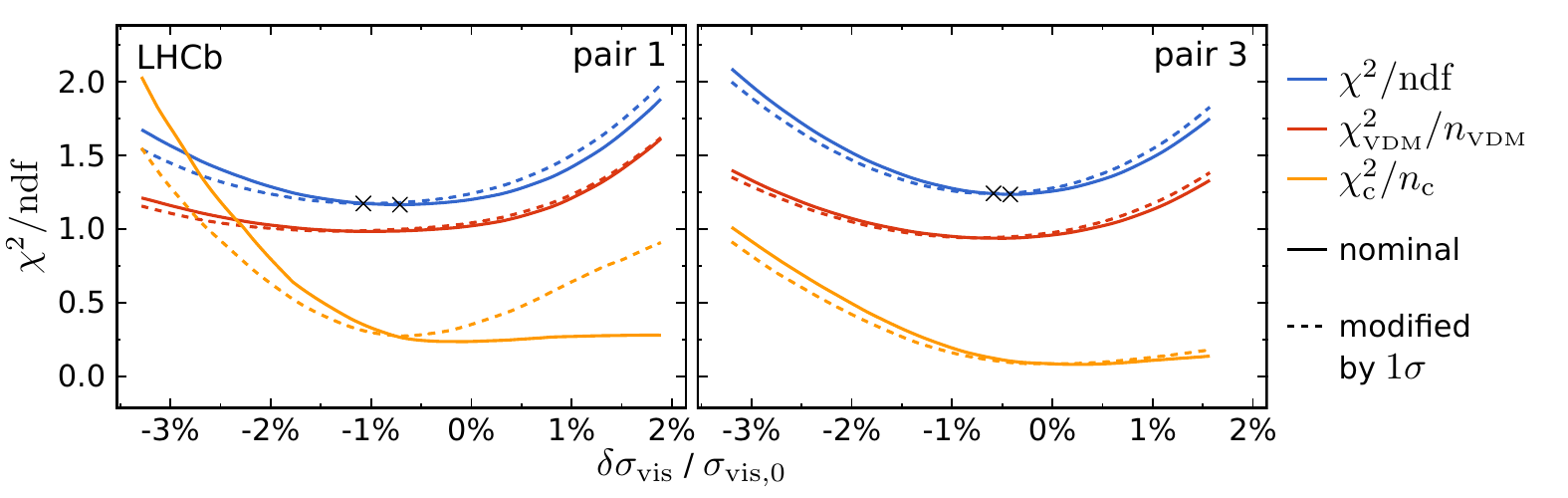}
    \includegraphics[width=0.98\linewidth]{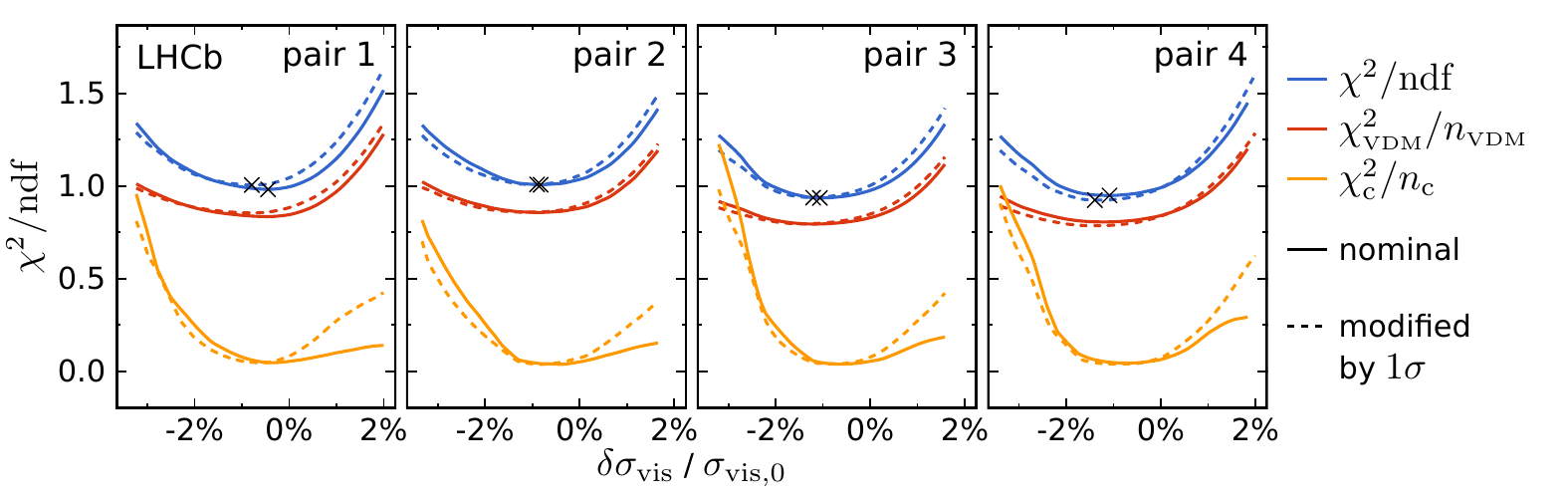}
    \vspace*{-0.5cm}
  \end{center}
  \caption{
    Average \chisqndf profiles as function of the relative difference to the baseline cross-section for the reference scan pairs in (top) April and (bottom) July shown with the top pair of lines (blue).
    The profiles corresponding to the nominal and modified constraints are shown with solid and dashed lines, respectively.
    The minima of the \chisqndf profiles are marked with a cross.
    The contribution of the \vdm data, $\chi^2_{\mbox{\tiny VDM}}$, normalized to the number of points in the VDM scan pair $n_{\mbox{\tiny VDM}}$ is shown with the middle pair of lines (red).
    The contribution of the constraints, $\chi_\mathrm{c}^2$, normalized to the number of constraints $n_\mathrm{c}$ is shown with the bottom pair of lines (yellow).
  }
  \label{fig:vdm_syst_chi2_avg}
\end{figure}

Another systematic uncertainty arises from the uncertainty in the central values of the constraints obtained from BGI analysis.
To estimate this uncertainty, the constraints on $f_{1,2}$ and $v_{1,2}$ are
varied by one standard deviation.
The cross-section obtained with the modified constraints is compared to that obtained with the nominal constraints.
The difference of $0.3\%$ (for both April and July calibrations) is taken as an additional systematic uncertainty, which is considered fully correlated with the BGI result.
The average \chisqndf profiles for the reference scan pairs are shown in Fig.~\ref{fig:vdm_syst_chi2_avg} together with the relative contributions of the \vdm data and the constraints.

\subsection{Reproducibility}
\label{sec:VDM:Reproducibility}
An important source of non-reproducibility of \vdm scan measurements is the drift of the beam orbits.
In addition, in the presence of a non-zero crossing angle, a drift in the
longitudinal bunch crossing point, $\zrf$,
can influence the beam separation.
While the effect of the latter is found to be negligible, the former has potentially a sizeable effect, which cannot be reliably corrected.
For the following discussion data are averaged over bunch pairs since these effects are common for all bunches.

The relation between the position of the luminous region and the beam separation from Eq.~\eqref{eq:zrf_muzl_deltazl} can be used to estimate orbit drifts and $\zrf$ values in the case of non-zero crossing angle.
In the general case of a crossing angle in both $xz$ and $yz$ planes, the luminous region centre for Gaussian beams is
\begin{equation}
\label{eq:zlr_tilted_crossing}
\xilz = \zrf + k \Dx' \, ,
\end{equation}
where $k$ contains the beam widths and angles and $\Dx'$ is the separation in the crossing plane, which is given by
\begin{equation}
\Dx' = \Dx \cos\psi + \Dy \sin\psi \,,\quad
\tan\psi = \frac{\tan\phi_y}{\tan\phi_x} \, .
\end{equation}
Here $\psi$ is the tilt of the crossing plane, which is approximately $-21.5\degrees$ in April 2012 and negligible for all other calibrations.

\begin{figure}[ptb]
  \begin{center}
    \includegraphics[width=0.49\linewidth]{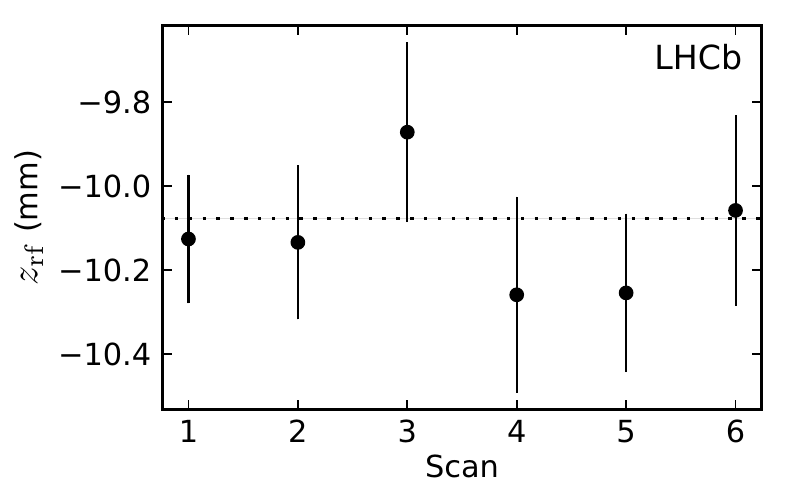}
    \includegraphics[width=0.49\linewidth]{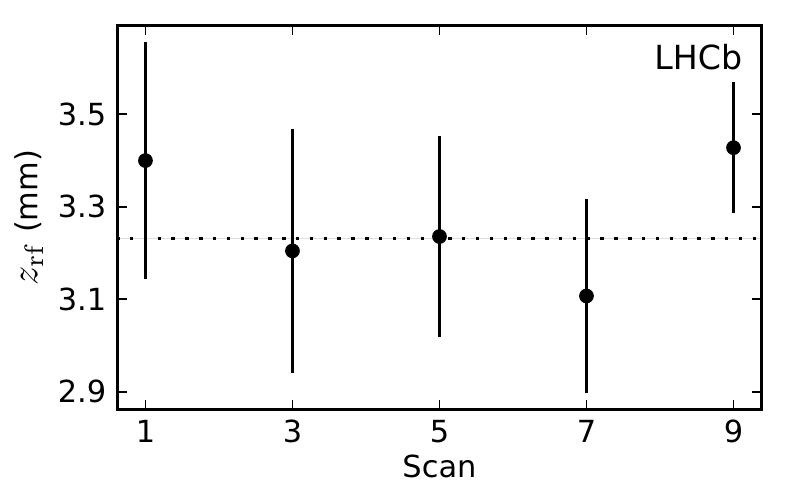}
    \vspace*{-0.5cm}
  \end{center}
  \caption{
    Values of $\zrf$ in (left) April and (right) July scans.
    The implied assumptions are that $\zrf$ does not change and there is no orbit drift during a scan pair.
    No measurements are performed for offset scans.
    The deviations of the measurements from the weighted average (horizontal dotted line) are not statistically significant.
  }
  \label{fig:cond_zrf}
\end{figure}
In accordance with Eq.~\eqref{eq:zrf_muzl_deltazl}, the value of $\zrf$ is measured by interpolating the longitudinal position of the luminous region $\xilz$ at the separation that corresponds to the maximum luminosity.
It is implicitly assumed that $\zrf$ does not change and that there is no orbit drift during a scan pair.
When the scanning axis is orthogonal to the beam crossing plane (\eg $y$ scans in July) this measurement cannot be made,
as $\xilz$ is not expected to change during the corresponding scan.
The measured $\zrf$ values, which are shown in Fig.~\ref{fig:cond_zrf}, are found to be consistent between scans and with those obtained with the BGI analysis in the same fills.
Therefore, it is assumed that $\zrf$ values do not vary during a scan and the associated uncertainty is negligible.

The overall drift of the beam separation between scan pairs is automatically taken into account in the \vdm analysis, since it is effectively only a shift of the nominal separation where the luminosity is maximal.
However, the drift during a scan pair can introduce a bias to the measurement.
It is useful to split the drift into two components: a ``slow'' one that corresponds to the time scale of a scan pair and a ``fast'' one that corresponds to the time scale of a step.

\begin{figure}[ptb]
  \begin{center}
    \includegraphics[width=0.49\linewidth]{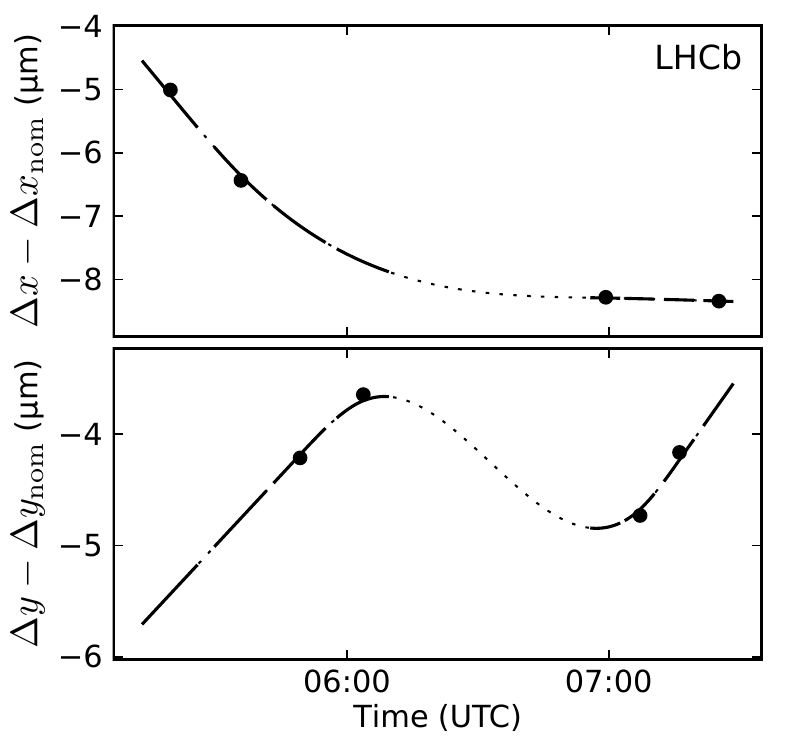}
    \includegraphics[width=0.49\linewidth]{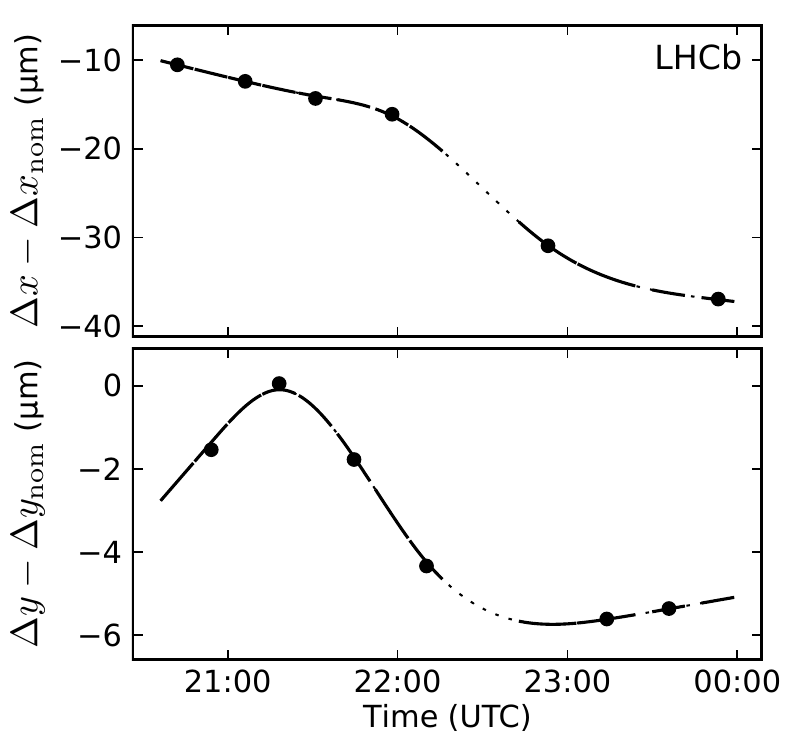}
    \vspace*{-0.5cm}
  \end{center}
  \caption{
    Estimated ``slow'' drift of the beam separation in the (left) April and (right) July \vdm sessions.
    The nominal separation where the luminosity is maximal gives a single     
    estimate per scan of the drift in the corresponding direction (circles).
    The uncertainties are smaller than the marker size.
    The data for each coordinate are fitted with a smooth function (curve) and a prediction is made at each step of every scan (solid curve).
  }
  \label{fig:cond_slow_drift}
\end{figure}
The slow beam separation drift can be estimated from the fitted position of the \vdm profile maximum.
One estimate per coordinate is obtained for each scan pair, as shown in Fig.~\ref{fig:cond_slow_drift}.

\begin{figure}[ptb]
  \begin{center}
    \includegraphics[width=0.49\linewidth]{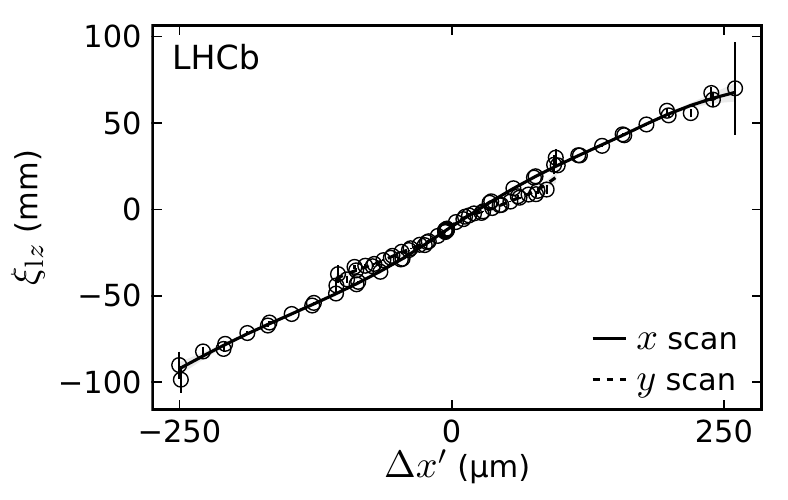}
    \includegraphics[width=0.49\linewidth]{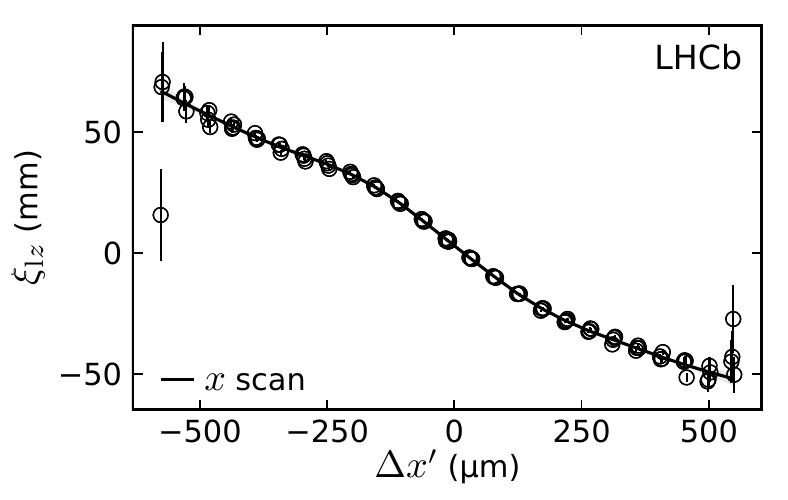}
    \vspace*{-0.5cm}
  \end{center}
  \caption{
    Longitudinal position of the luminous region as function of separation in the crossing plane for reference scans in the (left) April and (right) July \vdm sessions.
    It is assumed that there is no significant orbit and $\zrf$ drift 
    during the reference scans.
    The data for each scan direction are fitted with a smooth function.
    In July the crossing plane is orthogonal to the $y$ axis, thus 
    $\xilz$ is not expected to change during the $y$ scans and the data are omitted.
    The non-linearity of the curves is due to second order effects, which are not expected for pure Gaussian beams.
  }
  \label{fig:cond_zlr_ref}
\end{figure}
\begin{figure}[ptb]
  \begin{center}
    \includegraphics[width=0.49\linewidth]{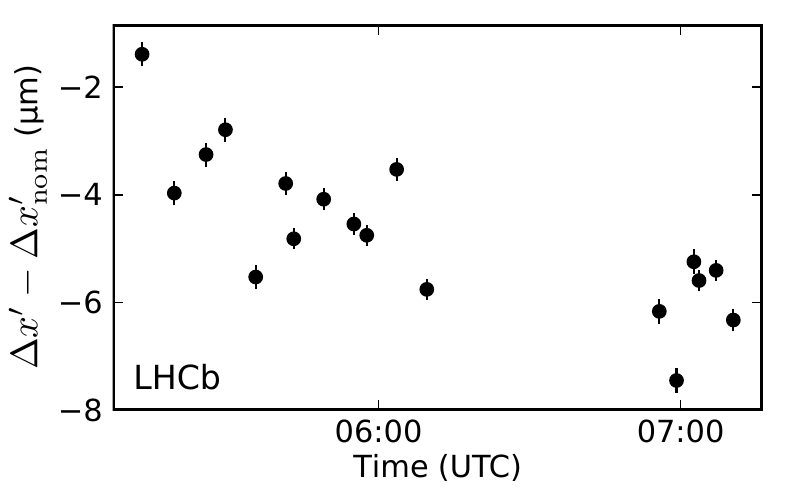}
    \includegraphics[width=0.49\linewidth]{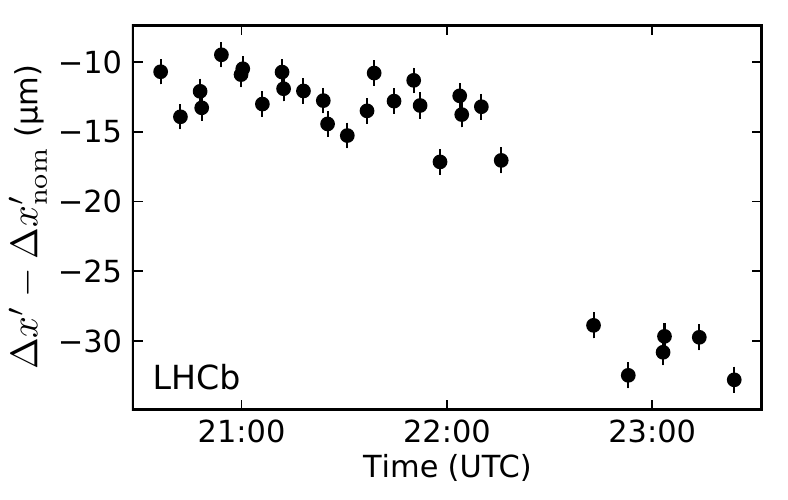}
    \vspace*{-0.5cm}
  \end{center}
  \caption{
    Estimated drift of the beam separation in the crossing plane $\Dx'$
    in the (left) April and (right) July \vdm sessions.
    The drift is estimated only for each nominally head-on step. 
  }
  \label{fig:cond_fast_drift}
\end{figure}
The fast component of the beam separation drift is more difficult to estimate.
The presence of a crossing angle enables an estimation of the drift in the beam separation in the crossing plane $\Dx'$.
This is possible because of the correlation between the $\Dx'$ and the $z$ position of the luminous region as seen in Eq.~\eqref{eq:zrf_muzl_deltazl}.
While the latter is strictly true only for Gaussian beams, the exact form of the function $\xilz(\Dx')$ can be estimated by fitting the measured $\xilz$ as function of $\Dx'$ with a smooth function.
The data and the estimated dependence are shown in Fig.~\ref{fig:cond_zlr_ref}.
The deviations of the data points $\xilz$ from the curve give an estimate of the beam drift.
This approach is only reliable when the reference curve is obtained by averaging enough measurements.
Therefore, the drift is estimated only for the three nominally head-on steps for each scan, as shown in Fig.~\ref{fig:cond_fast_drift}.

\begin{figure}[ptb]
  \begin{center}
    \includegraphics[width=0.49\linewidth]{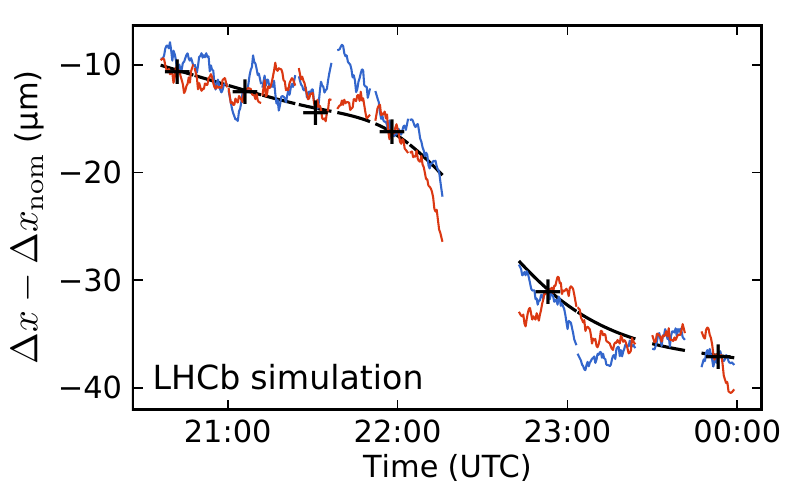}
    \vspace*{-0.5cm}
  \end{center}
  \caption{
    Two example simulations of the beam separation drift in the $x$ coordinate for the July \vdm session.
    The separation drift is modelled with a Brownian motion that is constrained to the measured slow component of the drift (smooth black line) at the measurement points (crosses).
  }
  \label{fig:cond_drift_sim}
\end{figure}
\begin{figure}[ptb]
  \begin{center}
    \includegraphics[width=0.49\linewidth]{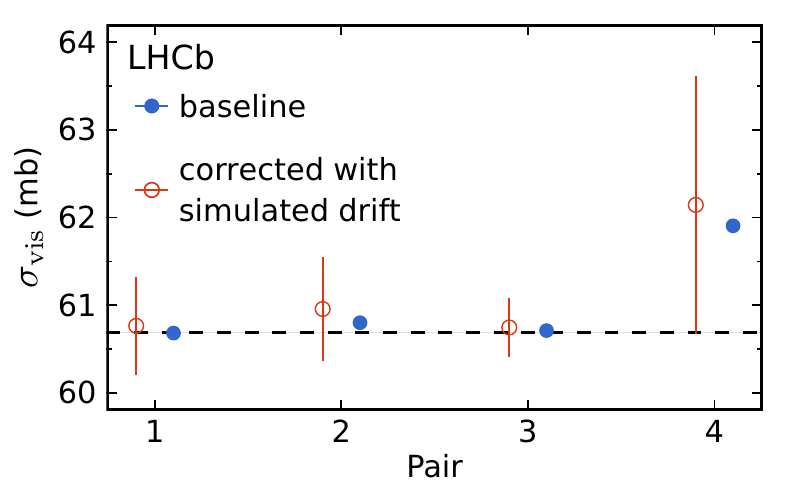}
    \includegraphics[width=0.49\linewidth]{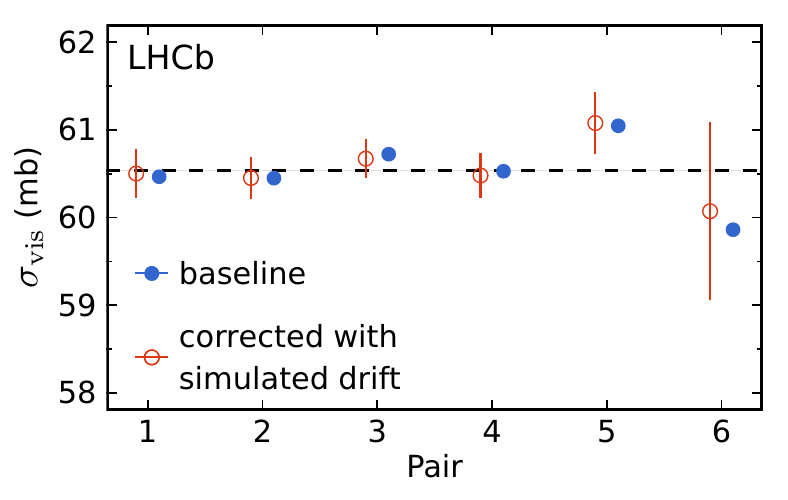}
    \vspace*{-0.5cm}
  \end{center}
  \caption{
    Cross-section bias caused by beam orbit drifts 
    in the (left) April and (right) July \vdm sessions.
    The baseline cross-section for each pair is shown with a solid circle and the average calibration cross-section (using the reference pairs) is shown with a horizontal dashed line.
    The average biased cross-section obtained with simulation of the beam drifts is shown with open circles.
    The error bars indicate the \rms of the biased cross-section of 400 independent simulations.
    The larger error bars for the offset scan pairs indicate that they are more sensitive to random drifts.
  }
  \label{fig:cond_drift_uncertainty}
\end{figure}
To estimate the effect of the orbit drifts on the cross-section measurement a simulation approach is used.
The fast component of the drift in each coordinate is modelled with a probabilistic process according to a Brownian motion.
The diffusion coefficient, which is the parameter of the model, is estimated from the measured fast component of the drift in $\Dx'$.
The value is found to be $0.005\,\mathrm{\micro m^2 s^{-1}}$ and
it is assumed to be equal for both coordinates and constant during a fill.
The separation drift model is constrained to the measured slow component of the drift at the measurement points.
Examples of the result of the simulation are shown in Fig.~\ref{fig:cond_drift_sim}.
The systematic uncertainty is evaluated by analysing
400 statistically independent simulations for each scan session,
where the simulated drift is added to the nominal beam separation.
The average bias on the cross-section and its \rms are summarized on Fig.~\ref{fig:cond_drift_uncertainty}.
It is seen that the average bias, which is mainly due to the slow component of the drift, is small.
On the other hand, the uncertainty in the bias, which is represented by the \rms of all simulations and is driven by the fast drift component, is sizeable.

Since the fast drifts are modelled to be independent for each scan pair, the uncertainty on the average cross-section is reduced.
The uncertainty due to the slow component is assumed to be correlated, thus the estimates for individual scan pairs are averaged.
The two uncertainties are added in quadrature to obtain a total uncertainty of 0.54\% and 0.21\% in April and July, respectively.
The non-reproducibility can be estimated directly by the deviations from the average of the cross-section from individual scan pairs.
The maximum deviations for reference scan pairs observed in April and July are $0.03\%$ and $0.30\%$, respectively.
Assuming that the deviations are mainly due to the drift, and in order to avoid double counting, only the larger value of the drift estimate and the maximum deviation from the average is taken as a systematic uncertainty.

It is seen in Fig.~\ref{fig:cond_drift_uncertainty} that the scan pairs with an offset working point are more sensitive to random beam drifts, which can be explained by the large derivative of the \vdm profile at non-zero beam separation.
This is the main reason for excluding offset scans from the cross-section measurement.
Moreover, since scan pair number five in July had a working point offset by about $30\mum$ in $x$, it is also more sensitive to beam drifts.
Therefore, it is not included in the determination of the central value of the cross-section.
Considering the reference scan pairs (one to four) in July, the uncertainties of the drift bias are too large compared to the observed fluctuations.
This suggests that the value used for the simulation parameter, 
on which these uncertainties depend directly,
is overestimated or the employed model does not describe well the beam drifts.
Therefore, the higher cross-section measured from scan pair number five (see Fig.~\ref{fig:xsec_fit_xsec}) cannot be explained by random beam drifts
and the deviation from the average cross-section (using pairs one to four) of $0.8\%$ is taken as a systematic uncertainty to account for a potential unknown source of non-reproducibility.
The latter systematic uncertainty is assigned to all \PPP calibrations and is considered to be correlated among calibrations.

The effects of the applied corrections are summarized in Table~\ref{tab:vdm_corrections}.
The values for the other energies will be discussed below (Sec.~\ref{sec:VDM:Results 2}).
\begin{table}[ptb]
  \centering
  \caption{
    Effect of corrections on the result of the \vdm scan calibrations at $\sqrt{\sNN}$ values of 8, 7 and \Five\unit{TeV} (in \%).
    To obtain the effects, each correction is excluded, the fits are redone and the result is compared with the baseline.
  }
  \begin{tabular}{lrrrrr}
    & \multicolumn{2}{c}{8\tev} & 7\tev & \multicolumn{2}{c}{\Five\tev} \\
    Source & April & July &      & \PPB & \PBP \\
    \midrule
    Ghost charge         & $+0.93$ & $+0.68$ & $+1.59$ & $+0.80$ & $+0.97$ \\
    Satellites charge    & $+0.85$ & $+0.43$ & $+0.78$ & $+0.19$ & $+0.19$ \\
    Efficiency of the observable  & $+0.32$ & $-0.03$ & $-0.04$ & $+0.08$ & $+0.11$ \\
    Length scale         & $-1.39$ & $+0.39$ & $+0.06$ & $-2.14$ & $-1.03$ \\
    Dynamic $\beta$      & $-0.39$ & $-0.37$ & $-0.35$ & $-0.05$ & $-0.05$ \\
    Beam-beam deflection & $+1.12$ & $+1.12$ & $+1.00$ & $+0.15$ & $+0.16$ \\
    Linear correlation   & $+0.30$ & n.a. & n.a. & n.a. & n.a. \\
  \end{tabular}
  \label{tab:vdm_corrections}
\end{table}

%% file: vdm_results.tex
\subsection{Results}
\label{sec:VDM:Results}
The reference cross-section for \PPP collisions at $\sqrt{s}=8\tev$ for the \emph{Track} observable is determined by computing a weighted average of the results from the calibrations in the April and July fills.
It is assumed that most of the systematic uncertainties are fully correlated in order to avoid underestimating the uncertainty on the combined result.
The individual calibration results and the average reference cross-section are shown in Table~\ref{tab:vdm_results_pp8}.
A list of all uncertainties for the July and April calibrations is provided in Table~\ref{tab:vdm_syst} along with estimates of their correlation.
The values for the other energies are discussed below.
\begin{table}[ptb]
  \centering
  \caption{
    Individual calibration results and average of the $8\unit{TeV}$ \PPP \vdm scan sessions.
    The part of the relative uncertainty that is correlated between the calibrations is shown in the fourth column.
    The weights used to obtain the average are given in the last column.
  }
  \begin{tabular}{lcccc}
    \mulrow{2}{*}{Session} & \mulrow{2}{*}{$\sigma_\mathrm{Track}$ (mb)} & Relative & Correlated & \mulrow{2}{*}{Weight} \\
     && uncertainty & part & \\
    \midrule
   April 2012   & $60.70\pm0.92$ & $1.52\%$ & $1.34\%$ & 0.53  \\
   July 2012   & $60.54\pm0.93$ & $1.54\%$ & $1.51\%$ & 0.47   \\
   \midrule
   Average & $60.62\pm0.89$ & $1.47\%$ &          &  \\    
  \end{tabular}
  \label{tab:vdm_results_pp8}
\end{table}
\begin{table}[ptb]
  \centering
  \caption[Systematic uncertainties on the luminosity calibration using the van der Meer method.]{
    Relative systematic uncertainties on the reference cross-section for the \vdm calibrations at $\sqrt{\sNN}$ values of 8, 7 and \Five\tev (in \%).
    The uncertainties are divided into groups affecting the description of the \vdm profile
    (Secs.~\ref{sec:VDM:Length scale}--\ref{sec:VDM:Reproducibility}),
    the measurement of the rate (\sect~\ref{sec:VDM:Rate measurement}) and the bunch population product measurement (\sect~\ref{sec:Bunch current normalization}).
    The fourth (eight) column indicates whether the uncertainties are correlated between the two \PPP calibrations at $\sqrt{s}=8\tev$ (the \PPB and \PBP calibrations).
    Empty cells indicate that the corresponding source is not applicable or the uncertainty is negligible.
    For the \PPB and \PBP calibrations, the uncertainties due to scan variation and drift, and non-reproducibility cannot be estimated separately as only two \vdm scan pairs per calibration were performed.
  }
  \begin{tabular}{lccccccc}
       & \multicolumn{3}{c}{8\tev} & 7\tev & \multicolumn{3}{c}{\Five\tev} \\
Source & April & July & Corr. &      & \PPB & \PBP & Corr. \\
\midrule
\multicolumn{8}{c}{\vdm profile description}\\
Fit model                   & 0.50 & 0.50 & yes & 0.50 & 1.31 & 1.34 & yes \\
Fit bias                    & 0.20 & 0.20 & yes & 0.20 &      &      &     \\
Linear correlation          & 0.15 &      & no  &      &      &      &     \\
Parameter assumptions       & 0.60 & 0.90 & yes & 0.90 &      &      &     \\
Constraints from BGI        & 0.30 & 0.30 & yes & 0.80 &      &      &     \\
Length scale                & 0.50 & 0.50 & yes & 0.50 & 1.10 & 1.10 & no  \\
VELO transverse scale       & 0.05 & 0.05 & yes & 0.05 & 0.05 & 0.05 & yes \\
Beam-beam effects           & 0.27 & 0.30 & yes & 0.29 & 0.06 & 0.05 & yes \\
Scan variation and drift    & 0.54 & 0.30 & no  & 0.20 & 0.67 & 1.31 & no  \\
Non-reproducibility         & 0.80 & 0.80 & yes & 0.80 &      &      &     \\
(fifth scan pair in July)   & & & & & & & \\
Statistical                 & 0.06 & 0.05 & no  & 0.11 & 0.26 & 0.25 & no  \\
\midrule
\multicolumn{8}{c}{Rate measurement}\\
Beam-gas background         & 0.03 & 0.07 & yes & 0.08 & 0.34 & 0.21 & yes \\
Beam-beam background        & 0.09 & 0.18 & yes & 0.09 & 0.65 & 0.71 & yes \\
Efficiency of rate observable & 0.13 & 0.07 & yes & 0.17 & 0.04 & 0.06 & yes \\
\midrule
\multicolumn{8}{c}{Bunch population uncertainties} \\
DCCT population product     & 0.22 & 0.24 & yes & 0.24 & 0.31 & 0.34 & no  \\
FBCT offset                 & 0.04 & 0.06 & yes & 0.01 & 0.21 & 0.09 & no  \\
BPTX cross-check            & 0.13 & 0.06 & yes &      & 0.14 & 0.14 & no  \\
Ghost charge                & 0.05 & 0.03 & yes & 0.07 & 0.14 & 0.19 & no  \\
Satellites charge           & 0.02 & 0.03 & yes & 0.25 & 0.07 & 0.09 & no  \\
No satellite measurements   & 0.43 &      & no  &      &      &      &     \\
\midrule
Total                       & 1.52 & 1.54 &     & 1.71 & 2.05 & 2.36 & \\
  \end{tabular}
  \label{tab:vdm_syst}
\end{table}

\subsection{Summary of other van der Meer scan calibrations}
\label{sec:VDM:Results 2}

The \vdm analysis presented in this paper focuses on the calibration of the reference cross-section at $\sqrt{s}=8\tev$ (\PPP).
Comparable results are obtained with the \vdm method for the 2011 \PPP calibration at $\sqrt{s}=7\tev$ and the 2013 \PPB and \PBP calibrations at $\sqrt{\sNN}=\Five\tev$.

A \vdm scan session was performed in fill 2234 for the calibration of the reference cross-section at $\sqrt{s}=7\tev$ (\PPP).
The operational procedure and the relevant trigger configuration were very similar to the one employed in 2012.
The fill conditions and the scan parameters are listed in Tables~\ref{tab:fills} and \ref{tab:vdm_scans}.
Three symmetric $x$-$y$ scan pairs were performed.
The average decay time of the bunch population product amounts to 70 hours and the luminosity drop caused by emittance growth is negligible.

The \vdm analysis at 7\tev is performed in exactly the same manner as for the 8\tev calibrations, which are discussed in detail.
Given the similarity of the conditions, the systematic uncertainties on the fit model, the non-reproducibility and the length scale calibration are directly translated from the 2012 calibration.
The corrections and the systematic uncertainty due to beam-beam effects are found to be similar, owing to the similar beam energy and bunch intensities.
The dominating systematic uncertainties are related to the constraints on the beam factorizability as obtained from the BGI analysis (Sec.~\ref{sec:intermediate_energy}).
By using these constraints instead of making parameter assumptions (see Sec.~\ref{sec:Cross-section determination}), the measurement is shifted by 0.9\%.
The systematic uncertainty arising from the uncertainty in the central values of the constraints is found to be 0.8\% and it is fully correlated with the BGI result.

Individual calibrations of the \PPB and \PBP reference cross-sections were performed in separate fills (3505, 3542).
Due to the limited time of the ion runs, normal fills for physics data taking were used.
Therefore, the filling schemes (see Table~\ref{tab:fills}) were not optimized as for dedicated luminosity calibration fills.
The number of colliding bunches was 38, instead of usually 16 in dedicated fills.
Moreover, the peak $\mu_\eff$ value was very low, $<0.02$.
In order to decrease statistical uncertainties, the random trigger rate was doubled to 45\kHz during the \PPB calibration.
The statistical uncertainties on the cross-section measurement are significantly larger compared to \PPP calibrations and amount to $0.25\%$.

The statistical uncertainty on the cross-section measurement per colliding bunch pair and per scan pair is about 2\%.
Therefore, a systematic structure in the fit residuals, which leads to a measurement bias of that order, may not be revealed by the \chisqndf values of individual fits.
The fit quality of a model is estimated from the sum of the residuals of all bunch pairs as function of beam separation.
The aggregated residuals show a statistically significant structure if a pure Gaussian model is employed.
It is found that two factorizable empirical models (a double Gaussian function with a negative weight for the narrow component and a Gaussian function multiplied by an even fourth order polynomial) fit the \vdm scan profiles well and give similar cross-section values.
In addition, two non-factorizable empirical models for the two-dimensional luminosity profiles have similar fit quality.
The first non-factorizable model is a sum of two two-dimensional Gaussian functions with a negative weight for the narrow term.
The second non-factorizable model is a $xy$ rotationally symmetric Gaussian function multiplied by an even fourth order polynomial and scaled by the $x$ and $y$ profile widths.
The two non-factorizable models give similar cross-section values, which are about 2\% lower than the values from the factorizable models.
The \vdm scans performed provide no information on the factorizability.
Therefore, the full range of obtained cross-section values is considered by taking the central value at the middle and assigning half of the span as a fit model uncertainty.

An independent length scale calibration was performed for both fills.
The \lhc optics setup was almost identical in the two proton-lead beam configurations, thus the calibration constants are expected to be identical.
However, a difference of about 1\% is observed and is assigned as a systematic uncertainty to both calibrations.

The uncertainty due to the beam orbit drift is estimated to be $0.67\%$ and $0.88\%$ for the \PPB and the \PBP calibration, respectively.
The deviation of the visible cross-section measurement from the average in repeated \vdm scans amounts to $0.23\%$ and $1.31\%$ for the \PPB and the \PBP calibration, respectively.
Following the procedure from Sec.~\ref{sec:VDM:Reproducibility}, only the larger value of the uncertainty due to orbit drift and the deviation from the average is taken as a systematic uncertainty.
Beam-beam effects 
have a small impact on the \PPB and \PBP calibrations owing to the low bunch intensities of (1.3--1.7)$ \times 10^{10}$ elementary charges.
The low $\mu$ values effectively increase the fraction of beam-gas and beam-beam induced backgrounds.
The corresponding uncertainties are evaluated as described in Sec.~\ref{sec:VDM:Rate measurement} and amount to $~0.3\%$ and $~0.7\%$ for beam-gas and beam-beam background respectively.
It is assumed that there is no correlation between the uncertainties of the \PPB and \PBP calibrations when it is partial.
This is done to avoid underestimating the uncertainty on the ratio of the luminosities of the two data samples, which enters into ratios of cross-section measurements.
A summary of all uncertainties is provided in Table~\ref{tab:vdm_syst}.
The reference cross-section for the \emph{Track} observable is
$2.126 \pm 0.049\barn$ and $2.120 \pm 0.053\barn$
for the \PPB and \PBP mode, respectively.
The compatibility of the two results indicates that the \emph{Track} observable has similar efficiency for the two beam modes.

%% file: combination.tex
\section{Summary and conclusion}
\label{sec:Combination}

\begin{table}[t]
  \centering
  \caption{
    Systematic uncertainties of the BGI and \vdm methods for \pp interactions at $8\unit{TeV}$.
    The fourth column indicates whether the uncertainties are correlated between the two calibrations.
    All values are given in \%.
  \label{tab:combined_uncertainties}
  }
  \small
  \renewcommand{\arraystretch}{0.9}
  \begin{tabular}{lccc}
Source & BGI & VDM & Correlated \\    
\midrule
\multicolumn{4}{c}{Bunch population uncertainties (\sect~\ref{sec:Bunch current normalization})}\\ 
FBCT offset                 & 0.04 & 0.05 & yes \\
BPTX cross-check            & n.a. & 0.09 & yes \\
DCCT population product     & 0.22 & 0.23 & yes \\
Ghost charge                & 0.02 & 0.04 & yes \\
Satellite charge            & 0.06 & 0.02 & yes \\
Missing satellite measurements  & n.a. & 0.23 & no \\
\midrule
\multicolumn{4}{c}{Rate measurement}\\ 
Background subtraction      & 0.20 & 0.14 & yes \\  %
Ratio of observables \roc   & 0.20 & n.a. & no \\
Efficiency of rate observables & negl. & 0.09 & no \\
\midrule
Fit model                   & \multicolumn{2}{c}{0.50} & yes \\    
VELO transverse scale       & \multicolumn{2}{c}{0.05} & yes \\
\midrule
\multicolumn{4}{c}{BGI specific (\sect~\ref{sec:Beam-gas imaging method})}\\ 
Beam-beam resolution        & 0.93 &  & no \\
Beam-gas resolution         & 0.55 &  & no \\
Detector alignment          & 0.45 &  & no \\
Measurement spread          & 0.54 &  & no \\
Bunch length                & 0.05 &  & no \\
Reconstruction efficiency   & 0.04 &  & no \\
Pressure gradient           & 0.03 &  & no \\
\midrule
\multicolumn{4}{c}{\vdm specific (\sect~\ref{sec:Van der Meer scan method})}\\ 
Length scale                    &  & 0.50 & no  \\
Beam-beam effects               &  & 0.28 & no  \\
Fit bias                        &  & 0.20 & no  \\
Linear correlation              &  & 0.08 & no  \\
Parameter assumptions           &  & 0.74 & no  \\
Constraints from BGI            &  & 0.30 & yes \\
Scan variation and drift        &  & 0.32 & no  \\
Non-reproducibility             &  & 0.80 & no  \\
Statistical                     &  & 0.04 & no  \\
\midrule
Uncorrelated  & 1.31 & 1.32 & \\
Correlated    & 0.59 & 0.65 & \\
  \end{tabular}
  \renewcommand{\arraystretch}{1}
\end{table}

Since some of the luminosity calibrations have been performed with both the VDM and BGI method 
(\pp cross-section at $\sqrt{s}=8\tev$ and $\sqrt{s}=7\tev$),
the best result is obtained by computing an average of the two methods, taking into account the 
correlation between the systematic error sources.
A summary of the final reference cross-section results is presented in Table~\ref{tab:combined_results}.

\begin{table}[b]
  \centering
  \caption{
    Results of the luminosity calibration measurements.
    The total uncertainty on the luminosity calibration (last column) is the sum in quadrature of the absolute calibration uncertainty (fourth column) and the relative calibration uncertainty (fifth column).
    The weights used to obtain the average absolute calibration at 8 and 7\tevcap (\pp) are given in the third column.
    The part of the uncertainty that is correlated between \vdm and BGI calibrations is shown in parentheses (fourth column).
  }
  \setlength{\tabcolsep}{4pt}
  \begin{tabular}{l@{\qquad}ccccc}
    \toprule
    \multirow{2}{*}{Method} & \multicolumn{3}{c}{Absolute calibration} & Relative calibration & Total \\
& $\sigma_\mathrm{vis}$~(mb) & Weight & Uncertainty & uncertainty & uncertainty \\
    \midrule
    \multicolumn{6}{l}{\pp at $\sqrt{s}=8\tev$} \\
    BGI     & $60.62\pm0.87$ & 0.50 & 1.43\% (0.59\%) & & \\  %
    VDM     & $60.63\pm0.89$ & 0.50 & 1.47\% (0.65\%) & & \\  %
    Average & $60.62\pm0.68$ &      & 1.12\% \phantom{(0.00\%)} & 0.31\% & 1.16\% \\  %
    \midrule
    \multicolumn{6}{l}{\pp at $\sqrt{s}=7\tev$} \\
    BGI     & $63.00\pm2.22$ & 0.13 & 3.52\% (1.00\%) & &  \\  %
    VDM     & $60.01\pm1.03$ & 0.87 & 1.71\% (1.00\%) & &  \\  %
    Average & $60.40\pm0.99$ &      & 1.63\% \phantom{(0.00\%)} & 0.53\% & 1.71\% \\  %
    \midrule
    \multicolumn{6}{l}{\PPP at $\sqrt{s}=2.76\tev$} \\
    BGI     & $52.7\phantom{0}\pm1.2\phantom{0}$ & & 2.20\% \phantom{(0.00\%)} & 0.25\% & 2.21\% \\
    \midrule
    \multicolumn{6}{l}{\PPB at $\sqrt{\sNN}=\Five\tev$} \\
    VDM     & $\phantom{.}2126\pm49\phantom{.0}$ & & 2.05\% \phantom{(0.00\%)} & 1.03\% & 2.29\% \\
    \midrule
    \multicolumn{6}{l}{\PBP at $\sqrt{\sNN}=\Five\tev$} \\
    VDM     & $\phantom{.}2120\pm53\phantom{.0}$ & & 2.36\% \phantom{(0.00\%)} & 0.82\% & 2.50\% \\
    \bottomrule
  \end{tabular}
  \label{tab:combined_results}
\end{table}

The BGI and \vdm calibrations of the visible \pp cross-section at $\sqrt{s}=8\tev$ achieve very similar precision.
Therefore, for simplicity the two results are combined with equal weights; the correlated components of the 
uncertainties are averaged linearly and the others are averaged in quadrature.
The different sources of uncertainty are compared in Table~\ref{tab:combined_uncertainties}, 
indicating also whether they are correlated or not.
The result of the combination is given in Table~\ref{tab:combined_results}.
If the uncertainty on the propagation to physics data (0.31\%, see \sect~\ref{sec:InteractionRate}) 
is included, the total uncertainty on the luminosity is 1.16\%. 
The latter uncertainty is valid if the complete 2012 data set or a major part of it is used.  
In some cases, for small partial sets, the uncertainty may be different. 

A weighted average of the two \pp cross-section measurements at $\sqrt{s}=7\tev$ is computed, taking into account that the \vdm calibration is more precise.
The correlated part of the uncertainty due to the bunch population measurements and the factorizability is estimated to be 1\%.
A weight of 0.87 is given to the \vdm measurement based on the uncorrelated part of the uncertainties.
The final result given in Table~\ref{tab:combined_results} has a precision of 1.63\%.
If the uncertainty on the propagation to physics data (0.53\%, see \sect~\ref{sec:InteractionRate}) 
is also included, the total uncertainty on the luminosity is 1.71\%. 
As for the 2012 data set, the total uncertainty on the luminosity is valid if a major part of the 2011 data set is used.

\label{sec:Conclusion}

\begin{figure}[tb]
  \begin{center}
    \includegraphics[width=0.74\linewidth]{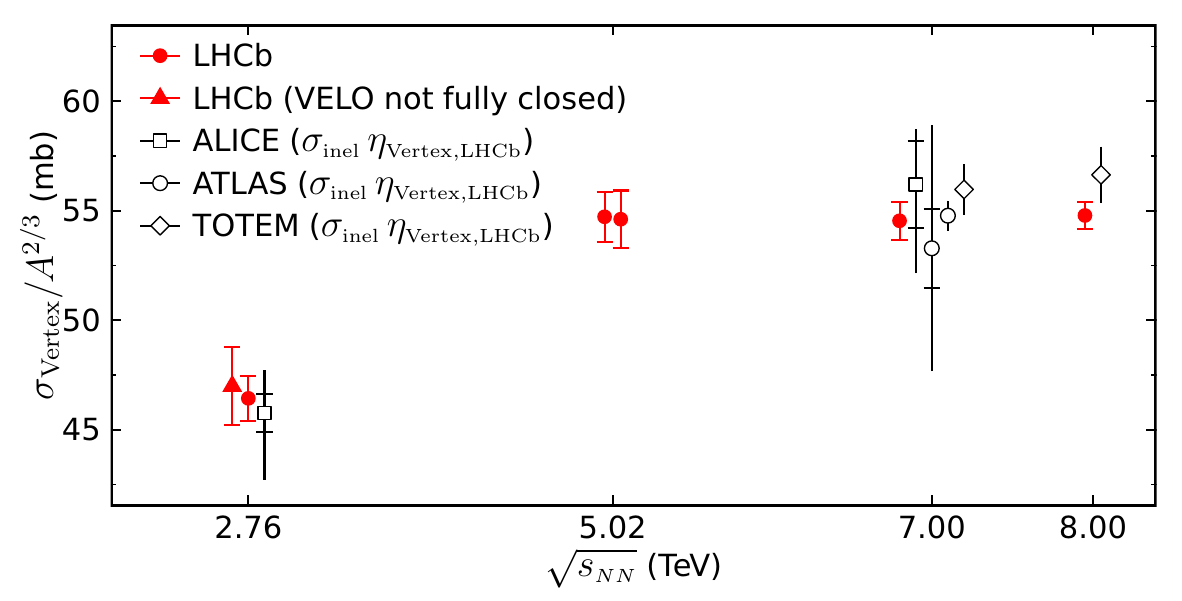}
    \vspace*{-0.5cm}
  \end{center}
  \caption{
    Measurements of the visible cross-section for the \emph{Vertex} observable as function of 
    centre-of-mass energy.
    The measurement at 2.76\tevcap \cite{LHCb-PAPER-2012-039} performed using the \vdm method 
    (solid triangle) is corrected 
    for the reduced efficiency due to the \velo being not fully closed.
    The visible cross-section for proton-lead collisions at \Five\tevcap is scaled by $A^{-2/3}$.
    A comparison is made with the luminosity-independent measurements of the \PPP inelastic 
    cross-section by the TOTEM collaboration \cite{TOTEM:7tev, Antchev:1495764} at 7 and 8\tevcap,
    with direct measurements by the ALICE \cite{Abelev:2012sea} and the ATLAS \cite{Aad:2011eu} experiments,
    and with a measurement by ATLAS from elastic \pp scattering \cite{ATLAS-CONF-2014-040}.
    The measurements from other experiments are scaled with the \lhcb efficiency for inelastic events 
    \EffVertex, obtained from simulation.
    The uncertainties of the direct measurements from ALICE and ATLAS are dominated by the extrapolation of 
    the visible cross-section to the total inelastic cross-section and are not to be compared with 
    the uncertainties of the \lhcb measurements.
    The tick marks represent the uncertainty due to the luminosity calibration only.
    Data points at the same centre-of-mass energy are displaced horizontally for clarity.
  }
  \label{fig:sigma_energy}
\end{figure}

Measurements of the visible cross-section for the \emph{Vertex} observable as function of centre-of-mass 
energy are shown in Fig.~\ref{fig:sigma_energy}.
The visible cross-section for proton-lead collisions at \Five\tevcap is scaled by $A^{-2/3}$
($A=208$ for lead).
A comparison is made with the luminosity-independent measurements of the \PPP inelastic cross-section by the 
TOTEM collaboration \cite{TOTEM:7tev, Antchev:1495764} at 7 and 8\tevcap,
and direct measurements by the ALICE \cite{Abelev:2012sea} and the ATLAS \cite{Aad:2011eu} experiments.
The measurements from other experiments are scaled with the \lhcb efficiency for inelastic 
events \EffVertex, which is obtained from simulation with a negligible statistical uncertainty.
The values of \EffVertex are 0.729, 0.768 and 0.758 for 2.76, 7 and 8\tev, respectively. 
A measurement from CMS \cite{Chatrchyan:2012nj} is not shown as it does not include an extrapolation 
to the total inelastic cross-section.
No systematic uncertainties are included to account for the \EffVertex scaling, nor for the scaling 
with $A^{-2/3}$.

Figure~\ref{fig:sigma_energy} also shows the result of a first luminosity calibration of the
\PPP cross-section at 2.76\tev which was performed using the \vdm method in 2011 
\cite{LHCb-PAPER-2012-039}.
During this data taking period, the \velo was positioned with its sensitive area at a minimum distance of 13\mm 
from the beam instead of the nominal 8\mm.
The corresponding drop in efficiency is estimated to be 5\% from simulation.
Taking this difference into account, as well as the unaccounted potential non-factorizability of the beams,
a good agreement is found when comparing to the more precise BGI measurement from 2013.

An earlier measurement of the \PPP visible cross-section based on the VDM and BGI methods for the \emph{Track} 
observable at $\sqrt{s}=7\tev$ was reported in Ref.~\cite{LHCb-PAPER-2011-015}.
The data available in the older publication were not sufficient to measure the factorizability 
of the beams and complete factorizability was assumed.
It was shown here that neglecting effects of non-factorizability may cause an underestimate of
the visible cross-section at the few percent level.
Nevertheless, the value (58.8~$\pm$~2.0\unit{mb}) is consistent with the significantly more precise result reported here.
The results reported here supersede those of Ref.~\cite{LHCb-PAPER-2011-015}.

In conclusion, several luminosity calibration measurements were performed at the LHC
using the LHCb detector and two experimental methods, the \vdm scan method and
the BGI method.
Ghost charge fractions were also measured using beam-gas interactions and the results
are used in several luminosity calibrations by other LHC experiments.
The LHCb luminosity calibrations were made for proton-proton collisions at three different 
centre-of-mass energies (2.76, 7 and 8\tev) and for proton-lead collisions at the equivalent
nucleon-nucleon centre-of-mass energy of \Five\tev.
The analysis strategies and the results of the calibrations were presented in detail.
Compared to the calibration performed in 2010, an improvement by an order of magnitude was achieved 
in the bunch population normalization, chiefly obtained by means of a thorough study
of the LHC beam current measuring devices.
This achievement opened the way to a global reduction of the systematic and statistical 
uncertainties for both the \vdm scan method and the BGI method.
A controlled gas injection into the LHC vacuum was employed to increase the beam-gas interaction 
rate by almost two orders of magnitude and
detailed systematic studies (including offset scans and reproducibility checks) 
were conducted with the \vdm scan method.
Modelling the non-factorizability in the transverse distribution of the bunch particles 
is required and, if neglected, would have changed the calibration results by up to 3\%.
In the case of proton-proton collisions at 8\tev, a precision of 1.47\% is
obtained with the \vdm scan method and 1.43\% with the BGI method.
When combining the results, %
the precision obtained on the reference visible cross-section is 1.12\%,
which constitutes to date the most precise luminosity
calibration at a bunched-beam hadron collider.
The precision of the calibrations for the other beams and beam energies is close to 2\%.
The luminosity calibration results are used to determine a reference cross-section,
which is employed in the LHCb physics data analysis to measure absolute cross-sections
of various processes.

%% file: acknowledgements.tex
\section*{Acknowledgements}
\noindent We express our gratitude to our colleagues in the CERN
accelerator departments for the excellent performance of the LHC. We
thank the technical and administrative staff at the LHCb
institutes. We acknowledge support from CERN and from the national
agencies: CAPES, CNPq, FAPERJ and FINEP (Brazil); NSFC (China);
CNRS/IN2P3 (France); BMBF, DFG, HGF and MPG (Germany); SFI (Ireland); INFN (Italy); 
FOM and NWO (The Netherlands); MNiSW and NCN (Poland); MEN/IFA (Romania); 
MinES and FANO (Russia); MinECo (Spain); SNSF and SER (Switzerland); 
NASU (Ukraine); STFC (United Kingdom); NSF (USA).
The Tier1 computing centres are supported by IN2P3 (France), KIT and BMBF 
(Germany), INFN (Italy), NWO and SURF (The Netherlands), PIC (Spain), GridPP 
(United Kingdom).
We are indebted to the communities behind the multiple open 
source software packages on which we depend. We are also thankful for the 
computing resources and the access to software R\&D tools provided by Yandex LLC (Russia).
Individual groups or members have received support from 
EPLANET, Marie Sk\l{}odowska-Curie Actions and ERC (European Union), 
Conseil g\'{e}n\'{e}ral de Haute-Savoie, Labex ENIGMASS and OCEVU, 
R\'{e}gion Auvergne (France), RFBR (Russia), XuntaGal and GENCAT (Spain), Royal Society and Royal
Commission for the Exhibition of 1851 (United Kingdom).

%% file: LHCb_HD_authorlist_2014-07-09.tex
\centerline{\large\bf LHCb collaboration}
\begin{flushleft}
\small
R.~Aaij$^{41}$, 
B.~Adeva$^{37}$, 
M.~Adinolfi$^{46}$, 
A.~Affolder$^{52}$, 
Z.~Ajaltouni$^{5}$, 
S.~Akar$^{6}$, 
J.~Albrecht$^{9}$, 
F.~Alessio$^{38}$, 
M.~Alexander$^{51}$, 
S.~Ali$^{41}$
% ...

{\footnotesize \it
% ...
$ ^{5}$Clermont Universit\'{e}, Universit\'{e} Blaise Pascal, CNRS/IN2P3, LPC, Clermont-Ferrand, France\\
$ ^{6}$CPPM, Aix-Marseille Universit\'{e}, CNRS/IN2P3, Marseille, France\\
% ...
$ ^{9}$Fakult\"{a}t Physik, Technische Universit\"{a}t Dortmund, Dortmund, Germany\\
% ...
$ ^{37}$Universidad de Santiago de Compostela, Santiago de Compostela, Spain\\
$ ^{38}$European Organization for Nuclear Research (CERN), Geneva, Switzerland\\
% ...
$ ^{41}$Nikhef National Institute for Subatomic Physics, Amsterdam, The Netherlands\\
% ...
$ ^{46}$H.H. Wills Physics Laboratory, University of Bristol, Bristol, United Kingdom\\
% ...
$ ^{51}$School of Physics and Astronomy, University of Glasgow, Glasgow, United Kingdom\\
$ ^{52}$Oliver Lodge Laboratory, University of Liverpool, Liverpool, United Kingdom\\
% ...
}
\end{flushleft}

%% file: lumicalib_PAPER.bbl
\ifx\mcitethebibliography\mciteundefinedmacro
\PackageError{LHCb.bst}{mciteplus.sty has not been loaded}
{This bibstyle requires the use of the mciteplus package.}\fi
\providecommand{\href}[2]{#2}
\begin{mcitethebibliography}{10}
\mciteSetBstSublistMode{n}
\mciteSetBstMaxWidthForm{subitem}{\alph{mcitesubitemcount})}
\mciteSetBstSublistLabelBeginEnd{\mcitemaxwidthsubitemform\space}
{\relax}{\relax}

\bibitem{Alves:2008zz}
LHCb collaboration, A.~A. Alves~Jr.\ {\em et~al.},
  \ifthenelse{\boolean{articletitles}}{\emph{{The \lhcb detector at the LHC}},
  }{}\href{http://dx.doi.org/10.1088/1748-0221/3/08/S08005}{JINST \textbf{3}
  (2008) S08005}\relax
\mciteBstWouldAddEndPuncttrue
\mciteSetBstMidEndSepPunct{\mcitedefaultmidpunct}
{\mcitedefaultendpunct}{\mcitedefaultseppunct}\relax
\EndOfBibitem
\bibitem{LHCb-PAPER-2012-008}
LHCb collaboration, R.~Aaij {\em et~al.},
  \ifthenelse{\boolean{articletitles}}{\emph{{Inclusive $W$ and $Z$ production
  in the forward region at $\sqrt{s}=7$ TeV}},
  }{}\href{http://dx.doi.org/10.1007/JHEP06(2012)058}{JHEP \textbf{06} (2012)
  058}, \href{http://arxiv.org/abs/1204.1620}{{\tt arXiv:1204.1620}}\relax
\mciteBstWouldAddEndPuncttrue
\mciteSetBstMidEndSepPunct{\mcitedefaultmidpunct}
{\mcitedefaultendpunct}{\mcitedefaultseppunct}\relax
\EndOfBibitem
\bibitem{LHCb-PAPER-2014-033}
LHCb collaboration, R.~Aaij {\em et~al.},
  \ifthenelse{\boolean{articletitles}}{\emph{{Measurement of the forward $W$
  boson production cross-section in $pp$ collisions at $\sqrt{s}=7$ TeV}},
  }{}\href{http://arxiv.org/abs/1408.4354}{{\tt arXiv:1408.4354}}, {submitted
  to JHEP}\relax
\mciteBstWouldAddEndPuncttrue
\mciteSetBstMidEndSepPunct{\mcitedefaultmidpunct}
{\mcitedefaultendpunct}{\mcitedefaultseppunct}\relax
\EndOfBibitem
\bibitem{LHCb-PAPER-2013-059}
LHCb collaboration, R.~Aaij {\em et~al.},
  \ifthenelse{\boolean{articletitles}}{\emph{{Updated measurements of exclusive
  $J/\psi$ and $\psi(2S)$ production cross-sections in $pp$ collisions at
  $\sqrt{s} = 7$ TeV}},
  }{}\href{http://dx.doi.org/10.1088/0954-3899/41/5/055002}{J.\ Phys.\
  \textbf{G41} (2014) 055002}, \href{http://arxiv.org/abs/1401.3288}{{\tt
  arXiv:1401.3288}}\relax
\mciteBstWouldAddEndPuncttrue
\mciteSetBstMidEndSepPunct{\mcitedefaultmidpunct}
{\mcitedefaultendpunct}{\mcitedefaultseppunct}\relax
\EndOfBibitem
\bibitem{ref:moller}
C.~M{\o}ller, \ifthenelse{\boolean{articletitles}}{\emph{{General properties of
  the characteristic matrix in the theory of elementary particles}}, }{}K.\
  Danske Vidensk.\ Selsk.\ Mat.\ -Fys.\ Medd.\  \textbf{23} (1945) 1\relax
\mciteBstWouldAddEndPuncttrue
\mciteSetBstMidEndSepPunct{\mcitedefaultmidpunct}
{\mcitedefaultendpunct}{\mcitedefaultseppunct}\relax
\EndOfBibitem
\bibitem{Napoly:1992kn}
O.~Napoly, \ifthenelse{\boolean{articletitles}}{\emph{{The luminosity for beam
  distributions with error and wake field effects in linear colliders}},
  }{}Part.\ Accel.\  \textbf{40} (1993) 181
  \href{https://cds.cern.ch/record/240071}{CERN-SL-92-34-AP,
  CERN-CLIC-NOTE-173}\relax
\mciteBstWouldAddEndPuncttrue
\mciteSetBstMidEndSepPunct{\mcitedefaultmidpunct}
{\mcitedefaultendpunct}{\mcitedefaultseppunct}\relax
\EndOfBibitem
\bibitem{Herr:941318}
W.~Herr and B.~Muratori, \ifthenelse{\boolean{articletitles}}{\emph{{Concept of
  luminosity}},
  }{}\href{http://dx.doi.org/10.5170/CERN-2006-002.361}{Proceedings of the CERN
  Accelerator School (CAS) (2003) 361}, Zeuthen, Germany\relax
\mciteBstWouldAddEndPuncttrue
\mciteSetBstMidEndSepPunct{\mcitedefaultmidpunct}
{\mcitedefaultendpunct}{\mcitedefaultseppunct}\relax
\EndOfBibitem
\bibitem{Antchev:1495764}
TOTEM collaboration, G.~Antchev {\em et~al.},
  \ifthenelse{\boolean{articletitles}}{\emph{{Luminosity-independent
  measurement of the proton-proton total cross section at $\sqrt{s}$ = 8 TeV}},
  }{}\href{http://dx.doi.org/10.1103/PhysRevLett.111.012001}{Phys.\ Rev.\
  Lett.\  \textbf{111} (2013) 012001}\relax
\mciteBstWouldAddEndPuncttrue
\mciteSetBstMidEndSepPunct{\mcitedefaultmidpunct}
{\mcitedefaultendpunct}{\mcitedefaultseppunct}\relax
\EndOfBibitem
\bibitem{ATLAS-CONF-2014-040}
ATLAS collaboration, \ifthenelse{\boolean{articletitles}}{\emph{{Measurement of
  the total cross section from elastic scattering in $pp$ collisions at
  $\sqrt{s}=7$ TeV with the ATLAS detector}},
  }{}\href{http://arxiv.org/abs/1408.5778}{{\tt arXiv:1408.5778}}\relax
\mciteBstWouldAddEndPuncttrue
\mciteSetBstMidEndSepPunct{\mcitedefaultmidpunct}
{\mcitedefaultendpunct}{\mcitedefaultseppunct}\relax
\EndOfBibitem
\bibitem{LHCb-PAPER-2011-015}
LHCb collaboration, R.~Aaij {\em et~al.},
  \ifthenelse{\boolean{articletitles}}{\emph{{Absolute luminosity measurements
  with the LHCb detector at the LHC}},
  }{}\href{http://dx.doi.org/10.1088/1748-0221/7/01/P01010}{JINST \textbf{7}
  (2012) P01010}, \href{http://arxiv.org/abs/1110.2866}{{\tt
  arXiv:1110.2866}}\relax
\mciteBstWouldAddEndPuncttrue
\mciteSetBstMidEndSepPunct{\mcitedefaultmidpunct}
{\mcitedefaultendpunct}{\mcitedefaultseppunct}\relax
\EndOfBibitem
\bibitem{vanderMeer:296752}
S.~van~der Meer, \ifthenelse{\boolean{articletitles}}{\emph{{Calibration of the
  effective beam height in the ISR}}, }{}
  \href{https://cds.cern.ch/record/296752}{CERN-ISR-PO-68-31} (1968)\relax
\mciteBstWouldAddEndPuncttrue
\mciteSetBstMidEndSepPunct{\mcitedefaultmidpunct}
{\mcitedefaultendpunct}{\mcitedefaultseppunct}\relax
\EndOfBibitem
\bibitem{Rubbia:1025746}
C.~Rubbia, \ifthenelse{\boolean{articletitles}}{\emph{{Measurement of the
  luminosity of $p\overline{p}$ collider with a (generalized) Van der Meer
  Method}}, }{}
  \href{https://cds.cern.ch/record/1025746}{CERN-p$\overline{p}$-Note-38}
  (1977)\relax
\mciteBstWouldAddEndPuncttrue
\mciteSetBstMidEndSepPunct{\mcitedefaultmidpunct}
{\mcitedefaultendpunct}{\mcitedefaultseppunct}\relax
\EndOfBibitem
\bibitem{2011NIMPA.654..634B}
V.~Balagura, \ifthenelse{\boolean{articletitles}}{\emph{{Notes on Van der Meer
  scan for absolute luminosity measurements}},
  }{}\href{http://dx.doi.org/10.1016/j.nima.2011.06.007}{Nucl.\ Instrum.\
  Meth.\  \textbf{A654} (2011) 634}\relax
\mciteBstWouldAddEndPuncttrue
\mciteSetBstMidEndSepPunct{\mcitedefaultmidpunct}
{\mcitedefaultendpunct}{\mcitedefaultseppunct}\relax
\EndOfBibitem
\bibitem{ref:vdm-LHC}
H.~Burkhardt and P.~Grafstr\"{o}m,
  \ifthenelse{\boolean{articletitles}}{\emph{{Absolute luminosity from machine
  parameters}}, }{}
  \href{https://cds.cern.ch/record/1056691}{CERN-LHC-PROJECT-Report-1019}
  (2007)\relax
\mciteBstWouldAddEndPuncttrue
\mciteSetBstMidEndSepPunct{\mcitedefaultmidpunct}
{\mcitedefaultendpunct}{\mcitedefaultseppunct}\relax
\EndOfBibitem
\bibitem{FerroLuzzi:2005em}
M.~Ferro-Luzzi, \ifthenelse{\boolean{articletitles}}{\emph{{Proposal for an
  absolute luminosity determination in colliding beam experiments using vertex
  detection of beam-gas interactions}},
  }{}\href{http://dx.doi.org/10.1016/j.nima.2005.07.010}{Nucl.\ Instrum.\
  Meth.\  \textbf{A553} (2005) 388}\relax
\mciteBstWouldAddEndPuncttrue
\mciteSetBstMidEndSepPunct{\mcitedefaultmidpunct}
{\mcitedefaultendpunct}{\mcitedefaultseppunct}\relax
\EndOfBibitem
\bibitem{Aad:1517411}
ATLAS collaboration, G.~Aad {\em et~al.},
  \ifthenelse{\boolean{articletitles}}{\emph{{Improved luminosity determination
  in pp collisions at $\sqrt{s}$ = 7 TeV using the ATLAS detector at the LHC}},
  }{}\href{http://dx.doi.org/10.1140/epjc/s10052-013-2518-3}{Eur.\ Phys.\ J.\
  \textbf{C73} (2013) 2518}, \href{http://arxiv.org/abs/1302.4393}{{\tt
  arXiv:1302.4393}}\relax
\mciteBstWouldAddEndPuncttrue
\mciteSetBstMidEndSepPunct{\mcitedefaultmidpunct}
{\mcitedefaultendpunct}{\mcitedefaultseppunct}\relax
\EndOfBibitem
\bibitem{CMS:2013gfa}
CMS collaboration, \ifthenelse{\boolean{articletitles}}{\emph{{CMS luminosity
  based on pixel cluster counting -- summer 2013 update}}, }{}
  \href{http://cds.cern.ch/record/1598864}{CMS-PAS-LUM-13-001}\relax
\mciteBstWouldAddEndPuncttrue
\mciteSetBstMidEndSepPunct{\mcitedefaultmidpunct}
{\mcitedefaultendpunct}{\mcitedefaultseppunct}\relax
\EndOfBibitem
\bibitem{Abelev:1700665}
ALICE collaboration, B.~Abelev {\em et~al.},
  \ifthenelse{\boolean{articletitles}}{\emph{{Measurement of visible cross
  sections in proton-lead collisions at $\sqrt{s_{NN}}$=5.02 TeV in van der
  Meer scans with the ALICE detector}},
  }{}\href{http://dx.doi.org/10.1088/1748-0221/9/11/P11003}{JINST \textbf{9}
  (2014) 11003}, \href{http://arxiv.org/abs/1405.1849}{{\tt
  arXiv:1405.1849}}\relax
\mciteBstWouldAddEndPuncttrue
\mciteSetBstMidEndSepPunct{\mcitedefaultmidpunct}
{\mcitedefaultendpunct}{\mcitedefaultseppunct}\relax
\EndOfBibitem
\bibitem{LHCMachine}
L.~Evans and P.~Bryant, \ifthenelse{\boolean{articletitles}}{\emph{{LHC
  Machine}}, }{}\href{http://dx.doi.org/10.1088/1748-0221/3/08/S08001}{JINST
  \textbf{3} (2008) S08001}\relax
\mciteBstWouldAddEndPuncttrue
\mciteSetBstMidEndSepPunct{\mcitedefaultmidpunct}
{\mcitedefaultendpunct}{\mcitedefaultseppunct}\relax
\EndOfBibitem
\bibitem{LHCb-PAPER-2010-001}
LHCb collaboration, R.~Aaij {\em et~al.},
  \ifthenelse{\boolean{articletitles}}{\emph{{Prompt $K^0_S$ production in $pp$
  collisions at $\sqrt{s}=0.9$ TeV}},
  }{}\href{http://dx.doi.org/10.1016/j.physletb.2010.08.055}{Phys.\ Lett.\
  \textbf{B693} (2010) 69}, \href{http://arxiv.org/abs/1008.3105}{{\tt
  arXiv:1008.3105}}\relax
\mciteBstWouldAddEndPuncttrue
\mciteSetBstMidEndSepPunct{\mcitedefaultmidpunct}
{\mcitedefaultendpunct}{\mcitedefaultseppunct}\relax
\EndOfBibitem
\bibitem{ref:balagura-moriond}
V.~Balagura, \ifthenelse{\boolean{articletitles}}{\emph{{Luminosity measurement
  at LHCb}}, }{}Procceedings of 45th Rencontres de Moriond (QCD and High Energy
  Interactions), La Thuile, Aosta Valley Italy, 13--20 March 2010,
  \href{http://cds.cern.ch/record/1260472}{LHCb-TALK-2010-021}\relax
\mciteBstWouldAddEndPuncttrue
\mciteSetBstMidEndSepPunct{\mcitedefaultmidpunct}
{\mcitedefaultendpunct}{\mcitedefaultseppunct}\relax
\EndOfBibitem
\bibitem{ref:plamen-moriond}
P.~Hopchev, \ifthenelse{\boolean{articletitles}}{\emph{{The beam-gas method for
  luminosity measurement at LHCb}}, }{}Proceedings of the 45th Rencontres de
  Moriond (Electroweak Interactions and Unified Theories), La Thuile, Aosta
  Valley Italy, 6--13 March 2010, \href{http://arxiv.org/abs/1005.4398}{\tt
  arXiv:1005.4398}\relax
\mciteBstWouldAddEndPuncttrue
\mciteSetBstMidEndSepPunct{\mcitedefaultmidpunct}
{\mcitedefaultendpunct}{\mcitedefaultseppunct}\relax
\EndOfBibitem
\bibitem{BCNWG1}
G.~Anders {\em et~al.}, \ifthenelse{\boolean{articletitles}}{\emph{{LHC bunch
  current normalisation for the April-May 2010 luminosity calibration
  measurements}}, }{}
  \href{https://cds.cern.ch/record/1325370}{CERN-ATS-Note-2011-004 PERF}
  (2011)\relax
\mciteBstWouldAddEndPuncttrue
\mciteSetBstMidEndSepPunct{\mcitedefaultmidpunct}
{\mcitedefaultendpunct}{\mcitedefaultseppunct}\relax
\EndOfBibitem
\bibitem{BCNWG2}
G.~Alici {\em et~al.}, \ifthenelse{\boolean{articletitles}}{\emph{{LHC bunch
  current normalisation for the October 2010 luminosity calibration
  measurements}}, }{}
  \href{https://cds.cern.ch/record/1333997}{CERN-ATS-Note-2011-016 PERF}
  (2011)\relax
\mciteBstWouldAddEndPuncttrue
\mciteSetBstMidEndSepPunct{\mcitedefaultmidpunct}
{\mcitedefaultendpunct}{\mcitedefaultseppunct}\relax
\EndOfBibitem
\bibitem{Barschel:1425904}
C.~Barschel {\em et~al.}, \ifthenelse{\boolean{articletitles}}{\emph{{Results
  of the LHC DCCT calibration studies}}, }{}
  \href{https://cds.cern.ch/record/1425904}{CERN-ATS-Note-2012-026 PERF}
  (2012)\relax
\mciteBstWouldAddEndPuncttrue
\mciteSetBstMidEndSepPunct{\mcitedefaultmidpunct}
{\mcitedefaultendpunct}{\mcitedefaultseppunct}\relax
\EndOfBibitem
\bibitem{BCNWG3}
G.~Anders {\em et~al.}, \ifthenelse{\boolean{articletitles}}{\emph{{Study of
  the relative LHC bunch populations for luminosity calibration}}, }{}
  \href{https://cds.cern.ch/record/1427726}{CERN-ATS-Note-2012-028 PERF}
  (2012)\relax
\mciteBstWouldAddEndPuncttrue
\mciteSetBstMidEndSepPunct{\mcitedefaultmidpunct}
{\mcitedefaultendpunct}{\mcitedefaultseppunct}\relax
\EndOfBibitem
\bibitem{BCNWG4}
A.~Alici {\em et~al.}, \ifthenelse{\boolean{articletitles}}{\emph{{Study of the
  LHC ghost charge and satellite bunches for luminosity calibration}}, }{}
  \href{https://cds.cern.ch/record/1427728}{CERN-ATS-Note-2012-029 PERF}
  (2012)\relax
\mciteBstWouldAddEndPuncttrue
\mciteSetBstMidEndSepPunct{\mcitedefaultmidpunct}
{\mcitedefaultendpunct}{\mcitedefaultseppunct}\relax
\EndOfBibitem
\bibitem{Antunes-Nobrega:630827}
LHCb collaboration, R.~Antunes-Nobrega {\em et~al.},
  \ifthenelse{\boolean{articletitles}}{\emph{{LHCb reoptimized detector design
  and performance: Technical Design Report}}, }{}
  \href{http://cds.cern.ch/record/630827}{CERN-LHCC-2003-030, LHCb-TDR-9}\relax
\mciteBstWouldAddEndPuncttrue
\mciteSetBstMidEndSepPunct{\mcitedefaultmidpunct}
{\mcitedefaultendpunct}{\mcitedefaultseppunct}\relax
\EndOfBibitem
\bibitem{LHCb-DP-2014-001}
R.~Aaij {\em et~al.}, \ifthenelse{\boolean{articletitles}}{\emph{{Performance
  of the LHCb Vertex Locator}},
  }{}\href{http://dx.doi.org/10.1088/1748-0221/9/09/P09007}{JINST \textbf{9}
  (2014) P09007}, \href{http://arxiv.org/abs/1405.7808}{{\tt
  arXiv:1405.7808}}\relax
\mciteBstWouldAddEndPuncttrue
\mciteSetBstMidEndSepPunct{\mcitedefaultmidpunct}
{\mcitedefaultendpunct}{\mcitedefaultseppunct}\relax
\EndOfBibitem
\bibitem{LHCb-DP-2012-004}
R.~Aaij {\em et~al.}, \ifthenelse{\boolean{articletitles}}{\emph{{The \lhcb
  trigger and its performance in 2011}},
  }{}\href{http://dx.doi.org/10.1088/1748-0221/8/04/P04022}{JINST \textbf{8}
  (2013) P04022}, \href{http://arxiv.org/abs/1211.3055}{{\tt
  arXiv:1211.3055}}\relax
\mciteBstWouldAddEndPuncttrue
\mciteSetBstMidEndSepPunct{\mcitedefaultmidpunct}
{\mcitedefaultendpunct}{\mcitedefaultseppunct}\relax
\EndOfBibitem
\bibitem{LHCb-PUB-2014-044}
M.~Kucharczyk, P.~Morawski, and M.~Witek,
  \ifthenelse{\boolean{articletitles}}{\emph{{Primary Vertex Reconstruction at
  LHCb}}, }{}
  \href{https://cds.cern.ch/record/1756296}{LHCb-PUB-2014-044}\relax
\mciteBstWouldAddEndPuncttrue
\mciteSetBstMidEndSepPunct{\mcitedefaultmidpunct}
{\mcitedefaultendpunct}{\mcitedefaultseppunct}\relax
\EndOfBibitem
\bibitem{P.Odier.LHC.DCCT}
P.~Odier, M.~Ludwig, and S.~Thoulet,
  \ifthenelse{\boolean{articletitles}}{\emph{{The DCCT for the LHC Beam
  Intensity Measurement}}, }{}Proceedings of DIPAC09 (Beam Diagnostics and
  Instrumentation for Particle Accelerators), Basel, Switzerland, 25--27 May
  2009, \href{http://cds.cern.ch/record/1183400}{CERN-BE-2009-019}\relax
\mciteBstWouldAddEndPuncttrue
\mciteSetBstMidEndSepPunct{\mcitedefaultmidpunct}
{\mcitedefaultendpunct}{\mcitedefaultseppunct}\relax
\EndOfBibitem
\bibitem{ref:fbct}
D.~Belohrad {\em et~al.}, \ifthenelse{\boolean{articletitles}}{\emph{{The LHC
  fast BCT system: a comparison of design parameters with initial
  performance}}, }{}Proceedings of the Beam Instrumentation Workshop, BIW'10,
  Santa Fe, New Mexico, United States Of America, 2--6 May 2010,
  \href{https://cds.cern.ch/record/1267400}{CERN-BE-2010-010}\relax
\mciteBstWouldAddEndPuncttrue
\mciteSetBstMidEndSepPunct{\mcitedefaultmidpunct}
{\mcitedefaultendpunct}{\mcitedefaultseppunct}\relax
\EndOfBibitem
\bibitem{ref:LHC-RF}
P.~Baudrenghien {\em et~al.}, \ifthenelse{\boolean{articletitles}}{\emph{{The
  LHC RF system -- experience with beam operation}}, }{}Proceedings of the 2nd
  International Particle Accelerator Conference, San Sebastian, Spain, 4--9
  September 2011, \href{https://cds.cern.ch/record/1378471}{CERN-ATS-2011-048,
  CERN-ATS-2011-039}\relax
\mciteBstWouldAddEndPuncttrue
\mciteSetBstMidEndSepPunct{\mcitedefaultmidpunct}
{\mcitedefaultendpunct}{\mcitedefaultseppunct}\relax
\EndOfBibitem
\bibitem{Jeff:2012zz}
A.~Jeff {\em et~al.}, \ifthenelse{\boolean{articletitles}}{\emph{{Longitudinal
  density monitor for the LHC}},
  }{}\href{http://dx.doi.org/10.1103/PhysRevSTAB.15.032803}{Phys.\ Rev.\ ST
  Accel.\ Beams \textbf{15} (2012) 032803}\relax
\mciteBstWouldAddEndPuncttrue
\mciteSetBstMidEndSepPunct{\mcitedefaultmidpunct}
{\mcitedefaultendpunct}{\mcitedefaultseppunct}\relax
\EndOfBibitem
\bibitem{Jeff:1513180}
A.~Jeff, {\em {A longitudinal density monitor for the LHC}}, PhD thesis,
  Liverpool University, 2012,
  \href{https://cds.cern.ch/record/1513180}{CERN-THESIS-2012-240}\relax
\mciteBstWouldAddEndPuncttrue
\mciteSetBstMidEndSepPunct{\mcitedefaultmidpunct}
{\mcitedefaultendpunct}{\mcitedefaultseppunct}\relax
\EndOfBibitem
\bibitem{Colin-thesis}
C.~Barschel, {\em {Precision luminosity measurements at LHCb with beam-gas
  imaging}}, PhD thesis, RWTH Aachen University, 2014,
  \href{https://cds.cern.ch/record/1693671}{CERN-THESIS-2013-301}\relax
\mciteBstWouldAddEndPuncttrue
\mciteSetBstMidEndSepPunct{\mcitedefaultmidpunct}
{\mcitedefaultendpunct}{\mcitedefaultseppunct}\relax
\EndOfBibitem
\bibitem{ATLAS-BPTX}
C.~Ohm and T.~Pauly, \ifthenelse{\boolean{articletitles}}{\emph{{The ATLAS beam
  pick-up based timing system}},
  }{}\href{http://dx.doi.org/10.1016/j.nima.2010.03.069}{Nucl.\ Instrum.\
  Meth.\  \textbf{623} (2010) 558}, \href{http://arxiv.org/abs/0905.3648}{{\tt
  arXiv:0905.3648}}\relax
\mciteBstWouldAddEndPuncttrue
\mciteSetBstMidEndSepPunct{\mcitedefaultmidpunct}
{\mcitedefaultendpunct}{\mcitedefaultseppunct}\relax
\EndOfBibitem
\bibitem{Plamen-thesis}
P.~Hopchev, {\em {Absolute luminosity measurements at LHCb}}, PhD thesis,
  Grenoble University, 2011,
  \href{https://cds.cern.ch/record/1433396}{CERN-THESIS-2011-210}\relax
\mciteBstWouldAddEndPuncttrue
\mciteSetBstMidEndSepPunct{\mcitedefaultmidpunct}
{\mcitedefaultendpunct}{\mcitedefaultseppunct}\relax
\EndOfBibitem
\bibitem{Boccardi:1556087}
A.~Boccardi, E.~Bravin, M.~Ferro-Luzzi, and S.~Mazzoni,
  \ifthenelse{\boolean{articletitles}}{\emph{{LHC luminosity calibration using
  the longitudinal density monitor}}, }{}
  \href{https://cds.cern.ch/record/1556087}{CERN-ATS-Note-2013-034 TECH}
  (2013)\relax
\mciteBstWouldAddEndPuncttrue
\mciteSetBstMidEndSepPunct{\mcitedefaultmidpunct}
{\mcitedefaultendpunct}{\mcitedefaultseppunct}\relax
\EndOfBibitem
\bibitem{Zaitsev:473383}
N.~Y. Zaitsev, {\em {Study of the LHCb pile-up trigger and $B_S \rightarrow
  J/\psi \phi$ decay}}, PhD thesis, Amsterdam University, 2000,
  \href{https://cds.cern.ch/record/473383}{CERN-THESIS-2000-043}\relax
\mciteBstWouldAddEndPuncttrue
\mciteSetBstMidEndSepPunct{\mcitedefaultmidpunct}
{\mcitedefaultendpunct}{\mcitedefaultseppunct}\relax
\EndOfBibitem
\bibitem{CERN-ACC-2013-0028}
R.~Alemany-Fernandez, F.~Follin, and R.~Jacobsson,
  \ifthenelse{\boolean{articletitles}}{\emph{{The LHCb online luminosity
  control and monitoring}}, }{}Proceedings of the 4th International Particle
  Accelerator Conference, IPAC13, 12--17 May 2013, Shanghai, China,
  \href{http://cds.cern.ch/record/1567250}{CERN-ACC-2013-0028}\relax
\mciteBstWouldAddEndPuncttrue
\mciteSetBstMidEndSepPunct{\mcitedefaultmidpunct}
{\mcitedefaultendpunct}{\mcitedefaultseppunct}\relax
\EndOfBibitem
\bibitem{decker_1994}
F.~J. Decker, \ifthenelse{\boolean{articletitles}}{\emph{{Beam distributions
  beyond RMS}}, }{}Proceedings of the Beam Instrumentation Workshop, Vancouver,
  Canada, 3--6 October 1994,
  \href{http://inspirehep.net/record/384364}{SLAC-PUB-6684}\relax
\mciteBstWouldAddEndPuncttrue
\mciteSetBstMidEndSepPunct{\mcitedefaultmidpunct}
{\mcitedefaultendpunct}{\mcitedefaultseppunct}\relax
\EndOfBibitem
\bibitem{hastie1990generalized}
T.~J. Hastie and R.~J. Tibshirani, {\em Generalized Additive Models}, Chapman
  and Hall, 1990\relax
\mciteBstWouldAddEndPuncttrue
\mciteSetBstMidEndSepPunct{\mcitedefaultmidpunct}
{\mcitedefaultendpunct}{\mcitedefaultseppunct}\relax
\EndOfBibitem
\bibitem{Freedman:6360487}
D.~Freedman and P.~Diaconis, \ifthenelse{\boolean{articletitles}}{\emph{{On the
  histogram as a density estimator: L2 theory}},
  }{}\href{http://dx.doi.org/10.1007/BF01025868}{Probability Theory and Related
  Fields \textbf{57} (1981) 453}\relax
\mciteBstWouldAddEndPuncttrue
\mciteSetBstMidEndSepPunct{\mcitedefaultmidpunct}
{\mcitedefaultendpunct}{\mcitedefaultseppunct}\relax
\EndOfBibitem
\bibitem{Herr:CAS2003}
W.~Herr, \ifthenelse{\boolean{articletitles}}{\emph{Beam-beam interactions},
  }{}\href{http://dx.doi.org/10.5170/CERN-2006-002.379}{Proceedings of the CERN
  Accelerator School (CAS) (2003) 379}, Zeuthen, Germany\relax
\mciteBstWouldAddEndPuncttrue
\mciteSetBstMidEndSepPunct{\mcitedefaultmidpunct}
{\mcitedefaultendpunct}{\mcitedefaultseppunct}\relax
\EndOfBibitem
\bibitem{Herr:lumidays12}
W.~Herr, \ifthenelse{\boolean{articletitles}}{\emph{{Beam-beam effects and
  dynamic $\beta^*$}}, }{}LHC Lumi Days: LHC Workshop on LHC Luminosity
  Calibration, 29 Feb -- 1 Mar 2012, CERN, Geneva, Switzerland\relax
\mciteBstWouldAddEndPuncttrue
\mciteSetBstMidEndSepPunct{\mcitedefaultmidpunct}
{\mcitedefaultendpunct}{\mcitedefaultseppunct}\relax
\EndOfBibitem
\bibitem{MADX}
MAD - Methodical Accelerator Design, see
  {\href{http://cern.ch/mad}{http://cern.ch/mad}}\relax
\mciteBstWouldAddEndPuncttrue
\mciteSetBstMidEndSepPunct{\mcitedefaultmidpunct}
{\mcitedefaultendpunct}{\mcitedefaultseppunct}\relax
\EndOfBibitem
\bibitem{Ziemann:1991sb}
V.~Ziemann, \ifthenelse{\boolean{articletitles}}{\emph{{Beyond Bassetti and
  Erskine: Beam-beam deflections for non-Gaussian beams}}, }{}Proceedings of
  the Beam Dynamics Workshop, Los Angeles, USA, 13--16 May 1991,
  \href{http://inspirehep.net/record/316705}{SLAC-PUB-5582}\relax
\mciteBstWouldAddEndPuncttrue
\mciteSetBstMidEndSepPunct{\mcitedefaultmidpunct}
{\mcitedefaultendpunct}{\mcitedefaultseppunct}\relax
\EndOfBibitem
\bibitem{Kozanecki:1581723}
W.~Kozanecki, T.~Pieloni, and J.~Wenninger,
  \ifthenelse{\boolean{articletitles}}{\emph{{Observation of beam-beam
  deflections with LHC orbit data}}, }{}
  \href{http://cds.cern.ch/record/1581723}{CERN-ACC-NOTE-2013-0006}
  (2013)\relax
\mciteBstWouldAddEndPuncttrue
\mciteSetBstMidEndSepPunct{\mcitedefaultmidpunct}
{\mcitedefaultendpunct}{\mcitedefaultseppunct}\relax
\EndOfBibitem
\bibitem{LHCb-PAPER-2012-039}
LHCb collaboration, R.~Aaij {\em et~al.},
  \ifthenelse{\boolean{articletitles}}{\emph{{Measurement of $J/\psi$
  production in $pp$ collisions at $\sqrt{s}=2.76$ TeV}},
  }{}\href{http://dx.doi.org/10.1007/JHEP02(2013)041}{JHEP \textbf{02} (2013)
  041}, \href{http://arxiv.org/abs/1212.1045}{{\tt arXiv:1212.1045}}\relax
\mciteBstWouldAddEndPuncttrue
\mciteSetBstMidEndSepPunct{\mcitedefaultmidpunct}
{\mcitedefaultendpunct}{\mcitedefaultseppunct}\relax
\EndOfBibitem
\bibitem{TOTEM:7tev}
TOTEM collaboration, G.~Antchev {\em et~al.},
  \ifthenelse{\boolean{articletitles}}{\emph{{Luminosity-independent
  measurements of total, elastic and inelastic cross-sections at $\sqrt{s}$ = 7
  TeV}}, }{}\href{http://dx.doi.org/10.1209/0295-5075/101/21004}{Europhys.\
  Lett.\  \textbf{101} (2013) 21004}\relax
\mciteBstWouldAddEndPuncttrue
\mciteSetBstMidEndSepPunct{\mcitedefaultmidpunct}
{\mcitedefaultendpunct}{\mcitedefaultseppunct}\relax
\EndOfBibitem
\bibitem{Abelev:2012sea}
ALICE collaboration, B.~Abelev {\em et~al.},
  \ifthenelse{\boolean{articletitles}}{\emph{{Measurement of inelastic, single-
  and double-diffraction cross sections in proton--proton collisions at the LHC
  with ALICE}},
  }{}\href{http://dx.doi.org/10.1140/epjc/s10052-013-2456-0}{Eur.\ Phys.\ J.\
  \textbf{C73} (2013) 2456}, \href{http://arxiv.org/abs/1208.4968}{{\tt
  arXiv:1208.4968}}\relax
\mciteBstWouldAddEndPuncttrue
\mciteSetBstMidEndSepPunct{\mcitedefaultmidpunct}
{\mcitedefaultendpunct}{\mcitedefaultseppunct}\relax
\EndOfBibitem
\bibitem{Aad:2011eu}
ATLAS collaboration, G.~Aad {\em et~al.},
  \ifthenelse{\boolean{articletitles}}{\emph{{Measurement of the Inelastic
  Proton-Proton Cross-Section at $\sqrt{s}=7$ TeV with the ATLAS Detector}},
  }{}\href{http://dx.doi.org/10.1038/ncomms1472}{Nature Commun.\  \textbf{2}
  (2011) 463}, \href{http://arxiv.org/abs/1104.0326}{{\tt
  arXiv:1104.0326}}\relax
\mciteBstWouldAddEndPuncttrue
\mciteSetBstMidEndSepPunct{\mcitedefaultmidpunct}
{\mcitedefaultendpunct}{\mcitedefaultseppunct}\relax
\EndOfBibitem
\bibitem{Chatrchyan:2012nj}
CMS collaboration, S.~Chatrchyan {\em et~al.},
  \ifthenelse{\boolean{articletitles}}{\emph{{Measurement of the inelastic
  proton-proton cross section at $\sqrt{s}=7$ TeV}},
  }{}\href{http://dx.doi.org/10.1016/j.physletb.2013.03.024}{Phys.\ Lett.\
  \textbf{B722} (2013) 5}, \href{http://arxiv.org/abs/1210.6718}{{\tt
  arXiv:1210.6718}}\relax
\mciteBstWouldAddEndPuncttrue
\mciteSetBstMidEndSepPunct{\mcitedefaultmidpunct}
{\mcitedefaultendpunct}{\mcitedefaultseppunct}\relax
\EndOfBibitem
\end{mcitethebibliography}
